\newif\ifprintedversion
\printedversionfalse  %

\ifprintedversion
    \documentclass[11pt,b5paper,twoside,fleqn,english,openany]{memoir}
\else
    \documentclass[12pt,a4paper,oneside,fleqn,english,openany]{memoir}
\fi

\usepackage{beamerarticle}           %

\usepackage{styles/vubimagepagestyle}
\usepackage{styles/ulbhousestyle}  %
\usepackage{styles/myListings}     %
\usepackage{styles/myPackages}     %
\usepackage{styles/myPageSetup}    %
\usepackage{styles/myBibSettings}  %
\usepackage{styles/myCommands}     %
\usepackage{styles/myDrawingStuff} %

\def\vubprogram{Thesis submitted in fulfilment of the requirements for the award of the degree of Doctor of Engineering Sciences}

\def\vubpretitle{\vubprogram}

\title{How to effectively manage the energy in residential systems} %

\def\vubsubtitle{Development and comparison of methods} %

\author{Julian Ruddick} %

\date{February 2026}

\def\vubNDA{} %

\def\vubpromoters{Prof. Maarten Messagie\linebreak
Prof. Thierry Coosemans}

\def\vubfaculty{Faculty of engineering }      %

\usepackage{cleveref}
\crefname{subsection}{subsection}{subsections}
\usepackage{subcaption}
\usepackage{float}
\usepackage[super]{nth}
\usepackage[noend]{algorithmic}
\usepackage{algorithm}
\usepackage[hyphens]{xurl}

\usepackage{eurosym}
\usepackage{siunitx}
\usepackage{tabularray}
\usepackage{multirow}
\usepackage{fmtcount}
\usepackage{xparse}
\usepackage[pass]{geometry}
\usepackage{ragged2e}

\makeatletter
\newcommand{\useopenany}{\@twosidefalse}
\newcommand{\useopenright}{\@twosidetrue}
\makeatother

\DeclareSIUnit{\EUR}{\text{\euro}}

\ifprintedversion
\setlrmarginsandblock{2cm}{2cm}{*}             %
\setulmarginsandblock{2cm}{2.5cm}{*}           %
\checkandfixthelayout
\fi

\appto{\bibsetup}{\sloppy}

\makeatletter\AddToHook{insertmark}{\let\label\@gobble}\makeatother
\begin{document}
    \ifprintedversion
    \raggedbottom
    \fi
    \ifprintedversion
    \makebookletcover
    \else
    \maketitleVUB                    %
    \makecopyrightpageVUBdigital
    \fi

    \frontmatter  
    \ifprintedversion
    \makeinnertitlepageVUBbooklet
    \else
    \makeinnertitlepageVUBdigital
    \fi
    
\chapter*{Members of the Jury}\addcontentsline{toc}{chapter}{Members of the Jury}
\label{sec:jury_members}

\vspace{0.5\baselineskip}

\begin{center}
\renewcommand{\arraystretch}{1.25}
\large
\begin{tabularx}{\linewidth}{@{}>{\raggedright\arraybackslash}p{0.25\linewidth}X@{}}
Promotor               & \textbf{Prof. Maarten Messagie}, Department of Electric Engineering and Energy Technology, Vrije Universiteit Brussel \\
Promotor               & \textbf{Prof. Thierry Coosemans}, Department of Electric Engineering and Energy Technology, Vrije Universiteit Brussel \\
President of the Jury  & \textbf{Prof. Ann Nowe}, Department of Computer Science, Vrije Universiteit Brussel \\
Vice President of the Jury & \textbf{Prof. Sebastiaan Eeltink}, Department of Chemical Engineering and Industrial Chemistry, Vrije Universiteit Brussel \\
Secretary of the Jury  & \textbf{Prof. Valery Ann Jacobs}, Department of Electric Engineering and Energy Technology, Vrije Universiteit Brussel \\
External Jury member   & \textbf{Prof. Andrea Michiorri}, Department of Energy and Processes, École Nationale Supérieure des Mines de Paris \\
External Jury member   & \textbf{Prof. Zoltan Nagy}, Department of the Built Environment, Technische Universiteit Eindhoven \\
\end{tabularx}
\end{center}

\mode<all> %

    \IfLanguageName{english} %
    { %
        \def\AcknowledgementsName{Acknowledgements}
    } 
    { %
        \def\AcknowledgementsName{Dankwoord}
    }

\chapter*{\AcknowledgementsName}\addcontentsline{toc}{chapter}{\AcknowledgementsName}

First of all, as the apple never falls far from the tree, thank you Maman and Daddy for always being there for me, giving me the taste and curiosity for science, being able to give me advice all along the way, being demanding enough to push myself but also trustful enough to let me try things on my own, and thank you for feeding me tasty food for 30 years, that also helped a lot. Nothing has ever been super difficult, but that's thanks to your love and education. Gros bisous, je vous aime <3

Merci à Papy, Mamy et Parrain, de vous être autant occupés de moi, vous m'avez vraiment aidé à passer une enfance heureuse, je garderai toujours un beau souvenir des moments passés avec vous.
And thanks to my family in England, it has always been a pleasure to cross the Channel for the past 30 years and spend some good time with you around a Christmas dinner, board games, or for a walk in the countryside.

Next up, thank you to the trio at the head of EVERGi: Maarten, Thierry, and Cedric. It was a real delight to work under your guidance for 6 years now. I had the best interview where, after a bit more than an hour of discussion, you invited me for some drinks with the rest of the team and voilà, that was the start of the journey. You really put in place a great atmosphere that stimulates discussion and collaboration while also working hard to get the projects that allow us to work on relevant challenges of our times. Thank you also for your trust, openness to discussion, and also extensions. It all really allowed me to do this thesis in great conditions and to enjoy doing it.

I would also like to thank the members of my jury for their time, expertise, and valuable feedback, which have helped improve the quality of this thesis.

And now a big one: thank you to all my EVERGi colleagues for making so many good moments over the years. Thank you: Luis for helping me start this thesis the best way possible and the nice BBQs, Gilles for the great friendship and getting me the job in the first place, Alex for the countless Friday beers and your warm and caring nature, Glenn for the countless fun hours spent at GEP and being my EMS buddy in this journey, Simin for being my day one partner by starting at the same time, Anne for your contagious joy and philosophical discussions, Maria for your positive energy and continuing to propose and come do things with us, Stella for always being there and caring about us, Maeva for the example you give and bringing some fun when you come up north, Shary for the nice lunch discussions and being the eyes and ears of the group, Andrei for your love of food and drinks with colleagues, Dominik for your enthusiasm, the chill computer debugging sessions and always checking in to see how things are going, Yuva for the stories, Ander for always proposing your help and enjoying the good moments of life, Evgenii for being my partner in competitions, travel, adventures, desk, and machine learning during this thesis, Daniele for the special discussions and positive attitude to life, Simon for being my Linux buddy and helping out on the computer science tasks, Lennert for all the help with GEP and the nice movie discussions, Léa for the kindness and being one of the latest to leave when we stay out, Wouter for the nice chats and help with the coding and algorithmic work, Farzad for the camaraderie and the crazy stories, Stas for the nice moments that came out of Optimesh, Rémy for your youth, smile and teaching me stuff about voltages, Andy and Majid for helping me out when needed, Saeed for the number of goals avoided and being our official professional event photographer, Stijn for our ancestral Geraardsbergen connection and the joy to see mattentaart on the kitchen table, and Bart for being the newest nice colleague that will help continue the good spirit of the team.

Thank you also to all MOBI colleagues, it's great to share the Z building and Complex with all of you, with hopefully many more memorable team buildings to come. 

Thank you also to the VUB and the institutions that financed the projects I worked on (VLAIO, FOD Economie, European Commission). Both the funding and Belgian working conditions for PhD students allowed me to do this thesis in very comfortable circumstances.

I was also able to start this thesis from a good basis thanks to everything I was taught at l'École Saint Joseph, Lycée Émile Jacqmain, the Maastricht science programme, the DISAL lab in the EPFL, l'École polytechnique de l'ULB et au labo IRIDIA de l'ULB. I learned and had a lot of fun during all these studies with some memorable teachers and tutors along the way.

Finally, thank you to all my friends for the laughs, the joy, the festivals, the holiday trips, the restaurant outings and many more. Among them, my friends for now 15 years and still going strong: Samy, Antoine, Ismaël, Romain, Simon, Jérôme, and Ayoub, les rois de la teuf and all the memorable moments we lived together, to the different lovely people I had the pleasure to live with during these 6 years, la colloc de l'amour, le Clos et son voisinage bâtisseur, et les aigles du 44, and to my Ixelles hockey teammates.

All this to say: life is good, work is good, let's continue to do good science and live nice experiences all together for the many years to come.

\hfill \theauthor %

\hfill\textit{Brussels, February 25, 2026}

\mode<all> %
    \chapter*{Abstract}\addcontentsline{toc}{chapter}{Abstract}

The transition from fossil fuels presents significant challenges in energy management. The era of simply activating gas power plants to meet electricity demand or refuelling vehicles with petrol in a matter of minutes is coming to an end. To ensure a sustainable energy future, it is crucial to reduce energy consumption when possible and to synchronize electricity consumption and storage with the production from sources with low greenhouse gas emissions. Effective energy management in residential settings can help achieve these goals. This thesis explores a variety of energy management methods used to manage batteries, heat pumps, and other energy-related systems. The thesis presents three simulation cases alongside one real-world experiment to compare energy management done through simple rules, optimization, and machine learning methods. Furthermore, this thesis introduces TreeC, a novel machine learning-based method that learns understandable yet efficient energy management strategies, applicable to many cases. 
The thesis concludes with a comprehensive analysis of the benefits of energy management methods in electrified residential settings.

The three key contributions of this thesis can be summarized as follows:
1) The developed TreeC method demonstrates competitive performance compared to a variety of energy management approaches, such as model predictive control and reinforcement learning, across a diverse set of energy management problems. 
The understandability TreeC's energy management strategies greatly improves the transparency and trustworthiness of machine learning-based energy management.
2) The real-world experiment highlights the need to validate energy management methods on actual hardware and not solely on simulations.
Both TreeC and model predictive control performed worse than in simulations, due to, respectively, insufficiently representative training data and the real-time input data not being reliable enough.
3) Energy management algorithms reduce costs in electrified Belgian dwellings, particularly in poorly insulated dwellings, electric vehicle chargers, and photovoltaic installations. However, on average, these cost savings do not yet outweigh the current costs and effort required to implement such algorithms.

\bigskip

\textbf{keywords}: energy management system, machine learning, decision tree, reinforcement learning, model predictive control, demand response

\mode<all> %
    \chapter*{Disclosure on the use of AI}\addcontentsline{toc}{chapter}{Disclosure on the use of AI}

\textbf{Declaration}

During the preparation of this work, the author(s) used the following tool(s) and service(s):

\begin{center}
    \renewcommand{\arraystretch}{1.3} %
    \begin{tabular}{p{4cm}lp{6cm}}
         \toprule
         name of tool/service             &              & reason         \\[+2ex]
         \emph{GitHub Copilot}      & in order to: & \emph{auto-complete code, improve the writing, make LaTeX tables and TikZ figures}  \\
         \bottomrule
    \end{tabular}
\end{center}

After using this tool(s) and/or service(s), the author(s) reviewed and edited the content as needed. The author(s) take(s) full responsibility for the content of this work.

\mode<all> %

    \tableofcontents\newpage         %
    \listoffigures\newpage           %
    \listoftables\newpage            %
    \IfLanguageName{english} %
    { %
        \def\nomenclatureName{Acronyms}
    } 
    { %
        \def\acronymSection{Acroniemen}
    }

\chapter*{\nomenclatureName}\addcontentsline{toc}{chapter}{\nomenclatureName}\label{sec:nomenclature}

\def\widthA{1.5cm} \def\widthB{9cm} 

\begin{tabular}{p{1.8cm}p{9cm}}
    \acronym{CMA-ES}{Covariance matrix adaptation evolution strategy}
    \acronym{EMS}{Energy management system}
    \acronym{EV}{Electric vehicle}
    \acronym{HVAC}{Heating, ventilation and air conditioning}
    \acronym{MPC}{Model predictive control}
    \acronym{PV}{Photovoltaic}
    \acronym{RBC}{Rule-based control}
    \acronym{RL}{Reinforcement learning}
    \acronym{RQ}{Research question}
    \acronym{SOC}{State of charge}
\end{tabular}

\mode<all> %

    \mainmatter

\chapter[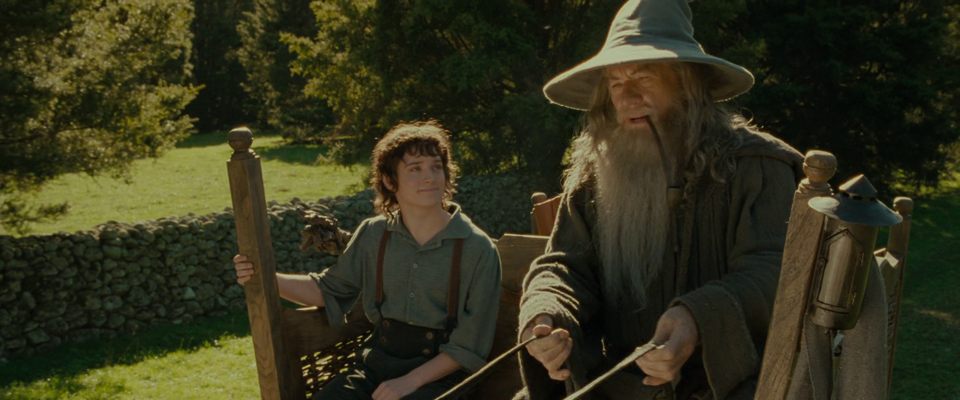]{Introduction\label{ch:introduction}}

\section{Context}
\subsection{Why do we need good energy management}

\begin{figure}
    \centering
    \includegraphics[width=1.0\textwidth]{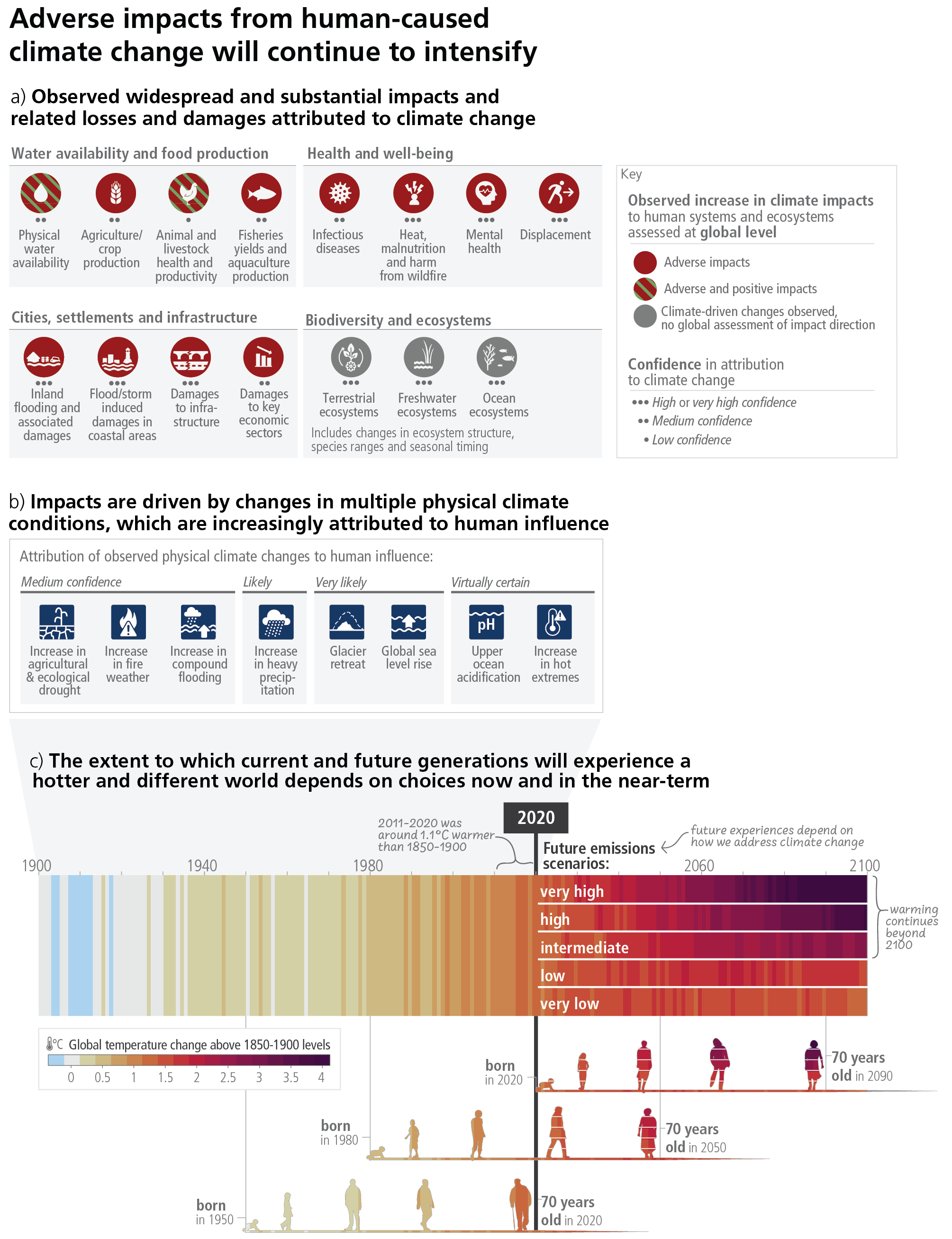}
    \caption{The IPCC's summary of a) Observed impacts of climate change b) Impacts attributed to human influence c) Future scenarios and consequences \cite{ipccSummaryPolicymakers2023}}
    \label{fig:climate_change_effects}
\end{figure}

Climate change is one of the most pressing challenges of our time, with adverse effects on water availability, food production, health, and infrastructure already evident and predicted to worsen in most projected scenarios (see summary in \cref{fig:climate_change_effects}). According to the Intergovernmental Panel on Climate Change (IPCC), ``Human activities, principally through emissions of greenhouse gases, have unequivocally caused global warming, [...] (high confidence)'' \cite{ipccSections2023}.

The primary contributing sectors to greenhouse gas emissions are energy supply (34\%), industry (24\%), agriculture, forestry and other land use (22\%), transport (15\%), and buildings (6\%). 
CO\textsubscript{2} emissions accounted for an estimated three quarters of human-induced greenhouse gas emissions in 2019, with emissions from fossil fuel combustion and industrial processes alone responsible for two thirds of the same greenhouse gas emissions total.
Achieving net-zero CO\textsubscript{2} energy systems requires ``a substantial reduction in overall fossil fuel use, [...]; widespread electrification; [...] (high confidence)'' according to the IPCC 2023 report.
The report also states that ``Several mitigation options, notably \textbf{solar energy}, \textbf{wind energy}, \textbf{electrification of urban systems}, urban green infrastructure, \textbf{energy efficiency}, \textbf{demand-side management}, improved forest and crop/grassland management, and reduced food waste and loss, are technically viable, are becoming increasingly cost-effective and are generally supported by the public. [...] (high confidence)''  \cite{ipccSections2023}. The bolded terms highlight the topics where energy management can play a significant role.

The European Green Deal sets the ambitious goal of achieving net-zero greenhouse gas emissions in the European Union by 2050 \cite{secretariat-generaloftheeuropeancommissionEuropeanGreenDeal2019}. In 2022, Belgium’s greenhouse gas footprint, including emissions from imported goods and services, stood at 12.4 tons of CO\textsubscript{2}-eq per person. This figure is higher than both the European Union average of 10.7 tons per person and the global average of 6.4 tons per person \cite{eurostatGreenhouseGasEmission2025,eurostatEUGHGEmissions2024}. Within Belgium, the main contributors to greenhouse gas emissions are industry (28\%), transport (24\%), energy supply (18\%), buildings (17\%), agriculture, forestry and other land use (11\%), and other sectors (2\%) \cite{climat.beEmissionsParSecteur2023}.

Progress has already been made in several sectors to reduce emissions. For example, wind and solar energy accounted for 30\% of Belgium’s electricity generation in 2024, up from 13\% in 2015. Over the same period, gas-based electricity generation declined from 33\% to 16\%, while nuclear increased slightly from 38\% to 42\% \cite{entso-eENTSOETransparencyPlatform2025}. Despite these improvements, electricity generated from gas still resulted in 417 kilograms of CO\textsubscript{2}-eq per person in 2024, whereas electricity from nuclear, wind, solar, and biomass combined contributed only 105 kilograms of CO\textsubscript{2}-eq per person (see \cref{fig:carbon_intensity_electricity_belgium}).

\newcommand{\carboninternsityfig}{\begin{figure}[H]
\centering
\includegraphics[width=0.8\textwidth]{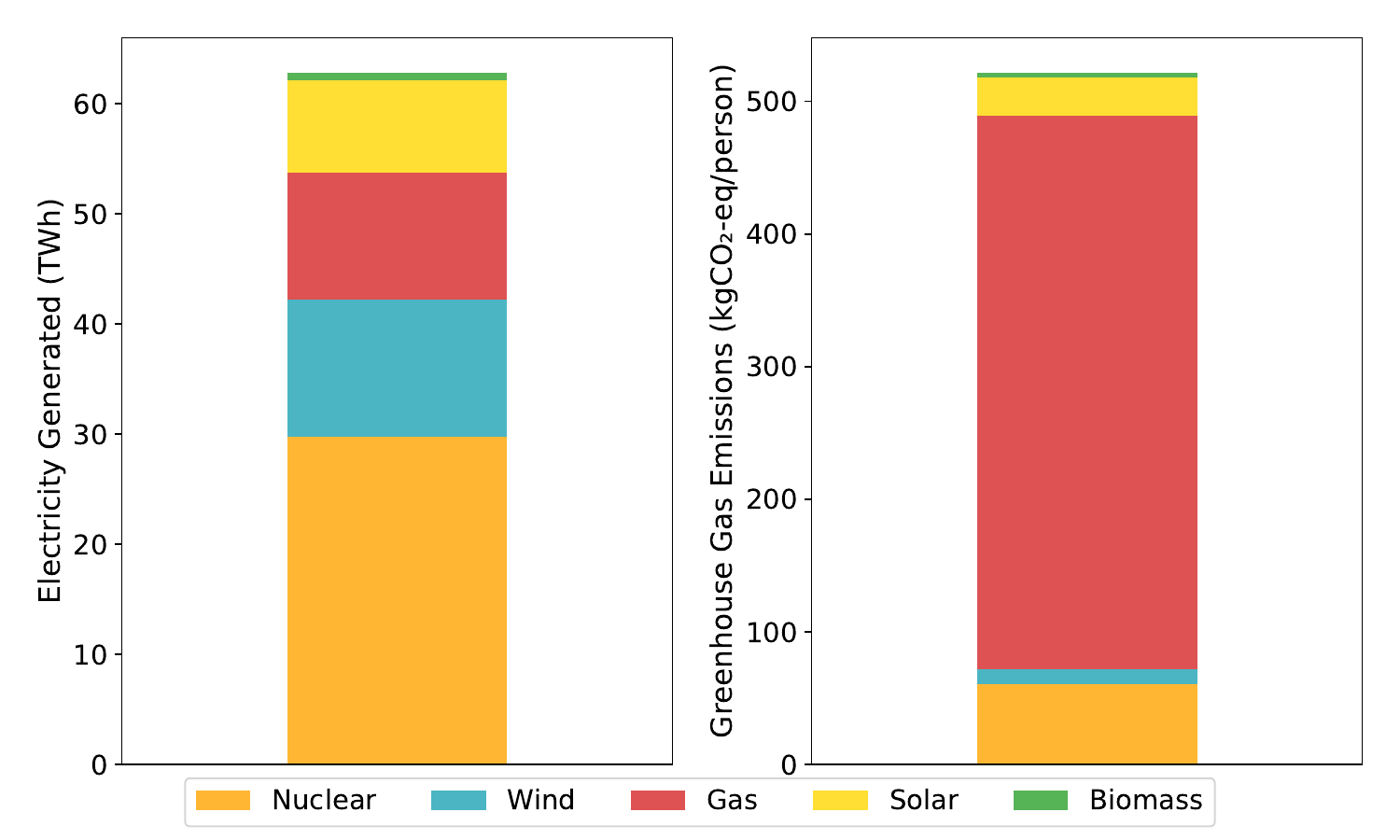}
\caption{Electricity generated using Nuclear, Wind, Gas, Solar and Biomass in Belgium for 2024 and the corresponding greenhouse gas emissions expressed in kg CO\textsubscript{2} equivalent per person. Data sources: \citet{entso-eENTSOETransparencyPlatform2025} for electricity generation and \citet{scarlatQuantificationCarbonIntensity2022} for carbon intensity factors.}
\label{fig:carbon_intensity_electricity_belgium}
\end{figure}
}

\ifprintedversion
\carboninternsityfig
\fi

Electrification of transport is also advancing: the share of electric vehicles in new car sales reached 28\% in 2024 \cite{vbofebBelgianMobilityDashboard2025}, and should approach 100\% by 2035 to meet the European Union’s targets \cite{europeanparliamentRevisionCO2Emission2025}.
In contrast, heat pump adoption remains low: only 2\% of households have them while gas boilers still account for 88\% of new heating system sales. This slow uptake is largely due to the electricity price in Belgium which has been over the past decade three to six times higher than the gas price (in \euro{}/kWh) while heat pumps become cost-competitive when the electricity price is lower than three times the gas price \cite{voltHeatPumpMarket2024}.

\ifprintedversion
\else
\carboninternsityfig
\fi

\newcommand{\elecgenfig}{\begin{figure}[H]
\centering
\includegraphics[width=0.8\textwidth]{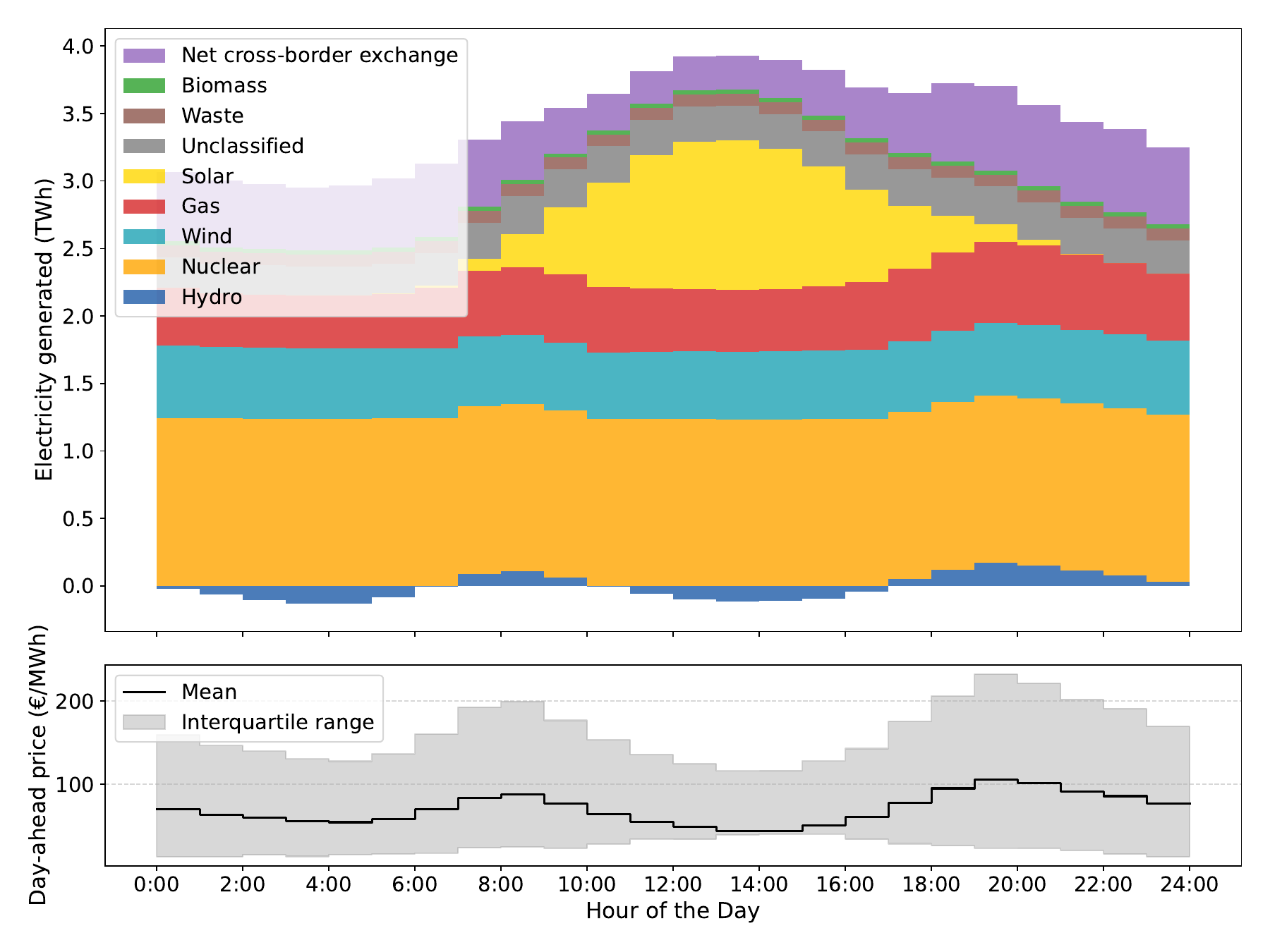}
\caption{(Top) Hourly breakdown of electricity sources in Belgium for 2024, illustrating the average contribution of each source throughout the day. (Bottom) Corresponding day-ahead electricity prices, which serve as a reference for various hourly pricing mechanisms. The solid line represents the mean price, while the shaded area indicates the range where 50\% of prices fall throughout the year. Prices are typically the lowest around 04:00 and 13:00, coinciding with periods of low demand and high solar generation, and peak around 08:00 and 19:00, aligning with high demand. Hydro and cross-border electricity exchanges contribute to balancing supply and demand during these periods. Data from \citet{entso-eENTSOETransparencyPlatform2025}.}
\label{fig:electricity_generation}
\end{figure}}

\ifprintedversion

\elecgenfig
\fi

In response to the evolving electricity generation mix and changing consumption patterns, a range of strategies are being adopted to better align electricity demand with periods of high renewable energy production (see \cref{fig:electricity_generation}). These measures include hourly electricity pricing made mandatory for large consumers and available voluntarily for residential users, the introduction of a capacity tariff that penalizes high peak demand for all consumers with a digital meter, and the development of various balancing markets. Such markets enable participants, such as operators of large battery energy storage systems, to shift their demand or generation to periods when the electricity grid is under stress.

\ifprintedversion
\else
\elecgenfig
\fi

These evolving pricing mechanisms, together with the growing use of energy assets such as batteries, heat pumps, and electric vehicles, demand more advanced energy management strategies. This marks a fundamental shift from the traditional use of fossil fuels, which could be stored and dispatched on demand.

This thesis looks into different algorithms for energy management systems (EMS), focusing on how they make decisions such as when to charge or discharge batteries, optimize heating schedules, or determine the best times to charge electric vehicles. 

This thesis focuses on EMSs applied in residential and commercial buildings, including homes, offices, and universities. It does not cover EMSs for large industrial facilities, power plants, or national grids.

\subsection{Energy management throughout history}

Before the 1970s, most efforts focused on generating and transmitting enough electricity to meet demand \cite{moranPowerSystemAutomatic1959}. Starting in the 1970s, there was a shift towards controlling or reducing electricity demand instead of just increasing generation capacity \cite{lampropoulosHistoryDemandSide2013}. For example, in 1978, the IEEE Power and Energy Society created the Load Management Working Group, which brought together many professionals working on ways to shift the electricity use on the consumer side \cite{gellingsIEEEPESLoad1981}.

Various strategies were developed and implemented to balance the electricity demand and production more effectively.  Since the 1960s, Germany introduced lower electricity prices at night and promoted the use of electric night storage heaters. These heaters stored heat during night hours, utilizing electricity generated by electricity power plants that could not be easily shut down overnight, such as coal or nuclear power plants \cite{quaschningLastmanagementZukuenftigenEnergieversorgung1999}. 
Since the 1970s in the United States, automatic frequency relays were deployed to disconnect parts of the grid when demand was too high in order to prevent large-scale blackouts \cite{maliszewskiFrequencyActuatedLoad1971}.
In 1978 \citet{garciaDemandSideManagement1987} did a pilot study in which electricity providers implemented higher prices for consumers during periods of peak demand. In this study, the utility could remotely switch off certain household appliances, such as air conditioners and water heaters, during high-demand periods.

The increasing integration of renewable energy sources, such as wind and solar, has made electricity production more dependent on weather conditions. As a result, effective energy management on the demand side has become increasingly important to maximize the use of renewable generation.
In 2011 the International Organization for Standardization released the first version of the ISO 50001 standard, which provides a definition and framework for energy management systems to help organizations improve their energy performance \cite{EnergyManagementSystems2018}. 
As computers became more powerful and cheaper, they were increasingly used to control electricity demand.
As an example, Home assistant, a free and open-source home automation software that enables energy management and can be run on a mini computer, is amongst the most popular open-source projects in the world and has been installed in over 2 million homes \cite{nijhof20255TwoMillion2025}.

While matching electricity production and consumption has always been essential, the growing share of renewable energy sources, the electrification of various sectors, and the decreasing cost of computing systems are making automated, computer-based energy management systems more relevant than ever.

\subsection{Control algorithms for energy management systems}

The concept of automated control in energy management is not new. For example the first record of a thermostat dates back to the 17th century. At the time, Cornelis Drebbel invented a device that could automatically regulate the temperature inside a chicken egg incubator by opening a lid when the temperature became too high through the expansion of alcohol \cite{tierieCornelisDrebbel157216331932}.

Most of our energy devices function using rule-based controllers, for example the default behaviour of many home batteries is to charge when the photovoltaic installation of the home produces more electricity than needed and to discharge when it produces less.
These simple rules are everywhere but in certain cases they can be improved by using more advanced control algorithms.

A control algorithm called model predictive control is often used to manage the heating of buildings.
This method uses a physical model of the controlled building and predictions of many variables such as weather conditions and occupancy to optimize the heating schedule of the building \cite{maDemandReductionBuilding2012,sirokyExperimentalAnalysisModel2011}.

A more recent and promising approach used in the field of energy management is reinforcement learning, which has achieved impressive results in fields like robotics and game playing \cite{liDeepReinforcementLearning2018}. 
Reinforcement learning is a machine learning technique where the controller learns by interacting with the system and receiving feedback as rewards or penalties. Through trial and error, it explores different actions and gradually improves its strategy to maximize long-term objectives, such as lowering energy costs or greenhouse gas emissions.
A key advantage of reinforcement learning is its ability to adapt to the specific characteristics of the system without needing a detailed physical model or simulation.

Both approaches have distinct limitations. Model predictive control requires significant effort to develop accurate physical models and install the necessary sensors. 
It also does not improve its control model by training on the controlled system. For example, if predictions are inaccurate and reduce the controller's performance, model predictive control will not adjust to this.
Reinforcement learning does improve its control model by training directly on the controlled system but often requires extensive training periods before achieving satisfactory performance, which can be impractical for real-world energy systems.

Regarding interpretability, model predictive control models are typically based on physical equations and remain understandable as long as the model complexity is manageable. In contrast, reinforcement learning methods overwhelmingly rely on neural networks, resulting in black-box controllers whose decision-making processes are difficult to interpret and may exhibit unpredictable behaviour.

Interpretable models have several advantages: errors in their decision-making can be identified more easily, unexpected behaviours are simpler to analyse, key inputs and actions are clearly highlighted, and their logic can be more easily transferred to similar systems.

This thesis, along with other recent research, seeks to address these limitations by exploring and evaluating alternative approaches to energy management control.

\section{Control algorithm concepts}

The following five concepts related to control algorithms are central to defining the research problems and questions in this thesis:

\textbf{Control model} The control model of an energy management method is the internal representation it uses to make decisions. This can be a set of rules, a mathematical model, a neural network, or any other structure that captures the relationships between inputs (such as sensor data and predictions) and outputs (control actions). The choice of control model has implications for the method's interpretability, and performance.

\textbf{Interpretable model} According to \citet{rudinStopExplainingBlack2019}, an interpretable model ``provides its own explanations, which are faithful to what the model actually computes.'' This contrasts with a black-box model defined as ``(1) a function that is too complicated for any human to comprehend or (2) a function that is proprietary''.
Examples of interpretable models include decision trees with a limited number of nodes, linear regression, and finite state machines with a limited number of states. Examples of black-box models include deep neural networks, boosted decision trees, and random forests.

\textbf{Improvable control model} A control model consists of the parameters and equations governing control actions. An improvable model allows these components to be refined through data training, enabling adaptation to the system over time. While machine learning methods typically feature such adaptive properties, rule-based control and model predictive control generally rely on fixed parameters that do not evolve with experience.

\textbf{Versatile method} A method is considered versatile if it can be applied to a wide range of cases with different objectives, constraints, and system characteristics without requiring significant modifications. for example, a method that can be used for both battery management and heating control without needing to change its core structure. While machine learning methods typically feature such versatile properties, rule-based control and model predictive control generally require defining specific rules or physical models for each new case.

\textbf{Proven technology} A method is considered a proven technology if it has been widely adopted in practice and has demonstrated its effectiveness in real-world applications. For example, model predictive control is a proven technology, it is adopted to do energy management by major industrial companies such as ABB \cite{abbABBAbilityAdvanced2026}, is used in the most popular energy management software for Home assistant \cite{davidusb-geekEnergyManagementHome2025}, and has a prominent place in the scientific literature on energy management systems \cite{balaliEnergyModellingControl2023,ahmadReviewMicrogridEnergy2023, binghuihanHomeEnergyManagement2023}.

\section{Research problems and questions}

\textbf{Research problem 1:} Lack of an existing energy management methods that uses interpretable control models, can improve their control model through training on data, and is versatile.

As presented in the introduction, existing energy management methods have limitations. Rule-based control is interpretable but often suboptimal and case specific, model predictive control requires detailed models and generally does not improve its control model through experience, and reinforcement learning can be data-intensive and lacks interpretability.
This lead to the first research question:

\begin{itemize}
\item \textbf{Research question 1:} Can an energy management system method use interpretable control models, improve its control model by training on data, and be versatile ?
\end{itemize}

Developing a energy management system with these three desirable properties would represent a significant advancement in the field of energy management systems but it is unlikely to be used if it performs poorly compared to proven technology such as model predictive control.
This lead to the second research question: 

\begin{itemize}
\item \textbf{Research question 2:} Is such an energy management system performant meaning that it obtains performances closer to proven technology such as model predictive control than to baseline methods such as rule-based control in a variety of cases ?
\end{itemize}

\textbf{Research problem 2:} Lack of validation of energy management methods on real hardware.

Many studies on energy management systems are performed in simulations without testing the methods on real hardware. This can lead to discrepancies between simulated and real-world performance due to factors such as unmodelled dynamics, sensor inaccuracies, and unrealistic assumptions in simulations.
As will be seen the literature review (see \cref{ch:literature}), there is a real demand for real world experiments in the research community.
In this context, validating the method to address the first research problem on real hardware is a crucial step to ensure its practical applicability.
This lead to the third research question:

\begin{itemize}
\item \textbf{Research question 3:} Can such an energy management system be implemented in a real case and how does it perform compared to baseline and proven technology methods ?
\end{itemize}

\textbf{Research problem 3:} Difficulty in quantifying the tangible benefits of deploying energy management systems in electrified dwellings.

Showing that an energy management system brings benefits when controlling a dwelling with specific characteristics is valuable but it is generally difficult to extrapolate these results to a broader that have different user behaviour, assets, and energy prices.
Nonetheless, as electrification of dwellings progresses in Belgium, it is important to understand if deploying energy management systems beneficial for most households, which characteristics lead to the highest benefits, and if the investment and effort required to implement such systems is justified from the end-user perspective.
This lead to the fourth research question:
\begin{itemize}
\item \textbf{Research question 4:} What are the tangible benefits in terms of cost savings, electric vehicle charge completion and thermal comfort of deploying energy management systems in electrified dwellings for residents in Belgium ?
\end{itemize}

\textbf{Research problem 4:} Missing machine learning paradigms to categorize algorithms used to solve control problems.

Traditional machine learning paradigms include supervised learning, unsupervised learning, and reinforcement learning. However, different machine learning algorithms used for control problems, including the one developed in this thesis, do not fit neatly into these categories (for example neuroevolution).
This leads to the sixth research question:

\begin{itemize}
    \item \textbf{Research question 5:} Can we define a new machine learning paradigm that categorizes algorithms used to solve control problems but do not fit in the three traditional machine learning paradigms ?
\end{itemize}

\section{Leap of knowledge}

To address these research questions, this thesis makes the following key leap of knowledge:

\begin{enumerate}
    \item Development of TreeC, a machine learning-based control algorithm for energy management that uses interpretable control models, can improve its control model by training on data, and is versatile.
    \item Comprehensive evaluation of TreeC on two public benchmarks and in a competition, comparing its performance to rule-based control, model predictive control, and reinforcement learning algorithms.
    \item Validation of TreeC on real home hardware, assessing its effectiveness relative to rule-based control, model predictive control, and reinforcement learning in practical settings.
    \item Analysis of the benefits of TreeC and model predictive control over rule-based control in a representative scenario of electrified dwellings in Belgium.
    \item Proposition of a new machine learning paradigm, termed holistic learning, to categorize methods that solve problems similarly.
\end{enumerate}

These leaps of knowledge have resulted in two first author journal articles, each publication comes with its own code repository to reproduce and build upon the results.

\begin{itemize}
    \item J. Ruddick, L. Ramirez Camargo, M. A. Putratama, M. Messagie, and T. Coosemans. “TreeC: A Method to Generate Interpretable Energy Management Systems Using a Metaheuristic Algorithm”. In: Knowledge-Based Systems (2025). \\
    Code: \url{https://github.com/EVERGi/treec-paper-results}\\
    Thesis sections: \Cref{sec:treec} and \cref{ch:simul_cases} \cref{sec:ANM6-Easy,sec:boptest}
    \item J. Ruddick, G. Ceusters, G. Van Kriekinge, E. Genov, C. De Cauwer, T. Coosemans, and M. Messagie. “Real-World Validation of Safe Reinforcement Learning, Model Predictive Control and Decision Tree-Based Home Energy Management Systems”. In: Energy and AI (2024). \\
    Code: \url{https://github.com/EVERGi/real_validation_saferl_treec_paper} \\
    Thesis chapter: \Cref{ch:summary_cases}
\end{itemize}

The work done in \cref{ch:ems_relevance} has been submitted to a journal under the title "Should Belgians install an advanced energy management system in their electrified dwellings ?".
The code repository for this work is available at \url{https://github.com/EVERGi/should_belgians_install_advanced_ems_paper}.

During this thesis, my colleague Evgeny Genov and I participated in two international competitions: the \textit{IEEE-CIS Technical Challenge on Predict+Optimize for Renewable Energy Scheduling}, where we achieved 3\textsuperscript{rd} place and received a cash prize of \$2000, and the \textit{CityLearn Challenge 2022}, where we placed 6\textsuperscript{th} and were awarded a community contribution prize worth \$1000 of travel grant to attend ACM BuildSys 2023.

Participation to both of these competitions as led to the following additional publications:

\begin{itemize}
    \item J. Ruddick, E. Genov, L. R. Camargo, T. Coosemans, and M. Messagie. “Evolutionary Scheduling of University Activities Based on Consumption Forecasts to Minimise Electricity Costs”. In: 2022 IEEE Congress on Evolutionary Computation (CEC) (2022) \\ 
    Code: \url{https://github.com/EVERGi/predict-then-schedule-university-activities}\\
    \item C. Bergmeir et al. “Predict+Optimize Problem in Renewable Energy Scheduling”. In: IEEE Access (2025).
    \item E. Genov, J. Ruddick, C. Bergmeir, M. Vafaeipour, T. Coosemans, S. Garcia, and M. Messagie. Balancing Forecast Accuracy and Switching Costs in Online Optimization of Energy Management Systems. In: Expert Systems with Applications (2025).
\end{itemize}

\section{Outline}

This thesis is structured into seven chapters:

\Cref{ch:introduction} introduces the context of the research, outlines the research questions, and presents the contributions of this thesis.

\Cref{ch:theoretical_foundations} provides the theoretical foundations used in this thesis to solve optimization problems.  An explanation is given on when certain optimization or machine learning methods should be used over others. It also introduces the concept of holistic learning, a paradigm useful to group methods that solve problems similarly. Finally, the chapter presents rule-based control and model predictive control, two of the main control algorithms used in energy management systems.

\Cref{ch:literature} provides an overview of the most important work done in the main fields of control algorithms used for energy management systems as well as look at real implementations of these algorithms and what are the public competitions and benchmarks in the field.

\Cref{sec:treec} describes in detail the developed TreeC method. 

\Cref{ch:simul_cases} presents the results of the simulation cases conducted in this thesis. The first case the \textit{Gym-ANM ANM6-Easy} benchmark which requires controlling large photovoltaic, wind, and battery systems to reduce energy loss and not overload the grid. The second case is the \textit{BOPTEST hydronic heat pump} benchmark which is a domestic heat pump control problem. The third case is the \textit{CityLearn Challenge 2022} competition where 5 domestic batteries need to be controlled in coordination to reduce electricity costs, CO\textsubscript{2} emissions, and grid power fluctuations. The three cases all evaluate the performance of TreeC against rule-based control, model predictive control, and reinforcement learning methods.

\Cref{ch:summary_cases} presents the results of the real experiment conducted on 4 emulated homes equipped with real batteries, photovoltaic panels, digital meters and controllable loads that mimic base electric consumption and the charging of an electric vehicle. TreeC is compared to rule-based control, model predictive control, and reinforcement learning using a real electricity contract with hourly prices and grid peak penalties.  

\Cref{ch:ems_relevance} aims to answer research question 2 more thoroughly by comparing the performance of TreeC, model predictive control, and rule-based control in a representative scenario of electrified dwellings in Belgium. 

\begin{table}[H]
    \centering
    \caption{Overview of the research problems (RP), research questions (RQ) and associated leaps of knowledge, publications, and chapters in this thesis.}
    \begin{tabular}{|c|c|c|c|c|}
        \hline
        RP & RQ & Leap of knowledge & Publication & Chapter \\
        \hline
        \multirow{2}{*}{1} & 1 & Development of TreeC & \multirow{2}{*}{\citet{ruddickTreeCMethodGenerate2025}} & \ref{sec:treec} \\
        \cline{2-3} \cline{5-5}
         & 2 & Evaluation TreeC in simulation & & \ref{ch:simul_cases} \\
        \hline
        2 & 3 & Real-world validation TreeC & \citet{ruddickRealworldValidationSafe2024} & \ref{ch:summary_cases} \\
        \hline
        3 & 4 & Analysis electrified Belgium & Submitted & \ref{ch:ems_relevance} \\
        \hline
        4 & 5 & Holistic learning & - & \ref{ch:theoretical_foundations} \\
        \hline
    \end{tabular}
\end{table}

\mode<all> %

    \chapter[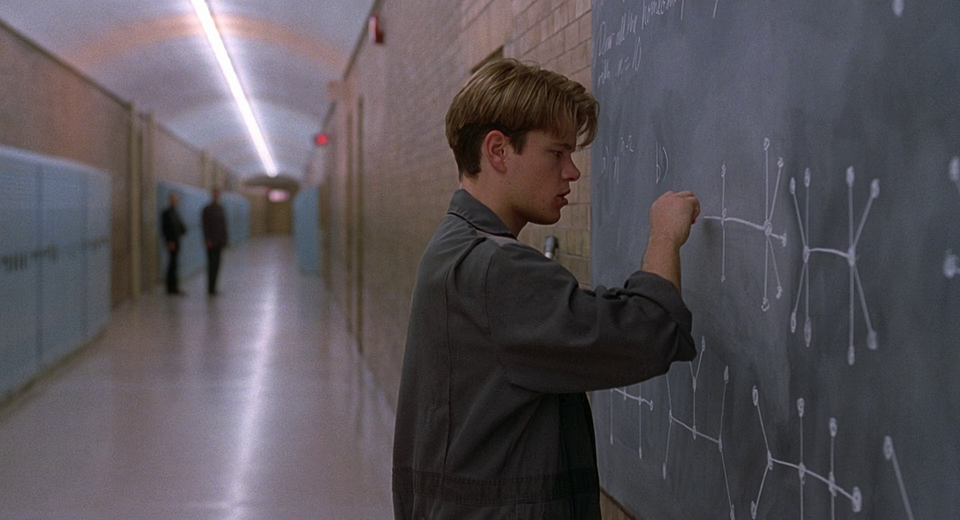]{Theoretical foundations\label{ch:theoretical_foundations}}

Energy management is fundamentally an optimization problem, typically aiming to minimize the operational cost of an energy system while satisfying constraints such as user comfort and device limitations.
Understanding the theoretical foundations of the methods used to address these problems is valuable.
This chapter defines the fundamental concepts related to optimization problems and provides an overview of the key theoretical concepts underlying the approaches employed in this thesis, focusing on optimization, machine learning, and their relevant subfields. It also discusses the criteria for selecting the most suitable subfield for a given energy management problem.

\section{Fundamental concepts in optimization problems}

Let's first define some terminology that will be used throughout this chapter.

\begin{itemize}
\item \textbf{Problem}: A problem is a task that needs to be solved.
\item \textbf{Variable}: A variable is a value that can change. In optimization, variables are the values that need to be determined to solve a problem.
\item \textbf{Solution}: A solution represents a potential resolution to a problem. It is typically a set of values assigned to the variables of the problem.
\item \textbf{Objective function}: An objective function is a function that evaluates the quality of a solution to a problem.
\item \textbf{Problem objective}: A problem objective is the goal of an optimization problem, which can be to minimize or maximize the objective function.
\item \textbf{Objective value}: An objective value is the value of the objective function for a given solution.
\item \textbf{Constraint}: A constraint is a condition that must be satisfied by a solution to a problem.
\item \textbf{Model}: A model represents a relationship that maps inputs to corresponding outputs.

\end{itemize}

Let's illustrate these concepts with a simple energy management problem.

Problem:
It is 20:00, and you want to charge an electric vehicle with 2 kWh in the next 4 hours while minimizing the cost of electricity.
The cost of electricity is 0.1 \euro/kWh between 8:00 and 22:00 and 0.05 \euro/kWh between 22:00 and 8:00.
The vehicle can not be charged more than 1 kWh per hour, so at a max power of 1 kW.

Variables:
The variables are the amount of energy charged at each hour of the next 4 hours.

Solution:
The solution is the values assigned to all variables, for example 0.5 kWh from 20:00 to 21:00, 1 kWh from 21:00 to 22:00, 0.5 kWh from 22:00 to 23:00 and 0 kWh at 23:00 to 0:00.

Objective function:
The objective function is the formula that calculates the total cost of electricity based on the values assigned to the variables.
In this case, the objective function is the sum of the cost of electricity for each hour multiplied by the amount of energy charged at that hour.

Problem objective:
The problem objective is to minimize the total cost of electricity given by the objective function.

Objective value:
The objective value is the total cost of electricity for the given solution.

Constraints:
The two constraints here are that the total amount of energy charged must be equal to 2 kWh and that the amount of energy charged during each hour must be less than or equal to 1 kWh.

Model:
Let's introduce a model that maps price of electricity to the amount of energy charged at each hour.
When the price of electricity is equal to 0.05 \euro/kWh, the model will charge 1 kWh at that hour, otherwise charge 0 kWh.

The problem is formulated mathematically in \cref{fig:energy_management_problem} which shows all the introduced concepts.

\begin{figure}[H]
    \centering
    \begin{tikzpicture}[
        every node/.style={align=left, font=\footnotesize},
        label/.style={font=\bfseries}
    ]
        \node[label] at (0, 2) {Time:};
        \node at (4, 2) {20:00-21:00};
        \node at (6, 2) {21:00-22:00};
        \node at (8, 2) {22:00-23:00};
        \node at (10, 2) {23:00-0:00};

        \node[label] at (0, 1) {Electricity price (\euro/kWh):};
        \node at (4, 1) {$e_1 = 0.1$};
        \node at (6, 1) {$e_2 = 0.1$};
        \node at (8, 1) {$e_3 = 0.05$};
        \node at (10, 1) {$e_4 = 0.05$};

        \node[label] at (0, 0) {Variable (kWh):};
        \node at (7, 0) {$x_1, x_2, x_3, x_4 \in \mathbb{R}$ };

        \node[label] at (0, -1) {Constraints:};
        \node at (7, -1) {
            Charge limit: $0 \text{ kWh} \leq x_i \leq 1$ kWh \\[4pt] 
            Total charge: $\sum_{i=1}^{4} x_i = 2$ kWh
        };

        \node[label] at (0, -2) {Objective function:};
        \node at (7, -2) {$\sum_{i=1}^{4} (x_i \times e_i)$};
        
        \node[label] at (0, -3) {Problem objective:};
        \node at (7, -3) {$\min \sum_{i=1}^{4} (x_i \times e_i)$};

        \node[label] at (0, -4) {Introduced model:};
        \node at (7, -4) {$x_i = 
        \begin{cases} 
        1 \text{ kWh}, & \text{if } e_i = 0.05 \text{ \euro/kWh} \\ 
        0 \text{ kWh}, & \text{otherwise} 
        \end{cases}$};

        \node[label] at (0, -5) {Solution (kWh):};
        \node at (4, -5) {$x_1 = 0$};
        \node at (6, -5) {$x_2 = 0$};
        \node at (8, -5) {$x_3 = 1$};
        \node at (10, -5) {$x_4 = 1$};

        \node[label] at (0, -6) {Objective value:};
        \node at (7, -6) {$\sum_{i=1}^{4} (x_i \times e_i)$ = 0.1 \euro};
    \end{tikzpicture}
    \caption{Mathematical formulation of the simple energy management problem described above.}
    \label{fig:energy_management_problem}
\end{figure}

\section{How to solve optimization problems}

\begin{figure}[H]
    \centering
    \begin{tikzpicture}[
        every node/.style={draw, rounded corners, font=\footnotesize, text width=2.5cm, align=center}
        ]
        \newlength{\xsep}
        \newlength{\ysep}
        \setlength{\xsep}{-1.2cm} %
        \setlength{\ysep}{0.8cm}    %
        \node (root) {Solving optimization problem(s)};
        
        \node[fill=blue!20, below left=\ysep and 0pt of root, xshift=-\xsep] (opt) {Optimization};
        \node[fill=green!20, below right=2*\ysep+0.6cm and 0pt of root, xshift=\xsep+1.6cm] (ml) {Machine learning};
        
        \node[below left=\ysep and 0pt of opt, xshift=-\xsep] (math) {Mathematical programming};
        \node[below right=\ysep and 0pt of opt, xshift=\xsep] (meta) {Metaheuristics};
        
        \node[fill=red!20, below left=\ysep and 0pt of ml, xshift=-\xsep-4.78cm] (indep) {Independent problems};
        \node[fill=yellow!20, below right=\ysep and 0pt of ml, xshift=\xsep-1.55cm] (control) {Control problems};
        
        \node[below left=\ysep and 0pt of indep, xshift=-\xsep] (super) {Supervised learning};
        \node[below right=\ysep and 0pt of indep, xshift=\xsep] (unsuper) {Unsupervised learning};
        
        \node[below left=\ysep and 0pt of control, xshift=-\xsep] (reinf) {Reinforcement learning};
        \node[below right=\ysep and 0pt of control, xshift=\xsep] (holis) {Holistic learning};
        
        \draw[->] (root) -- (opt);
        \draw[->] (root) -- (ml);
        \draw[->] (opt) -- (math);
        \draw[->] (opt) -- (meta);
        \draw[->] (ml) -- (indep);
        \draw[->] (ml) -- (control);
        \draw[->] (indep) -- (super);
        \draw[->] (indep) -- (unsuper);
        \draw[->] (control) -- (reinf);
        \draw[->] (control) -- (holis);
    \end{tikzpicture}
    \caption{Overview of the approaches discussed in this chapter for solving optimization problems. Each split represents a decision point, guiding the selection of the most suitable method based on the problem characteristics. The details on each decision point are provided in \cref{fig:decision_tree_optimization_vs_ml,fig:decision_tree_mathematical_vs_metaheuristic,fig:decision_tree_ml_independent_vs_control,fig:decision_tree_ml_supervised_vs_unsupervised,fig:decision_tree_ml_reinforcement_vs_holistic_learning} and their respective sections.}
    \label{fig:overview_theoretical_foundations}
\end{figure}
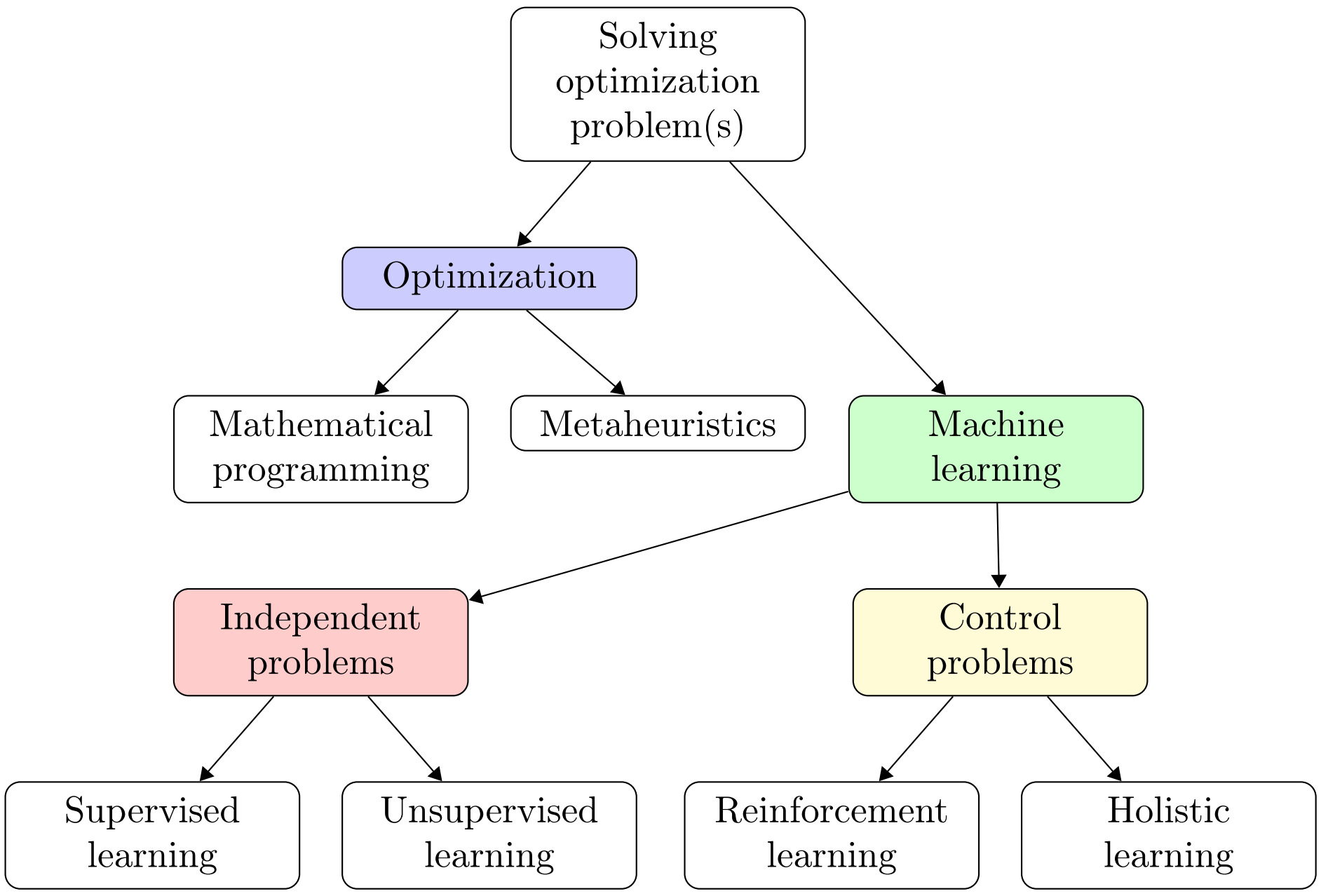

There are multiple approaches to solve optimization problems with some more suitable than others depending on the problem characteristics.
This section presents the main approaches encountered in this thesis and gives guidelines on how to choose the most appropriate one for a given problem.
\Cref{fig:overview_theoretical_foundations} provides an overview of the presented approaches to solving optimization problems and illustrates the decision process for selecting the most appropriate subfield.

\subsection{Difference optimization and machine learning}

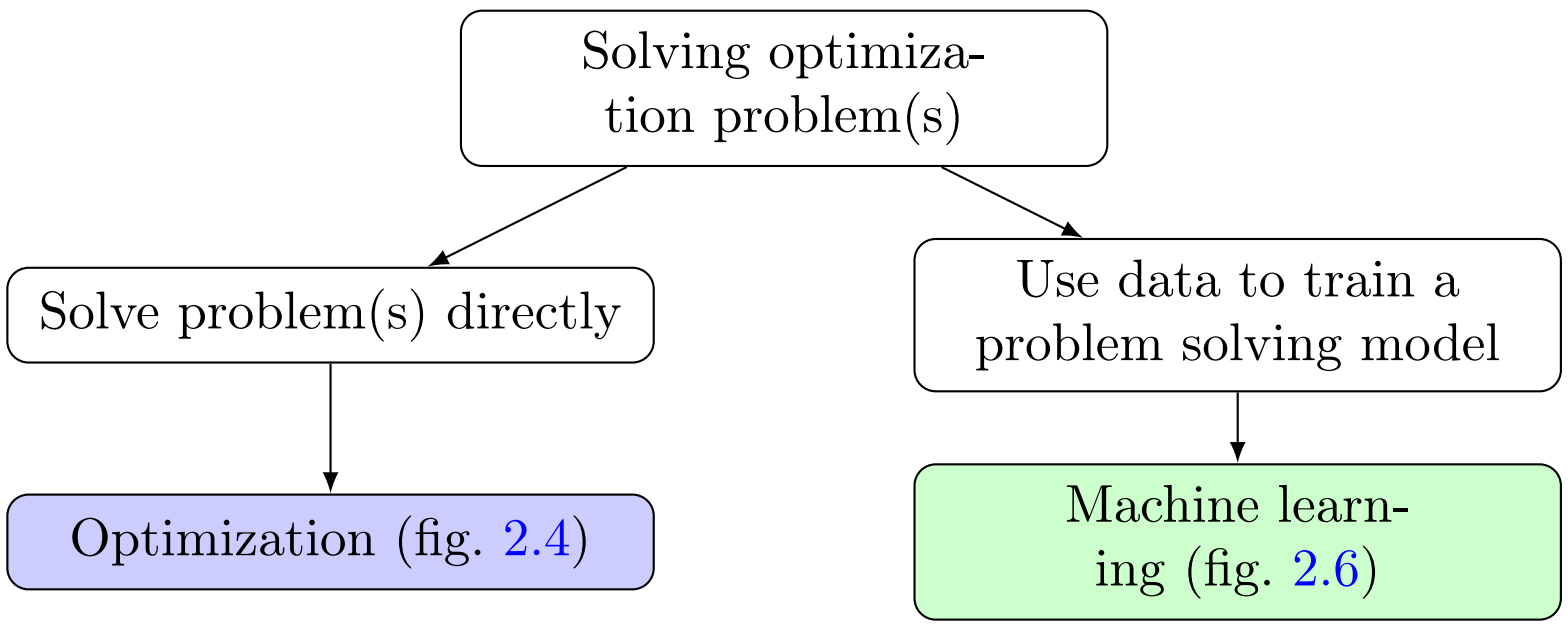
\begin{figure}[H]
    \centering
    \begin{tikzpicture}[
        level 1/.style={sibling distance=60mm},
        edge from parent/.style={draw, -latex},
        every node/.style={draw, align=center, rounded corners, text width=40mm, font=\footnotesize}
    ]
        \node {Solving optimization problem(s)}
            child {node {Solve problem(s) directly}
                child {node[fill=blue!20] {Optimization (\cref{fig:decision_tree_mathematical_vs_metaheuristic})}}
            }
            child {node {Use data to train a problem solving model}
                child {node[fill=green!20] {Machine learning (\cref{fig:decision_tree_ml_independent_vs_control})}}
            };
    \end{tikzpicture}
    \caption{Optimization vs machine learning.}
    \label{fig:decision_tree_optimization_vs_ml}
\end{figure}

Optimization and machine learning are two distinct approaches to solving problems. 
Optimization aims to determine the best possible solution for a problem, whereas machine learning develops models that can generate good solutions not only for the original problem but also for similar problems by learning from data (see \cref{fig:decision_tree_optimization_vs_ml}) \cite{burkovHundredPageMachineLearning2019}.
In machine learning, training refers to the process of learning a model.

The choice between optimization and machine learning depends on the problem to be solved.

Optimization typically requires to clearly define a problem by 
formulating the relationship between the solution and the objective function as a mathematical expression or a set of rules.
In machine learning, this is not always required because a human can provide optimal solutions embedded in the training data, allowing the model to learn how to connect data inputs to optimal solutions.

Consider the problem of recognizing digits (0 to 9) from an image. The optimization approach involves defining explicit rules to identify each digit, such as detecting specific pixel patterns unique to each digit.
However, for handwritten digits, defining such rules can become too complex, even though a human can still label the images with the correct digits.
In this case machine learning is a preferred approach because it will teach a model which patterns are relevant to identify the digits from the labelled images.

\subsection{Optimization}

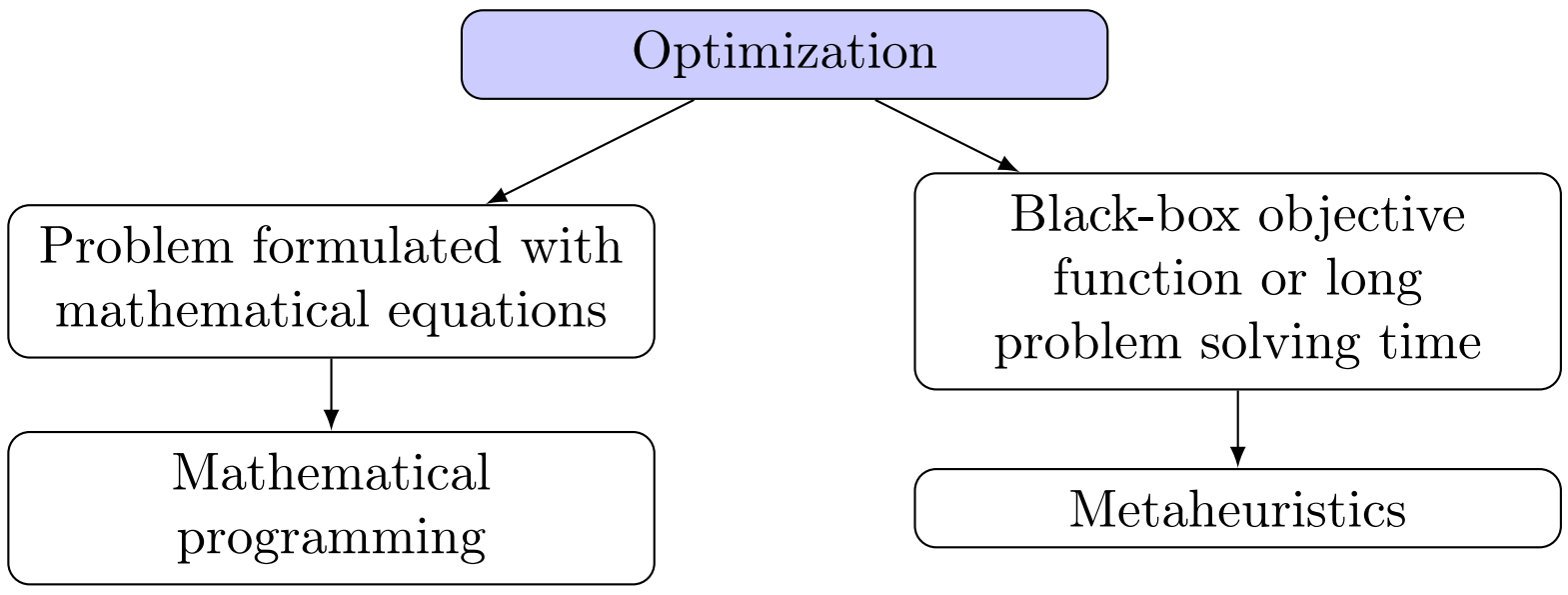
\begin{figure}[H]
    \centering
    \begin{tikzpicture}[
        level 1/.style={sibling distance=60mm},
        edge from parent/.style={draw, -latex},
        every node/.style={draw, align=center, rounded corners, text width=40mm, font=\footnotesize}
    ]
        \node[fill=blue!20] {Optimization}
            child {node {Problem formulated with mathematical equations}
                child {node {Mathematical programming}}
            }
            child {node {Black-box objective function or long problem solving time}
                child {node {Metaheuristics}}
            };
    \end{tikzpicture}
    \caption{Mathematical optimization vs Metaheuristics.}
    \label{fig:decision_tree_mathematical_vs_metaheuristic}
\end{figure}

There are two types of optimization algorithms: deterministic and stochastic.
Deterministic algorithms always give the same solution to a problem.
Stochastic algorithms involve some randomness that will give different solutions to a problem for different runs.

In the context of energy management system control strategies, the most common deterministic and stochastic algorithms are respectively mathematical programming and metaheuristic algorithms \cite{mariano-hernandezReviewStrategiesBuilding2021,salehiComprehensiveReviewControl2022}.

Mathematical programming algorithms use a mathematical formulation of the link between the variables and the objective function of a problem to search for the optimal solution. 
In contrast, metaheuristic algorithms sample the solution space by testing different solutions and obtaining their objective values. The optimization process indicates which solutions should be explored further to obtain better objective values.

\begin{figure}[H]
    \centering
    \includegraphics[width=0.8\textwidth]{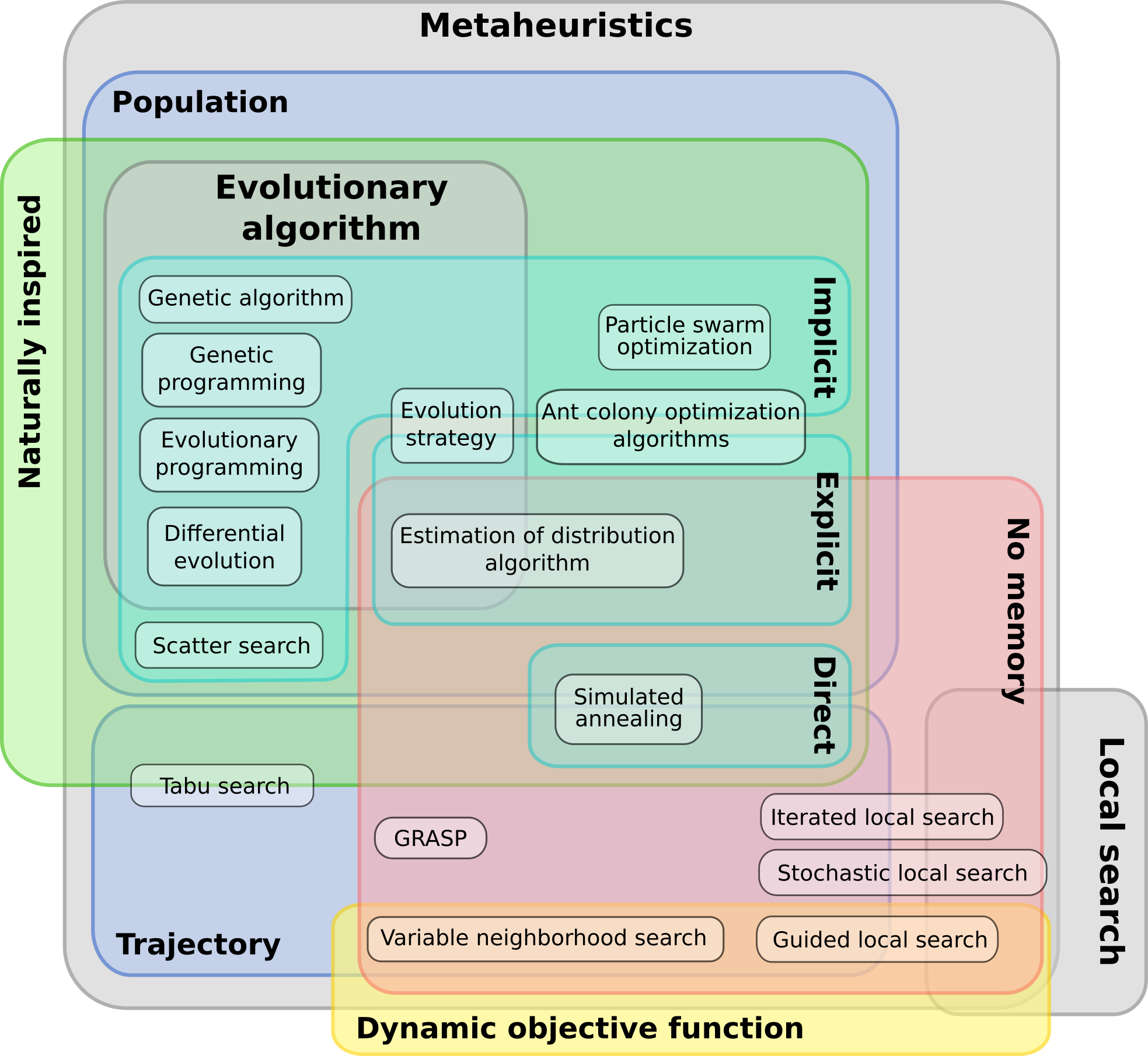}
    \caption{Classification of various metaheuristic algorithms \cite{dreoDifferentesClassificationsMetaheuristiques2007}.}
    \label{fig:classification_metaheuristic_algorithms}
\end{figure}

While many works focus specifically on evolutionary algorithms to describe relevant stochastic optimization methods, this thesis adopts the broader metaheuristic category. Metaheuristics encompass a wider range of stochastic algorithms, including evolutionary algorithms, that solve optimization problems with the same base principles.
\Cref{fig:classification_metaheuristic_algorithms} illustrates this with specific algorithm examples and the various ways to classify them (e.g. evolutionary algorithm, naturally inspired, population-based, etc.). 

The choice between mathematical programming and metaheuristic algorithms depends on the problem's characteristics. 
Mathematical programming is often more efficient and can guarantee an optimal solution if the problem can be clearly defined with mathematical equations. 
However, when problems become to long to solve because of too many variables, complex or non-linear constraints, or when the objective function is a black-box (i.e., it can be evaluated but not explicitly defined), metaheuristic algorithms are usually a better choice (see \cref{fig:decision_tree_mathematical_vs_metaheuristic}).

\subsection{Machine learning}
\subsubsection{Training approaches}
\label{sec:ml_training_approaches}

Machine learning traditionally has three training approaches: supervised learning, unsupervised learning, and reinforcement learning \cite{bishopPatternRecognitionMachine2006}.
This thesis employs a fourth approach used by many works but that could not be found explicitly defined in the literature: holistic learning.

\begin{itemize}
\item \textbf{Supervised learning} trains a model by giving it data containing inputs and associated outputs (often a good solution). 
This approach is often used in regression and classification applications such as time series forecasting and image recognition.
\item \textbf{Unsupervised learning} trains a model by giving it input data but without indicating what are good solutions.
The goal is to recognize patterns in the data for applications such as detecting possible anomalies and creating groups of similar data (clustering).
\item \textbf{Reinforcement learning} trains a model to find good solutions for a sequence of dependent problems. In training, each intermediate solution receives a reward score to assess the quality of the solution. The objective function in reinforcement learning is the sum of the all the individual rewards of the sequence of solutions. This approach is often used to control robots or play games.
\item \textbf{Holistic learning} trains a model to find good solutions for a sequence of dependent problems. In training, the model is evaluated through its objective function only for the entire sequence of solutions and not for each intermediate solution.
This approach is used for the same applications as reinforcement learning.
\end{itemize}

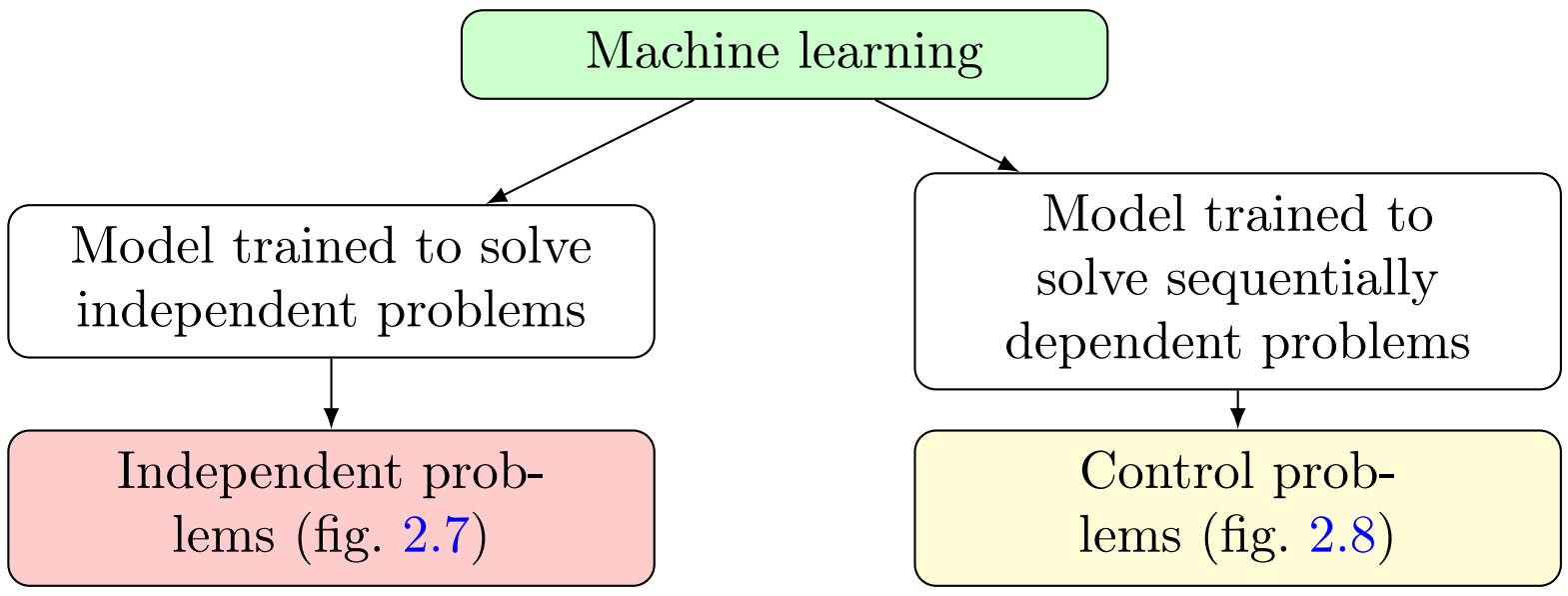
\begin{figure}[H]
    \centering
    \begin{tikzpicture}[
        level 1/.style={sibling distance=60mm},
        edge from parent/.style={draw, -latex},
        every node/.style={draw, align=center, rounded corners, text width=40mm, font=\footnotesize}
    ]
        \node[fill=green!20] {Machine learning}
            child {node {Model trained to solve independent problems}
                child {node[fill=red!20] {Independent problems (\cref{fig:decision_tree_ml_supervised_vs_unsupervised})}}
            }
            child {node {Model trained to solve sequentially dependent problems}
                child {node[fill=yellow!20] {Control problems
                (\cref{fig:decision_tree_ml_reinforcement_vs_holistic_learning})}}
            };
    \end{tikzpicture}
    \caption{Independent vs Control problems in machine learning.}
    \label{fig:decision_tree_ml_independent_vs_control}
\end{figure}

These approaches are trained to tackle different types of problems. 
Supervised and unsupervised learning are trained to solve independent problems in the sense that the quality of the trained model is evaluated independently for each problem.
With reinforcement and holistic learning, the model is trained to solve a sequence of dependent problems where the quality of the model is evaluated based on the entire sequence of solutions.
These problems are here called independent and control problems respectively (see \cref{fig:decision_tree_ml_independent_vs_control}).

An example of an independent problem is the recognition of handwritten digits from images. Recognizing a digit in an image is an independent problem because the quality of the solution depends only on the specific image being analysed, not on the solutions for other images. 

In contrast, playing chess is considered more of a control problem. Here, the machine learning model determines a move at each turn, but the overall outcome of the game depends on the sequence of all moves made. While it is possible to evaluate the quality of individual moves, the ultimate goal is to achieve a favourable outcome for the entire sequence of moves.

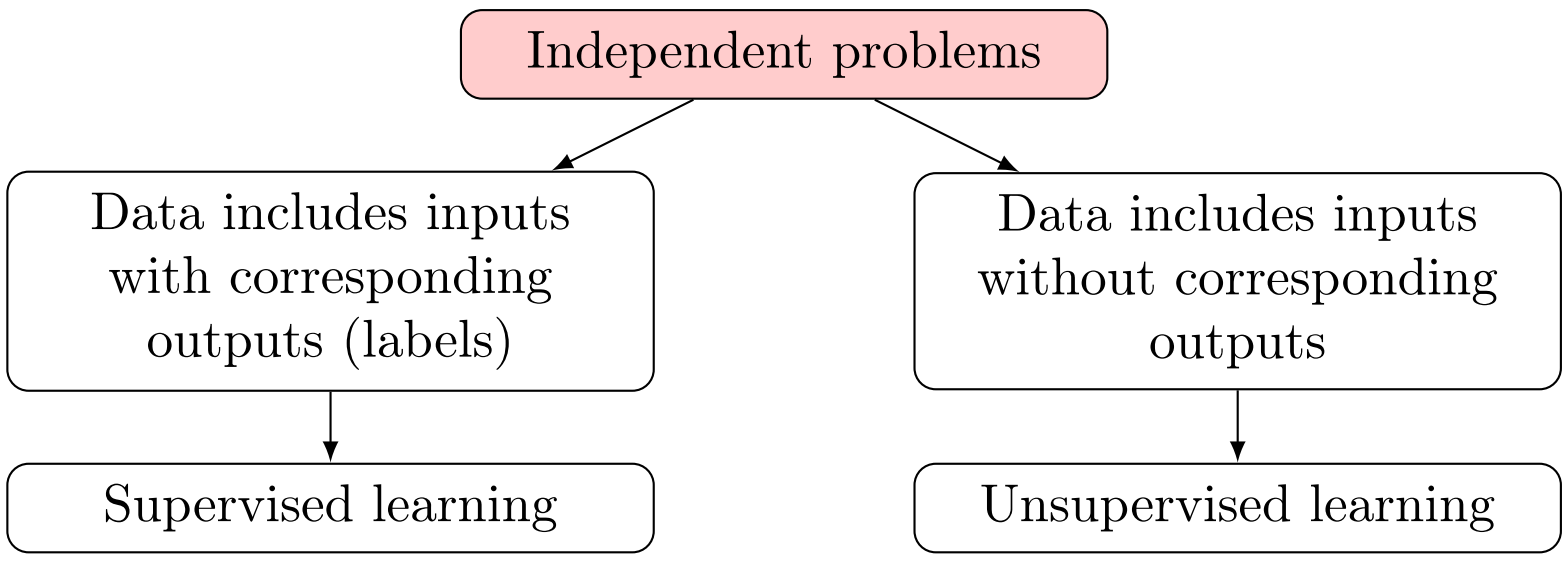
\begin{figure}[H]
    \centering
    \begin{tikzpicture}[
        level 1/.style={sibling distance=60mm},
        level 2/.style={sibling distance=40mm},
        edge from parent/.style={draw, -latex},
        every node/.style={draw, align=center, rounded corners, text width=40mm, font=\footnotesize}
    ]
        \node[fill=red!20] {Independent problems}
            child {node {Data includes inputs with corresponding\\ outputs (labels)}
                child {node {Supervised learning}}
            }
            child {node {Data includes inputs\\ without corresponding\\ outputs}
                child {node {Unsupervised learning}}
            };
    \end{tikzpicture}
    \caption{Supervised vs Unsupervised learning.}
    \label{fig:decision_tree_ml_supervised_vs_unsupervised}
\end{figure}

The difference between supervised and unsupervised learning is that in supervised learning, the data includes inputs with corresponding outputs (labels), while in unsupervised learning, the data includes inputs without corresponding outputs (see \cref{fig:decision_tree_ml_supervised_vs_unsupervised}).

In the example of the handwritten digit recognition problem, supervised learning uses labelled images, where each image is paired with its corresponding digit. 
A similar problem but tackled using unsupervised learning would be to group the images into 10 different clusters based on their similarities. 
Ideally, each cluster would represent a single digit, containing all images corresponding to that digit.
Both the supervised and unsupervised final models would probably have similar internal rules to recognize the digits, but the approach to obtain them is very different.

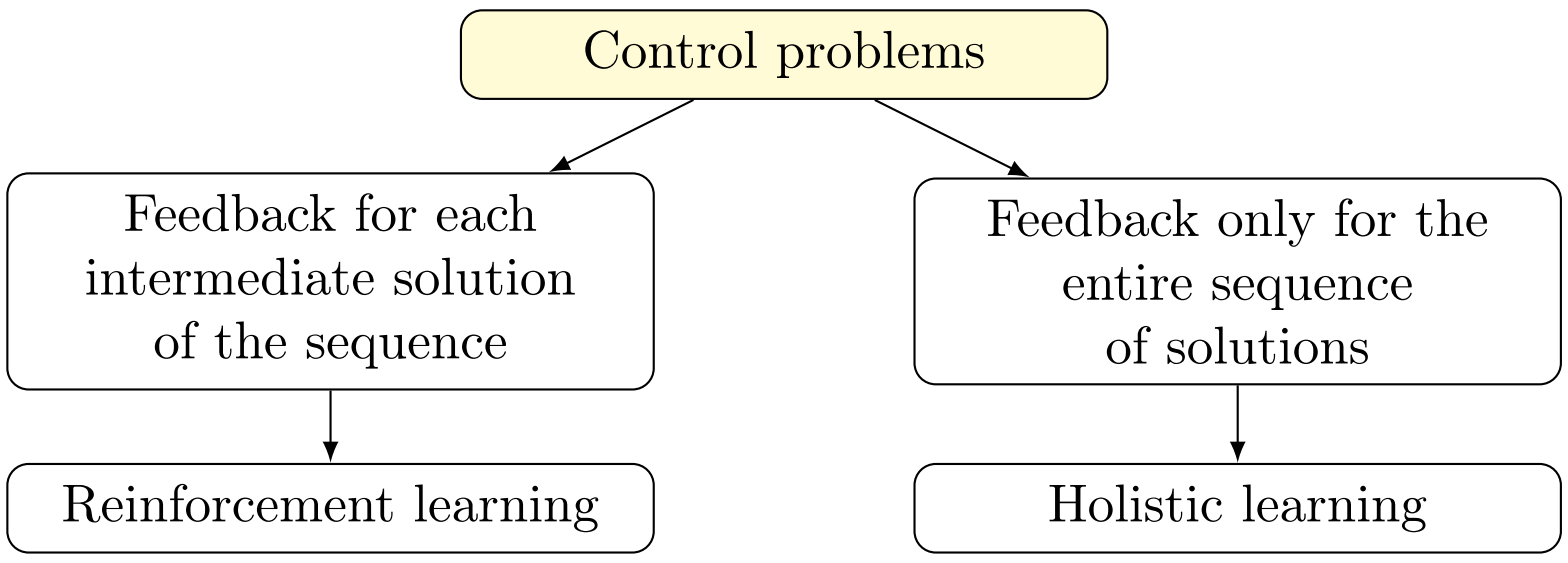
\begin{figure}[H]
    \centering
    \begin{tikzpicture}[
        level 1/.style={sibling distance=60mm},
        level 2/.style={sibling distance=40mm},
        edge from parent/.style={draw, -latex},
        every node/.style={draw, align=center, rounded corners, text width=40mm, font=\footnotesize}
    ]
        \node[fill=yellow!20] {Control problems}
        child {node {Feedback for each intermediate solution\\ of the sequence}
            child {node {Reinforcement learning}}
        }
        child {node {Feedback only for the\\ entire sequence of solutions}      
            child {node {Holistic learning}}
        };
    \end{tikzpicture}
    \caption{Reinforcement vs Holistic learning.}
    \label{fig:decision_tree_ml_reinforcement_vs_holistic_learning}
\end{figure}

The difference between reinforcement and holistic learning is that in reinforcement learning, the model receives feedback in the form of a reward score for each intermediate solution of the sequence, while in holistic learning, the model receives feedback only for the entire sequence of solutions (see \cref{fig:decision_tree_ml_reinforcement_vs_holistic_learning}).
Both reinforcement and holistic learning can be trained on same problems with same objective functions, but in reinforcement learning, the objective function is divided into individual rewards to provide relevant feedback for each intermediate solution in the sequence.

In the chess game example, reinforcement learning would provide feedback after each move, enabling the model to learn from individual moves. In contrast, holistic learning would provide feedback only at the end of the game, evaluating the entire sequence of moves as a whole.

To be clear, in reinforcement learning the model can receive neutral feedback for intermediate solutions within a sequence, while the final solution is evaluated with positive or negative feedback based on the overall outcome \cite{silverGeneralReinforcementLearning2018}.
For example in chess, a move that does not directly lead to a win or loss might receive a neutral reward score of 0, while a winning move receives a positive reward score (e.g., +1) and a losing move a negative reward score (e.g., -1).
Even thought the final outcome is the most important, the model still learns from the intermediate neutral feedback during the process.

Holistic learning can often be implemented by framing the model's parameters as variables in a metaheuristic optimization problem. The objective function evaluates the overall performance of the sequence of solutions generated by the model. This approach iteratively adjusts the model's parameters to improve the objective value of the entire sequence generated by the model.
This already exists in the literature applied to neural network models \cite{floreanoNeuroevolutionArchitecturesLearning2008}, fuzzy logic models \cite{thriftFuzzyLogicSynthesis1991} and decision tree models \cite{jankowskiEvolutionaryAlgorithmDecision2014}. 

In general, reinforcement learning is most effective when meaningful feedback can be provided for each intermediate solution, accurately reflecting its contribution to the overall objective. If such intermediate feedback is not available or does not give accurate quality assessment of the intermediate solutions, holistic learning would generally be more appropriate.

\subsubsection{Models}
\label{sec:ml_models}

Since machine learning essentially trains models to solve problems, the choice of model is important.
There are many types of models, each with its own strengths and weaknesses.

To understand better some main models used in machine learning, \cref{fig:ml_models_example} illustrates the exact same control strategy but implemented with three different models: a neural network, a linear regression, and a decision tree.

\newcommand{\mlmodelsfig}{
\begin{figure}[H]
    \centering
    \begin{subfigure}[b]{0.48\textwidth}
        \centering
        \begin{tikzpicture}[node distance=1.2cm, every node/.style={font=\footnotesize}]
            \node[draw, circle, fill=white] (s) {$s$};
            \node[draw, circle, fill=white, right=2cm of s] (b) {$b$};
            \node[draw, circle, fill=white, below=1.5cm of $(s)!0.5!(b)$] (sum) {$\Sigma$};
            \node[draw, circle, fill=white, below=0.5 cm of sum] (out) {$P_{batt}$};
            
            \draw[->, thick] (s) -- node[left] {$w_1=2$} (sum);
            \draw[->, thick] (b) -- node[right] {$w_2=-1$} (sum);
            \draw[->, thick] (sum) -- (out);
            
            \node[above=0.1cm of b] (biaslbl) {$bias=1$};
            \node[above=0.1cm of s] (slbl) {$sensor=\{0,1\}$};
            \node[below=0.1cm of out] (outlbl) {Battery power [kW]};
        \end{tikzpicture}
        \caption{Neural network}
    \end{subfigure}
    \hfill
    \begin{subfigure}[b]{0.48\textwidth}
        \centering
        \vspace{1.5em} %
        \begin{tikzpicture}
            \draw[->] (-1,0) -- (2,0) node[right] {$s$};
            \draw[->] (0,-2.5) -- (0,2) node[above] {$P_{batt}\ (kW)$};
            
            \draw (1,-0.1) -- (1,0.1) node[below=6pt] {$1$};
            \draw (-0.1,-1) -- (0.1,-1) node[left=8pt] {$-1$};
            \draw (-0.1,1) -- (0.1,1) node[left=8pt] {$1$};
            
            \draw[thick, blue] (-0.5,-2) -- (1.5,2);
            
            \filldraw[blue] (0,-1) circle (2pt) node[below right] {$(0,-1)$};
            \filldraw[blue] (1,1) circle (2pt) node[below right] {$(1,1)$};
            \node[fill=white, font=\footnotesize, text=blue] at (2.7,1.5) {$P_{batt} = 2s - 1$};
        \end{tikzpicture}
        \vspace{0.5em} %
        \caption{Linear regression}
        \end{subfigure}
    \begin{subfigure}[b]{0.6\textwidth}
        \centering
        \vspace{0.8em}
        \begin{tikzpicture}[
            level 1/.style={sibling distance=3cm, level distance=1.5cm},
            level 2/.style={sibling distance=2cm, level distance=1.5cm},
            every node/.style={draw, rounded corners, font=\footnotesize, align=center, fill=white}
        ]
            \node {Does sensor $s$ detect solar irradiation?}
                child {node {Battery power:\\1 kW} 
                    edge from parent node[left, draw=none, font=\footnotesize] {Yes}
                }
                child {node {Battery power:\\-1 kW}
                    edge from parent node[right, draw=none, font=\footnotesize] {No}
                };
        \end{tikzpicture}
        \caption{Decision tree}
    \end{subfigure}
    \caption{Three common models used in machine learning exhibiting exactly the same control strategy: charge the controlled battery at 1 kW when the solar sensor detects solar irradiation and discharge the battery by 1 kW otherwise.}
\label{fig:ml_models_example}
\end{figure}
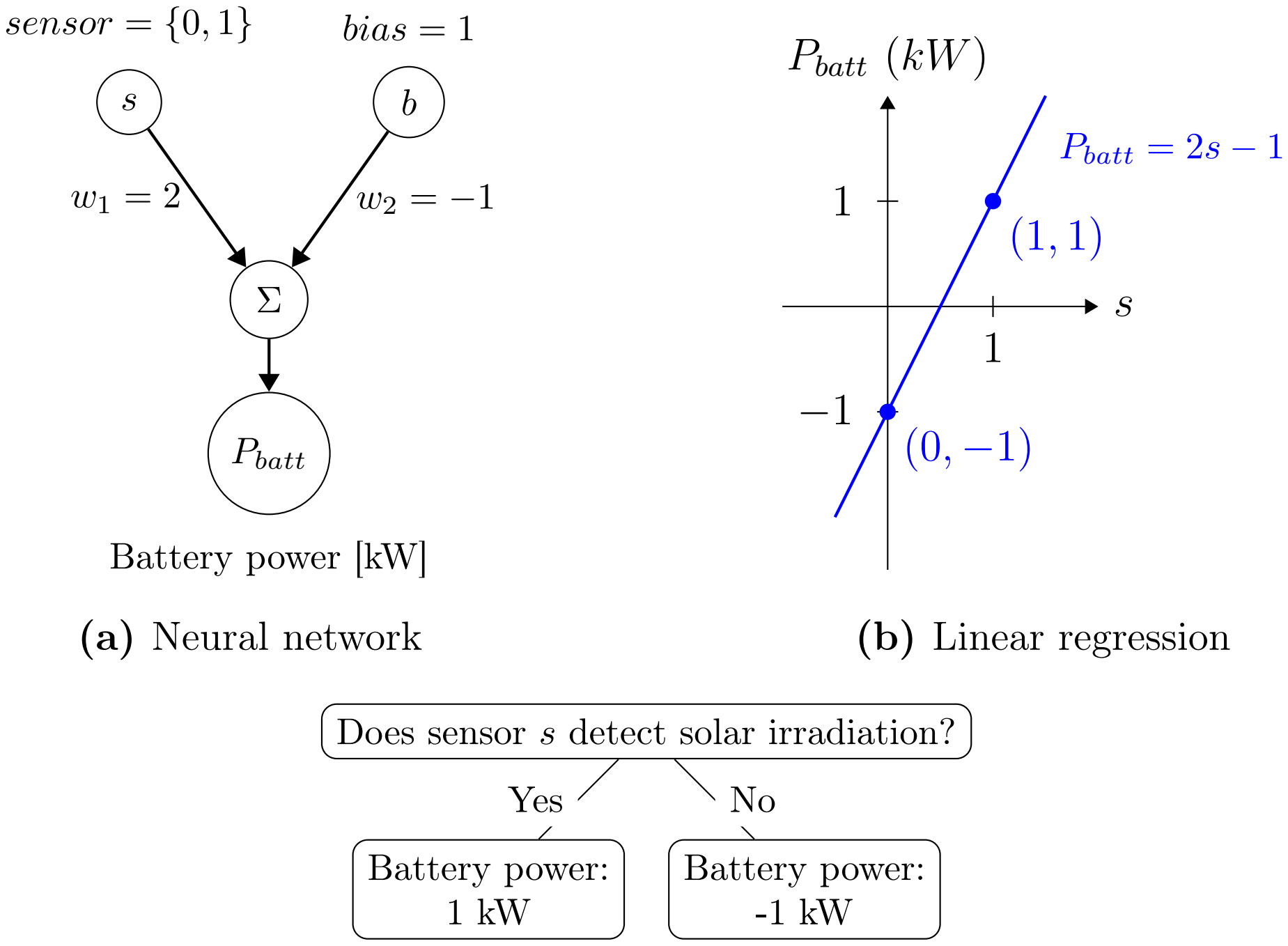
}

\ifprintedversion
\mlmodelsfig
\fi

The neural network consists of an input layer (here the sensor $s$ with values 0 or 1 and a bias $b$ set to 1), an output layer (the battery power in kW), and weighted connections ($w_1$, $w_2$) between them. Each input is multiplied by its corresponding weight, and the outputs are summed to produce the final value. In this example, the output is therefore given by $P_{batt} = w_1\times s + w_2 \times b = 2s - 1$.

The linear regression model is an equation of the form $y = a \times x + b$, where $a$ is the slope of the line and $b$ is the y-intercept. In this case, the linear regression model has a slope of 2 and a y-intercept of -1, which gives the same equation as the neural network, $P_{batt} = 2s - 1$.

The decision tree model is a flowchart-like structure that makes decisions based on the value of the sensor $s$. If the sensor detects solar irradiation (i.e. $s=1$), the battery power is set to 1 kW, otherwise it is set to -1 kW. 

\ifprintedversion
\else
\mlmodelsfig
\fi

During training, machine learning learns the parameters of the model, such as the weights in the neural network, the slope and y-intercept in the linear regression, or the decision rules in the decision tree, to obtain a good objective value for the trained problem.

Neural network models are generally not interpretable, as they often contain a vast number of parameters (e.g. the ChatGPT model GPT-3 has 175 billion parameters \cite{brownLanguageModelsAre2020}). In contrast, linear regression and decision tree models are considered interpretable, provided they remain sufficiently simple, with a limited number of parameters or rules.

\Citet{rudinStopExplainingBlack2019} recommends to use interpretable models when possible, as interpretability brings many benefits such as trust, transparency, and accountability.
It is important to note that the complexity of a problem does not always require a complex model. In some cases, simple models can achieve good objective values even for complex problems. However, there are problems for which only more complex models are capable of delivering satisfactory performance.

\section{Specific methods used for energy management problems}

This section presents two other common methods used in energy management systems: rule-based control and model predictive control (MPC).

\subsection{Rule based control}

Rule-based control is the most straightforward control strategy. The EMS directs each asset to operate according to predefined rules, without mathematical programming or machine learning. This approach typically reflects the default behaviour of energy assets.

Here below are examples of rule-based control strategies used in this thesis:

\begin{itemize}
\item \textbf{Battery control:} The battery operates in self-consumption mode. It charges when photovoltaic (PV) production exceeds the local load, storing excess energy, and discharges to supply the load when PV production is insufficient.
\item \textbf{Electric vehicle charger control:} The charger operates at the maximum allowed power to charge the electric vehicle as quickly as possible.
\item \textbf{Heat pump control:} Follow indoor comfort temperature setpoints set by the user.
\end{itemize}

Rule-based control strategies are easy to implement and understand, but can be limited in terms of performance.

\subsection{Model predictive control}

MPC is a method that models a problem for a defined horizon, solves it to obtain a control setpoint then regularly re-evaluates the problem to adapt to model inaccuracies or prediction errors made over the horizon.

If both the model of the system and predictions are perfectly accurate and the optimization is run over the entire time of the problem, then the MPC will find the optimal control strategy.
In practice, however, achieving this is challenging because real systems cannot be perfectly forecasted or modelled.

In energy management problems, MPC typically uses mathematical equations to model the physical behaviour of the energy system. For forecasting, common predicted inputs include the expected power of each asset, weather conditions, and energy prices.

There are three categories of MPC describing how the model of the system is implemented \cite{deconinckToolboxDevelopmentValidation2016, drgonaAllYouNeed2020}: 

\begin{itemize}
    \item White-box MPC: The equations and parameters of the model are all based on the knowledge of the system without estimating them through data driven methods.
    \item Grey-box MPC: Part of the model is estimated through data driven methods and part is based on the knowledge of the system.
    \item Black-box MPC: The model is entirely approximated through data driven methods.
\end{itemize}

If even one of the parameter is estimated through data driven methods the MPC is considered grey-box and not white-box \cite{deconinckToolboxDevelopmentValidation2016}.
Doing the forecasting using a data driven method does not disqualify the MPC from being white-box as this predicts the inputs to the model and does not change the model itself.

The advantage of white-box MPC is that if the model is accurate the MPC will be very reliable and it is possible to track meaningful parameters of the system \cite{drgonaAllYouNeed2020}.
The disadvantage is that describing the system accurately can be very difficult, time-consuming and take time to solve when the description is mathematically complex.

The advantage of black-box MPC is that no knowledge of the system is needed to implement the model and the solving time is generally much faster.
The disadvantage is that data is necessary to estimate the model and if the data does not cover the complete behaviour of the system it can lead to very inaccurate models.

What distinguishes MPC from other optimization-based methods used in EMSs is that the optimization is re-executed on a smaller time-period compared to the horizon of the optimization.
Where other optimization methods would calculate every 24 hours the control schedule for the entire day and stick to it, MPC would calculate the control schedule for the next 24 hours, apply the given control for one hour and then re-calculate for the following 24 hours.
Recalculating more regularly than the horizon allows to be more reactive to unexpected events.

This thesis implements mainly white-box MPCs with a grey-box MPC used in \cref{ch:ems_relevance}.

\mode<all> %

\chapter[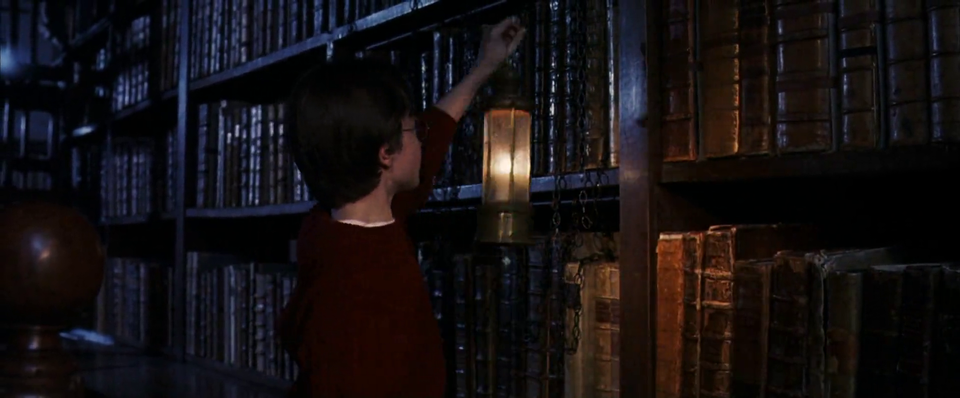]{Literature review\label{ch:literature}}

This chapter reviews the literature on energy management systems (EMS) and aims to: 1) define the scope and key terms relevant to the thesis, 2) provide an overview of the control algorithm fields used in EMSs and relevant to this thesis, 3) identify machine learning methods using decision tree models similar to the TreeC method developed in this thesis, 4) summarize real-world EMS implementations, 5) present publicly available benchmarks and competitions for EMS performance comparison, and 6) goes into recent literature reviews to identify the state of the art in energy management control algorithms.

The chapter is structured as follows:

\begin{itemize}
    \item Scope definition: outlines the scope and key terms, and describes the literature search strategy.
    \item Control algorithm fields: reviews key contributions in mathematical programming, metaheuristics, deterministic and stochastic model predictive control, reinforcement learning, and holistic learning.
    \item Decision tree based machine learning methods: summarizes literature on machine learning based EMSs using decision tree models.
    \item Real implementations: highlights studies where EMSs have been tested on real hardware.
    \item Benchmarks and competitions: lists major energy management benchmarks and competitions for EMS evaluation.
    \item Conclusion: summarizes the main findings.
\end{itemize}

\section{Scope definition}

\subsection{Term definition}

The key terms used in this literature review are defined in this section.

An \textbf{energy management system} is a set of interrelated or interacting elements that are organized to achieve objectives and energy targets \cite{EnergyManagementSystems2018}.
This definition is very general and not restricted to computer-aided systems meaning an EMS can be solely based on human decisions.
This thesis will focus on computer-aided EMSs which implement an automated control strategy.

A computer-aided EMS needs to gather inputs, process them and then control one or more elements of the system.
Relevant inputs can be the energy prices, the power consumption and production of different energy assets, the weather forecast, etc.
The control is usually done by giving setpoints to energy assets. 

In this context, a \textbf{control strategy} is the algorithm that defines how the inputs are processed to obtain the control setpoints.
Energy management strategy is another term found in literature for the control strategy of an EMS.

The goal of the control strategy is to adjust the system's energy demand to achieve specific objectives, most commonly to minimize energy costs.

\textbf{Demand-side management} refers to all measures that modifies the energy demand of a system.
This includes demand response and energy efficiency measures \cite{lauby2011NorthAmerican2013}.

\textbf{Demand response} refers to all measures that changes the energy demand of a system without the explicit goal of reducing the energy consumption \cite{wohlfarthEnergyEfficiencyDemand2020}.
This often means increasing the power demand when there is good power supply and reducing it when the supply is low.
In literature the term load management is often used to describe the same concept \cite{lampropoulosHistoryDemandSide2013}.

\textbf{Energy efficiency} refers to all measures aimed at reducing the energy consumption of a system.
This is mainly done through building insulation and energy efficient appliances but can also be done through better control of the system \cite{wohlfarthEnergyEfficiencyDemand2020}.

\subsection{Scope of the literature review}

The scope of the literature review is the studies done on \textbf{automated control strategies used in computer-aided EMSs}.
This section describes the basic search queries used in SCOPUS to find the literature that fits this scope.
This basic search query is used by all subsections of this literature review, with additional search queries added to restrict the results further.

The first query restricts the search to energy related optimization in a broad sense.

Query: "energy management system" or "demand side management" or "demand-side management" or "demand response" or "energy efficiency"

The second query restricts the search to management related topics removing for example energy efficiency studies done through building insulation.

Query: management or control or scheduling

The third query restricts the search to assets or similar concepts that can be controlled by an EMS relevant to the scope of this thesis.

Query: battery or microgrid or "heat pump" or grid or "electric vehicle" or boiler or building or chp or hvac or lighting or "energy storage" or "renewable energy" or "heating" or "cooling" or "photovoltaic"

Without any additional filters, the search returns 80\ 808 documents.
Subsection of the literature review will add search queries to restrict the results.

The literature review focuses on control strategies working with time-steps of 1 minute to 1 hour therefore excluding research done on lower \cite{barklundEnergyManagementAutonomous2008,songEnergyManagementStrategies2014,tanCentralizedControlParallel2012,xiePontryaginsMinimumPrinciple2019,zhangModelPredictiveControl2017} granularity time-steps. Checking that each paper fits this criterion as well as the overall scope of the literature review is done manually.

\section{Control algorithm fields}

This section reviews the main algorithmic fields used in EMS control strategies. Each subsection summarizes the most cited articles in its field, covering mathematical programming, metaheuristics, deterministic and stochastic model predictive control, reinforcement learning, and holistic learning. Article selection criteria and key contributions are briefly presented for each field, providing a comparative overview of their strengths and limitations.

Given the vast number of publications in each algorithmic field, ranging from 230 to 24\,161 articles per query depending on the field, this review focuses on summarizing the five most cited papers in each field. This approach provides a representative overview of key contributions and trends without attempting exhaustive coverage.

\subsection{Mathematical programming}

Mathematical programming, introduced in the previous chapter, is a very common method used in EMSs.

\subsubsection{Article selection criteria}

For this section, the additional search query adds optimization to programming as some papers simply refer to their mathematical programming based EMS as optimization.

Query: programming or optimization

This additional query restricts the results to 24\ 161 documents.

\subsubsection{Key articles and their contributions}

\citet{conejoRealTimeDemandResponse2010} proposes an EMS method that schedules the consumption of a residential and commercial consumers with hourly real time prices.
The EMS predicts the upper and lower bounds of the future electricity prices and the paper shows this method is more efficient than predicting the most likely price.

\citet{palma-behnkeMicrogridEnergyManagement2013} implemented an EMS that controls a diesel generator, an energy storage, a water supply, photovoltaics and electric loads in a microgrid.
The paper compares in simulation the EMS when re-evaluating the strategy every hour and every 24 hours showing the economic benefits of the hourly re-evaluation.

\citet{sortommeOptimalSchedulingVehicletoGrid2012} demonstrates in simulation an EMS that controls 10\,000 vehicle-to-grid capable electric vehicles to do ancillary services and optimize on hourly electricity prices.

\citet{duApplianceCommitmentHousehold2011} proposes a load scheduling EMS and demonstrates its performance on a water heater optimized on hourly real time electricity prices and comfort.
The paper's novelty is its new multiloop algorithm shown to solve the appliance commitment problem in a fast, robust, and flexible way.

\citet{moraisOptimalSchedulingRenewable2010} presents an EMS for optimal scheduling of a wind turbine, solar unit, fuel cell, and storage battery in an isolated microgrid. The EMS is tested in simulation using data from a real small laboratory microgrid, demonstrating its effectiveness in managing isolated microgrid operations.

\subsection{Metaheuristics}

Metaheuristic optimization, also introduced in the previous chapter, is another common method used in EMSs.

\subsubsection{Article selection criteria}

For metaheuristic optimization, many papers use the term evolutionary algorithm or simply refer to the algorithm they use, so the additional search query adds evolutionary and other names of common metaheuristic algorithms.

Query: "metaheuristic" or "evolutionary" or "ant colony" or "particle swarm" or "genetic" or "differential evolution" or "simulated annealing" or "tabu search"

This additional query restricts the results to 4\ 386 documents. 

\subsubsection{Key articles and their contributions}

\citet{logenthiranDemandSideManagement2012} introduces an EMS designed to manage the load shifting of thousands of electrical devices, including home appliances, heating, ventilation and air conditioning (HVAC) systems, lighting, and motors. The EMS achieved cost reductions of 5\% in residential and commercial cases and 10\% in an industrial case.

\citet{chenSmartEnergyManagement2011} introduces an EMS designed to optimize load shifting in a microgrid comprising photovoltaics, energy storage systems, fuel cells, microturbines, and various loads. The proposed EMS achieved a cost reduction of 27.92\% compared to a scenario without load shifting.

\citet{zhaoOptimalPowerScheduling2013} introduces an EMS for optimizing the power usage of various home appliances, including air conditioners, electric radiators, kettles, and water heaters. The proposed approach achieved a 26\% reduction in electricity costs and a 35\% decrease in the peak-to-average ratio compared to scenarios without load shifting.

\citet{moghaddamMultiobjectiveOperationManagement2011} presents an EMS for optimizing the operation of a microgrid consisting of a microturbine, fuel cell, photovoltaics, wind turbine, and battery, serving residential, commercial, and industrial loads. The proposed method outperforms other metaheuristic based EMSs in terms of cost reduction and emissions minimization.

\citet{luoOptimalSchedulingRenewable2020} presents an EMS for a grid-connected microgrid incorporating photovoltaics, wind turbines, fuel cells, microturbines, and battery energy storage systems. The proposed EMS achieves a cost reduction of 0.5\% to 2\% compared to other metaheuristic based EMSs, while also reducing simulation time by 2\% to 8\%.

\subsection{Deterministic model predictive control}

Model predictive control (MPC) is a method that models a problem for a defined horizon, solves it to obtain a control setpoint then regularly re-evaluates the problem to adapt to model inaccuracies or prediction errors made over the horizon.

Deterministic MPC is the classical approach. It uses the most useful model and prediction of the system to perform good control.

\subsubsection{Article selection criteria}
For this section, the additional search query is the following:

Query: "model predictive control" or "MPC" or "model-based predictive control"

This additional query restricts the results to 2\ 800 documents.

\subsubsection{Key articles and their contributions}

\citet{sirokyExperimentalAnalysisModel2011} implemented an MPC based EMS to control the HVAC of three real buildings.
The experiment showed that the EMS saved ~20\% energy compared to a heating curve based EMS.

\citet{olivaresCentralizedEnergyManagement2014} proposes an MPC based EMS that controls a diesel generator, a battery and a microturbine in an isolated three-phase microgrid equipped with non-controllable photovoltaics, wind turbines and loads.
The simulations show the importance of modelling the microgrid in the EMS as a three-phase system and provides a performant approximation to linearize the originally non-linear problem.

\citet{maDemandReductionBuilding2012} applied an MPC based EMS to control the HVAC of a building with 5 zones optimizing on and off-peak electricity prices.
The MPC based EMS was compared in simulation to 3 different rule based EMSs and obtained a 3\% to 25\% energy and cost reduction.

\citet{sturzeneggerModelPredictiveClimate2016} implemented an MPC based EMS in a real office building controlling the thermally activated building system (TABS), the air handling unit and the blinds.
The MPC based EMS saved 17\& energy compared to a rule based EMS, but the authors calculated that with current technology and energy prices the investment in the MPC based EMS would not be profitable.

\citet{smarraDatadrivenModelPredictive2018} introduces DPC, a black-box MPC method where the model of the controlled system is learned and not defined using mathematical equations.
The method is evaluated on three case studies.
In the first case study DPC is compared to a white-box MPC and controls blind position, electric lighting, evaporative coolers and radiators in building to minimize costs and keep the room temperature comfortable.
The MPC based EMS obtains a 44\% lower cost compared to DPC.
In the second case DPC controls temperature setpoints for a corridor, guest rooms, kitchen and water supply in a large hotel too complex to model using a white-box MPC.
It is shown that DPC manages to change appropriately the temperature setpoints to follow a desired electric power signal for the hotel.
In the third case DPC controls the radiators of an off-grid house modelled in simulation after a real house.
The DPC based EMS obtains a 25\% to 50\% lower energy consumption compared to a rule based EMS depending on how much DPC is permitted to violate the comfort constraints.

\subsection{Stochastic model predictive control}
\label{sec:stoc_mpc_literature}

Stochastic MPC is a newer approach to MPC.
It incorporates the prediction errors by either optimizing over multiple prediction scenarios or internalizing the uncertainties in the constraints of the problem.

\subsubsection{Article selection criteria}
This section adds to the query of Deterministic MPC the following query:

Query: "stochastic" or "scenario-based" 

This additional query restricts the results to 230 documents.

\subsubsection{Key articles and their contributions}
\citet{oldewurtelEnergyEfficientBuilding2010} proposes a stochastic MPC based EMS that controls HVAC, blind positioning and electric lighting to maintain comfortable room temperature, lighting and CO\textsubscript{2} levels.
The Stochastic MPC obtains better performances in simulation compared to a rule based EMS and a deterministic MPC based EMS in both \cite{oldewurtelEnergyEfficientBuilding2010} and \cite{oldewurtelUseModelPredictive2012}.

\citet{bazmohammadiHierarchicalEnergyManagement2019} introduces a hierarchical stochastic MPC based EMS for interconnected microgrids equipped with batteries, generators, wind turbines, and photovoltaics. The upper-level EMS coordinates the operation of three microgrids, while the lower-level EMS manages individual microgrid operations, such as battery charging/discharging and generator control. 
The study compares a chance-constrained stochastic MPC approach to a deterministic MPC for the lower-level EMS, both with and without communication between microgrids. Results indicate that the stochastic approach significantly reduces the average power exchanged with the main grid by 50\% to 75\%, while adding communication between microgrids achieves a smaller reduction of 4\% to 15\%. Conversely, in terms of total costs, the stochastic approach yields less than a 1\% reduction, whereas incorporating communication reduces costs by 9\%.

\citet{olivaresStochasticPredictiveEnergyManagement2015} introduces a stochastic MPC based EMS for isolated microgrids, managing diesel generators, batteries, and microturbines to meet energy demand. The EMS incorporates multiple wind power scenarios into the MPC formulation to address uncertainty. Simulation results demonstrate a 13\% cost reduction compared to a deterministic MPC approach.

\citet{kouDistributedEMPCMultiple2017} presents an EMS for coordinating 10 interconnected microgrids, each equipped with wind turbines, batteries, and microturbines. The EMS operates in two stages: the first stage optimizes the electricity trading schedule between the microgrids and the main grid, while the second stage optimizes the internal power flow within each microgrid to adhere to the trading schedule.
The stochastic nature of wind power and load forecasts is incorporated into the MPC formulation using chance constraints. Simulation results show that the proposed solution adheres to the given schedules within acceptable deviations, achieving a 12\% cost reduction compared to an uncoordinated approach. While the coordinated approach is 2\% more expensive than a fully centralized method, it offers faster computation and eliminates the need for fully centralized control.

\citet{tobajasResilienceorientedScheduleMicrogrids2022} presents a stochastic MPC EMS for a microgrid with a hybrid energy storage system composed of batteries and hydrogen storage, wind-turbines, electric loads, and photovoltaics.
The EMS is evaluated on different resilience scenarios where the EMS successfully maintains the microgrid's critical operation during grid outages.

\subsection{Reinforcement learning}

Reinforcement learning is a machine learning technique introduced in the previous chapter and increasingly used in EMSs.

\subsubsection{Article selection criteria}
For this section, the additional search query is the following:

"reinforcement learning"

This additional query restricts the results to 2\ 029 documents.

\subsubsection{Key articles and their contributions}

\citet{mocanuOnLineBuildingEnergy2019} evaluates two reinforcement learning methods to control the electric vehicle charging, air conditioning, and dishwasher of a building. 
The proposed Deep Q-learning and Deep Policy Gradient methods obtained peak reduction of 15\% to 25\% and energy cost reduction of 14\% to 30\% compared to no load shifting.

\citet{wanModelFreeRealTimeEV2019} proposes a reinforcement learning method to charge a home electric vehicle capable of charging and discharging to the grid.
The proposed method obtained cost reductions of 77\% and 9\% compared to normal charging and MPC respectively.

In a follow-up paper with a similar setup, \citet{liConstrainedEVCharging2020} proposed a new reinforcement learning EMS called constrained policy optimization. This method achieves a good balance between reaching the target battery state of charge at the end of the charging session and minimizing electricity costs, outperforming other reinforcement learning methods and MPC when taking both into account. 

\citet{zhangWholeBuildingEnergy2019} implemented a reinforcement learning EMS to control the HVAC of a real building and obtained a 16.7\% reduction in heating demand compared to a rule-based control EMS.

\citet{chenOptimalControlHVAC2018} proposes a reinforcement learning EMS to control the HVAC and window systems of a building to minimize energy consumption and thermal discomfort.
The proposed method outperforms a rule-based EMS in two different climates, achieving 13\% and 23\% lower HVAC energy consumption, 62\% and 80\% lower discomfort degree hours, and 63\% and 77\% fewer high humidity hours.

\subsection{Holistic learning}

Holistic learning is the new category proposed in the previous chapter to describe machine learning methods that update the model only based on the final training objective value and not on intermediate values like in reinforcement learning.

\subsubsection{Article selection criteria}

The main field of holistic learning applied to EMSs seems to be learned fuzzy logic controllers through metaheuristic optimization.

To find articles using fuzzy logic controllers the following search query was used:

Query: fuzzy

To find articles doing holistic learning, different metaheuristic optimization algorithms were added to the query as well as the word learning:

Query: metaheuristic or evolutionary or "ant colony" or "particle swarm" or "genetic" or "differential evolution" or "simulated annealing" or "tabu search" or learning

No articles were found when searching for neuroevolution which is a common holistic learning field in robotics.

These additional queries restrict the results to 697 documents.

\subsubsection{Key articles and their contributions}
\citet{athariOperationalPerformanceEnergy2016} introduces an EMS for managing a hybrid renewable energy system comprising a wind turbine, photovoltaic array, battery stack, and a hydrogen system with an electrolyser and fuel cell. 
The EMS employs a fuzzy logic controller trained using metaheuristic optimization to control the grid inverter and hydrogen system power setpoints efficiently.
The training is done on the daily or weekly forecasts of the system. 
Simulation results show that the daily optimized fuzzy logic controller achieves a cost reduction of 11\% compared to the weekly optimized fuzzy logic controller and 23\% compared to an unoptimized fuzzy logic controller.

\citet{hagrasHierarchicalFuzzyGenetic2003} introduces an EMS designed to manage room heating and illumination based on occupant preferences. The EMS employs a fuzzy logic controller whose rule base is optimized using metaheuristic techniques, guided by occupant feedback on comfort levels. The study compares this approach to a supervised learning method that trains the fuzzy logic controller using pre-collected occupant data. While both methods achieve comparable results, the proposed approach allows real-time interaction with occupants, enabling the EMS to iteratively improve its performance, unlike the supervised learning method, which requires training before deployment.

\citet{desantisHierarchicalGeneticOptimization2017} introduces a fuzzy logic based EMS for a microgrid equipped with renewable energy sources and a controllable energy storage system. The EMS employs a metaheuristic algorithm to optimize the fuzzy rule base, incorporating a mechanism to eliminate less effective rules during the learning process. This approach achieved a 67\% increase in profit while reducing the rule base size by 53\% compared to a similar method that does not eliminate rules.

\citet{teoOptimizationFuzzyEnergyManagement2021} presents a fuzzy logic based EMS for a grid-connected microgrid with renewable energy sources and a controllable energy storage system. The EMS aims to minimize operating costs and reduce peak load. Three approaches for generating the rule base are compared: expert-defined rules, a single-objective metaheuristic algorithm, and a multi-objective metaheuristic algorithm. The multi-objective approach achieved the best results, reducing costs by 2\% and peak load by 26\% compared to the expert-defined rules, while slightly outperforming the single-objective method. The proposed EMS was experimentally validated on a real microgrid and the measured behaviour was consistent with the parallel simulation with mean absolute percentage error less than 2\% for the state of charge and power of the energy storage system.

\citet{leonoriMicrogridEnergyManagement2020} conducted a comparative study of various EMS generation methods to optimize the profit and reduce grid stress in a microgrid equipped with renewable energy sources and a controllable energy storage system. The methods compared include: MPC with perfect forecasts, metaheuristic optimization combined with supervised learning using different models (Adaptive Neuro-Fuzzy Inference Systems, Support Vector Machines, Echo State Networks, and Multi-Layer Perceptrons), and metaheuristic algorithms for directly learning fuzzy rule bases. 
Model predictive control achieved the best overall performance. However, the Adaptive Neuro-Fuzzy Inference Systems based method outperformed MPC in one test case and delivered comparable results in three out of the six other test cases. Due to its significantly lower execution time (300 to 1900 times faster) and competitive performance, the authors recommend the Adaptive Neuro-Fuzzy Inference Systems based method as a practical alternative.

\subsection{Comparison different algorithmic fields}

This section provides a comparative overview of the most cited work in each algorithmic field discussed in this chapter. It summarizes their prevalence in the literature, citation impact, and typical application cases.

\begin{table}[H]
    \caption{Comparison of algorithmic fields in EMSs}
\centering
\begin{tabular}{|l|l|l|}
\hline
Algorithmic field & \begin{tabular}{@{}c@{}}Number of articles  \\ obtained from query \end{tabular} & \begin{tabular}{@{}c@{}}Number of citations \\ of reviewed articles \end{tabular}
\\ \hline
Mathematical programming & 24\,161 &  483 to 877  \\ \hline
Metaheuristics & 4\,640 & 281 to 1097\\ \hline
Deterministic MPC & 2\,800 & 286 to 653\\ \hline
Stochastic MPC & 230 & 108 to 963 \\ \hline
Reinforcement learning & 2\,029 & 252 to 506\\ \hline
Holistic learning & 697 & 79 to 121\\ \hline
\end{tabular}
\label{tab:comparison_algorithmic_fields}
\end{table}

\Cref{tab:comparison_algorithmic_fields} compares the different algorithmic fields in terms of literature prevalence and citation impact.
These numbers suggest that mathematical programming is the most applied EMS method in research, followed by metaheuristics. Deterministic MPC and reinforcement learning have comparable representation, while stochastic MPC and holistic learning are the least prevalent approaches.

\begin{table}[H]
    \caption{Algorithmic fields and their application cases in EMS literature}
    \centering
    \begin{tabular}{|l|l|l|l|l|}
        \hline
        \begin{tabular}{@{}l@{}}Algorithmic \\ field\end{tabular} & \begin{tabular}{@{}l@{}}Building \\ heating\end{tabular} & \begin{tabular}{@{}l@{}}Building \\ electricity \end{tabular} & Microgrid & \begin{tabular}{@{}l@{}}Electric \\ vehicles\end{tabular} \\
        \hline
        \begin{tabular}{@{}l@{}}Mathematical \\ programming\end{tabular} &  & \cite{conejoRealTimeDemandResponse2010}, \cite{duApplianceCommitmentHousehold2011}
        & \cite{palma-behnkeMicrogridEnergyManagement2013}, \cite{moraisOptimalSchedulingRenewable2010}
        & \cite{sortommeOptimalSchedulingVehicletoGrid2012} \\
        \hline
        Metaheuristics &  & \cite{logenthiranDemandSideManagement2012}, \cite{zhaoOptimalPowerScheduling2013} & \begin{tabular}{@{}l@{}}\cite{chenSmartEnergyManagement2011}, \cite{moghaddamMultiobjectiveOperationManagement2011}, \\ \cite{luoOptimalSchedulingRenewable2020}\end{tabular} &  \\
        \hline
        \begin{tabular}{@{}l@{}}Deterministic \\ MPC\end{tabular} & \begin{tabular}{@{}l@{}}\cite{sirokyExperimentalAnalysisModel2011}, \cite{maDemandReductionBuilding2012}, \\ \cite{sturzeneggerModelPredictiveClimate2016}, \cite{smarraDatadrivenModelPredictive2018}\end{tabular} &  & \cite{olivaresCentralizedEnergyManagement2014} &  \\
        \hline
        \begin{tabular}{@{}l@{}}Stochastic \\ MPC\end{tabular} &\cite{oldewurtelEnergyEfficientBuilding2010}, \cite{oldewurtelUseModelPredictive2012} &  & \begin{tabular}{@{}l@{}}\cite{bazmohammadiHierarchicalEnergyManagement2019}, \cite{olivaresStochasticPredictiveEnergyManagement2015}, \\ \cite{kouDistributedEMPCMultiple2017}, \cite{tobajasResilienceorientedScheduleMicrogrids2022}\end{tabular} &  \\
        \hline
        \begin{tabular}{@{}l@{}}Reinforcement \\ learning\end{tabular} & \cite{zhangWholeBuildingEnergy2019}, \cite{chenOptimalControlHVAC2018} & \cite{mocanuOnLineBuildingEnergy2019} &  & \cite{wanModelFreeRealTimeEV2019}, \cite{liConstrainedEVCharging2020} \\
        \hline
        \begin{tabular}{@{}l@{}}Holistic \\ learning\end{tabular} & \cite{hagrasHierarchicalFuzzyGenetic2003} &  & \begin{tabular}{@{}l@{}}\cite{athariOperationalPerformanceEnergy2016},  \cite{desantisHierarchicalGeneticOptimization2017}, \\ \cite{teoOptimizationFuzzyEnergyManagement2021},  \cite{leonoriMicrogridEnergyManagement2020}\end{tabular} &  \\
        \hline
    \end{tabular}
    \label{tab:algorithmic_subfields_cases}
\end{table}

\Cref{tab:algorithmic_subfields_cases} summarizes the application cases of each algorithmic field in the reviewed literature.
The most common applications among the top cited articles is microgrid energy management followed by building heating management, building electricity management, and electric vehicle charging management.
Stochastic MPC and holistic learning are predominantly applied to microgrid energy management, whereas deterministic MPC is mainly used for building heating control. Notably, among the reviewed articles, reinforcement learning has not been applied to microgrid energy management, in contrast to other algorithmic fields. While these observations suggest certain trends in the suitability of each algorithmic field for specific EMS applications, a more comprehensive review would be required to draw definitive conclusions.

This section highlights the prevalence of each algorithmic field in EMS literature, with mathematical programming being the most widely applied and stochastic MPC and holistic learning the least represented. The application cases of the most cited articles show that deterministic MPC is primarily used for building heating, while stochastic MPC and holistic learning are mainly applied to microgrid energy management.

However, most reviewed articles do not directly compare different algorithmic fields on the same application case, making it challenging to identify which approach works best for a given scenario. More meaningful comparisons are typically achieved through energy management benchmarks and competitions, with the most notable examples presented in \cref{sec:benchmarks}.

\section{Decision tree based machine learning methods}

Decision trees are widely recognized in machine learning for their interpretability and strong performance across various applications \cite{barredoarrietaExplainableArtificialIntelligence2020,quinlanInductionDecisionTrees1986,jordanMachineLearningTrends2015}. However, few studies use machine learning methods to generate decision tree based EMSs, with only two notable studies identified in the literature.

\citet{huoDecisionTreebasedOptimization2021} proposed a decision tree based EMS for microgrid management. The approach first employs an MPC method to generate optimal policies and then uses supervised learning to train decision trees that replicate the MPC's behaviour.

Similarly, \citet{daiDecipheringOptimalMixedmode2023} developed a reinforcement learning EMS for ventilation control, utilizing a neural network model. To enhance interpretability, decision trees were subsequently trained via supervised learning to approximate the neural network's decision-making process.

Both studies highlight the potential of decision trees to create interpretable or explainable EMSs. However, the decision trees in these cases serve as approximations of optimized control generated by other methods, rather than being directly optimized themselves.
This indirect approach may face performance challenges, as the decision tree model may not be suited to reproduce accurately the control behaviour found with the original model.
However, a decision tree optimized directly on the controlled system could very well be performant through a different control behaviour than the one found with the original model.

\section{Real implementations}

This part of the literature review comes from \citet{ruddickRealworldValidationSafe2024} that reviewed the literature on real-world implementations of EMSs.

One observation, and also specifically highlighted by \citet{rathorEnergyManagementSystem2020} and \citet{allwynComprehensiveReviewEnergy2023} in their reviews, is that most work is on simulated case studies and only a few show practical real-world implemented results. 
Nevertheless, notable examples include \citet{zhaoMPCbasedOptimalScheduling2015} who implemented a non-linear MPC in the Hong Kong Zero Carbon Building that optimized the usage of an electric chiller and combined cooling and power unit, considering a stratified chilled water storage tank as thermal energy storage (and thus source of flexibility) and a PV system under a day-ahead pricing mechanism. 
\citet{viotModelPredictiveControl2018} implemented an MPC in the SYNERGI building of the University of Bordeaux - France, which optimized the Thermally Activated Building Systems and reported significant energy savings (40\%) with an improved or equivalent thermal comfort. 
\citet{arroyoComparisonOptimalControl2022} then later compared white-, grey- and black-box modelling paradigms in an MPC based home EMS for the Thermally Activated Building Systems of the Vliet building of the KU Leuven University in Heverlee, Belgium. 
\citet{elkazazEnergyManagementSystem2020} implemented a two-layer structured EMS in a laboratory based grid-connected microgrid, including a non-controllable load, a solar PV system, wind turbine production and a battery. 
The mathematical program in the upper layer then determined the optimal reference values for the lower level MPC and reported a daily cost reduction of up to 30\%. 
\citet{zhangModelPredictiveControl2022} then more recently presented the results of a year-long evaluation of an MPC for the energy management in a small commercial building, located in California, United States, equipped with solar PV and controllable space conditioning, commercial refrigeration, and a battery while testing two types of demand flexibility applications: (1) electricity cost minimization under time-of-use tariffs and (2) responses to grid flexibility events. 
Their results showed 12\% annual electricity cost savings and 34\% peak demand reduction against a baseline, trained using supervised learning, while preserving thermal comfort and food safety. 
\citet{yangExperimentStudyMachinelearningbased2021} then also used supervised learning, yet to approximate an MPC and conducted an experimental study on two real testbeds (office floor and lecture theatre) in Singapore. 
The EMS controlled the air-conditioning systems and still managed to save 52\% of cooling energy consumption compared to 58\% of the original MPC but reduced the computational load by a factor of 100. 

When looking for real-world implemented machine learning EMS results, the options become even more scarce, as also mentioned by \citet{fuApplicationsReinforcementLearning2022} in their review. Even though multiple real-world challenges are yet to be verified and thus remain open, as e.g. stipulated by \citet{nweyeRealworldChallengesMultiagent2022}. 
This, together with recent advancements in imitation learning, makes it a more viable practical option \cite{deyReinforcementLearningBuilding2023}. 
Nonetheless, notable implemented examples include \citet{costanzoExperimentalAnalysisDatadriven2016} who demonstrated a model-assisted batch RL agent controlling heating, ventilation and air conditioning units, subjected to dynamic pricing, for the room climate control of a living lab and confirmed that within 10 to 20 days sensible control policies, that are further shaped by domain knowledge, could be obtained. \citet{kazmiGigawatthourScaleSavings2018} then later demonstrated a model based RL agent optimizing domestic hot water production, controlling reheating cycles of a domestic hot water storage vessel coupled to an air source heat pump in 19 real houses in the Netherlands. 
They reported a 20\% energy efficiency gain without any loss of user comfort (observed on a subset of 5 houses). 
Around the same time, \citet{zhangPracticalImplementationEvaluation2018} practically implemented and evaluated a deep RL agent for the optimal control of a radiant heating system in a real office building. 
The RL agent was trained offline on a physics-based model that was calibrated with real operational data before deploying online on the actual heating system (without taking additional exploratory actions). 
A heating demand savings of 16.6\% to 18.2\% was reported compared to a rule-based control strategy over a three-month-long evaluation period. Also considering distributed energy resources, \citet{touzaniControllingDistributedEnergy2021} demonstrated a deep RL agent optimizing the heating, ventilation and air conditioning and battery, in the presence of solar PV, of the FLEXLAB at Lawrence Berkeley National Laboratory in Berkeley, California, United States. 
The RL agent was also trained offline on a calibrated physics-based simulation model before the online deployment of the policy on the real system. 
They reported significant cost savings (up to 39.6\%) while maintaining similar thermal comfort levels over a three-week-long testing period. 
\citet{naugDeepReinforcementLearning2022} then presented a deep RL agent for a 5-zone simulated testbed, but also a real three-story building at the Vanderbilt University, United States - controlling heating setpoints in the air handling units for a full year. 
Here, again, the RL agent was trained offline and afterwards deployed online. 
If a performance degradation was detected, machine learning models of the building dynamics were updated and an offline relearning loop then also updated the RL agent (using these offline dynamic system models). 
As a final example, \citet{gokhaleRealWorldImplementationReinforcement2023} conducted a 4-week experimental case study for the demand response of a cluster of 8 residential buildings using an RL based energy coordinator together with a Proportional-Integral control based power dispatcher - that also would override actions that could lead to thermal comfort violations, given they utilize the thermal mass of the buildings as cost-free source of flexibility.

Overall the literature has examples of MPC and RL based EMSs implemented in real cases, even though the options are still scarce. Some of these studies compared their method to a rule-based EMS, but none compared different algorithmic fields on the same case.

\section{Energy management benchmarks and competitions}
\label{sec:benchmarks}

Open energy management benchmarks and competitions provide valuable platforms for evaluating and comparing the performance of various EMSs.
This section highlights several notable benchmarks and competitions, including those used in this thesis for method evaluation. Detailed descriptions and analyses of each benchmark and competition will be provided in subsequent chapters.

Several benchmark cases offer open datasets to test EMSs:
\begin{itemize}
    \item \textbf{Boptest} \cite{blumBuildingOptimizationTesting2021}: 8 buildings with HVAC systems.
    \item \textbf{Synergym} \cite{campoy-nievesSinergymVirtualTestbed2025}: 9 buildings with HVAC systems.
    \item \textbf{Energym} \cite{scharnhorstEnergymBuildingModel2021}: 14 buildings with HVAC systems, 4 of which include electric vehicles and batteries.
    \item \textbf{ProHMo} \cite{zinsmeisterBenchmarkingFrameworkEnergy2024}: 6 buildings with HVAC systems and domestic hot water control.
    \item \textbf{Gym-ANM} \cite{henryGymANMOpensourceSoftware2021}: 1 power system case.
\end{itemize}

These benchmarks enable researchers to compare their methods against others without needing to create custom cases tailored to their specific approaches.

Competitions also serve as an effective way to compare EMSs, as participants operate under similar time and resource constraints. Notable competitions include:

\begin{itemize}
    \item \textbf{CityLearn Challenge 2020} \cite{vazquez-canteliCityLearnChallenge20202020}: 9 buildings with HVAC and domestic hot water systems, exclusively for RL methods.
    \item \textbf{CityLearn Challenge 2021} \cite{nagyCitylearnChallenge20212021}: 9 buildings with domestic hot water, chilled water storage, and battery systems, exclusively for RL methods.
    \item \textbf{CityLearn Challenge 2022} \cite{nweyeCityLearnChallenge20222023}: 17 buildings with battery systems, open to all methods.
    \item \textbf{IEEE-CIS Technical Challenge on Predict+Optimize for Renewable Energy Scheduling} \cite{bergmeirPredict+OptimizeProblemRenewable2025}: 6 buildings with battery systems, requiring month-long scheduling of class activities and battery operations, open to all methods.
\end{itemize}
 
These benchmarks and competitions represent important steps toward establishing standardized evaluation frameworks for EMSs. As they continue to evolve, they could become reference benchmarks for the field, similar to the Makridakis competitions in forecasting \cite{makridakisAccuracyExtrapolationTime1982} and the CEC benchmark competitions in metaheuristic optimization \cite{liBenchmarkFunctionsCEC20132013}. Such initiatives have been instrumental in identifying the most effective methods in their respective domains. Such a level of comparative insight is still lacking in EMS research.

This thesis compares TreeC to methods of the Gym-ANM and Boptest benchmarks in \cref{sec:ANM6-Easy} and \cref{sec:boptest}, respectively, and the participation to the CityLearn Challenge 2022 is reported in \cref{sec:citylearn}.

\section{State-of-the-art energy management control methods}

While the previous sections highlighted key contributions within specific algorithmic fields, determining the overall state-of-the-art for energy management control methods requires a broader perspective. This section examines three reviews published in 2023 covering microgrids, buildings, and home energy management to identify potential leading control methodologies.

Regarding microgrids, \citet{ahmadReviewMicrogridEnergy2023} identifies five main fields of energy management optimization: mathematical, heuristic and metaheuristic, fuzzy, MPC, and multi-agent systems. The review does not explicitely identify a specific state-of-the-art method. Instead, it remarks that: ``In this study, it was noticed that not many papers have considered analyzing computational complexity and or compared performance with other existing algorithms. Therefore, there is a need for factoring in computational complexity and comparing various parameters with other proposed methods in the literature to enable a more concerted and productive effort at incrementing the microgrid EMCS research.''

Regarding building energy and HVAC system control, \citet{balaliEnergyModellingControl2023} identifies three main fields: MPC, RL, and reinforced-MPC. The authors conclude by emphasizing the need for standardized evaluation, stating that ``Finally, the development of a unified metric for performance verification and validation of simulated models would simplify comparisons across different studies.''

Regarding home energy management, \citet{binghuihanHomeEnergyManagement2023} identifies four main fields of optimization methods: mathematical, nature-inspired meta-heuristic, artificial intelligence, and model predictive control. While the study does not explicitly cite a lack of unified metrics or comparative studies as a challenge, it also does not identify a single definitive state-of-the-art method.

Overall, these reviews do not explicitly identify a single state-of-the-art energy management control method. Instead, they highlight the diversity of approaches and the need for more comparative studies and standardized evaluation metrics to better understand the relative performance of different methods in the field.
This analysis also shows that model predictive control is central in the field, as it is identified in all three reviews. 

Based on the reviewed literature and understanding of the methods, it is hypothesized that the state-of-the-art lies with stochastic model predictive control methods. These approaches combine the performance of model predictive control with the ability to handle uncertainty effectively, justifying their classification as a separate algorithmic field in \cref{sec:stoc_mpc_literature}. However, this has not been established in the literature, and more mature benchmarking for energy management control algorithms is necessary to confirm this hypothesis.

\section{Conclusion}

The literature review reveals several key insights regarding energy management systems (EMSs) literature:

\begin{enumerate}
    \item Mathematical programming is the most prevalent algorithmic field in EMS research, followed by metaheuristics, deterministic model predictive control, and reinforcement learning. Stochastic MPC and holistic learning are less commonly explored.
    \item Holistic learning methods, particularly those involving metaheuristic optimization of fuzzy logic controllers, have demonstrated some relevant results but remain underexplored. Notably, neuroevolution which is commonly used in robotics but has not yet been applied to EMSs.
    \item Decision tree based EMSs are largely unexplored, with only two studies identified that use supervised learning to approximate control strategies generated by other methods.
    \item Real-world implementations of MPC and reinforcement learning based EMSs exist, but such studies are still scarce. No found work has directly compared methods from different algorithmic fields on the same real-world case.
    \item Benchmarks and competitions for EMS evaluation are emerging, providing standardized platforms for method comparison. However, these initiatives are not yet as established as those in other fields such as forecasting and metaheuristic optimization.
    \item Relevant reviews of energy management control methods do not explicitly identify a single state-of-the-art approach, instead highlighting the diversity of methods. Model predictive control is commonly identified as a central method in the field.
\end{enumerate}

Overall, while significant progress has been made in EMS research, there remain notable gaps in the exploration of machine learning based methods using decision tree models, as well as in the direct comparison of algorithmic approaches on real-world implementations.

This thesis participates in addressing these gaps by developing TreeC, a machine learning based method using decision trees, and comparing it to other algorithmic fields using established benchmarks and a real-world case study.

\mode<all> %

    \chapter[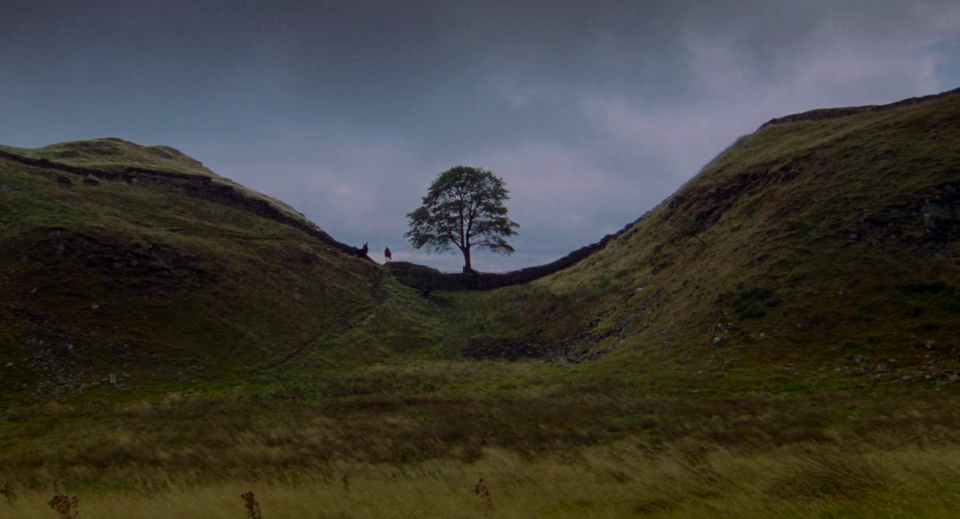]{The new TreeC EMS method\label{sec:treec}}

This chapter introduces TreeC, a novel machine learning method used to solve energy management problems. TreeC leverages metaheuristic optimization to automatically generate interpretable control strategies tailored to specific energy systems. The following sections describe the algorithm's structure, encoding, optimization process, parallel training capabilities, and enhancements made to the base implementation.

\section{Decision-tree generation}

TreeC uses metaheuristic optimization to generate decision trees for control. It generates and tests many decision tree based strategies on the system, selecting as final control strategy the one with the best objective value. This approach allows TreeC to learn an effective and interpretable control strategy tailored to the specific system it is trained on.

TreeC generates a model (the decision tree) by training on data (simulations of the energy system). The parameters of the model are improved by using metaheuristic optimization. This follows the definition of machine learning given in the theoretical foundations chapter (\cref{ch:theoretical_foundations}). More specifically, TreeC is a holistic learning method because it solves sequentially dependant problems and improves the model only by using the final objective value of the training problem as feedback.

The decision tree model used in TreeC is widely used in machine learning for making decisions based on input data. It consists of nodes, features, and leaves. Each \textbf{node} in the tree represents a decision point where a specific \textbf{feature} (an input variable or attribute) is evaluated against a threshold or condition. Depending on the outcome of this evaluation, the process follows a branch to the next node or to a leaf. A \textbf{leaf} is a terminal node that assigns an action or output value, representing the final decision for a given set of input features. Decision trees are interpretable models as long as the number of nodes is kept manageable and the meaning of the features and leafs are understandable.

TreeC has three important operations: 

\begin{itemize}
\item Encoding (see \cref{fig:encoding}): to translate a list of numbers to the tree EMS.
 \item Optimisation (see \cref{fig:optimisation_proc}): to successively improve the trees.
 \item Pruning (see \cref{fig:pruning}): to reduce the size of the tree by removing unused leafs and nodes.
\end{itemize}

\ifprintedversion
\newcommand{\widthfigtreecmeth}{0.9\linewidth}
\else
\newcommand{\widthfigtreecmeth}{1.0\linewidth}
\fi

\begin{figure}
    \centering
    \begin{subfigure}[b]{\textwidth}
        \centering
        \includegraphics[width=\widthfigtreecmeth]{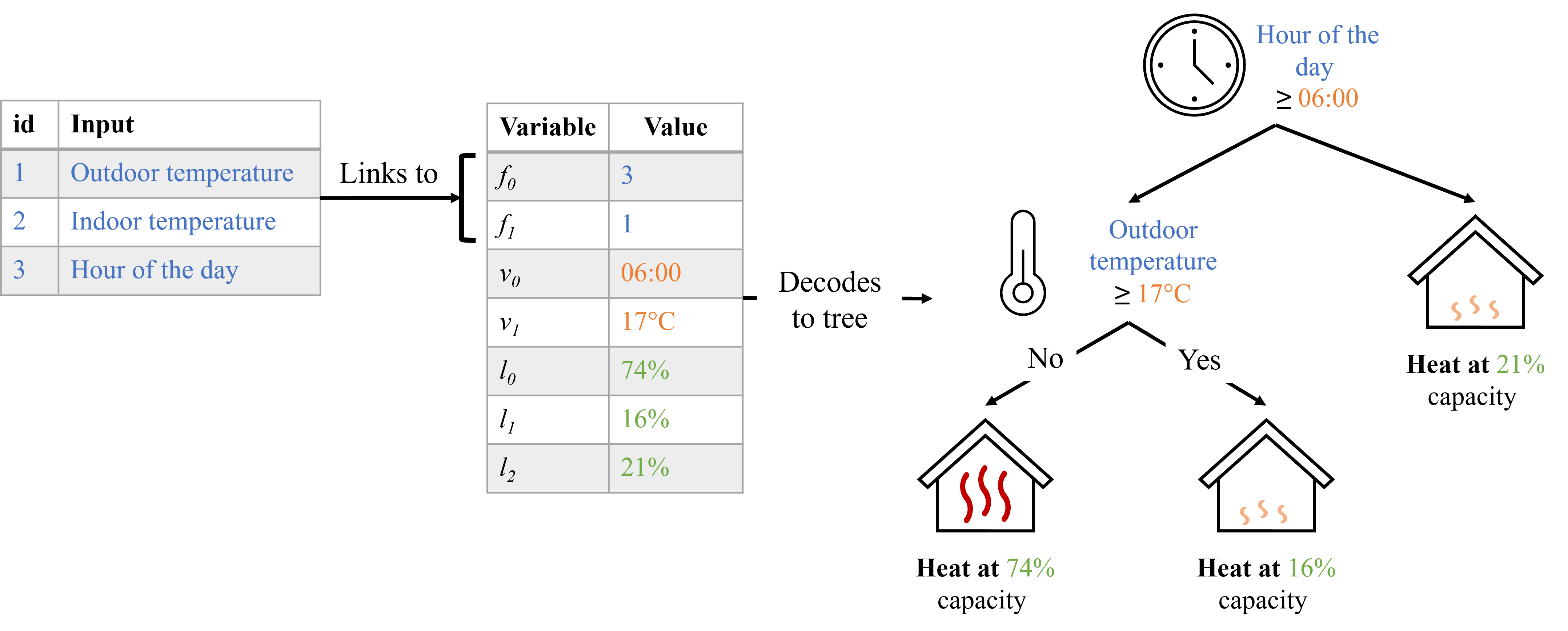}
        \caption{Encoding}
        \label{fig:encoding}
    \end{subfigure}
    \hfill
    \begin{subfigure}[b]{\textwidth}
        \centering
        \includegraphics[width=\widthfigtreecmeth]{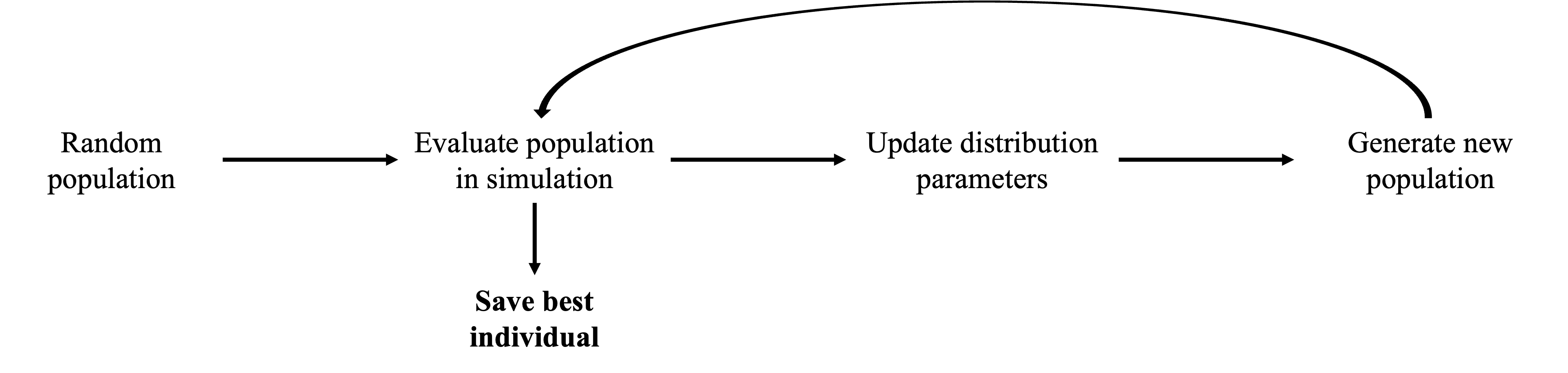}
        \caption{Optimisation}
        \label{fig:optimisation_proc}
       \end{subfigure}
    \hfill
    \begin{subfigure}[b]{0.9\textwidth}
        \centering
        \includegraphics[width=\widthfigtreecmeth]{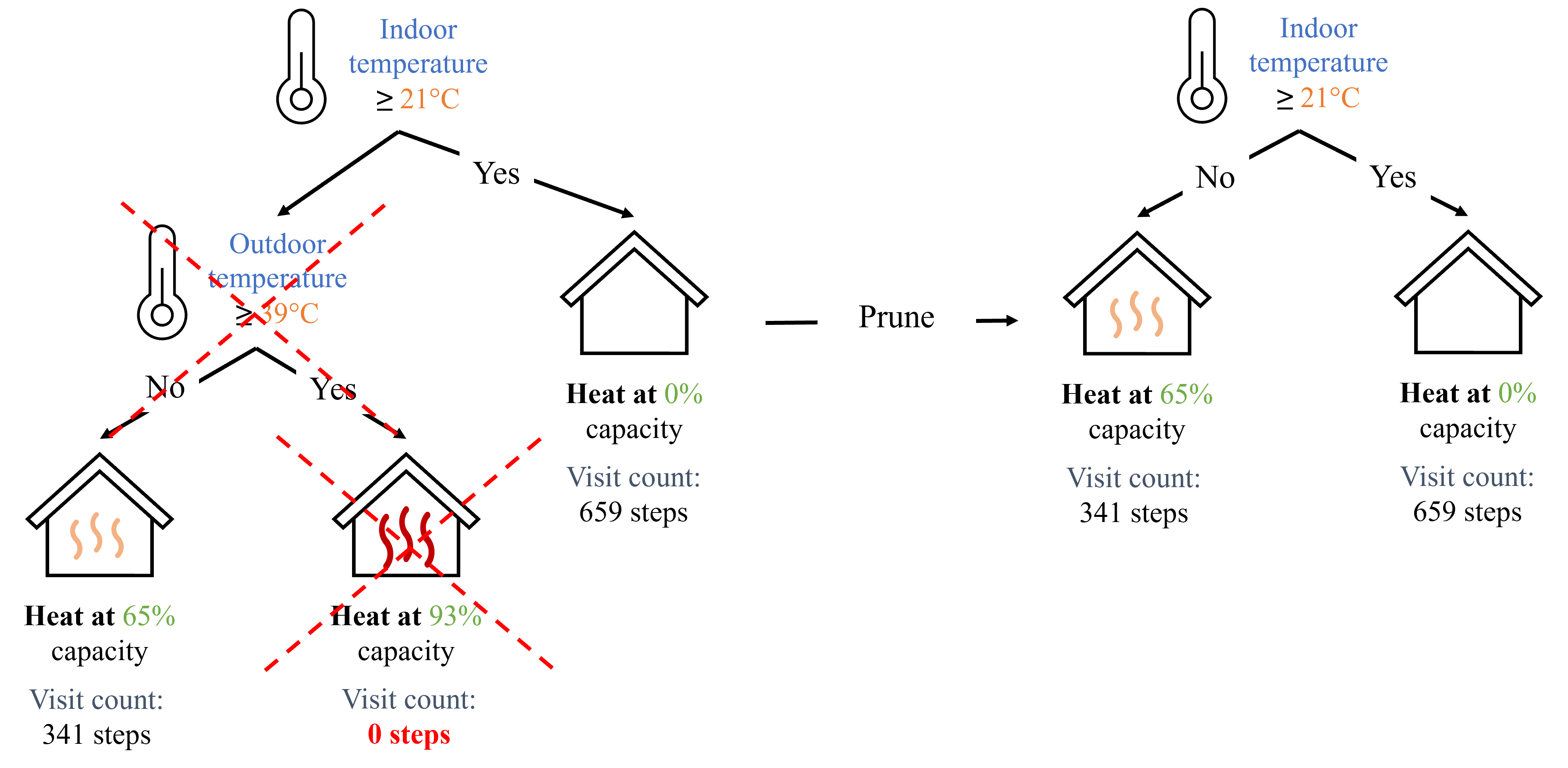}
        \caption{Pruning}
        \label{fig:pruning}
    \end{subfigure}
    \caption{Overview of the TreeC method for a household heating case. 
    \Cref{fig:encoding} illustrates how a decision tree energy management system (EMS) is encoded as a list of numerical variables, where each variable corresponds to a feature, split value or leaf action in the tree structure. 
    \Cref{fig:optimisation_proc} depicts the optimization process using the covariance matrix adaptation evolution strategy (CMA-ES), where multiple candidate trees are generated, evaluated, and iteratively improved based on their objective values. 
    \Cref{fig:pruning} demonstrates the pruning procedure, in which unused leaves and their parent nodes are removed from the tree to simplify the model and retain only the parts of the tree that are actually used during training.}
    \label{fig:method_viz}
\end{figure}

The decision tree used is a complete binary tree. 
The tree is encoded on $3N+1$ variables with $N$ number of parent or split nodes and $N+1$ number of leafs. 
The first $2N$ variables represent the features ($f_0, f_1,...,f_N$) and the values ($v_0,v_1,...,v_N$) of the node split. 
The remaining $N+1$ variables ($l_0,l_1,...,l_{N+1}$) are the values that represent which actions should be taken on each leaf as illustrated in \cref{fig:encoding}. 
\label{ch:var_explanation}
The $3N+1$ variables defining the tree EMS are the variables of the energy management problem and are modified to optimize the performance of the EMS on this problem.
This encoding approach is commonly used in many applications with different variations as shown in the review by \citet{barrosSurveyEvolutionaryAlgorithms2012}.

Note that the number of parent nodes $N$ is an important hyperparameter that influences the complexity of the behaviour and how interpretable the decision tree is. 
In this method, the $N$ value is set to 20 as this allows a relatively complex behaviour while keeping the tree interpretable after pruning.

The proposed method uses the covariance matrix adaptation evolution strategy (CMA-ES) \cite{hansenReducingTimeComplexity2003}, a state-of-the-art metaheuristic algorithm. 
CMA-ES was chosen due to its high performance in different optimization problems \cite{hansenComparingResults312010} and needing almost no user-defined hyperparameters. 
In CMA-ES, an \textbf{individual} is one possible solution to the problem, in this case a specific decision tree with certain parameter values. A \textbf{population} is a group of these individuals. In each round of optimization, also called generations, the algorithm creates and tests a population of different individuals, compares how well they perform using the objective function, and then uses the best ones to guide the search for even better solutions in the next round.
This process is repeated multiple times in order to obtain individuals with a better objective value over the generations. 
The final selected EMS is the individual that obtained the best objective value over the whole optimization process  (see \cref{fig:optimisation_proc}). 

The default CMA-ES population size of $4+3\lfloor ln(n)\rfloor $ is used here with $n$ being the number of variables defining the decision tree \cite{hansenPrincipledDesignContinuous2014}. 

It is important that the objective value obtained by each individual are comparable meaning that the evaluation conditions should be exactly the same (i.e. same start time, evaluation period, ...). 

Finally, the decision trees generated have leafs that are never reached during the training phase and therefore should be pruned. The pruning methods checks if each leaf has been reached and if not removes the leaf and replaces the parent node by its other child node. 
This process is repeated until all remaining leaves have been visited at least once (see \cref{fig:pruning}).

\section{Parallel training and multi-threading properties}
\label{sec:par_train}

Metaheuristics algorithms are good to solve complex problems but are non-deterministic so two separate trainings will not find the same solution. To obtain a better solution, five different trainings are executed with the same problem conditions but different initial population for the CMA-ES algorithm. The tree with the best score in training is then evaluated to obtain the results. Doing this also allows observing if the trees stemming from different trainings have recognizable similar behaviours or which are the most used features. 

This method can be executed using many parallel threads. 
The CMA-ES algorithm allows the evaluation of the individuals within a population to be done in parallel. 
This means the maximum number of possible threads is equal to the number of individuals within a population used for the training multiplied by the number of different parallel training (five in this case).

\begin{figure}[H]
	\centering
	\includegraphics[width=1.0\linewidth]{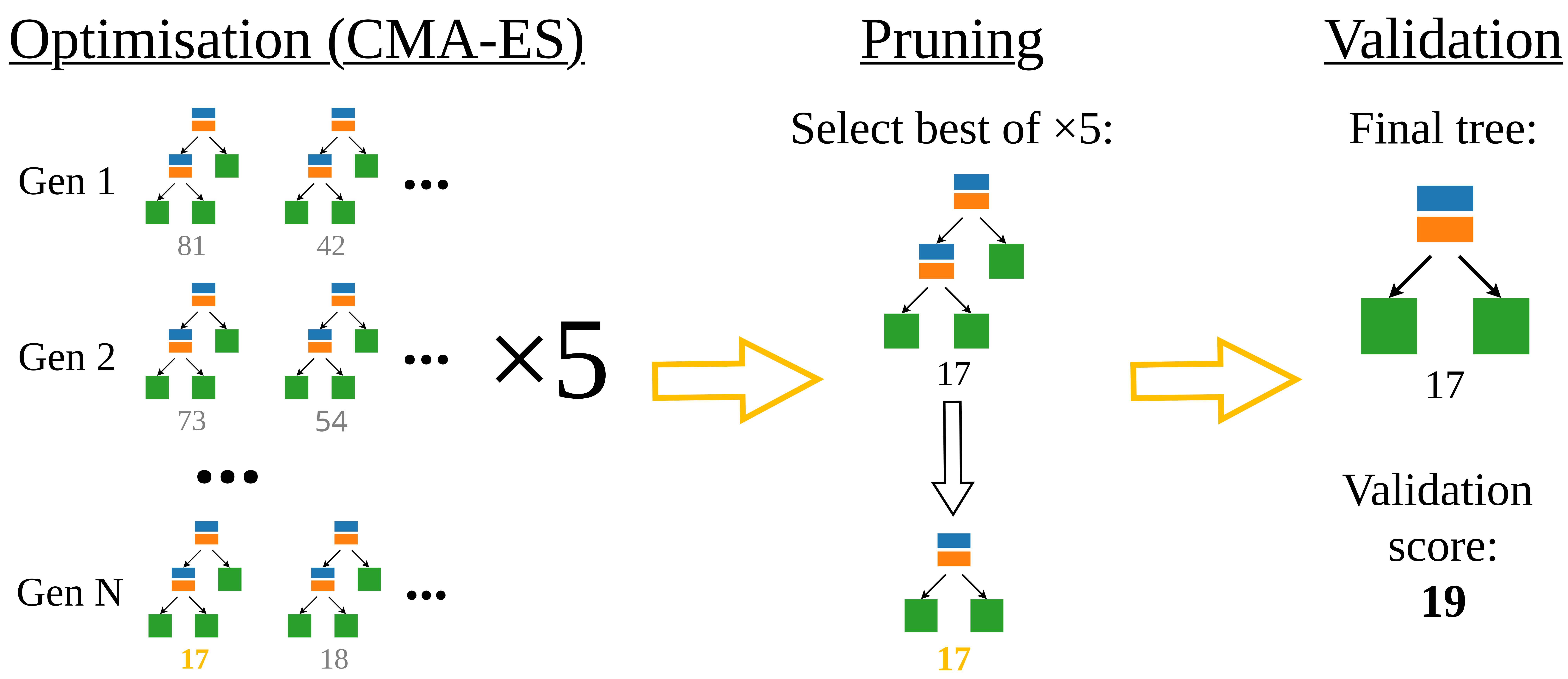}
	\caption{Summary of the optimization/training, pruning and validation method. 
	The decision trees represent different evaluated candidates with the score they obtained beneath them and uses the same colour code for variables as in \cref{fig:method_viz}.
	The decision trees with yellow scores are the best performing ones and taken to the next step.
	The optimization is executed 5 times in parallel as described in \cref{sec:par_train}.}
	\label{fig:summary_method}
\end{figure}

\Cref{fig:summary_method} shows a summary of the training, pruning and validation process including the five parallel training described in this section.

The final obtained decision tree model can be categorized as rule-based control but since it is generated by training on data, the method as a whole is a machine learning method.

The TreeC method is the main algorithmic contribution of this thesis and fits into holistic learning, the new machine learning paradigm proposed in \cref{ch:theoretical_foundations}.

\section{Enhancements and extensions}

The simulation cases in \cref{ch:simul_cases} uses the base TreeC implementation as described above. However, \cref{ch:summary_cases,ch:ems_relevance} improve on this base implementation in two ways:

\begin{itemize}
\item The CMA-ES metaheuristic algorithm frequently converged to local minima during training runs.
To address this limitation, CMA-ES was replaced with particle swarm optimization \cite{kennedyParticleSwarmOptimization1995}, another metaheuristic algorithm, which achieved better objective values during training.
This thesis used the canonical particle swarm optimization algorithm implementation described in \citet{poliParticleSwarmOptimization2007}. 
\item The original pruning method only removed unused leaves, which improved interpretability but kept leaves with minimal impact on EMS performance, making the decision tree unnecessarily complex. To address this, a more aggressive pruning approach was introduced: after training, the best decision tree is further pruned by removing any leaf whose absence does not worsen the objective value obtained after training by more than a small threshold (1\% for \cref{ch:summary_cases} and 5 \euro{} for \cref{ch:ems_relevance}). Leaves are tested for removal in order of least usage. This process eliminates both unused and minimally impactful leaves, resulting in a more interpretable decision tree.
\end{itemize}

\section{Interpretability}

The decision tree structures of TreeC are kept relatively simple by limiting the maximum number of splitting nodes to maximum 20 and by pruning unused and minimally impactful leaves. After pruning, the number of splitting nodes is generally much lower than 20, the different decision trees EMSs showcased in this thesis have between 0 and 4 splitting nodes (see \cref{fig:anm_trees,fig:boptest_tree,fig:house1_decision_trees,fig:house2_decision_trees,fig:house3_decision_trees,fig:house4_decision_trees,fig:complete_dwelling}).
This simplicity allows the decision trees to be easily visualized and understood by users, as they can follow the branches of the tree to see how different input features lead to specific actions.

\section{Conclusion}

TreeC is a novel machine learning method for energy management that uses metaheuristic optimization to generate interpretable decision tree-based control strategies. It encodes decision trees as numerical variables, optimizes them using CMA-ES, and prunes unused branches to simplify the model. TreeC can be trained in parallel, allowing multiple independent trainings to improve training speed. Enhancements such as replacing CMA-ES with particle swarm optimization and implementing more aggressive pruning further improve performance and interpretability in later chapters.

Overall, TreeC is a machine learning-based control algorithm for energy management that uses interpretable control models, is capable of improving its performance by training on data, and is versatile in its application.

\mode<all> %

\chapter[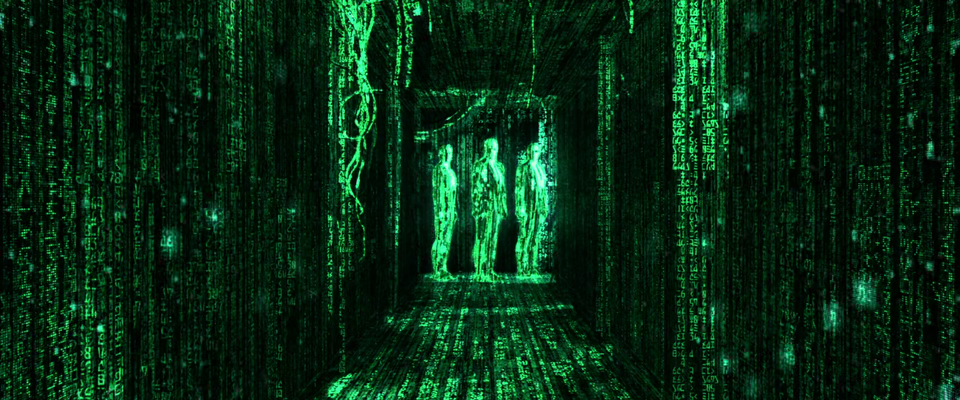]{Simulation cases\label{ch:simul_cases}}

This chapter presents three simulation cases—ANM6-Easy, BOPTEST, and CityLearn 2022, selected to evaluate and compare the developed TreeC method. These cases include one international competition and two benchmarks, covering a range of energy systems: grid management, residential heating, and district coordination. Together, they demonstrate the versatility and effectiveness of the TreeC method across diverse energy management problems.

The cases as well as the TreeC implementation are all open-source allowing other researchers to reproduce the results and compare their methods with the ones presented in this thesis.

The three cases are summarized as follows:

\begin{itemize}
    \item \textbf{Case 1: ANM6-Easy} The \textit{Gym-ANM ANM6-Easy} benchmark requires controlling large photovoltaic, wind, and battery systems to increase energy efficiency and not overload the local electrical grid.
    The evaluated EMSs are TreeC, model predictive control(MPC) and reinforcement learning (RL).
    \item \textbf{Case 2: BOPTEST} The \textit{BOPTEST hydronic heat pump} benchmark requires controlling a hydronic heat pump system for a home to minimize energy costs and thermal discomfort.
    The evaluated EMSs are TreeC, MPC, RL, and rule-based control (RBC).
    \item \textbf{Case 3: Citylearn 2022} The \textit{CityLearn Challenge 2022} is a competition on energy management in a district of 5 buildings with solar panels and batteries where the EMS should coordinate the batteries to minimize the electricity cost, CO\textsubscript{2} emissions, and grid power variability.
    TreeC is evaluated against stochastic and deterministic MPC, hybrid RL, holistic learning, and supervised learning approaches.
\end{itemize}

Cases 1 and 2 are based on \citet{ruddickTreeCMethodGenerate2025} and the section for case 3 is not based on a publication.

\section{Case 1: ANM6-Easy}
\label{sec:ANM6-Easy}

\subsection{Case description}

\begin{figure}[H]
\centerline{\includegraphics[width=0.8\linewidth]{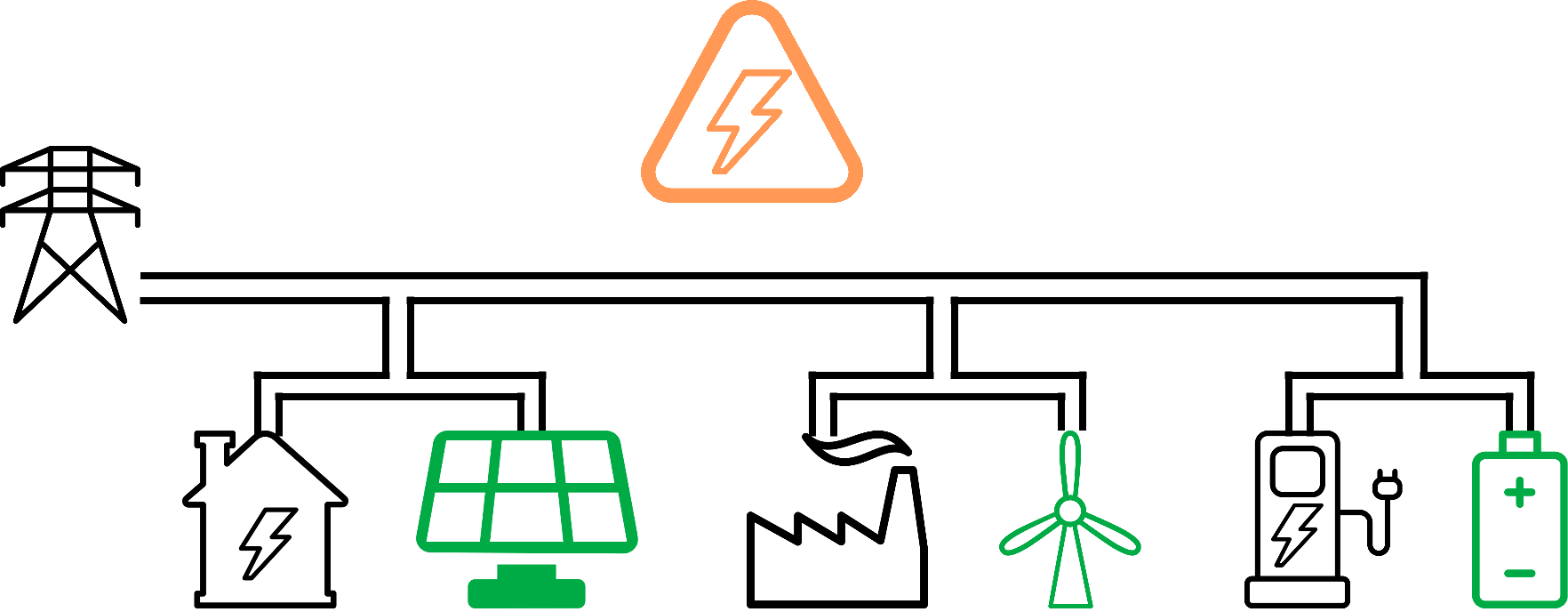}}
\caption{Visualization of the ANM6-Easy electrical grid benchmark. The controlled photovoltaic plant, wind-turbine farm and large electrical storage unit are shown in green. The optimized metric indicated in orange evaluates grid safety and energy loss.
}
\label{fig:anm6_scheme}
\end{figure}

\begin{figure}[H]
\centering
\includegraphics[width=1.0\linewidth]{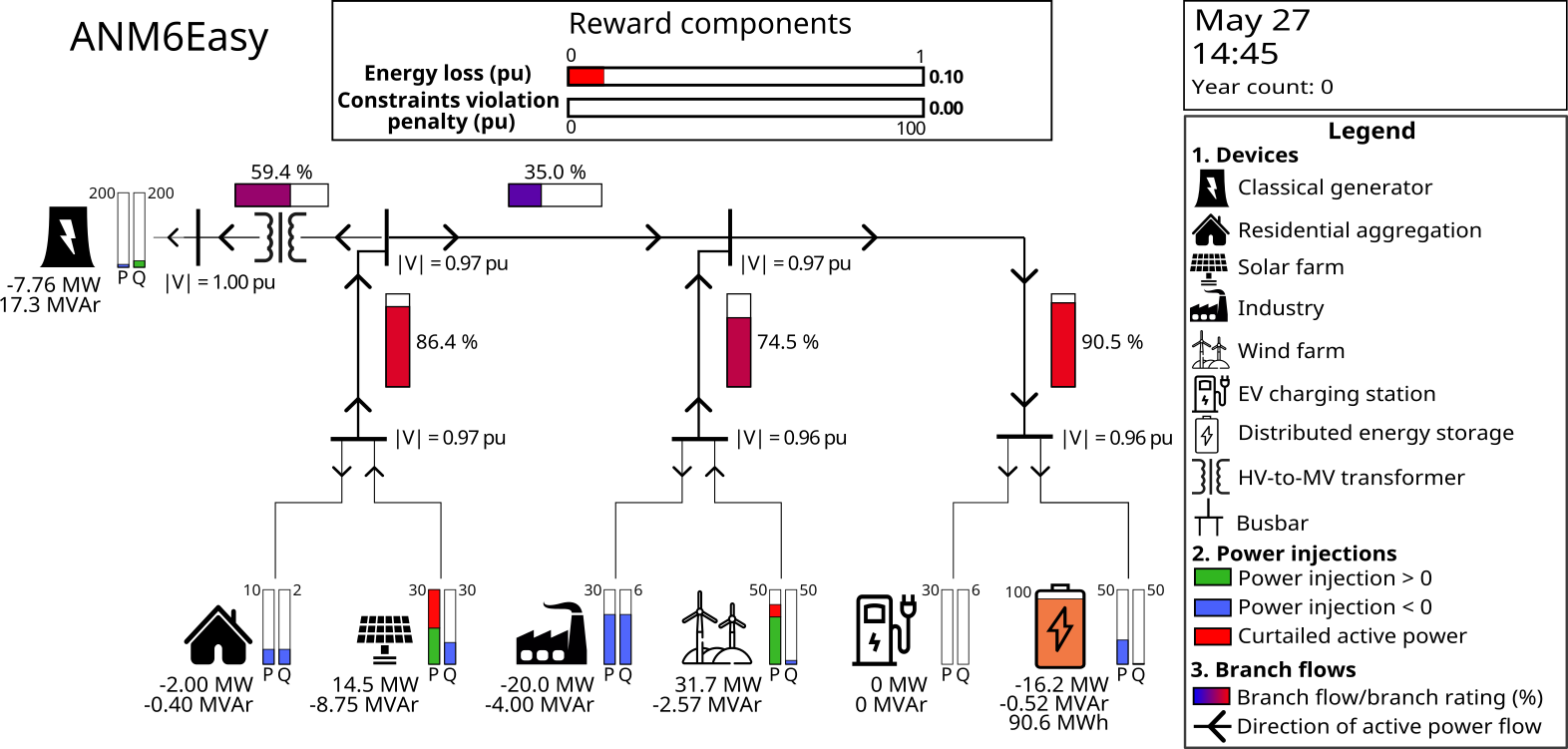}
\caption{ANM6-Easy case visualization for a typical working hour period using the EMS presented in \cref{fig:anm_trees}. All the assets are shown with their respective real and reactive power as well as the constraints on the grid lines and voltage magnitude of the busbars \cite{henryGymANMOpensourceSoftware2021}. }
\label{fig:ANM case}
\end{figure}

The second case study is ANM6-Easy, a test case proposed by \citet{henryGymANMReinforcementLearning2021} and provided in the gym-anm simulator. 
The version 1.0.2 of gym-anm was used in this work. ANM6-Easy simulates an electrical grid as shown in \cref{fig:ANM case}. 
The grid has three passive loads: a residential area (average power: 3 MW), an industrial complex (average power: 10 MW),  and an electric vehicle charging garage (average power: 10 MW). The controllable assets of the grid are a photovoltaic (PV) plant (average power: 8 MW), a wind-turbine farm (average power: 28 MW) and one large electrical storage unit (storage capacity: 100 MWh). A slack fossil fuel generator ensures that the grid is always supplied with enough power (peak power: 200 MW).

In this benchmark, the EMS has direct control over the renewable energy sources and the large electrical storage unit. Specifically, it determines the real power curtailment and reactive power set-points for both the photovoltaic and wind-turbine plants, as well as the real and reactive power set-points for the storage unit. These control actions are computed every 15 minutes, with each action affecting the subsequent time step.

The simulation environment is structured as a repeating daily cycle, characterized by three distinct periods that reflect typical variations in demand and renewable generation throughout the day.

\begin{itemize}
    \item From 23:00 to 06:00: Windy night with very low demand from the three passive loads.
    \item From 08:00 to 11:00 and from 18:00 to 21:00: Typical arrival time at work or home. Very high demand from the electric vehicle charging garage due to people plugging in their vehicles to charge. A low amount of power is produced by both renewable asset plants. The demand from the residential area is high while for the industrial complex it is moderate.
    \item From 13:00 to 16:00: Typical working hours. The demand for the industrial area is high while the residential area demand is low. Both renewable asset plants produce a lot of power. There is no power demand from the electric vehicle charging garage. 
\end{itemize}

During the hours not included in the three characteristic periods, the non-controllable powers shift linearly from the powers of the preceding period to the powers of the next period.

The electrical grid can collapse if mismanaged, most often this occurs due to a voltage collapse problem \cite{chiangVoltageCollapseElectric1990}.

The objective of the EMS in this test case is to control the real and reactive powers of the renewable asset plants and large electrical storage unit in order to: (1) avoid exceeding the maximum power allowed on the branches, (2) maintain the voltage within nominal range with a tolerance of $\pm 5\%$ and (3) minimize energy loss. 

The performance metric for this case depends on the reward $r_t$ calculated using Eq. (\ref{eq:reward_ANM}):

\begin{equation}
	r_{t}=clip(-100,\Delta E_{t:t+1}+1000\phi (s_{t+1}),100)
	\label{eq:reward_ANM}
\end{equation}

where:

\begin{tabular}{ll}
    $\Delta E_{t:t+1}$ [pu] & total energy loss between time-step $t$ and $t+1$ \\
    $\phi (s_{t+1})$ [pu] & penalty term for violating the constraints of the grid \\
\end{tabular}

The penalty term $\phi (s_{t+1})$ increases when the maximum power of a branch is exceeded and when a bus voltage magnitude is not kept within the 0.95 to 1.05 per unit range. 
Both $\Delta E_{t:t+1}$ and $\phi (s_{t+1})$ are expressed in per unit with the base power of the system set to 100 MVA. When the electrical grid collapses, $r_t$ is set to 20\,000.

The energy loss and constraint violation score $S$ is the same as defined in the original paper \cite{henryGymANMReinforcementLearning2021}. 
It is calculated over 3\,000 time-steps and uses an additional discount factor $0.995^t$ (see \cref{eq:obj_valid_anm}).

\begin{equation}
	S=\sum_{t}^{3000}r_t*0.995^t
\label{eq:obj_valid_anm}
\end{equation}

\subsection{EMS methods}

\subsubsection{TreeC training and validation}

During the training, six trees are evolved simultaneously. 
A different tree is used for the reactive power and real power of each of the three controllable assets. 
There are 18 possible split features to select from. 
These features are: the real and reactive power of each asset \label{ch:cascading_comment} from the previous time-step (14 features), the real power of the renewable asset plants from the previous time-step if they were not curtailed (two features), the current state of charge of the large electrical storage unit and the current time of the day. 
The values of the leafs sets the power for the next time-step. 

For the results, 20 sets of six trees are generated using the method described in \cref{sec:treec}. 
The problem has 366 variables, a population of 21 is used and the training was done over 1500 generations. 
During training, the population's individual are evaluated over a period of 300 time-steps (~3 days). 
The starting time of each training is randomly set but stays the same for all evaluations of the same training to ensure that the comparison between individuals is fair.

The objective function of the training and validation both use the reward $r_t$ provided by the simulator. 
\Cref{eq:reward_ANM} shows how $r_t$ is calculated.

The objective function for training $O_{train}$ is a sum of the rewards for 300 time-steps (see \cref{eq:obj_train_anm}). 

\begin{equation}
O_{train}=\sum_{t}^{300}r_t
\label{eq:obj_train_anm}
\end{equation}

The discount factor used in the energy loss and constraint score $S$ (see \cref{eq:obj_valid_anm}) was removed for the training objective function $O_{train}$ because it favours earlier time-steps of the simulation.
Removing the discount factor makes the reward of all time-steps equally important.

\subsubsection{Comparison reinforcement learning, model predictive control and random}

The validation results show for each generated EMS the average energy loss and constraint score $S$ (see \cref{eq:obj_valid_anm}) calculated over 10 different random start times (seeds 0 to 9 in gym-anm simulator). 

The original paper provides benchmark results and code implementations for two MPC variants (with perfect and constant forecasts), two reinforcement learning methods (Proximal Policy Optimization (PPO) and Soft Actor-Critic (SAC)) and a random EMS that samples actions from the allowed range for each controllable asset \cite{henryGymANMReinforcementLearning2021}. For each of these benchmark EMSs, the average constraint score $S$ over the 10 different random start times is computed using the EMS implementations from the original paper.-

\subsection{Results and discussion}

The training for the results was obtained on a high performance computer composed of 880 Intel\textregistered Xeon\textregistered Gold 6148 processors

\begin{figure}[H]
\centering
\includegraphics[width=0.75\linewidth]{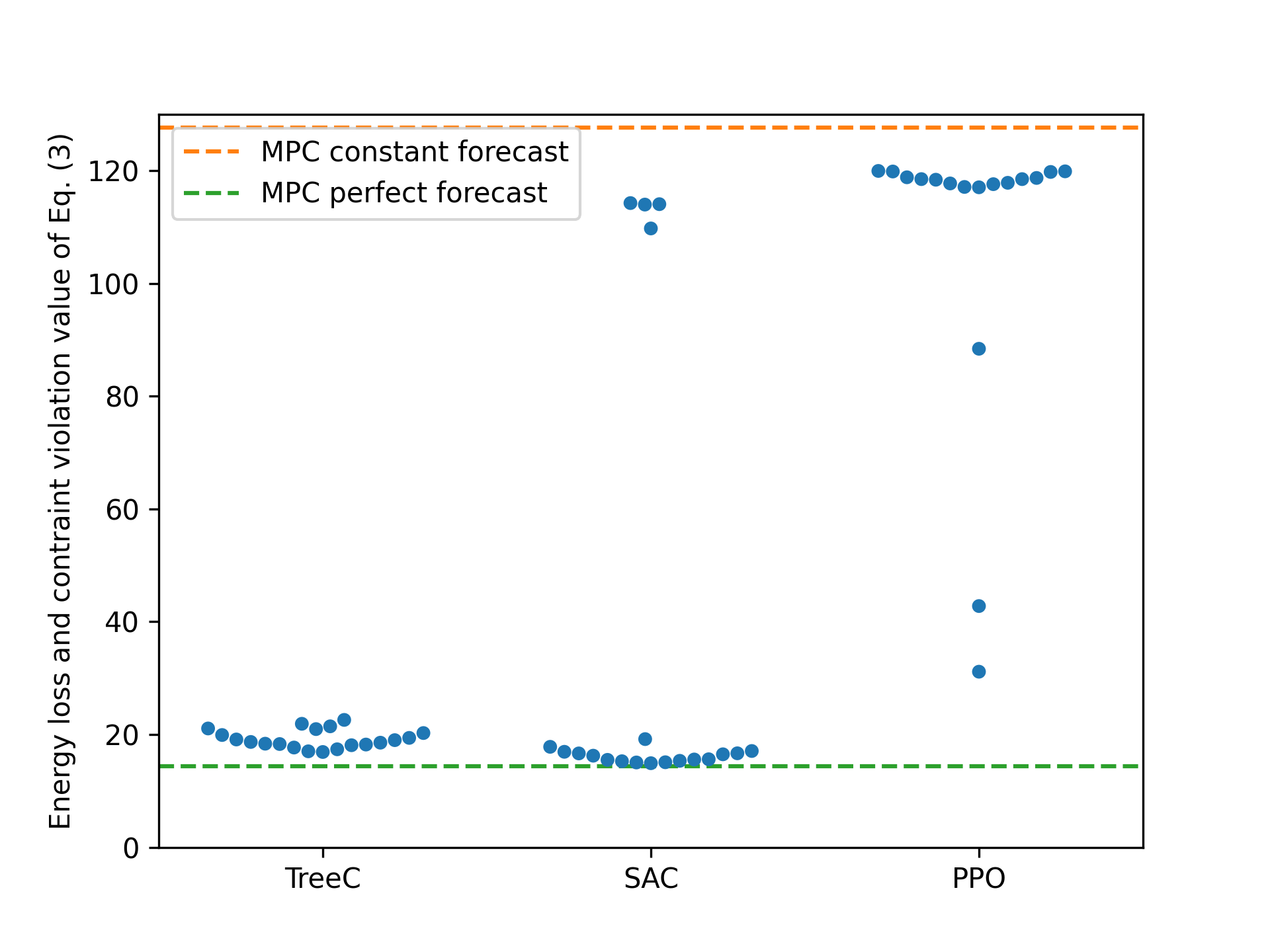}
\caption{Swarm plot representation of the results obtained for the ANM6-Easy case with each point being the validation of independent EMSs and the horizontal lines being the validation of the MPC EMSs. The TreeC and SAC EMSs both obtain results close to the MPC with perfect forecast. Disregarding its outliers, SAC performs slightly better than TreeC. PPO performs worse than the two other methods. Three outliers for PPO with a score higher than 130 are not shown in the plot.}
\label{fig:ANM_all_box}
\end{figure}

In this case study, running the evaluation of one individual simulation of 300 time-steps lasted ~10 seconds. 
The time to update the CMA-ES optimization model is negligible here. 
The trees generated for the results shown in \cref{fig:ANM_all_box} all ran for  157\,500 evaluation which amounts to ~18 days 5 hours and 30 minutes of training using a single thread and to ~4 hours and 10 minutes using the maximum number of 105 threads (population size of 21 $\times$ 5 parallel training).

In this case, the TreeC EMS perform much better than PPO and MPC with constant forecast methods and slightly worse than the MPC with perfect forecast and SAC method (see \cref{fig:ANM_all_box}).
The average score of SAC method is worse than TreeC due to outliers but it would be unfair to claim that TreeC performs better since it selects the best tree form five trainings to avoid such outliers. 
Implementing such a mechanism in the SAC method would improve its average score greatly. 

All methods perform much better than the random EMS that samples random actions from the possible action range of each controllable asset. 
This random EMS obtains an average score of 12\,875 over 200 runs. 
The electrical grid collapsed in each of the 200 runs after 181 steps on average.
All EMS methods presented in \cref{fig:ANM_all_box} avoid collapses consistently. 

\begin{table}[H]
	\centering
	\caption{Average and standard deviation of the EMSs scores for the ANM6easy case.}

\begin{tabular}{|c|c|}
	\hline	
	EMS & Energy loss and constraint violation score  \\ \hline
	MPC perfect forecast & \textbf{14.4 $\pm$ 0.0}\\ \hline
	SAC no outlier & 16.2 $\pm$ 1.2\\ \hline
	TreeC & 19.2 $\pm$ 1.7\\ \hline
	SAC & 35.6 $\pm$ 39.7 \\ \hline
	PPO & 120.9 $\pm$ 44.9  \\ \hline
	MPC constant forecast & 127.7 $\pm$ 0.0 \\ \hline
	Random EMS & 12\,875 $\pm$ 1308.8\\ \hline

	\end{tabular}
	\label{tab:ANM_table_results}
	
\end{table}

\Cref{tab:ANM_table_results} provides the average score obtained by each EMS method from best to worst.
The table also includes the performance of SAC without the four outliers that obtained a score higher than 100. 
\Cref{tab:comp} uses the average SAC score without outliers as the reference score for RL on the ANM6easy case.
We believe this is a more fair comparison because the TreeC method selects the best tree from five trainings to avoid such outliers.
Such a mechanism was not implemented in the SAC method to keep the exact same implementation as the original paper \cite{henryGymANMReinforcementLearning2021}.

\newcommand{\anmtreesfig}{
\begin{figure}[H]
    \centering
    \begin{subfigure}[b]{0.45\textwidth}
        \centering
        \includegraphics[width=\textwidth]{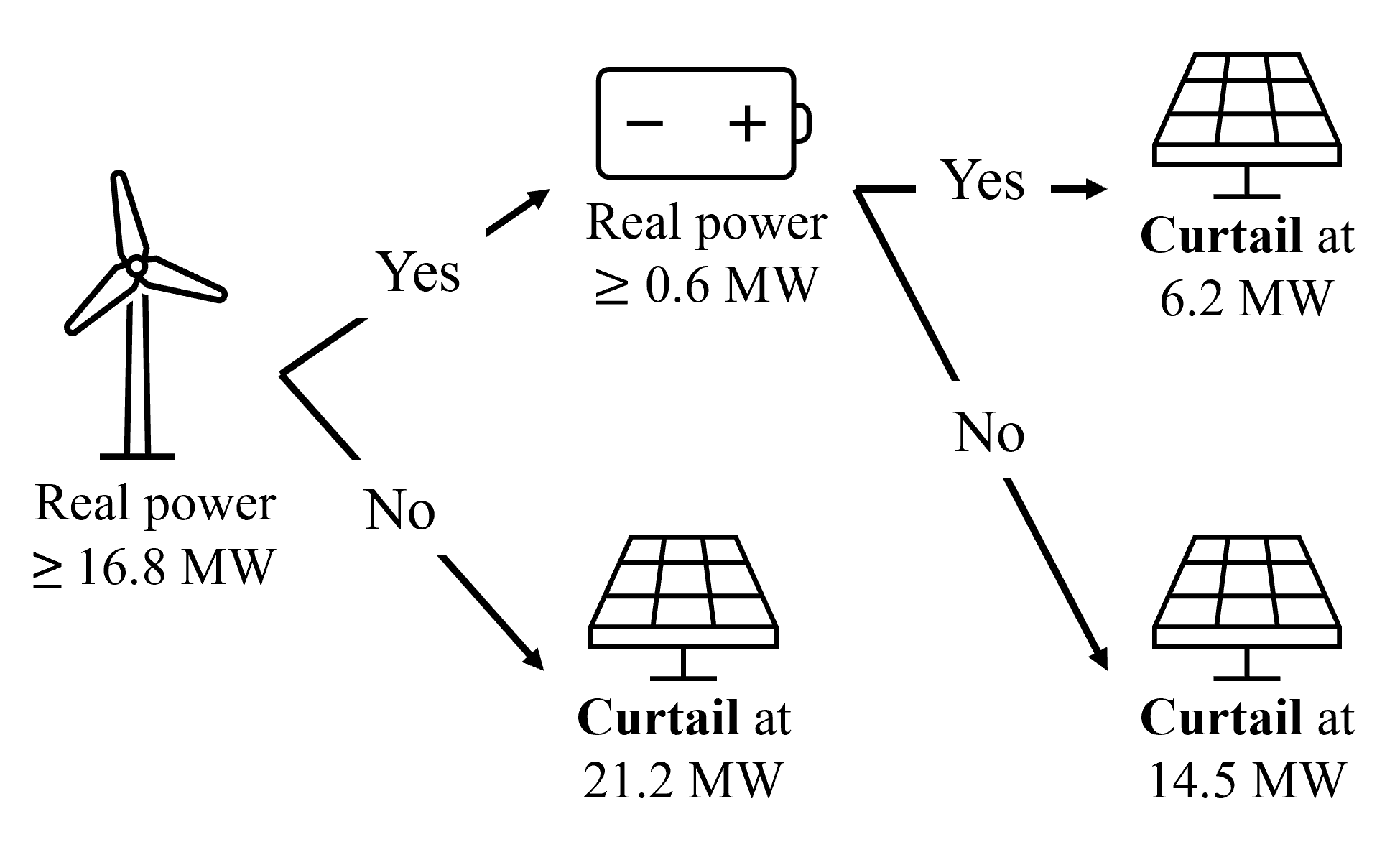}
        \caption{Curtailment photovoltaics}
        \label{fig:curt_pv}
    \end{subfigure}
    \hfill
    \begin{subfigure}[b]{0.45\textwidth}
        \centering
        \includegraphics[width=\textwidth]{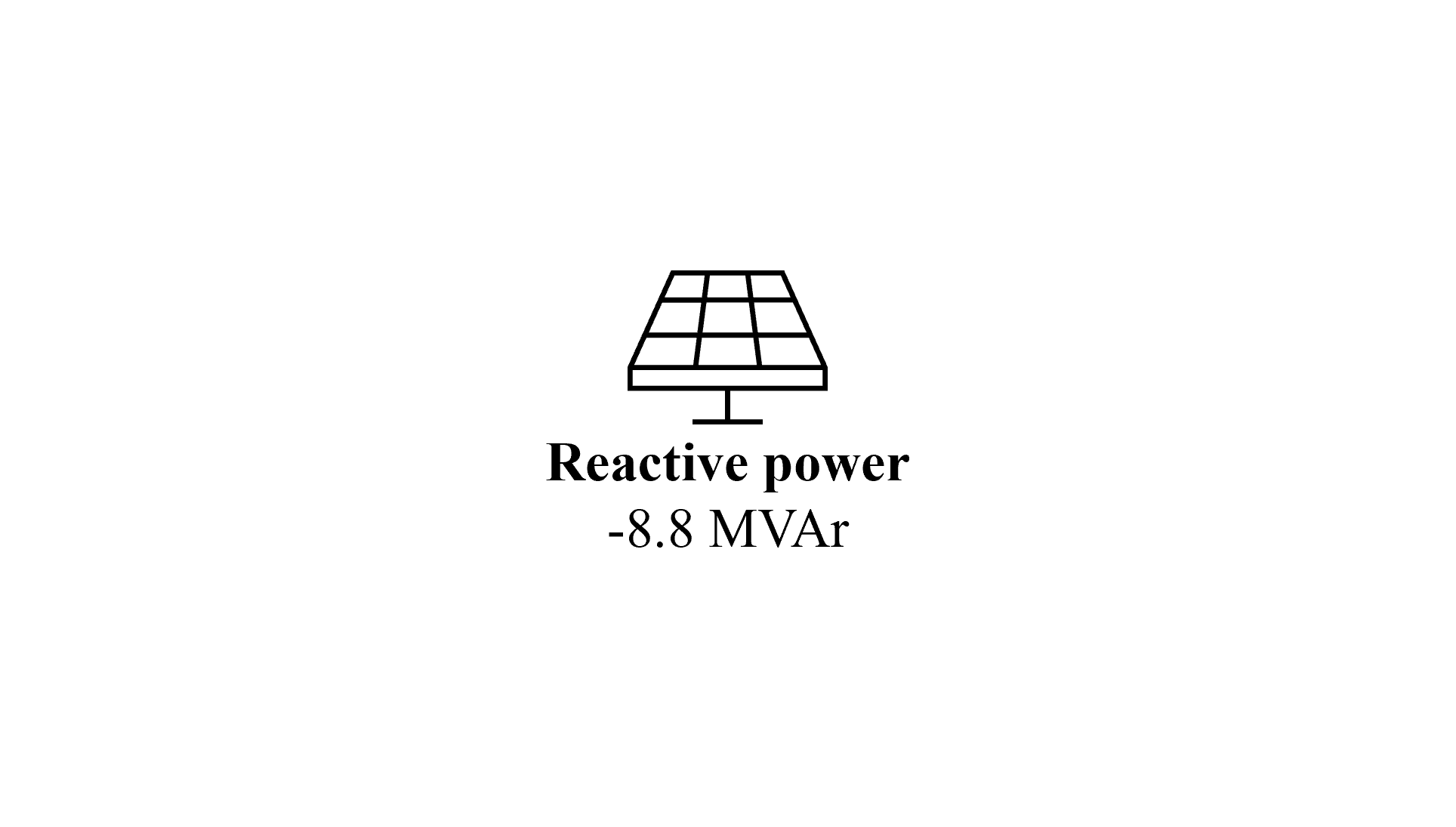}
        \caption{Reactive power photovoltaics}
        \label{fig:reac_pv}
       \end{subfigure}
    \hfill
    \begin{subfigure}[b]{0.45\textwidth}
        \centering
        \includegraphics[width=\textwidth]{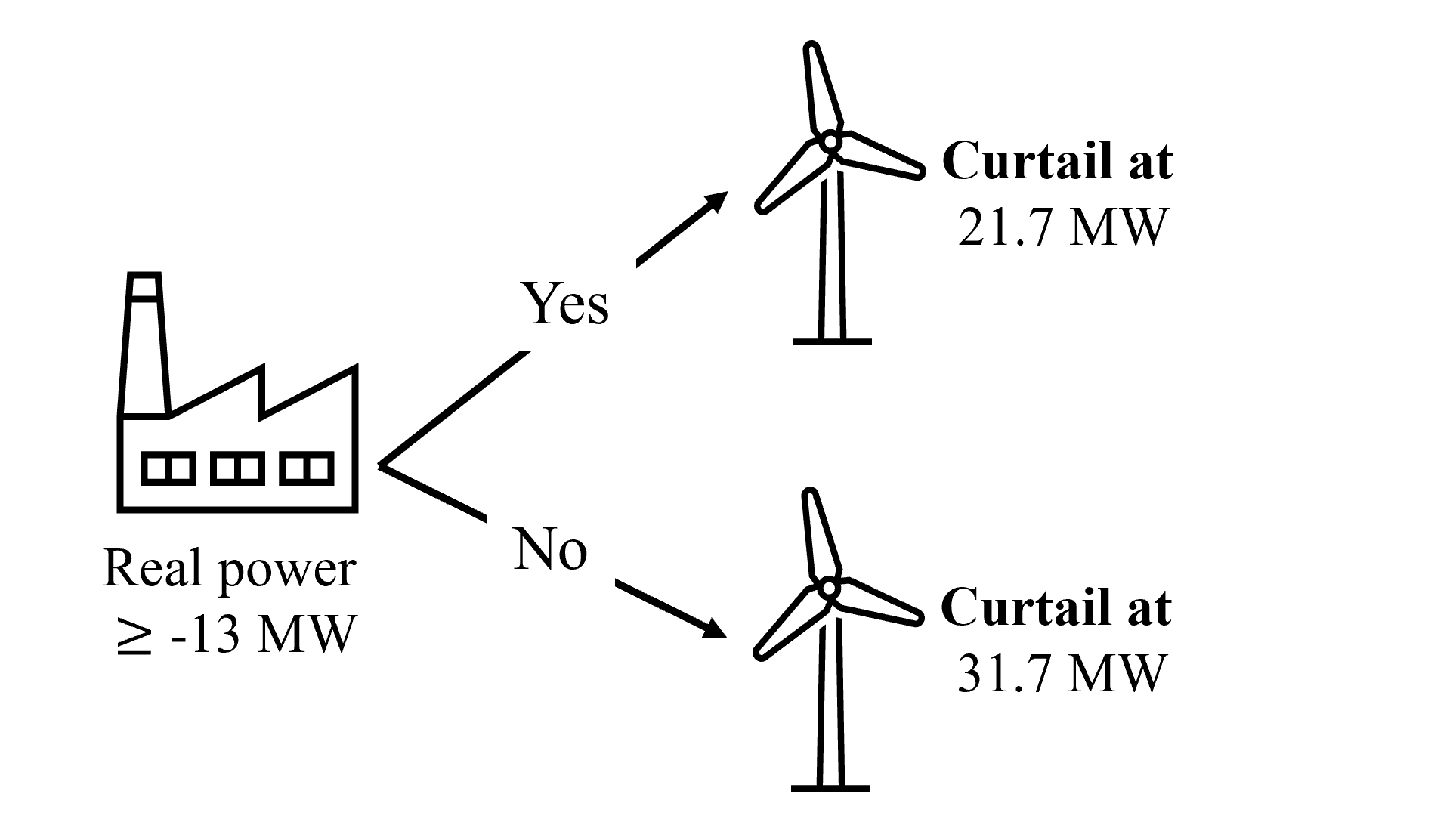}
        \caption{Curtailment wind-turbines}
        \label{fig:curt_wind}
    \end{subfigure}
    \hfill
    \begin{subfigure}[b]{0.45\textwidth}
        \centering
        \includegraphics[width=\textwidth]{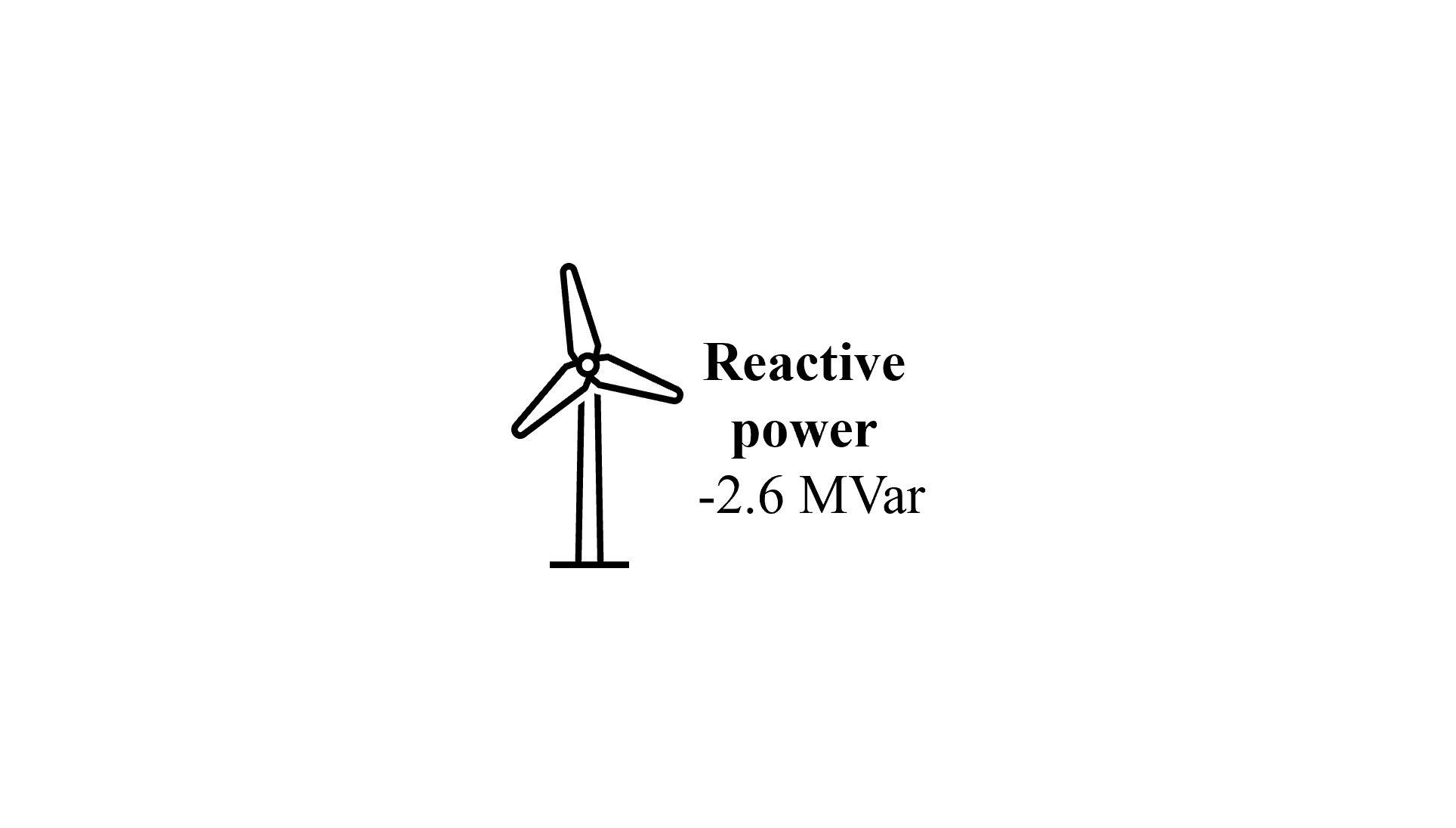}
        \caption{Reactive power wind-turbines}
        \label{fig:reac_wind}
    \end{subfigure}
    \hfill
    \begin{subfigure}[b]{0.45\textwidth}
        \centering
        \includegraphics[width=\textwidth]{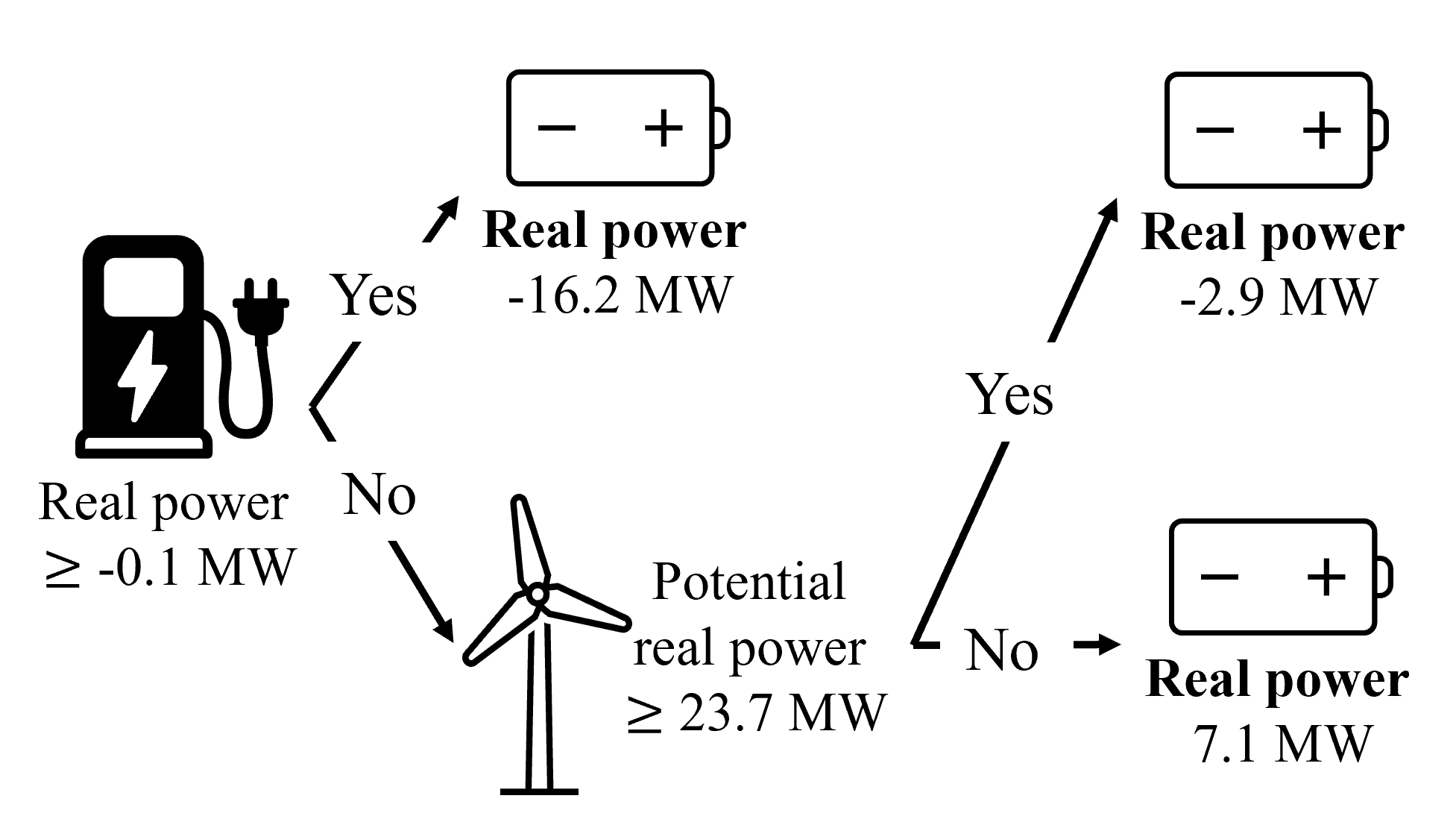}
        \caption{Real power large electrical storage}
        \label{fig:real_batt}
    \end{subfigure}
    \hfill
    \begin{subfigure}[b]{0.4\textwidth}
        \centering
        \includegraphics[width=\textwidth]{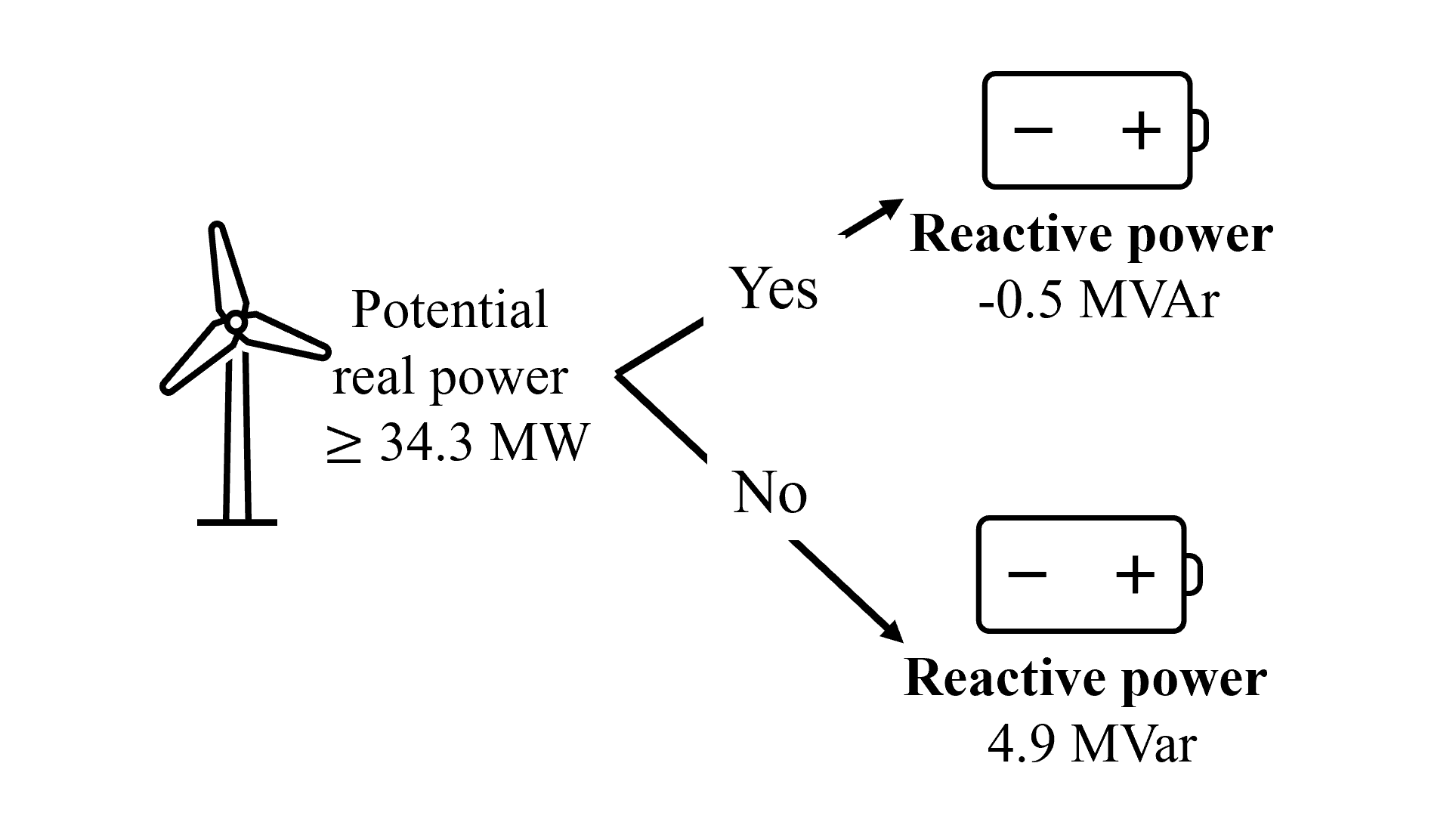}
        \caption{Reactive power large electrical storage}
        \label{fig:reac_batt}
    \end{subfigure}
    
       \caption{TreeC EMS which obtained the best score in \cref{fig:ANM_all_box}. Each tree controls the real power or reactive power of each controllable asset. To keep them simple for visualization, nodes that were visited 10 times or less during evaluation were removed.}
       \label{fig:anm_trees}
\end{figure}
}

The trees that obtained the best performance shown in \cref{fig:anm_trees} provide a simple and interpretable EMS shown in \cref{fig:anm_trees} that obtains performances only slightly worse than an MPC based EMS with perfect forecast. 
The tree in \cref{fig:curt_pv} curtails PV at 21.2 MW when the power of the wind-turbines is low and otherwise curtails at 14.5 MW when the large electrical storage charges and at 6.2 MW when the large electrical storage discharges. 
The tree in \cref{fig:reac_pv} keeps the reactive power of the PV at -8.8 MVAr. The tree in \cref{fig:curt_wind} curtails wind-turbines at 31.7 MW when the industry load exceeds 13 MW and curtails at 21.7 MW otherwise. 
The tree in \cref{fig:reac_wind} keeps the reactive power of the wind-turbines at -2.6 MVAr. 
The tree in \cref{fig:real_batt} charges the large electrical storage at 16.2 MW when there is no demand from the electric vehicle charging garage otherwise charges at 2.9 MW when the wind-turbines can produce more than 23.7 MW and discharges at 7.1 MW when they can produce less. 
The tree in \cref{fig:reac_batt} sets the reactive power of the large electrical storage to -0.5 MVAr when the wind-turbines can produce more than 34.3 MW and to 4.9 MVAr otherwise. 

\ifprintedversion
\anmtreesfig
\fi

The interpretability of the trees gives interesting insight on the problem itself, for example in this case the reactive power is very easy to control efficiently since the trees obtained have a maximum of two different power settings for each asset (see \cref{fig:reac_pv,fig:reac_wind,fig:reac_batt}). 

\ifprintedversion
\else
\anmtreesfig
\fi

\ifprintedversion
\pagebreak
\else
\pagebreak
\fi

\section{Case 2: BOPTEST}
\label{sec:boptest}

\subsection{Case description}

\begin{figure}[H]
    \ifprintedversion
    \centerline{\includegraphics[width=0.67\linewidth]{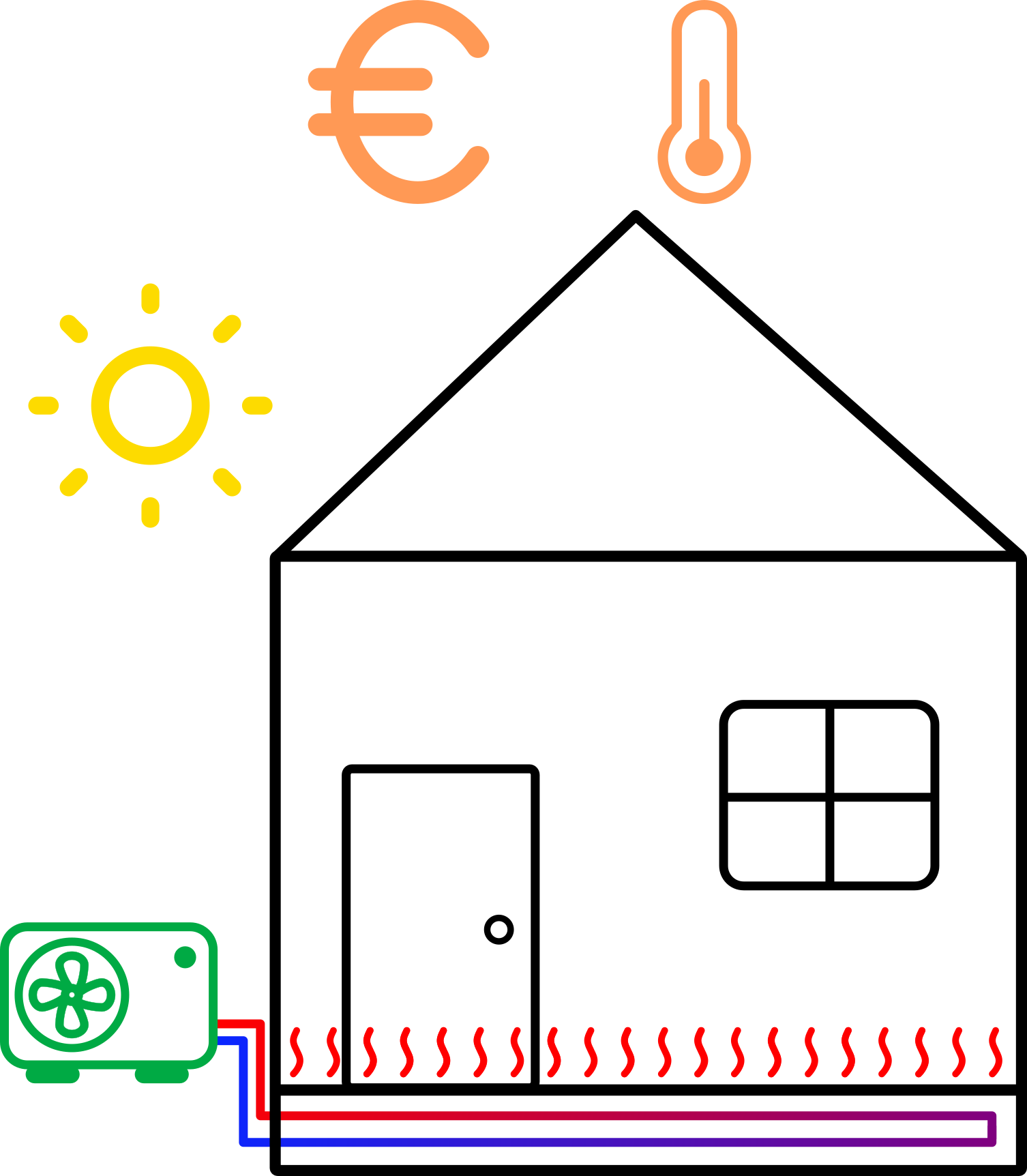}}
    \else
    \centerline{\includegraphics[width=0.8\linewidth]{figures/simulation_cases/boptest_scheme.png}}
    \fi
    \caption{Visualization of the BOPTEST Hydronic Heat Pump case. The EMS controls the air-to-water heat pump, in green, to minimize electricity costs while maintaining indoor comfort, in orange. The simulation considers weather conditions and the thermal properties of the house.}
        \label{fig:boptest_scheme}
\end{figure}
The third case study is BOPTEST Hydronic Heat Pump a test case provided in the BOPTEST simulator \cite{blumPrototypingBOPTESTFramework2019}. 
The version 0.1.0. of BOPTEST was used in this work. 
The case is a heating management problem in a household. 
In particular, this test case consist of a five-person house with a rectangular floor of 12 by 16 meters and a height of 2.7 meters. 
The house is equipped with water based floor heating, the water is heated by an air-to-water modulating heat pump of 15 kW nominal heating capacity. 
The interior temperature is impacted by the outdoor weather conditions. 
The walls, floor, roof and windows have thermal properties described extensively in the documentation of the test case \cite{blumDocumentationBestest_hydronic_heat_pumpCase2021}. 
The simulation uses one year of weather data from Brussels provided by the simulator \cite{blumPrototypingBOPTESTFramework2019}. 

Three different energy pricing scenarios are made available by the simulator:

\begin{itemize}
    \item The constant electricity price: Fixed electricity price of 0.2535 \euro/kWh.
    \item The dynamic electricity price: During on-peak hours between 07:00 and 22:00 the electricity price is  0.2666 \euro/kWh and between 22:00 and 07:00 the electricity price is 0.2383 \euro/kWh.
    \item The highly dynamic price: Hourly dynamic electricity price taken from the Belgian day-ahead energy prices of 2019. The median electricity price is 0.2389 \euro/kWh with a lower quartile price of 0.2317 \euro/kWh and an upper quartile price of 0.2392 \euro/kWh.
\end{itemize}

The simulator also provides two validation periods of 14 days each. 
The first spans from the \nth{17} to the \nth{31} of January and is centred around the peak heat day of the year, the second period spans from the \nth{19} April to the \nth{3} May and is centred around the typical heat day of the year.

The house is considered occupied before 7:00 and after 20:00 every weekday and at all times during the weekend. 
The comfort range is between 21°C and 24°C during occupied hours and between 15°C and 30°C during unoccupied hours. 

The objective of the case is to control the heat pump modulation signal for compressor frequency in order to keep the temperature of the house within the comfort range when occupied and minimize the electricity costs.

The two performance metrics for this case are the total electricity cost and the total discomfort.

The total electricity cost $E$ is calculated using Eq. (\ref{eq:elec_cost_boptest}):

\begin{equation}
E=\sum_{t}^{T}p_t*e_t
\label{eq:elec_cost_boptest}
\end{equation}

where:

\begin{tabular}{ll}
    $T$ & set of simulated time-steps \\
    $p_t$ [\euro/kWh] & price of electricity at time-step $t$ \\
    $e_t$ [kWh] &  \begin{tabular}[c]{@{}l@{}} electrical consumption of the heat pump, evaporator fan and \\ the circulation pump at time-step $t$ \end{tabular} \\
\end{tabular}

The total discomfort $D$ is calculated using Eq. (\ref{eq:discomfort_boptest}):

\begin{equation}
D=\sum_{t}^{T}\delta_t
\label{eq:discomfort_boptest}
\end{equation}

where:

\begin{tabular}{ll}
    $\delta_t$ [Kh] & discomfort metric at time-step $t$ \\
\end{tabular}

$\delta_t$ has two different possible values: (1) when the indoor house temperature is out of the comfort range it equals the absolute difference between the indoor house temperature and the closest comfort temperature range bound, (2) when the indoor house temperature is within the comfort range, $\delta_t=0Kh$.

\subsection{EMS methods}

\subsubsection{TreeC training and validation}
For this case, only one tree is necessary to control the heat pump. 
The tree has the same inputs and outputs as the SS2 RL based EMS presented in \cite{arroyoComparisonOptimalControl2022}. 
There are 5 possible split features which are the price of electricity, the indoor house temperature, the lower and upper comfort range bounds and the time of the week. 
The values of the leafs sets the heat pump modulation signal for compressor frequency for the next time-step. They are discretised to 11 values between 0 and 1 with a step of 0.1.

Two different trees are generated for each pricing scenario and validation period. The problem has 61 variables, a population of 16 is used and the training was done over 150 generations. 
The evaluation of the individuals was done over 14 days starting 15 days before the start of the validation period and with a warm-up period of 1 day performed by the simulator. 
The heat pump is controlled every 15 minutes.

The objective function $O_{train}$ used for training is the same used by \citet{arroyoComparisonOptimalControl2022} to obtain their results also shown in this paper. 
A weighted objective function is used to take into account the discomfort and electricity cost aspects of the case. 
The following \cref{eq:obj_train_boptest} shows how $O_{train}$ is calculated:

\begin{equation}
O_{train}=w_1*D+w_2*E
\label{eq:obj_train_boptest}
\end{equation}

where:

\begin{tabular}{ll}
    $w_1$ [1/Kh] & weight given to the total discomfort $D$ \\
    $\delta_t$ [Kh] & discomfort metric at time-step $t$ \\
    $w_2$ [\euro{}/kWh] & weight given to the total electricity cost $E$ \\
    $p_t$ [\euro/kWh] & price of electricity at time-step $t$ \\
    $e_t$ [kWh] & \begin{tabular}[c]{@{}l@{}}
    electrical consumption of the heat pump, evaporator fan and the\\ circulation pump at time-step $t$ \end{tabular}\\
\end{tabular}

$w_1$ and $w_2$ are set to 100 and 192 respectively to match the weights used in \cite{arroyoComparisonOptimalControl2022}.

\subsubsection{Comparison reinforcement learning, model predictive control and rule-based control}

The MPC, RL, and RBC results shown in \cref{fig:boptest_tree_res} are sourced from \cite{arroyoComparisonOptimalControl2022}. All MPC controllers use perfect forecasts and a grey-box resistance-capacitance thermal model of the house. The range of MPC results reflects different prediction horizons (3–48 hours) and control step durations (15–60 minutes). RL results include several algorithms (DDQN, SAC, A2C, PPO), with variations in control step (15 or 60 minutes), input features (weather, indoor temperature, occupancy), and action types (setpoint or heat pump power). The RBC controller is a Proportional-Integral controller tuned to provide adequate indoor comfort without excessive energy use. The RL and MPC hybrid approach is also included as presented in the original paper.

\subsection{Results and discussion}

The training for the results was obtained on a high performance computer composed of 880 Intel\textregistered Xeon\textregistered Gold 6148 processors

\newcommand{\figboptestresults}{\begin{figure}[H]
\centering
\includegraphics[width=0.85\linewidth]{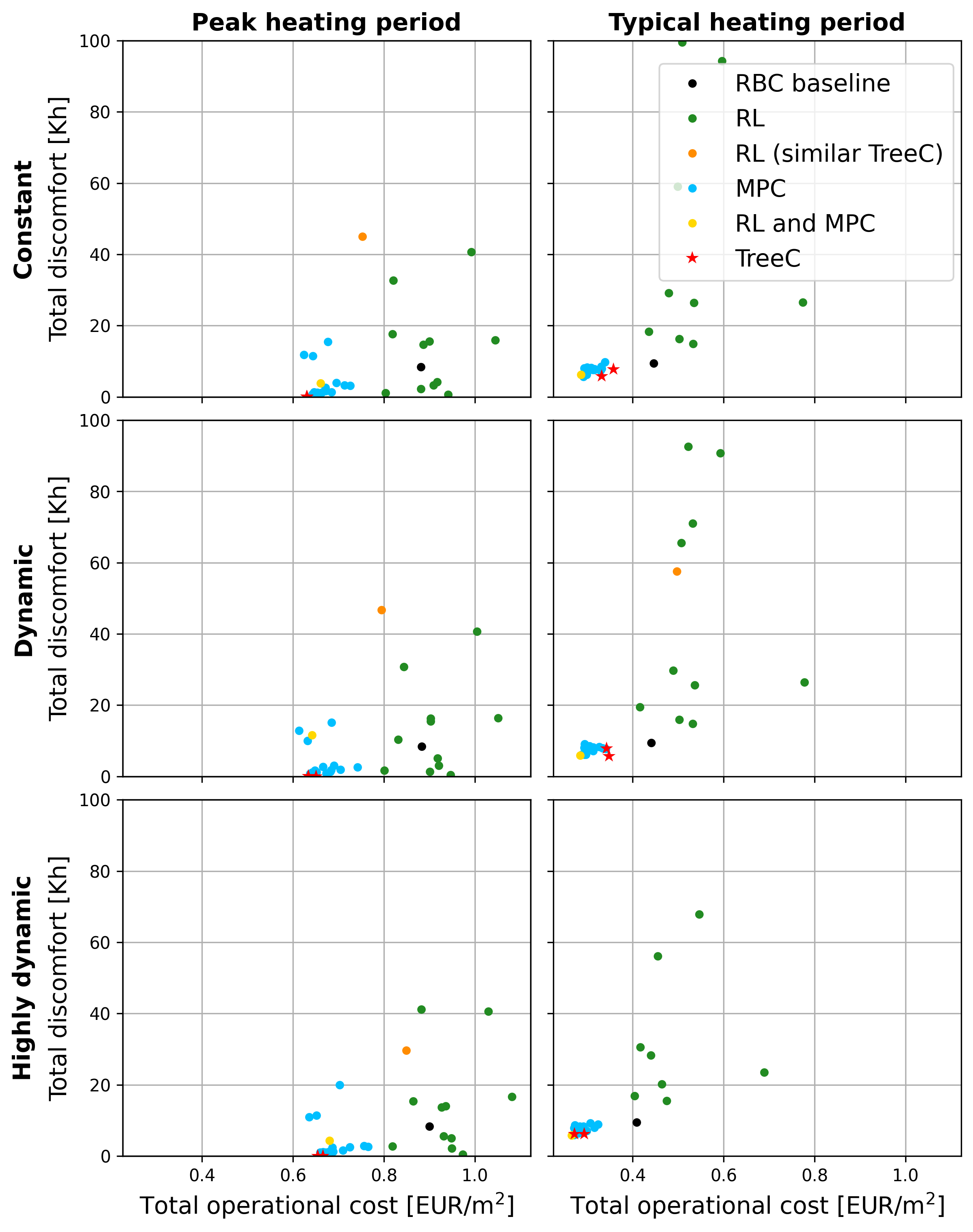}
\caption{BOPTEST results for the different pricing scenarios and validation time periods. The red stars are the results of the TreeC EMSs. The orange dot is the RL based EMSs that had exactly the same inputs and actions as the TreeC EMSs. For MPC and RL based EMSs, the different dots use different combination of parameters, inputs and actions. Some RL based EMSs have scores with a worse operational cost and total discomfort than the visualization range selected here (see \cite{arroyoComparisonOptimalControl2022} for full and detailed results). }
\label{fig:boptest_tree_res}
\end{figure}}

\ifprintedversion
\else
\fi

Running the evaluation of one individual simulation of two weeks lasts ~150 seconds. 
The time to update the CMA-ES optimization model is negligible here. 
The trees generated for the results shown in \cref{fig:boptest_tree_res} all ran for  12\,000 evaluation which amounts to ~20 days and 20 hours of training using a single thread and to ~6 hours and 15 minutes using the maximum number of 80 threads (population size of 16 $\times$ 5 parallel training).

As shown in \cref{fig:boptest_tree_res} the TreeC EMS obtains similar performances to the best performing MPC based EMS for all price scenarios of the peak heating period and the highly dynamic price scenario of the typical heating period. 
For the dynamic and constant price scenarios of the typical heating period, the TreeC EMS performs similarly in terms of comfort and slightly worse in terms of cost than the MPC based EMS. 
In all scenarios the TreeC EMS performed better than the RBC baseline EMS and the RL EMS. 
This is also the case for the RL based EMS that has exactly the same inputs and actions as the TreeC EMS. 

\ifprintedversion
\figboptestresults
\fi

\ifprintedversion
\else
\figboptestresults
\fi

\newcommand{\boptesttable}{

\begin{table}[H]
	\centering
	\caption{Average and standard deviation of the adjusted performances of the EMSs for the BOPTEST case.}
	\begin{tabular}{|c|c|c|}
		
	\hline
	
	EMS & Electricity cost [EUR/m\textsuperscript{2}] & Total discomfort [Kh]  \\ \hline
	RL and MPC & \textbf{0.018 $\pm$ 0.019} & 3.42 $\pm$ 4.04 \\ \hline
	TreeC & 0.033 $\pm$ 0.021 & \textbf{0.42 $\pm$ 0.77} \\ \hline
	MPC & 0.037 $\pm$ 0.029 & 2.91 $\pm$ 3.67 \\ \hline
	RBC & 0.208 $\pm$ 0.056 & 6.05 $\pm$ 2.34 \\ \hline	
	RL & 0.266 $\pm$ 0.085 & 24.41 $\pm$ 23.47 \\ \hline

	\end{tabular}
	
	\label{tab:boptest_table_results}
\end{table}
}

\Cref{tab:boptest_table_results} shows the adjusted and averaged electricity cost and total discomfort obtained by each EMS category in \cref{fig:boptest_tree_res}.
The electricity cost and total discomfort are adjusted in this table by subtracting the value of the best performing EMS for each metric on each of the 6 different scenarios (different price and heating periods).
This shows how far on average each EMS category is from the best performing EMS for each metric over the 6 scenarios.

\ifprintedversion
\boptesttable
\fi

\ifprintedversion
\else
\boptesttable
\fi

In terms of electricity cost, the RL and MPC hybrid EMS, TreeC and MPC based EMSs all perform similarly, while the RBC and RL based EMSs have a significantly higher electricity cost.

In terms of discomfort, a value of 7 Kh over one week corresponds to the indoor temperature being outside the comfort range by 1°C for an additional hour each day, compared to the best-performing EMS in each scenario.
Discomfort values close to and above 7 Kh may begin to noticeably decrease the occupant's comfort. The reinforcement learning approach yields a value of 24 Kh, which is substantially higher.

\begin{figure}[H]
\centering
\includegraphics[width=0.5\linewidth]{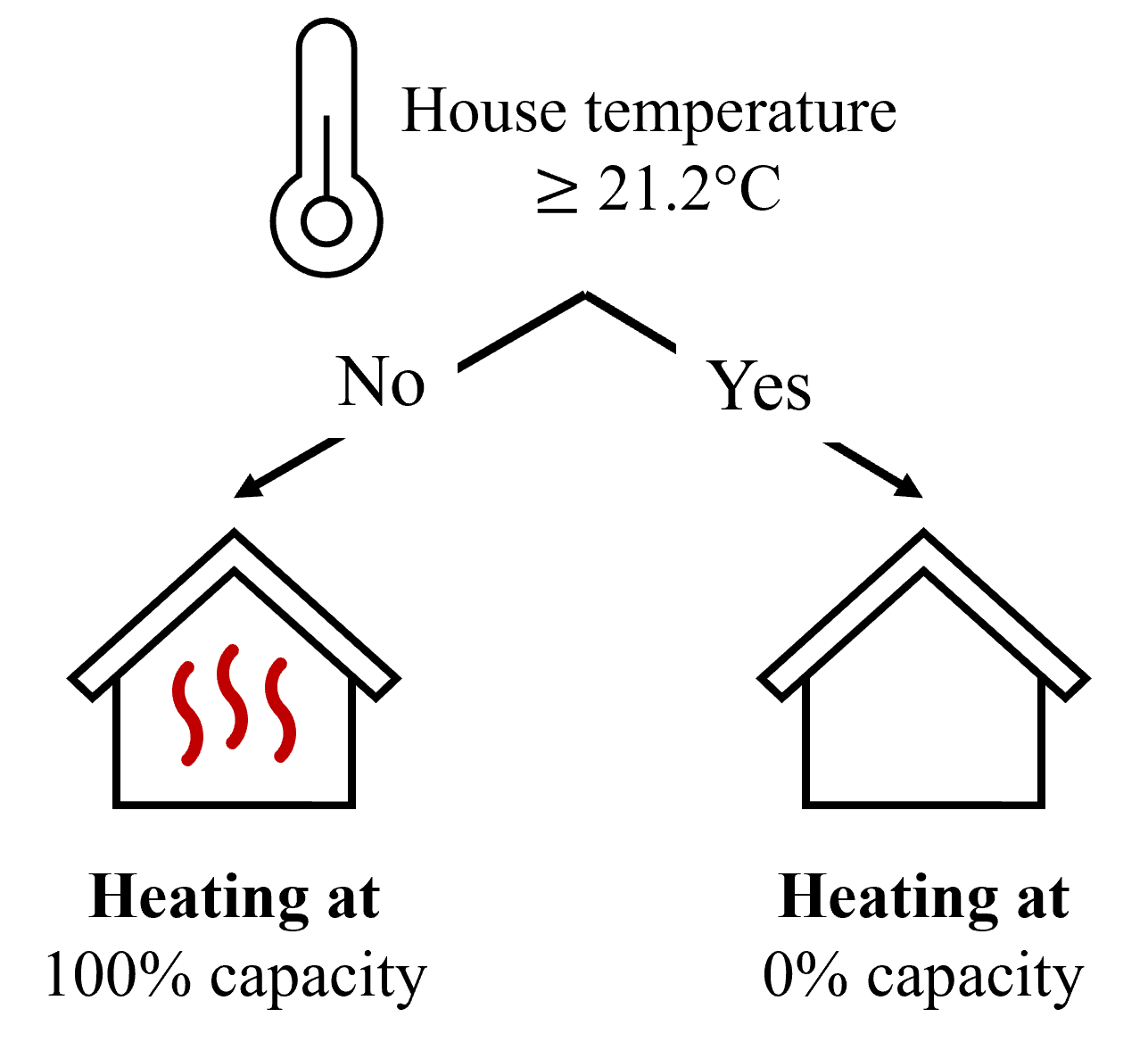}
\caption{TreeC EMS with the best score obtained on the BOPTEST case for the constant price and peak heating period scenario. }
\label{fig:boptest_tree}
\end{figure}

\newcommand{\figbotestvis}{
    \begin{figure}[H]
\centering
\includegraphics[width=0.9\linewidth]{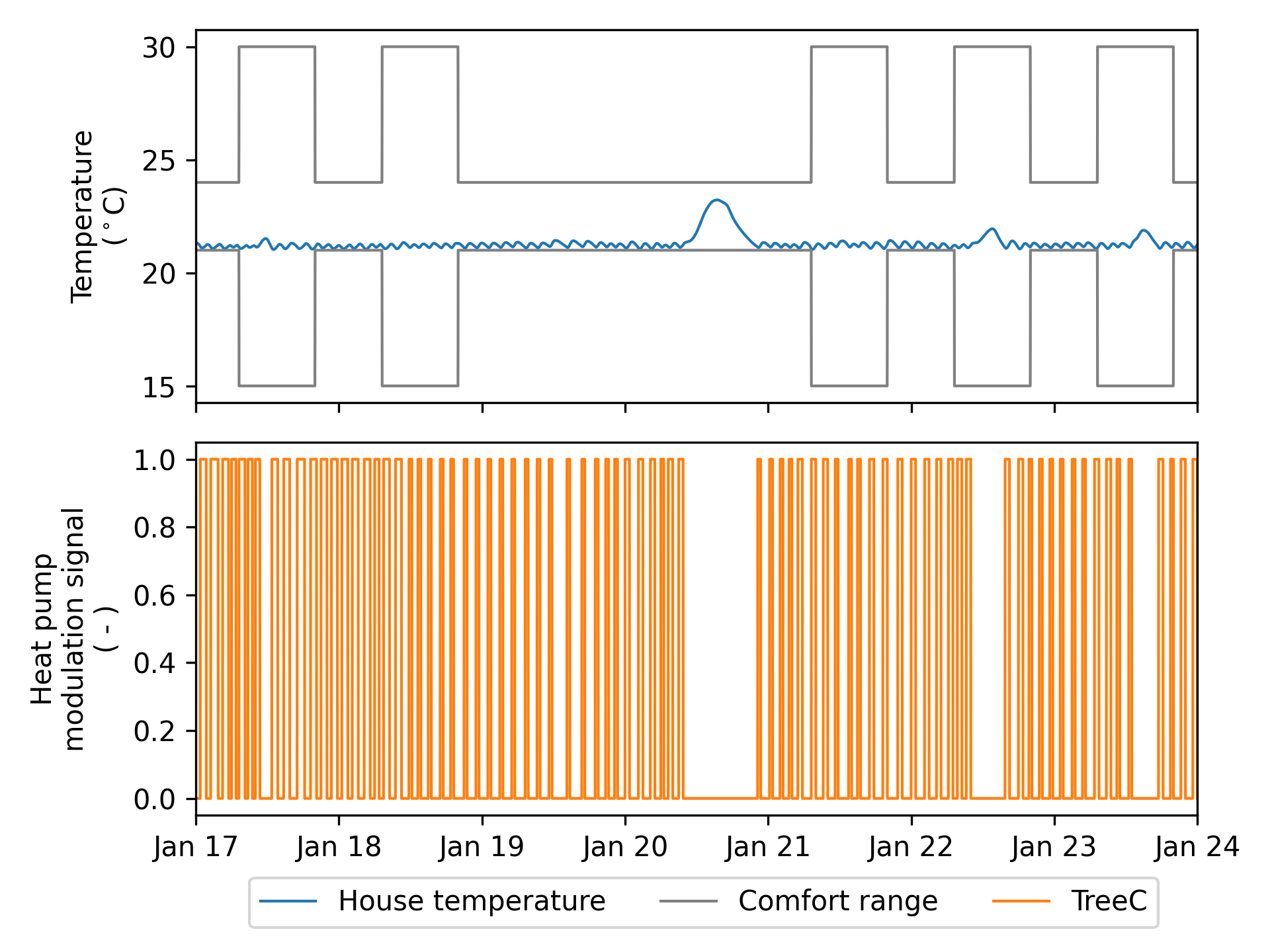}
\caption{TreeC EMS presented in \cref{fig:boptest_tree} controlling the heat pump for the first week of the peak heating validation period. The top plot shows that the indoor temperature always stays just above the lower bound of the comfort range during occupied hours. The increase in indoor temperature particularly noticeable on the \nth{20} of January is mainly caused by solar irradiation.}
\label{fig:boptest_vis}
\end{figure}
}

\ifprintedversion
\figbotestvis
\else
\fi

The trees obtained in training were all very similar to the one shown in \cref{fig:boptest_tree}. 
It is a  very simple tree that maintains the room temperature around 21.2\textdegree C which is just above the 21\textdegree C lower bound of the comfort range when the house is occupied. 
This behaviour ensures that the house temperature is always above the lower bound of the comfort range while keeping a relatively low temperature and therefore not having to buy too much electricity (see \cref{fig:boptest_vis}). 
The MPC behaviour is similar during occupied time periods but sometimes goes close to 20\textdegree C during unoccupied time periods when calculated as profitable. 
The TreeC EMS results show that going close to 20\textdegree C during the unoccupied time periods does not improve the score compare to simply keeping the temperature just above 21\textdegree C.

In this case TreeC found an EMS that is very performant and much more practical to implement than an MPC or RL based EMS. 

\ifprintedversion
\pagebreak
\else
\figbotestvis
\pagebreak
\fi

\section{Case 3: CityLearn 2022}
\label{sec:citylearn}

\subsection{Case description}
\begin{figure}[H]
\centering
\includegraphics[width=0.8\linewidth]{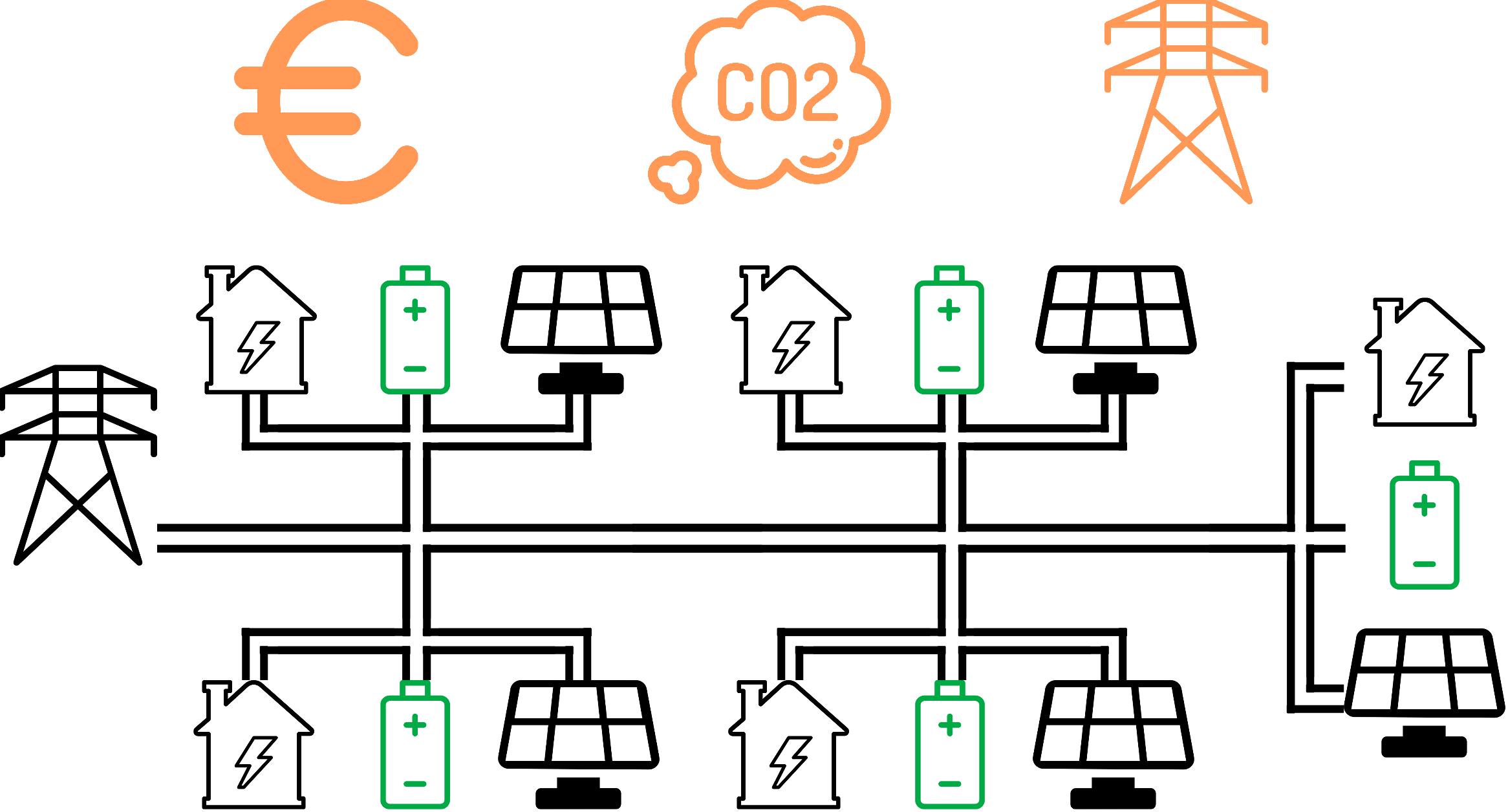}
\caption{Visualization of the CityLearn Challenge 2022. The EMS manages the batteries (in green) of residential buildings to jointly minimize electricity costs, CO\textsubscript{2} emissions, and grid stability metrics (in orange).}
\label{fig:citylearn_scheme}
\end{figure}

We participated with my colleague Evgeny Genov in the CityLearn Challenge 2022 \cite{nweyeCityLearnChallenge20222023} under the group name EVERGi\_team.

The objective of the challenge was to jointly minimize electricity cost, CO\textsubscript{2} emissions, and grid cost for a set of residential buildings equipped with battery and photovoltaic (PV) panels. The competition was structured in three phases: (1) a training phase with 5 houses and full data disclosure, (2) a validation phase introducing 5 additional houses with hidden data but accessible EMS performance evaluation, and (3) a test phase with 7 additional houses, for which no data was provided and only final EMS performance was assessed.

The total building power is defined in \cref{eq:citylearn_power}:
\begin{equation}
P^{build}_{t,b} = P^{base}_{t,b} + P^{batt}_{t,b} - P^{PV}_{t,b} \quad \forall\, b = 1,\ldots,B,\, t = 1, \ldots, 8760
\label{eq:citylearn_power}
\end{equation}
where:

\begin{tabular}{ll}
    $t$ & time step \\
    $b$ & building index \\    
    $P^{build}_{t,b}$ [kW] & total power of building $b$ at time $t$ \\
    $P^{base}_{t,b}$ [kW] & base uncontrollable load of building $b$ at time $t$ \\
    $P^{batt}_{t,b}$ [kW] & Battery power of building $b$ at time $t$ \\
    $P^{PV}_{t,b}$ [kW] & PV power of building $b$ at time $t$ \\
    $B$ & number of buildings \\

\end{tabular}

The electricity cost ($EC$) is defined with \cref{eq:citylearn_cost}:
\begin{equation}
EC = \sum_{t=1}^{8760} \max\left(\sum_{b=1}^{B} P^{build}_{t,b}\times E_{t},\, 0\right)
\label{eq:citylearn_cost}
\end{equation}

where:

\begin{tabular}{ll}
    $E_{t}$ [\euro/kWh] & electricity price at time $t$ \\
\end{tabular}

The collective pricing mechanism does not compensate for surplus power exported to the grid; therefore, the $\max$ function is applied to the aggregated building power to ensure that only net positive consumption is considered in the electricity cost calculation.

The CO\textsubscript{2} emissions ($CE$) are defined with \cref{eq:citylearn_co2}:

\begin{equation}
CE = \sum_{t=1}^{8760} \sum_{b=1}^{B} \max\left(P^{build}_{t,b} \times C_{t},\, 0\right)
\label{eq:citylearn_co2}
\end{equation}

where:

\begin{tabular}{ll}
    $C_{t}$ [gCO\textsubscript{2}/kWh] & CO\textsubscript{2} emissions factor at time $t$ \\
\end{tabular}

Injecting power into the grid does not result in negative emissions for individual buildings, so the $\max$ function is applied on the individual building power.

The ramping rate ($RR$) is defined with \cref{eq:citylearn_ramping}:

\begin{equation}
RR = \sum_{t=2}^{8760} \left|\sum_{b=1}^{B}  P^{build}_{t,b} - \sum_{b=1}^{B} P^{build}_{t-1,b} \right|
\label{eq:citylearn_ramping}
\end{equation}

The ramping rate quantifies how much the total power injected into or drawn from the grid by all buildings changes from one time-step to the next. Lower ramping rates are desirable because they indicate smoother power profiles, which reduce the reliance on fast-ramping generation units to maintain grid stability and balance supply with demand.

The load factor ($LF$) is defined with \cref{eq:citylearn_load_factor}:
\begin{equation}
LF = \sum_{m=0}^{11} \sum_{b=1}^{B} \frac{ \frac{1}{730} \sum_{k=0}^{729} P^{build}_{730m+k,\,b} }{ \max\limits_{0 \leq k \leq 729} P^{build}_{730m+k,\,b} }
\label{eq:citylearn_load_factor}
\end{equation}

where:

\begin{tabular}{ll}
    $m$ & month index \\
    $k$ & time-step increment \\
\end{tabular}

The load factor is calculated as the ratio of the average demand to the maximum demand within each month for each building. A higher load factor indicates that the average demand is closer to the monthly peak, reflecting a more consistent and less peaky load profile.

The number of time-steps in a month is defined as 730.

The grid cost ($GC$) is calculated as the sum of the ramping rate and one minus the load factor, as shown in \cref{eq:citylearn_grid_cost}:
\begin{equation}
GC = RR + (1 - LF)
    \label{eq:citylearn_grid_cost}
\end{equation}

The objective function is a weighted sum of the electricity cost, CO\textsubscript{2} emissions, with the EMS score in the numerator and the score without the battery in the denominator to normalize the three objectives. The objective function $O$ is defined with \cref{eq:citylearn_objective}:

\begin{equation}
O = \frac{1}{3} \frac{EC^{EMS}}{EC^{no\,bat}} + \frac{1}{3} \frac{CE^{EMS}}{CE^{no\,bat}} + \frac{1}{3} \frac{GC^{EMS}}{GC^{no\,bat}}
\label{eq:citylearn_objective}
\end{equation}

The score of the challenge is a weighted sum of the objective functions for the three phases giving a 20\% weight to the training phase, 30\% to the validation phase, and 50\% to the test phase as shown in \cref{eq:citylearn_final_score}:

\begin{equation}
Score = 0.2\times  O^{train} + 0.3\times O^{valid} + 0.5\times O^{test}
\label{eq:citylearn_final_score}
\end{equation}

\subsection{Challenge results}
Out of the 105 teams that participated in the challenge, 41 team submitted their solutions for the final evaluation \cite{nweyeCityLearnChallenge20222023}.
Each team was allowed to submit two solutions; we submitted one based on MPC and another using the TreeC approach.

Upon evaluation, the MPC and TreeC based solutions obtained a score of 0.840 and 0.853 respectively which ranks them 6\textsuperscript{th} and 7\textsuperscript{th} in the final leaderboard.
\citet{nweyeCityLearnChallenge20222023} reported the methods used by the top 5 teams in the challenge which is summarized in \cref{tab:citylearn_leaderboard}.

\begin{table}[H]
\ifprintedversion
\fontsize{10}{9}\selectfont
\else
\fontsize{10.95}{10}\selectfont
\fi
\renewcommand{\arraystretch}{1.2}
 
\centering
\caption{Final rank of the CityLearn Challenge 2022 with a short method categorization, score of the three objectives $EC$, $CE$, and $GC$, and the final score. This thesis' team result is indicated in bold \cite{nweyeCityLearnChallenge20222023}.}

\begin{tabular}{|l|l|l|l|l|l|l|}
\hline
Rank & Team & Method &$EC$&$CE$&$GC$& $Score$ \\ \hline

1\textsuperscript{st}                & Together       & \begin{tabular}[c]{@{}l@{}}Multi-scenario\\MPC+RL\end{tabular}&0.718&0.864&0.821& 0.801          \\ \hline
2\textsuperscript{nd}                & ambitiousengineers & Neuroevolution &0.748&0.863&0.827& 0.812          \\ \hline
3\textsuperscript{rd}                & CUFE           & MPC &0.750&0.859&0.872& 0.827          \\ \hline
4\textsuperscript{th}                & DivMARL        & RBC + RL &0.737&0.859&0.905& 0.834          \\ \hline
5\textsuperscript{th}                & Greener        & \begin{tabular}[c]{@{}l@{}}Supervised learning on\\ optimization solution\end{tabular} &0.740&0.879&0.902& 0.840          \\ \hline
6\textsuperscript{th}                & \textbf{EVERGi\_team} & MPC            &0.718&0.910&0.893& 0.840          \\ \hline
7\textsuperscript{th}                &  \textbf{EVERGi\_team} & TreeC          &0.732&0.882&0.946& 0.853          \\ \hline
8\textsuperscript{th}                & LA\_gang       & Unknown        &0.734&0.914&0.936& 0.861          \\
\hline
... & ...           & ...            & ...     &...&...&...       \\ \hline
\end{tabular}

\label{tab:citylearn_leaderboard}
\end{table}

The results in \cref{tab:citylearn_leaderboard} show that the first-place team achieved its leading score primarily through a significant reduction in the grid cost objective, outperforming both the MPC and TreeC EMSs significantly in this aspect.

\subsection{Discussion}

The top solutions in the challenge employed a variety of approaches. MPC was the most prevalent, used by the 1\textsuperscript{st}, 3\textsuperscript{rd}, and 6\textsuperscript{th} place teams. Holistic learning methods, such as neuroevolution and TreeC, achieved strong results, ranking 2\textsuperscript{nd} and 7\textsuperscript{th}, respectively, despite being less common in the literature (see \cref{ch:literature}). While no purely RL solutions appeared among the top entries, RL was integrated with MPC in the winning solution and with RBC in the 4\textsuperscript{th} place team, highlighting the value of hybrid approaches. A supervised learning method also reached the 5\textsuperscript{th} place. 

These results indicate that MPC remains highly competitive for this class of problems. Holistic learning approaches, though less established, demonstrated promising performance. RL, while not dominant on its own, proved effective when combined with other strategies. The success of the first-place team, which used multi-scenario stochastic MPC, suggests that explicitly addressing uncertainty through scenario-based optimization can provide a significant advantage.

\section{General discussion}

This chapter presented the results obtained from TreeC and different energy management methods for two benchmark cases and a competition.

Drawing from the results presented in this chapter, the main strengths and limitations of TreeC, MPC, RL and RBC are summarized in \cref{tab:comp}. 

\begin{table}[H]
\fontsize{8}{10}\selectfont
\caption{Relevant comparison criteria for the TreeC, MPC, RL and RBC methods. The colour scheme indicates the quality of the method for each criterion: light green is excellent, dark green is good, orange is moderate, and red is poor.
 }
\label{tab:comp}
\resizebox{1.0\linewidth}{!}{%
\begin{tblr}{
colspec={|l|l|l|l|l|},
cell{2}{2} = {green},
cell{2}{3} = {green},
cell{2}{4} = {green},
cell{2}{5} = {},
cell{3}{2} = {green},
cell{3}{3} = {green},
cell{3}{4} = {red},
cell{3}{5} = {orange},
cell{4}{2} = {green},
cell{4}{3} = {green},
cell{4}{4} = {red},
cell{4}{5} = {orange},
cell{5}{2} = {orange},
cell{5}{3} = {green},
cell{5}{4} = {teal7},
cell{5}{5} = {orange},
cell{6}{2} = {red},
cell{6}{3} = {green},
cell{6}{4} = {red},
cell{6}{5} = {green},
cell{7}{2} = {red},
cell{7}{3} = {green},
cell{7}{4} = {green},
cell{7}{5} = {orange},
cell{8}{2} = {teal7},
cell{8}{3} = {red},
cell{8}{4} = {green},
cell{8}{5} = {red},
cell{9}{2} = {green},
cell{9}{3} = {red},
cell{9}{4} = {green},
cell{9}{5} = {green},
cell{10}{2} = {green},
cell{10}{3} = {teal7},
cell{10}{4} = {red},
cell{10}{5} = {green},
cell{11}{2} = {green},
cell{11}{3} = {red},
cell{11}{4} = {green},
cell{11}{5} = {green},
cell{12}{2} = {red},
cell{12}{3} = {orange},
cell{12}{4} = {orange},
cell{12}{5} = {green},
}
\hline
\textbf{Criterion} & \textbf{TreeC} & \textbf{MPC} & \textbf{RL} & \textbf{RBC} \\ \hline
ANM6-Easy performance & 19.2 & 14.4 & 16.2 & None \\ \hline
BOPTEST performance & 0.03 EUR/m\textsuperscript{2} & 0.04 EUR/m\textsuperscript{2} & 0.27 EUR/m\textsuperscript{2} & 0.21 EUR/m\textsuperscript{2} \\ 
& 0.42 Kh & 2.91 Kh & 24.41 Kh & 6.05 Kh \\ \hline
CityLearn 2022 ranking & 7\textsuperscript{th} & 1\textsuperscript{st}, 3\textsuperscript{rd}, 6\textsuperscript{th} & 1\textsuperscript{st}, 4\textsuperscript{th} (hybrid) & 4\textsuperscript{th} (hybrid) \\ \hline
Common real implementations & No & Yes & No & Yes \\ \hline
Many possible output values & No & Yes & Yes & Moderate \\ \hline
Adapts model to problem & In simulation & No & In real case & No \\ \hline
Scales well with case complexity & Yes & No & Yes & Yes \\ \hline
Interpretability of model & Very good & Good & Bad & Very good \\ \hline
Relies on good forecasting & No & Yes & No & No \\ \hline
Training time & Very high & Forecast & High & No training \\ \hline

\end{tblr}}
\end{table}

The outstanding advantages of the MPC method are that the performance is good and has proven itself in several real use cases. 
The disadvantages are that the method is deterministic and does not have a learning behaviour. 
MPC relies on good forecasting of the system's various inputs and needs to model the system it controls using mathematical programming. 
This last point has the disadvantage that one must understand the system very well in order to be able to model it, and in almost all cases model simplifications are necessary in order to be able to solve the mathematical programming problem in the required time. 
MPC does not automatically adapt the control model to the problem by for example changing model parameters when the forecast is not accurate.
The interpretability is good (i.e. the MPC controls the system perfectly assuming all given models and inputs are accurate) but in most cases a lot of inputs and model constraints need to be taken into account to completely interpret the decision process of the MPC. 

On the other hand, the TreeC and RL methods both have the advantage of learning and adapting their control model to the controlled system. 

The RL method is capable of learning online meaning that RL can adapt the EMS while it is controlling the system. This is a real advantage as it would receive a very accurate feedback from the system and not have to rely on a less accurate simulation of the system. Nonetheless, many training steps, corresponding to multiple years, of bad performance are needed for the RL model to obtain a good performance \cite{ceustersModelpredictiveControlReinforcement2021}. Hence, either training in simulation or a transfer learning method would still be necessary to implement this in a real case. 

The RL method uses a neural network as control model. This allows the model to have very complex behaviours sometimes necessary to solve certain problems. On the other hand it is a black box model which is not interpretable and therefore could lead to unexpected behaviours which is not desirable for an EMS. 

For the three studied cases, RL fails to obtain consistently a good performance. In the ANM6-Easy case, SAC obtained a good performance but not PPO. In the BOPTEST case, all RL algorithm performed poorly even compared to the simple RBC baseline. In the CityLearn competition, RL performed well but only when combined with MPC or RBC. The trade-off of RL for using a model that is not interpretable should be that it is capable of solving simple and complex cases equally but there is no clear methodology in literature yet shown to obtain consistently good results over different benchmark cases. 

The main advantage of the TreeC method is the interpretability of its models as shown in \cref{fig:anm_trees,fig:boptest_tree}. This allows total confidence in the EMS when implementing it in a real case, the possibility to recognize some possible wrong behaviour that can be caused by the difference between the simulation and real case and makes it easier to transfer the knowledge gained on how to control this system to another similar system.

Another advantage is that the method does not require much customization other than the necessary ones for any machine learning problem (i.e. choosing good model inputs and outputs, which training steps to use and for how long should the training process run). The CMA-ES algorithm performs very well with default hyperparameters and the number of nodes of the tree can be defined to make the tree more or less interpretable. TreeC obtained consistently good performance over both the ANM6-Easy and BOPTEST cases changing only the part of the methodology specific to the problem (i.e. inputs, outputs, training steps and training length). 

One main disadvantage is that it cannot learn online and therefore requires an accurate simulation of the system it needs to control. This simulator also needs to be relatively fast as TreeC needs to simulate many steps to find a good solution. The CMA-ES algorithm allows to parallelize the evaluation of individuals within a generation but a fast simulator is still important to be able to do the individual's evaluation over many steps and the training over many generations. 

Another disadvantage is that the complexity of the behaviour of the model is limited by the size of the tree, one tree has a maximum number of output values limited to the number of leafs of the tree and the tree should remain small to keep it interpretable. For example, in the CityLearn competition, the neuroevolution approach, which can represent more intricate behaviours, outperformed TreeC, even though both are holistic learning methods with similar training methodologies.

\section{Conclusion}

This chapter the results from three simulation cases, including  two benchmark problems and a competition.

The TreeC method demonstrated good performances in the ANM6-Easy and BOPTEST benchmarks, and achieved a competitive ranking in the CityLearn 2022 Challenge. MPC consistently delivered high performance across all cases even though it can't adapt its model to the problems. RL showed variable results: it performed well in ANM6-Easy and in hybrid approaches for CityLearn, but performed poorly in BOPTEST.

Overall, TreeC demonstrated strong performance compared to other proven technology methods, proving its applicability across diverse energy management problems. Its main advantage lies in producing interpretable models, though this comes at the expense of longer training times.

\mode<all> %

\chapter[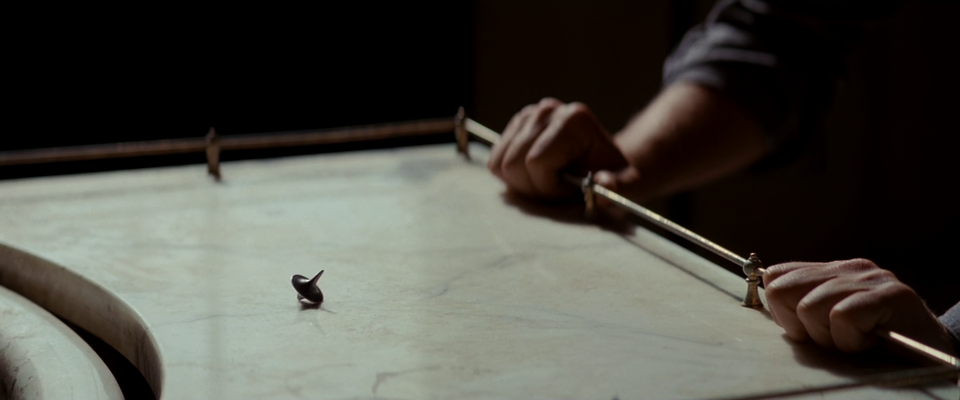]{Real-world experimental validation\label{ch:summary_cases}}

In the previous chapter, we presented three simulation cases to evaluate the performance of TreeC against other proven technology EMS methods.
Showing good performance in simulation is important, but the final goal is to implement these methods in real-world applications. This chapter presents a comparison of four EMSs in an experimental setup with real energy assets. The four implemented EMSs are TreeC, reinforcement learning (RL), model predictive control (MPC) and rule-based control (RBC). Each EMS manages a home battery and emulated EV charger and aim to minimize household electricity costs under a dynamic tariff with prices changing every hour. Implementing EMSs in a real-world setting is important to identify potential issues that do not arise in simulations. This chapter is an adaptation of the published journal article ``\textit{Real-world validation of safe reinforcement learning, model predictive control and decision tree-based home energy management systems}'' \cite{ruddickRealworldValidationSafe2024}.

The main contribution of this chapter are:
\begin{itemize}
    \item Implementation of TreeC in a real-world experimental setup.
    \item Comparison of TreeC, RL, MPC and RBC on a real experimental case with batteries, solar PV, electric loads, emulated electric vehicle charging and hourly electricity based pricing.
\end{itemize}

\ifprintedversion
\pagebreak
\else
\pagebreak
\fi

\section{Method}
\label{sec:method}

\subsection{Hardware setup}
\label{sec:hardware}

\begin{figure}[H]
    \centering
    \includegraphics[width=0.9\textwidth]{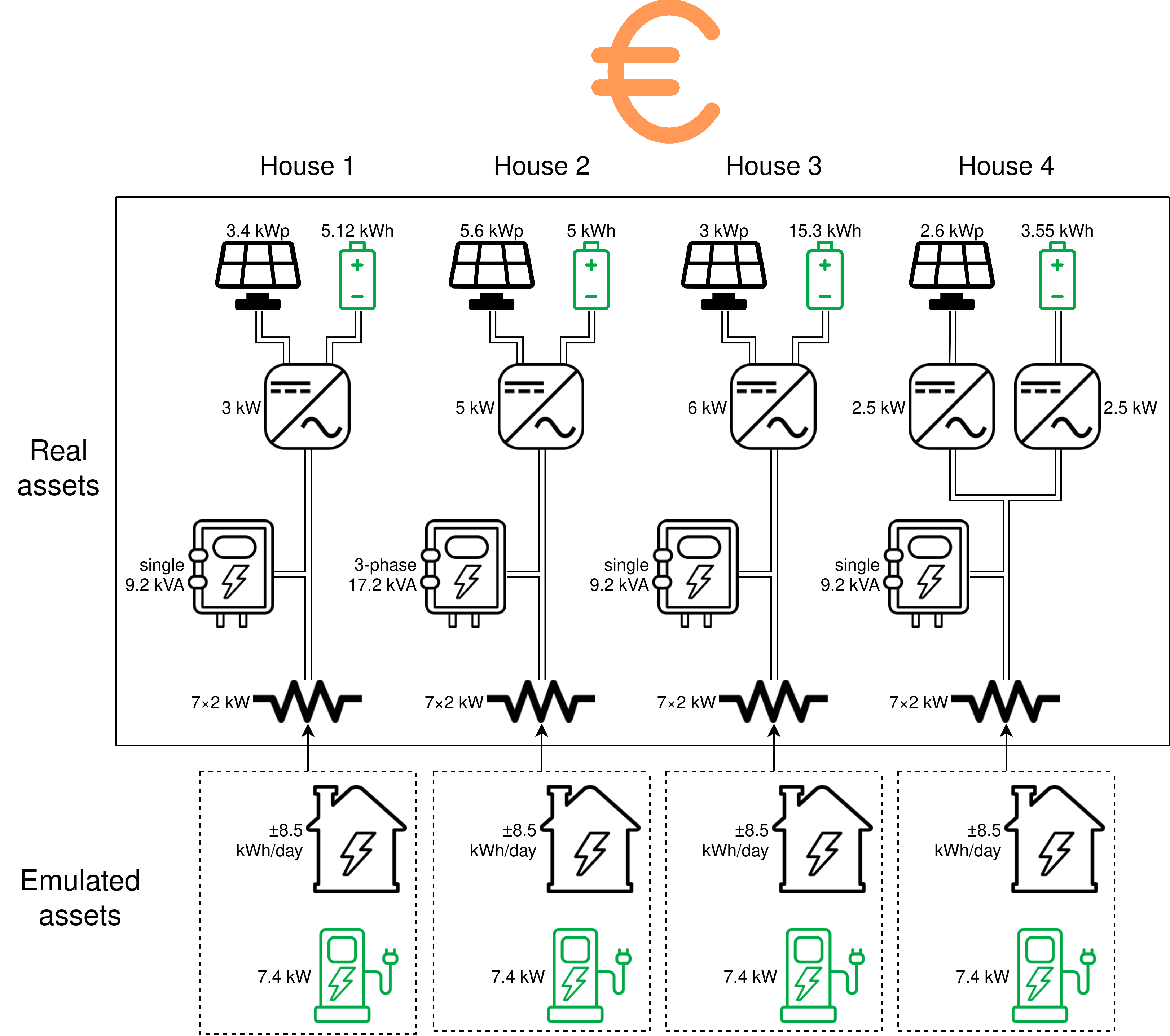}
    \caption{Schematic of this chapter's experimental setup for each house. From top to bottom: PV installation (with peak power), Battery (with capacity), inverter (with maximum power), grid connection (with phase specification and maximum power), dynamic loads (with maximum power), emulated house consumption (with average daily consumption), and EV charger (with maximum charge power). Assets highlighted in green are controlled by the EMS, optimization is performed on the electricity cost, indicated in orange.}
    
    \label{fig:setup_diagram}
\end{figure}

The experimental setup consists of four real reproductions of a house's electrical circuit. Each house has a battery, a PV installation, an emulated electric load consumption and emulated EV charging sessions. The houses are connected to the grid and have official digital meters from the local Belgian distribution system operator that measures energy imported from and exported to the grid. The home batteries and PV installations are different for each house, whereas the emulated electric load consumption and charging sessions are identical for all houses.  This setup replicates an electrical configuration typical of a modern home with any non-electrical heating system. The experiment was conducted in the Smart Village lab on the Green Energy Park site in Zellik, Belgium.

The electrical loads used for the consumption and charging sessions come from a dynamic load system composed of seven 2000W Perel\textregistered PHP2000 heaters controlled through Showtec Single DP-1 dimmers.
The electrical loads are calibrated so that the setpoints match the real measured active power consumption more closely (we expect this difference to come from measurement\footnote{No voltage and therefore power factor, measurement \textit{per} circuit was operational} inaccuracies and manufacturing tolerances of the loads). 

Measurement devices for power, voltage and current are present for the PVs, home batteries and grid connections.
The power of the electrical loads and other power losses in the setup is calculated via the PV, battery and grid measurements.

\begin{figure}
    \centering
    \includegraphics[width=0.9\textwidth]{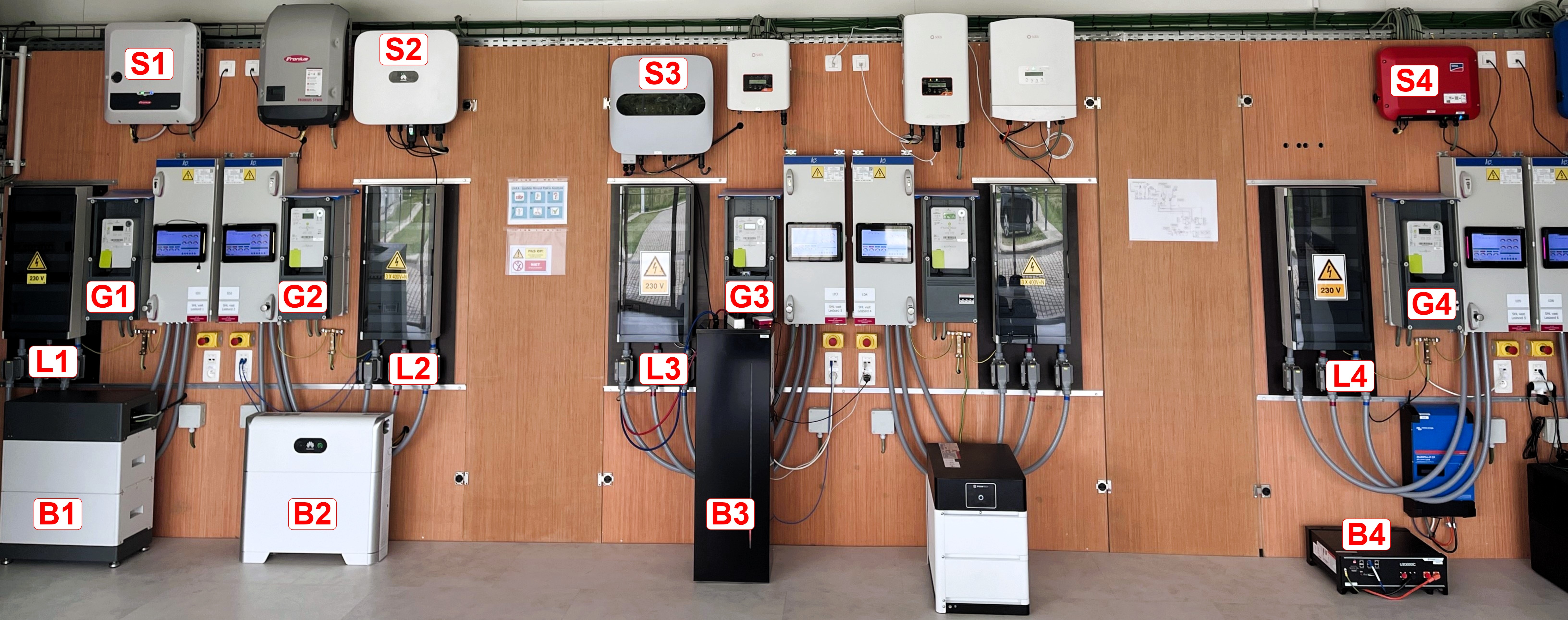}
    \caption{Photograph of the experimental setup: with \texttt{G} being the grid connection equipped with the official digital meter of the local distribution system operator, \texttt{B} the battery, \texttt{S} the solar PV installation inverters (solar panels equipped on the roof) and \texttt{L} the connectors to the dynamic loads (dimmers in separate cabinet and electrical loads stationed outdoors). The number represents the house numbers.
    }
    \label{fig:svl_picture}
\end{figure}

A schematic and a picture of the setup are displayed in \cref{fig:setup_diagram,fig:svl_picture}.

The hardware specifications of the home batteries, inverters and PV installations are summarized in \cref{tab:batt_spec,tab:inverter_spec,tab:pv_spec}.

\begin{table}[H]
    \caption{Battery specifications}
    \centering
    \footnotesize\begin{tabular}{c|l|l|c|c}
        \shortstack{\textbf{House}} & \shortstack{\textbf{Battery model}} & \shortstack{\textbf{Capacity}} & \shortstack{\textbf{Max} \\ \textbf{charge}} & \shortstack{\textbf{Max} \\ \textbf{discharge}} \\
        \hline
        1 & BYD HVS 5.1 & 5.12 kWh & 3.2 kW & 3.2 kW \\ 
        2 & Huawei LUNA2000-5-EO & 5 kWh & 2.5 kW & 2.5 kW \\ 
        3 & D-Centralized LF2 & 15.3 kWh & 3.0 kW & 4.0 kW \\ 
        4 & Pylontech US2000B & 3.55 kWh & 1.7 kW & 2.5 kW \\ 
    \end{tabular}
    
    \label{tab:batt_spec}
\end{table}

\begin{table}[H]
    \caption{Inverter specifications}
    \centering
    \footnotesize\begin{tabular}{c|l|l|c}
        \shortstack{\textbf{House}} & \shortstack{\textbf{Inverter model}} & \shortstack{\textbf{Assets}} & \shortstack{\textbf{Max} \\ \textbf{AC power}} \\
        \hline
        1 & Fronius Primo GEN24 3.0 & PV + battery & 3.0 kW \\ 
        2 & Huawei SUN2000-5KTL-M1 & PV + battery & 5.0 kW \\ 
        3 & Sermatec SMT-5K-TL-LV & PV + battery & 6.0 kW \\ 
        4 & SMA Sunny Boy SB2.5-1VL-40 & PV & 2.5 kW \\ 
        4 & Victron Multiplus-II GX 48/3000/35-32 & battery & 2.5 kW \\ 
    \end{tabular}
    
    \label{tab:inverter_spec}
\end{table}

\begin{table}[H]
    \caption{Additional specifications}
    \centering
    \footnotesize\begin{tabular}{l|c|c|c}
        \shortstack{\textbf{House}} & \shortstack{\textbf{PV peak power}} & \shortstack{\textbf{AC circuit}} & \shortstack{\textbf{House breaker}}  \\
        \hline
        1 & 3.4 kWp & mono phase & 9.2 kVA\\ 
        2 & 5.6 kWp & three phase & 17.2 kVA\\ 
        3 & 3 kWp & mono phase & 9.2 kVA \\ 
        4 & 2.6 kWp & mono phase & 9.2 kVA \\ 
    \end{tabular}
    
    \label{tab:pv_spec}
\end{table}

\subsection{Experimental setup}

\subsubsection{Consumption profile}
\label{sec:consumption}
We utilize the real-time (10 seconds) residential electricity \enquote{household} consumption data from \citet{schlemmingerDatasetElectricalSinglefamily2022}, focusing specifically on the 2019 dataset, which predates the COVID-19 pandemic. Our experiments employ data from single-family house (SFH) 19, which does not include solar PV, battery, or EV charging. Additionally, this household dataset excludes heat pump loads. SFH 19 is selected due to its high data availability and quality, along with its typical consumption metrics, averaging 3425 kWh per year.

\begin{figure}[H]
    \centering
    \begin{subfigure}{0.45\textwidth}
        \centering
        \includegraphics[width=\linewidth]{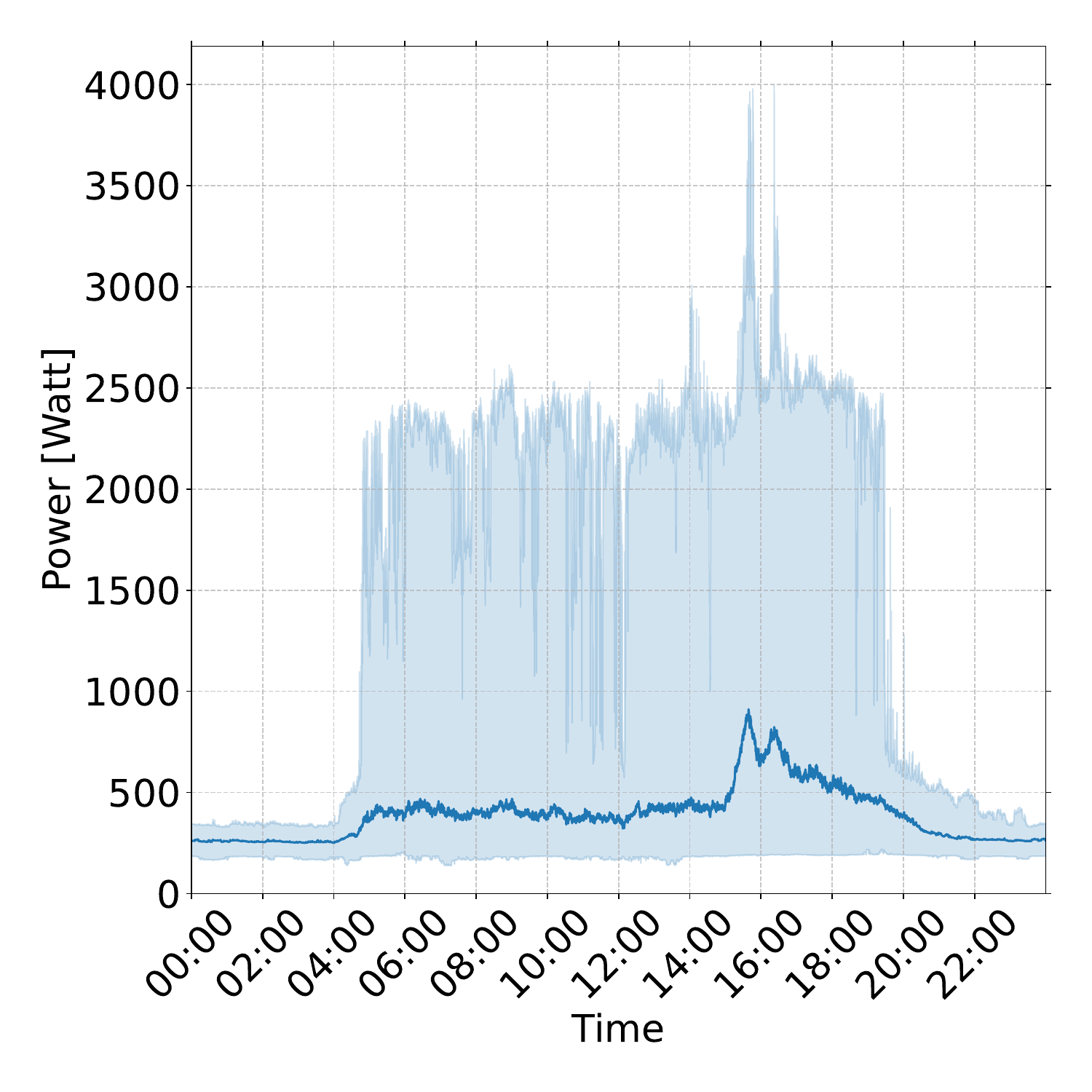}
        \caption{Time series with 2.5th and 97.5th percentile}
        \label{subfig:load lineplot}
    \end{subfigure}
    \hfill
    \begin{subfigure}{0.45\textwidth}
        \centering
        \includegraphics[width=\linewidth]{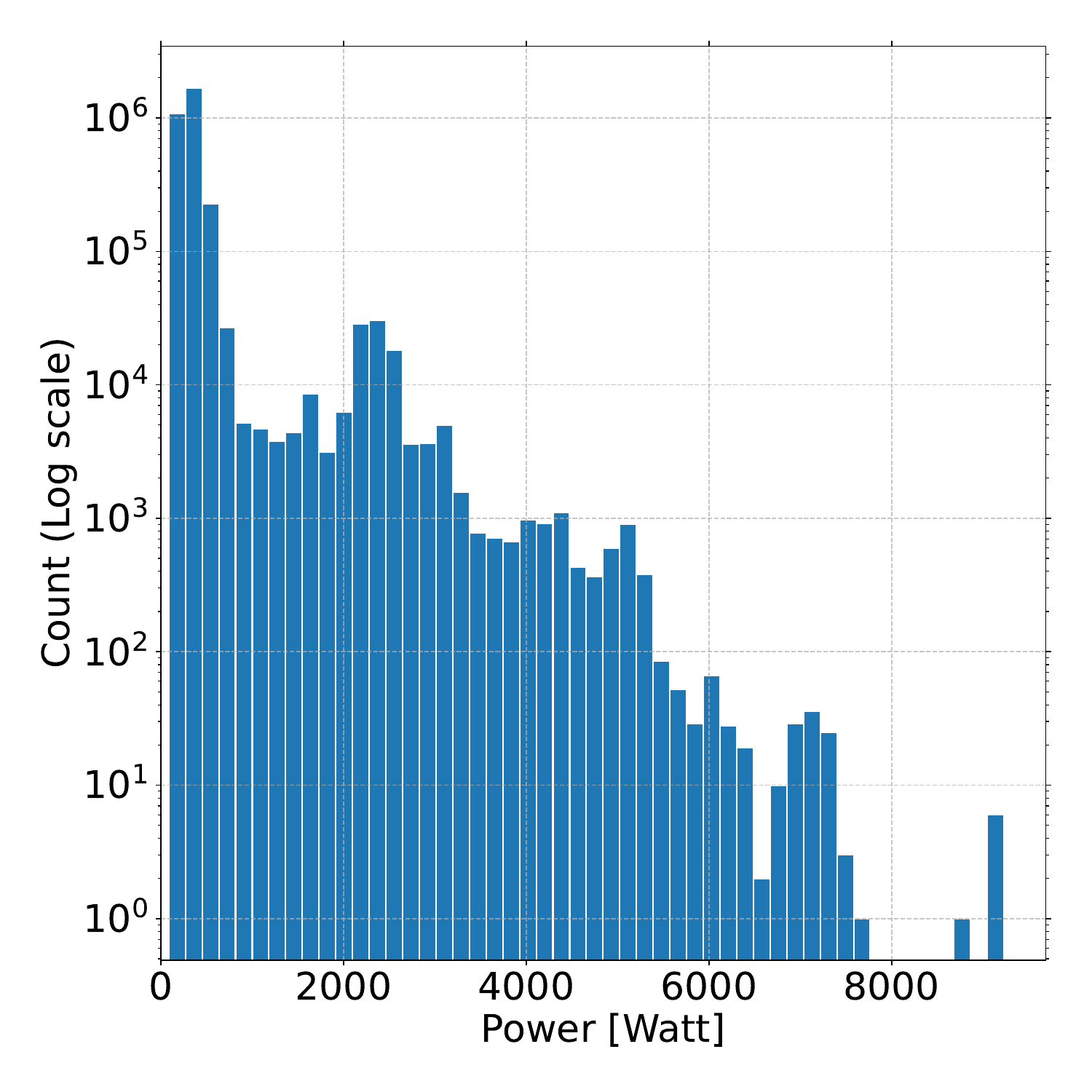}
        \caption{Histogram}
        \label{subfig:load histogram}
    \end{subfigure}
    \caption{Household consumption profile: 10 seconds SFH 19 data \cite{schlemmingerDatasetElectricalSinglefamily2022} without PV, battery, EV charging or heat pump having 3425.75 kWh/year, Std: 445.85 W, max: 9229 W.}
    \label{fig:consumption profile}
\end{figure}

We shifted the dataset so that weekdays align (that a Monday in runtime is also a Monday in the dataset) and used this for all our hardware-in-the-loop simulated dynamic loads so that each house has the same real-time non-controllable load.

\subsubsection{Electric vehicle and driver charging behaviour}
\label{sec:ev_sessions}
The residential electric mobility demand consists of two parts: a) a first part specific to the EV user charging behaviour (i.e. arrival time, departure time and energy need), and b) a second part specific to the EV charging power profile.

Firstly, the EV user charging behaviour is simulated directly using an existing open-source dataset available in \cite{sorensenResidentialElectricVehicle2021}. This dataset comprises data from 97 EV users, encompassing 6878 charging sessions at residential locations spanning from December 2018 to January 2020. 
Among these users, \enquote{Bl2-2} has been selected for the experiment due to its alignment with the average EV user in terms of arrival time, parking duration, and energy requirements. 
However, it distinguishes itself by exhibiting a notably high frequency of charging sessions, rendering it particularly interesting for the experiment as it allows for more interactions with the EMS developed in this chapter. From a statistical standpoint, this EV user has undergone 315 charging sessions, with an aggregate energy consumption of 3274.74 kWh across 386 days.

Secondly, the EV charging power profile is built based on certain variables and constraints. The EV charging power is constrained by (\ref{eq_ev_power_charge}).

\begin{equation}
    0 \leq P^{ev}_{t} \leq P^{ev}_{max}
\label{eq_ev_power_charge} 
\end{equation}

where: 

\begin{tabular}{ll}
$P^{ev}_{t}$ [kW] & EV charging power at time step $t$ \\
$P^{ev}_{max}$ [kW] & Maximum charging power \\
\end{tabular}

$P^{ev}_{max}$ is the maximum charging power in $[kW]$ set to 7.4 kW, following the IEC 61851 charging standards \cite{internationalelectrotechnicalcommissionIEC61851120172017}. 
Additionally, the maximum charging power follows the Constant-Current/Constant-Voltage (CC-CV) uncoordinated charging method defined in \citet{vagropoulosOptimalBiddingStrategy2013}. Such a method stipulates that the power is a function of the state of charge (SOC) as a piece-wise linear function summarized in (\ref{ch2_sc3_eq_p_final_soc}).

\begin{equation}
\ifprintedversion
\footnotesize
\fi
P^{ev}_t = \left\{
\begin{aligned}
&P^{ev}_{max}   &\textrm{if} \quad SOC^{ev}_{t} \leq SOC^{ev}_{cc,cv}  \\
&P^{ev}_{max}  - (P^{ev}_{max}  - P^{ev}_{min}) * \frac{SOC^{ev}_{t} - SOC^{ev}_{cc,cv}}{1 - SOC^{ev}_{cc,cv}}  &\textrm{if} \quad SOC^{ev}_{t} > SOC^{ev}_{cc,cv} 
\end{aligned}
\right.
\label{ch2_sc3_eq_p_final_soc} 
\end{equation}

where:

\begin{tabular}{ll}
$SOC^{ev}_{t}$ [\%] & SOC of the EV at time step $t$ \\
$SOC^{ev}_{cc,cv}$ [\%] & \begin{tabular}{l}    
Transition SOC between the Constant-Current \\ 
and Constant-Voltage phases, set to 90\% 
\end{tabular} \\
$P^{ev}_{min}$ [kW] & Maximum charging power at 100\% SOC, set to 1 kW \\
\end{tabular}

Finally, both charging power and SOC variables are linked using the equality constraint (\ref{eq_ev_soc_balance}).

\begin{equation}
SOC^{ev}_{t+1} = SOC^{ev}_{t} + \eta^{charge} \times \frac{P^{ev}_{t} \times \Delta t}{E^{ev}}  \label{eq_ev_soc_balance} 
\end{equation}

where:

\begin{tabular}{ll}
$SOC^{ev}_{t}$ [\%] & SOC at time step $t$ \\
$E^{ev}$ [kWh] & capacity of the EV battery, set to 60 kWh \\
$\eta^{charge}$ [\%] & charging efficiency, set to 95\% \\
\end{tabular}

The charging efficiency is set to 95\% as it is typical for lithium-ion batteries \cite{farhadIntroducingEnergyEfficiency2019}.

\subsubsection{Tariff}
The real-time pricing used in this study is based on a one-year dynamic contract of a major Belgian energy supplier, as if it were taken in April 2024\footnote{Full contract available on the paper's GitHub repository at \url{https://github.com/EVERGi/real_validation_saferl_treec_paper/}}.
The total cost is composed of four costs with different pricing mechanisms: the day-ahead cost, the offtake extras cost, the peak cost and the fixed yearly cost.

The day-ahead cost depends on the prices of \cref{eq:injection_price,eq:offtake_price}.

\begin{equation}
    Vo_t= 
\begin{cases}
    (Vd_t+0.011\frac{\SI{}{\EUR}}{kWh})*1.06& \text{if } (Vd_t+0.011\frac{\SI{}{\EUR}}{kWh})\geq 0\\
    (Vd_t+0.011\frac{\SI{}{\EUR}}{kWh}),              & \text{otherwise}
\end{cases}
\label{eq:injection_price}
\end{equation}

\begin{equation}
    Vi_t=Vd_t-0.009\frac{\SI{}{\EUR}}{kWh}
\label{eq:offtake_price}
\end{equation}

where:

\begin{tabular}{ll}
$Vo_t$ [\euro{}/kWh] & offtake price at time step $t$ \\
$Vi_t$ [\euro{}/kWh] & injection price at time step $t$ \\
$Vd_t$ [\euro{}/kWh] & Belgian day-ahead hourly price at time step $t$ \\
\end{tabular}

A value-added tax of 6\% is added to the offtake price when it is positive.

Extra costs that occur for energy taken from the grid for transport, distribution and taxes are calculated together and labelled as offtake extras cost.
The offtake price for all these additional offtake costs $Vf$ is 0.114 \euro{}/kWh. 

The monthly peak offtake cost $Cp_m$ is calculated by \cref{eq:peak_cost}.
\begin{equation}
Cp_m = \frac{|B_m|}{N_m}Vp*\max_{{t\in B_m}} (\frac{Eo_t}{0.25h}, 2.5 kW)
\label{eq:peak_cost}
\end{equation}

where:

\begin{tabular}{ll}
$B_m$ & set of all billed time steps in month $m$ \\
$N_m$ & total number of time steps in the month \\
$Vp$ [\euro{}/kW] & peak price of  3.5 \euro{}/kW \\
$Eo_t$ [kWh] & energy taken out of the grid at time step $t$ \\    
\end{tabular}

The $Cp_m$ cost adds a penalty to the highest offtake power averaged over a 15-minute time step that occurred for each month. 
In the context of this study, the monthly peak offtake cost is proportional to the billed period of each month to also include this cost for unfinished months. 
The offtake energy $Eo_t$ is divided by the length of a time step to obtain the power in kW.
A minimum monthly peak power of 2.5 kW is enforced in the cost, even if the measured peak power is lower.

The fixed yearly cost is calculated using the yearly price $Vy$ of 115.84 \euro{}\ and is proportional to the billing period.
The total electricity cost is calculated by \cref{eq:tot_cost}.

\begin{equation}
C_{tot}=\underbrace{\sum_{t\in T}{(Eo_tVo_t-Ei_tVi_t)}}_{\text{day-ahead cost}}+\underbrace{\sum_{t\in T}Eo_tVf}_{\text{offtake extras cost}}+\underbrace{\sum_{m \in M}{Cp_m}}_{\text{peak cost}}+\underbrace{\frac{|T|}{N_y}Vy}_{\text{yearly cost}}
\label{eq:tot_cost}
\end{equation}

where:

\begin{tabular}{ll}
$T$ & Set of all billed time steps \\
$M$ & Set of billed months \\
$N_y$ & Total number of time steps in the year \\
$Ei_t$ [kWh]& Energy injected to the grid at time step $t$ \\
\end{tabular}

\subsubsection{House switching}
As described in \cref{sec:hardware}, the four houses have different home batteries and PV installations. 
Assigning a different EMS to each house and comparing the performance by running them simultaneously would not yield comparable results.
One house could obtain better performances because it has a better battery or PV installation.
To avoid this bias, the EMSs are switched every 24 hours to a different house.

Since there are 4 houses and 4 EMSs to test, there are a total of 24 possible house and EMS combinations. 
Each combination is tested twice, resulting in 48 days of testing in total. 
The 48-day schedule is obtained by randomly shuffling the 24 combinations until two distinct 24-day schedules are generated. This ensures no consecutive days have the same RL or MPC EMS assigned to the same house. 
This rule is applied for practical reasons, as the RL and MPC EMSs behaviours depend on how they operated a house previously. 
Failed test days can occur, and not having an MPC or RL EMS on the same house for consecutive days provides more time to identify a failure while preventing its propagation to the following day.

The switching is conducted daily at 15:00. By this time, the batteries of all houses are charged to 100\% to ensure consistent starting conditions post-switch. 
The EV schedule is adjusted to prevent charging sessions from overlapping with the 15:00 switching time. 
This is achieved by shortening the duration of overlapping sessions, either by starting or finishing them at 15:00, with the option of the longest duration being selected. 
The start and end SOC of the EV batteries are maintained according to the original schedule.

These switching rules have been selected because 15:00 is a time when there are fewer charging sessions compared to the morning and evening hours. 
At 15:00, the home battery also had the opportunity to charge using the excess energy generated by the PV installation in the preceding hours.
Additionally, at this time, the day-ahead prices for the next day are available, eliminating the need to forecast them for the MPC EMS.

\subsubsection{Enforced charging behaviour}
\label{sec:enforced_charging}

An enforced charging behaviour is enforced in order to reach 100\% SOC at the switching time for the home batteries and to reach the imposed final SOC of an EV battery at the end of a charging session.
The batteries are charged at maximum power to reach their SOC goal at the latest possible time minus a buffer time $Btime$.
This buffer time is used to avoid not reaching the SOC goal in time due to non-perfect battery models, safety layer activation (which could reduce the battery setpoints, see \cref{sec:safety}) or other unexpected event.

The buffer time value is higher when the difference between the SOC goal and the current SOC of the battery is higher because more events could prevent the battery from reaching its SOC goal in time. This relation is described in \cref{eq:time_buffer}.

\begin{equation}
    B_{time} = (SOC_g-SOC)*(B_{max}-B_{min})+B_{min}
    \label{eq:time_buffer}
\end{equation}

where:

\begin{tabular}{ll}
$B_{time}$ [min] & time buffer in minutes \\
$SOC_g$ [\%] & SOC goal of the battery \\
$SOC$ [\%] & current SOC of the battery \\
$B_{max}$ [min] & time buffer if 100\% of the battery still needs charging \\
$B_{min}$ [min] & minimum time buffer\\
\end{tabular}

Every 5 seconds, the battery controller calculates with \cref{eq_ev_soc_balance,eq:batt_model} whether the battery can reach the SOC goal at the intermediate time, defined as the target time minus the time buffer. 
If not, the battery is charged at maximum power for 1 minute.
For both the home batteries and the EV batteries, $B_{min}$ is set to 3 minutes. 
For the home batteries $B_{max}$ is set to 1 hour and for the EV batteries $B_{max}$ is set to 4 hours. 
The $B_{min}$ and $B_{max}$ values were obtained by testing in simulation what values would consistently meet the SOC goal in time throughout the 3 months of training data (see \cref{sec:trainingdata}).

\subsection{Safety layer}
\label{sec:safety}

A safety layer is implemented for all EMSs to ensure the maximum power limits of the main circuit breaker is not exceeded.
The safety layer is run every 5 seconds and can override the setpoints of the EMSs if the power limits would be exceeded.
The method used in this study is the \texttt{OptLayerPolicy} safety layer from \citet{ceustersAdaptiveSafetyLayer2023} - as it is considered one of the state-of-the-art methods. 

\subsection{Simulation}
\label{sec:simulation} 

A simulation of the homes is implemented in order to train the TreeC EMS and compare the real experiment with the simulation.
A simulation time step of 15 minutes was used.
The simulation models the behaviour of the home battery, EV charging, PV, electrical load and grid connection and how they react to different control strategies from EMSs.

The simulation model of the home battery is described in \cref{eq:batt_model}.

\begin{equation}
    SOC^{batt}_{t+1} = SOC^{batt}_t + \frac{P^{ch}_{t}*\eta^{batt} - P^{di}_{t}/\eta^{batt}}{E^{batt}}\Delta t
    \label{eq:batt_model}
\end{equation}

where :

\begin{tabular}{ll}
$SOC^{batt}_{t}$ [\%] & SOC of the home battery at time step $t$ \\
$P^{ch}_{t}$ [kW] & Charge power at time step $t$ \\
$P^{di}_{t}$ [kW] & Discharge power at time step $t$ \\
$E^{batt}$ [kWh] & Energy capacity of the home battery \\
$\eta^{batt}$ [\%] & Charge and discharge efficiency of the home battery \\
$\Delta t$ [h] & Time step duration in hours \\
    
\end{tabular}

The efficiency of the home battery $\eta^{batt}$ is set to 95\% \cite{farhadIntroducingEnergyEfficiency2019} for home batteries 1,2 and 4 and 96\% for battery 3 as stated in the documentation of the battery. 
The energy capacity of each home battery is stated in \cref{tab:batt_spec}.

The EV is modelled using  \cref{eq_ev_power_charge,ch2_sc3_eq_p_final_soc,eq_ev_soc_balance}. The PV and electric loads come from the measurements and are included in the simulation. The grid power is calculated with \cref{eq:grid_power}.

\begin{equation}
    P^{grid}_{t} = - P^{load}_{t} - P^{ev}_{t} + P^{pv}_{t} + P^{batt}_{t} \forall t
    \label{eq:grid_power}
\end{equation}
To simplify the modelling of the simulations, the limit imposed by the hybrid inverter on the maximum combined power from the PV installation and the home battery is not enforced (see \cref{tab:inverter_spec}).
This simplification is not expected to significantly impact the results of the simulations because the hybrid inverter limits are very rarely reached in practice for the setups used in this study.

The enforced charging behaviour described in \cref{sec:enforced_charging} is closely approximated in the simulation.
The time buffer is calculated using \cref{eq:time_buffer} at each time step.
If following the battery charge or discharge power given by the EMS does not allow reaching the SOC goal at the intermediate time, the battery is charged at the minimum power necessary to meet that goal.

\subsection{Energy management systems}

This section presents the four EMS methods used and the data used to train the MPC, RL and TreeC EMSs.
All EMSs send a setpoint to the home battery and the EV charger every 15 minutes; how they calculate this setpoint is described in this section.

\subsubsection{Training data}
\label{sec:trainingdata}
The MPC, RL, and TreeC EMSs required training data and used three months of data spanning from 2024/01/01 at 00:00 to 2024/04/01 at 00:00.
The dataset included real PV power production, household power consumption described in \cref{sec:consumption} and EV charging sessions from \cref{sec:ev_sessions}.
The MPC utilized this data to train its forecasting models, while the RL models were pre-trained on the training data with additional power profiles for the home batteries and EV chargers. 
The power profiles of the home batteries and EV chargers were generated as if an RBC EMS was controlling these assets. 
The TreeC EMS used this data in the simulator, described in \cref{sec:simulation}, to construct its decision tree. 

\subsubsection{Rule-based control}

The RBC EMS implemented in this setup consists of default rules commonly found in many devices.
The primary rule for the home battery focuses on self-consumption \enquote{behind-the-meter}. It charges when excess power is available from the PV installation and discharges to meet load demands.
The rule for the EV charger is to always charge the EV immediately at maximum allowed power.
This EMS serves as a benchmark to evaluate the performance of the other EMSs.

\subsubsection{Model predictive control}

The MPC EMS in this setup is implemented as a Mixed-Integer Linear Program, which is solved at each (15-minute) time step.
The horizon of the MPC is variable and ends at the next switching time. The MPC model is solved every 15 minutes.
The variables of the MPC are the power profiles of the home battery and the EV charger, which are the assets it needs to control.

The objective function of the MPC is described in \cref{eq:mpc_obj}.

\begin{equation}
O=\underbrace{\sum_{t\in H}{(Eo_tVo_t-Ei_tVi_t)}}_{\text{day-ahead cost}}+\underbrace{\sum_{t\in H}Eo_tVf}_{\text{offtake extras cost}}+\underbrace{Vp*\max_{{t\in H}} (\frac{Eo_t}{0.25h}, Pp)}_{\text{peak cost}}
\label{eq:mpc_obj}
\end{equation}

where:

\begin{tabular}{ll}
$H$ & Set of time steps in the MPC horizon \\
$Pp$ [kW] & Previous grid peak power for that house when running MPC \\
\end{tabular}

At the beginning of the experiment, $Pp$ is set to 2.5 kW.
This objective function penalizes heavily exceeding the previous grid peak power $Pp$.

The MPC system models take as inputs the time of the day, the SOC of the home battery and EV battery, the Belgian day-ahead hourly prices $Vd_t$ until the next switching time, the forecasted power profiles until the next switching time of the PV installation and the electric load and the forecasted final SOC and departure time of the connected EV.

The MPC constraints for the home battery, grid power, and EV are consistent with the simulation models described in \cref{eq_ev_power_charge,eq_ev_soc_balance,eq:grid_power,eq:batt_model}. However, the EV charging behaviour in the MPC formulation differs from the simulator model in \cref{ch2_sc3_eq_p_final_soc}, as detailed in \cref{eq:mpc_change_ev}. 

\begin{equation}
\ifprintedversion
\scriptsize
\fi
P^{ev}_t \leq \left\{
\begin{aligned}
&P^{ev}_{max}   &\textrm{if} \quad SOC^{ev}_{\textcolor{red}{t+1}} \leq SOC^{ev}_{cc,cv}  \\
&P^{ev}_{max}  - (P^{ev}_{max}  - P^{ev}_{min}) * \frac{SOC^{ev}_{\textcolor{red}{t+1}} - SOC^{ev}_{cc,cv}}{1 - SOC^{ev}_{cc,cv}}  &\textrm{if} \quad SOC^{ev}_{\textcolor{red}{t+1}} > SOC^{ev}_{cc,cv} 
\end{aligned}
\right.
\label{eq:mpc_change_ev} 
\end{equation}

Calculating the average maximum power of the EV charger over the next time step when the $SOC^{ev}_{cc,cv}$ is exceeded would introduce an exponential relationship to the MPC model.
The linearity of the MPC model is kept by limiting the maximum power of the EV based on the SOC of the EV at the following time step $t+1$ instead.

A constraint requires the modelled SOC of the home battery to be charged to 100\% at the next switching time. A similar constraint is imposed to ensure the EV battery reaches the forecasted final SOC at the forecasted departure time.
This forecasted SOC and departure time can be underestimated, leading to the activation of the enforced charging behaviour (see \cref{sec:enforced_charging}).
For this reason, the approximated enforced EV charging power for the next time step (see \cref{sec:simulation}) is set as the minimum EV power setpoint in the MPC model.
Finally, a constraint requires the modelled grid power not to exceed the maximum grid power of 9.2 kW.

Forecasting of electrical load and PV generation are accomplished using a Light Gradient Boosting Machine (LightGBM) model \cite{keLightGBMHighlyEfficient2017}. The model predicts the next 96 quarter-hourly values (totalling 24 hours) using the previous 96 quarter-hourly values, the time of the day and the day of the week. 
An optimal combination of hyperparameters is found using the Tree-structured Parzen Estimator \cite{bergstraAlgorithmsHyperParameterOptimization2011}. 
The ranges of hyperparameters used for the optimization are shown in Table \ref{tab:forec_hyperparameters}. 
We refrain from using the weather data provided by external services as an input feature for the PV forecast to avoid possible errors, such as the external server being down for maintenance.
For electrical load prediction, the model is trained using the original data (see \cref{sec:consumption}) and therefore the same for the four houses while for PV generation the model is trained on a building-by-building basis. 
The forecasting model for EV charging duration and demands employs k-Nearest Neighbors (k-NN) \cite{coverNearestNeighborPattern1967} regression. The k-NN models utilize the initial SOC and the hour of the day as features, leveraging cosine similarity to find the most similar historical patterns. This model has previously shown superiority in data-scarce applications, such as the prediction of EV charging session parameters \cite{genovForecastingFlexibilityCharging2024}. 

\begin{table}[H]
    \caption{Hyperparameter Ranges for LightGBM Model}
    \centering
    \begin{tabular}{@{}l||l@{}}  %
        \textbf{Parameter} & \textbf{Range} \\ 
        \hline
        Learning Rate & 0.001, 0.01, 0.1, 0.2, 0.3 \\ 
        Num Leaves & 32, 64, 128, 256 \\ 
        Max Depth & 2, 4, 6, 8, 10 \\ 
        ColSample ByTree & 0.6, 0.8, 1.0 \\ 
        Min Child Samples & 1, 2, 3, 4, 5 \\ 
    \end{tabular}
    \label{tab:forec_hyperparameters}
\end{table}

\subsubsection{Reinforcement learning}
\label{sec:rl}

The specific reinforcement learning algorithm used in this case is the twin delayed deep deterministic policy gradient (TD3) \cite{fujimotoAddressingFunctionApproximation2018}, with its implementation taken from the stable baseline \cite{raffinStableBaselines3ReliableReinforcement2021} library.

Every 15 minutes, the algorithm receives the following inputs from the system (instantaneous values at the start of each time step):

\begin{itemize}
    \item Non-controllable electrical load [kW]
    \item PV power [kW]
    \item State of charge of the home battery [\%]
    \item State of charge of the EV [\%]
    \item Belgian day-ahead hourly price ($Vd_t$) [\euro{}/kWh]
    \item Hour of the day [h]
    \item Day of the week [-]
\end{itemize}

At each time step, the RL algorithm selects an action for both the home battery and the EV charger. This action is a continuous value representing the power setpoint in kW, constrained by the respective maximum charge and discharge limits of the home battery and the maximum charge limit of the EV charger.

At the end of each time step, the RL algorithm receives a reward $R_t$ based on the previous time step $t$.
The reward is computed as follows:

\begin{equation}
R_t = - ( \underbrace{Eo_tVo_t-Ei_tVi_t}_{\text{day-ahead cost}} + \underbrace{Eo_tVf}_{\text{offtake extras cost}} ) - z_t
\end{equation}

Where $z_t = 1\euro$ if the safety layer corrected the setpoint of either the home battery or the EV charger, and $z_t = 0\euro$ otherwise. This gives a significant penalty to unsafe behaviours, encouraging the RL model to avoid them.

The RL agent is initially pre-trained using Behavioural Cloning \cite{bainFrameworkBehaviouralCloning1999}, with the simulated data of the RBC EMS over the three-month training period to reproduce the RBC's behaviour. This approach reflects a practical scenario where only historical rule-based EMS data is available, and no simulator is present, highlighting the RL algorithm's ability to learn directly from real-world data without requiring a system model.
The pre-training was performed using an 80\%-20\% train-test split and trained over a period of 7 months and 6 days of data (i.e., three times the training data without the test data).

The TF3 algorithm was implemented with the hyperparameters described in \cref{tab:RL_hyperp}.

\begin{table}[H]
    \caption{TD3 hyperparameters. The parameters are the result of an optimization study from \citet{ceustersSafeReinforcementLearning2023}, yet with an increased \texttt{train\_freq} and reduced \texttt{learning\_starts}.}
    \centering
    \begin{tabular}{l||c}
    \textbf{Parameters} & \textbf{TD3} \\
    \hline
    gamma           &  0.7 \\
    learning\_rate  &  0.000583 \\
    batch\_size     &  16 \\
    buffer\_size    &  1e6 \\
    train\_freq     &  (4, \enquote{step}) \\
    noise\_type     &  normal \\
    noise\_std      &  0.183 \\
    learning\_starts & 96
    \end{tabular}
    
    \label{tab:RL_hyperp}
\end{table}

\subsubsection{TreeC}

Using the TreeC method, one decision tree is generated to control the home battery and another one to control the EV charger for each house.
The decision trees used in the experiment are generated using the training data of \cref{sec:trainingdata}, the simulator of \cref{sec:simulation} and the objective function described in \cref{eq:tot_cost}.

Two changes have been made to the original TreeC method described in \cref{sec:treec}:

\begin{itemize}
\item The metaheuristic algorithm Covariance matrix adaptation evolution strategy seemed to end up in local minima during different training runs regularly.
The covariance matrix adaptation evolution strategy was therefore replaced by the metaheuristic algorithm particle swarm optimization \cite{kennedyParticleSwarmOptimization1995}, which obtained better training results.
This study used the particle swarm optimization algorithm implemented in the Python library pygmo \cite{biscaniParallelGlobalMultiobjective2020} with default parameters which corresponds to the canonical particle swarm optimization algorithm described in \citet{poliParticleSwarmOptimization2007}.
\item At the end of the training, the best decision tree is pruned by removing all leaves that do not make the objective value worse than 1\% compared to the unpruned decision tree.
The least used leaves are tested for removal first.
\end{itemize}

The splitting features of the decision trees are the following:

\begin{itemize}
    \item Non-controllable electrical load ($P^{load}_{t-1}$) [kW]
    \item PV power ($P^{pv}_{t-1}$) [kW]
    \item Home battery power ($P^{batt}_{t-1}$) [kW]
    \item EV charging power ($P^{ev}_{t-1}$) [kW]
    \item State of charge of the home battery ($SOC_t^{batt}$) [\%]
    \item State of charge of the EV ($SOC_t^{ev}$) [\%]
    \item Belgian day-ahead hourly price ($Vd_t$) [\euro{}/kWh]
    \item Hour of the day [h]
    \item Day of the week [-]
    \item Difference with minimum day-ahead price [\euro{}/kWh]
\end{itemize}

All power values are the average power over the previous 15-minute time step (i.e., from $t-1$ to $t$). The SOC, price and time values are instantaneous values at the start of the time step $t$.
The "difference with minimum day-ahead price" feature represents the day-ahead price shifted to have the minimum price of the ongoing experiment day equal to 0.
This additional feature is calculated by subtracting to the current day-ahead price the minimum price of the experiment day period ranging from the previous switching time to the next switching time.

The actions of the TreeC EMS include charging or discharging the home battery at different power levels and also the possibility to do self-consumption.
The action space is normalized between 0 and 1. 
If the action value is less than 0.1, then it does self-consumption; otherwise, it charges or discharges the home battery at the power level corresponding to the denormalized action value.

The TreeC EMS is trained in simulation with the particle swarm optimization algorithm over 1000 generations and a population of 1000 individuals.
This means 1\,000\,000 decision tree based EMSs are evaluated in simulation on the three months of training data and the best one is then selected.
The population size of 1000 was chosen as it obtained better results for the real-world and non-unimodal benchmarks of the IEEE Competitions in Evolutionary Computation compared to the more historically implemented population sizes of 20-50 individuals \cite{piotrowskiPopulationSizeParticle2020}.
As in classical TreeC, five different EMSs are generated for each house through training then pruning.
The best-performing EMS of the five is then used in the experiment.

The TreeC EMS obtained after training for house 1 is presented in \cref{fig:house1_decision_trees}.

\ifprintedversion
    \newcommand{\widthfigdecisiontreereal}{0.45\linewidth}
\else
    \newcommand{\widthfigdecisiontreereal}{0.45\linewidth}
\fi

\begin{figure}[H]
    \centering
    \begin{subfigure}[b]{\widthfigdecisiontreereal}
        \centering
        \includegraphics[width=\textwidth]{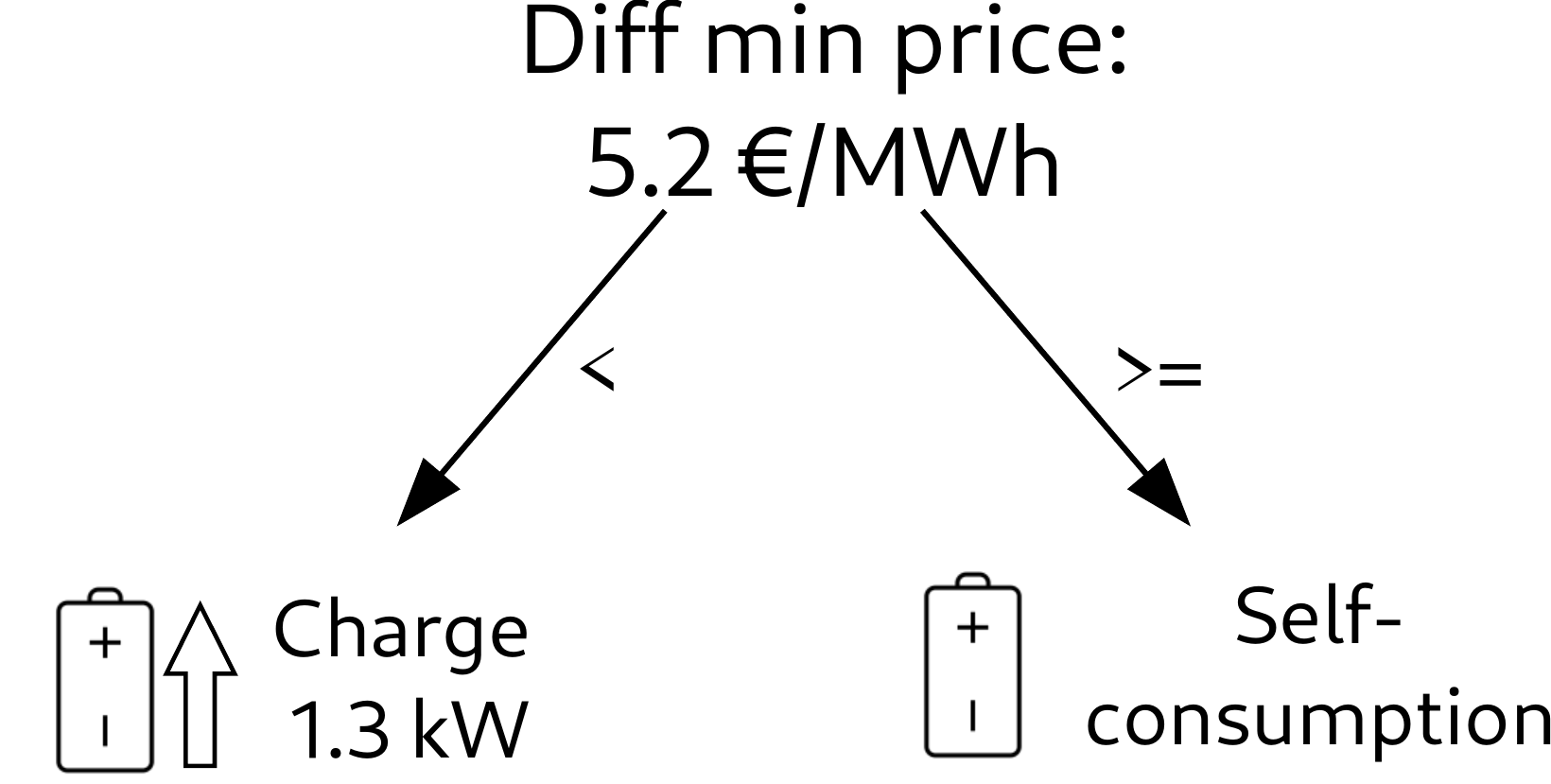}
        \caption{Decision tree for the home battery}
        \label{fig:house1_bess}
    \end{subfigure}
    \hfill
    \begin{subfigure}[b]{\widthfigdecisiontreereal}
        \centering
        \includegraphics[width=\textwidth]{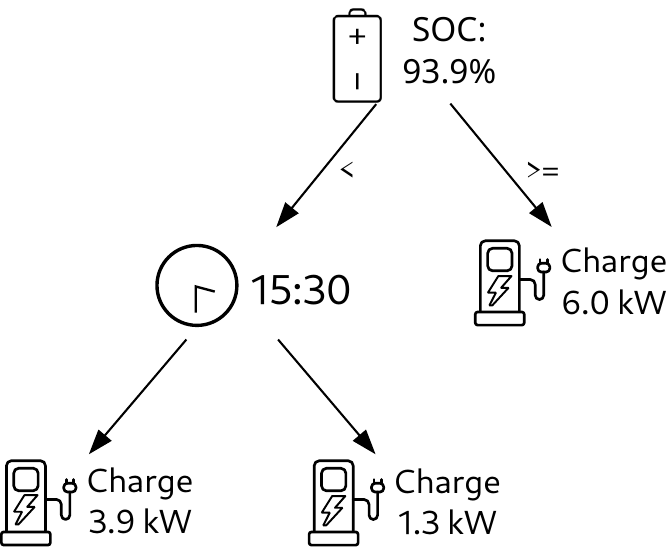}
        \caption{Decision tree for the EV charger}
        \label{fig:house1_ev}
    \end{subfigure}
    \caption{The TreeC EMS obtained after training and pruning for house 1.}
    \label{fig:house1_decision_trees}
\end{figure}

The obtained EMS is interpretable.
The home battery charges when the electricity price is close to the minimum price for the experiment day and otherwise does self-consumption. 
The EV charger sets a high charging power when the home battery is almost full; otherwise, it sets a low charging power after 15:30 and an intermediate charging power before 15:30.

Here below are the EMSs obtained for the other houses.

\begin{figure}[H]
    \centering
    \begin{subfigure}[b]{\widthfigdecisiontreereal}
        \centering
        \includegraphics[width=\linewidth]{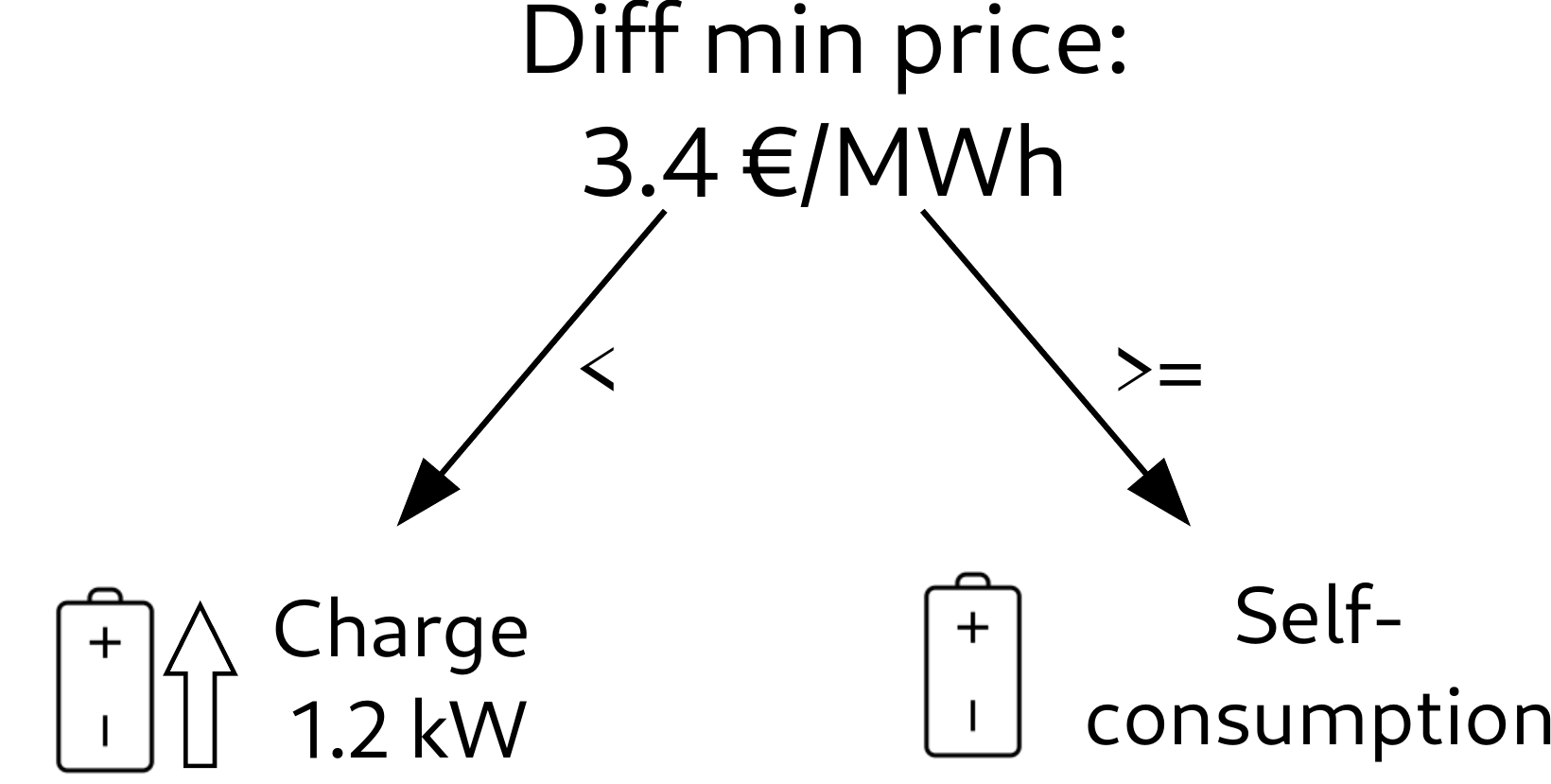}
        \caption{Decision tree for the home battery}
        \label{fig:house2_bess}
    \end{subfigure}
    \hfill
    \begin{subfigure}[b]{\widthfigdecisiontreereal}
        \centering
        \includegraphics[width=\linewidth]{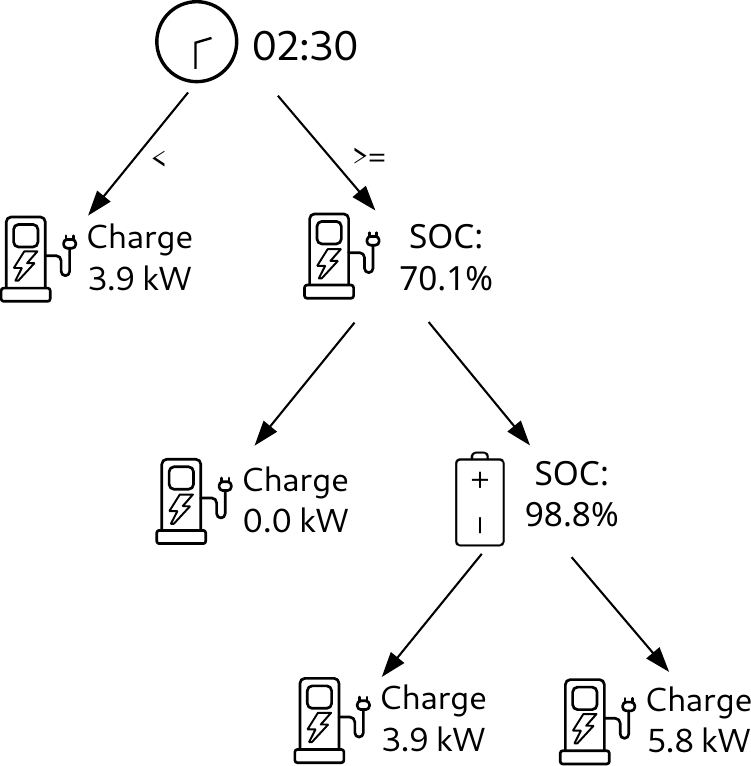}
        \caption{Decision tree for the EV charger}
        \label{fig:house2_ev}
    \end{subfigure}
    \caption{The TreeC EMS of house 2.}
    \label{fig:house2_decision_trees}
\end{figure}
\begin{figure}[H]
    \centering
    \begin{subfigure}[b]{\widthfigdecisiontreereal}
        \centering
        \includegraphics[width=\linewidth]{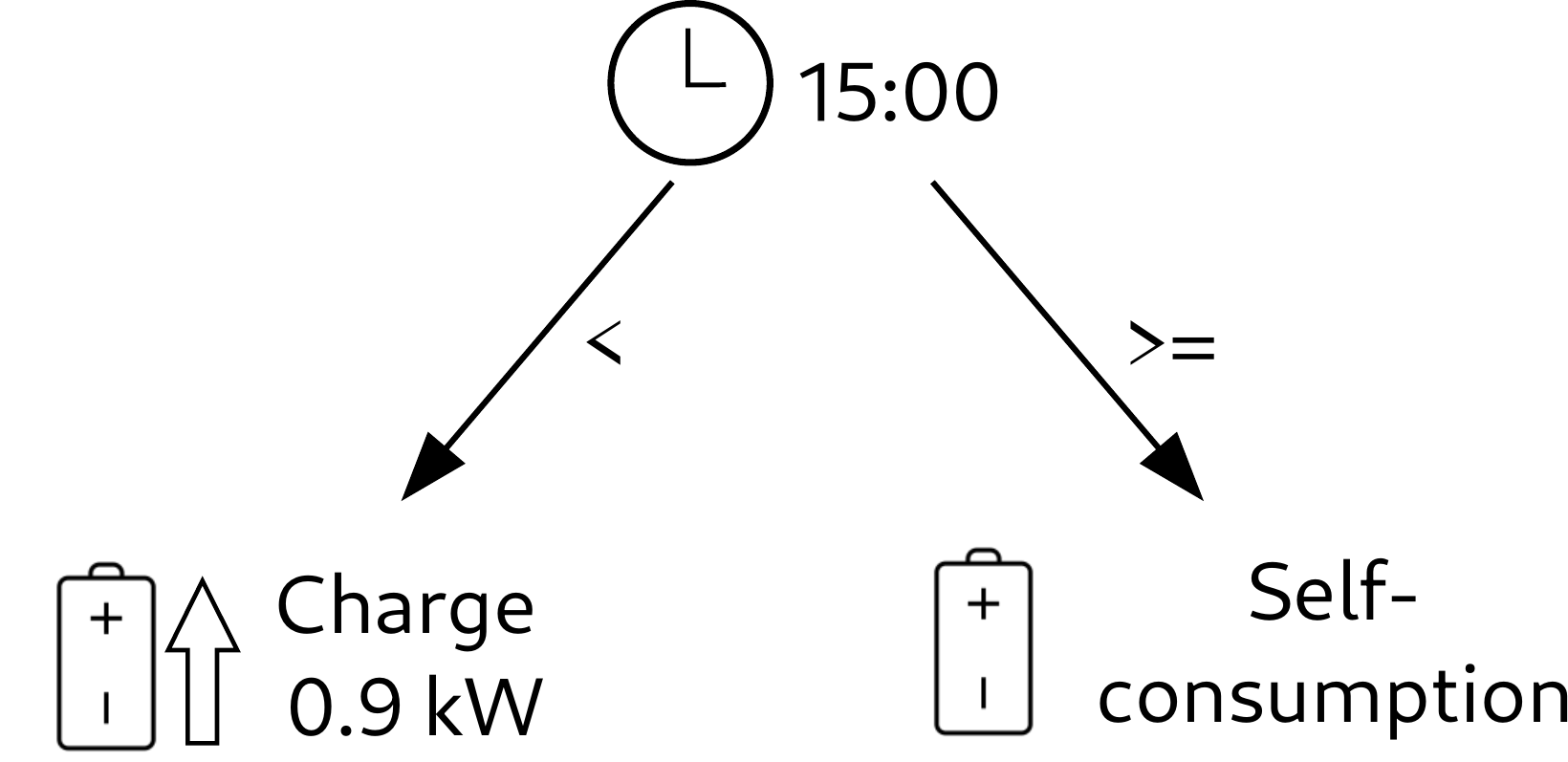}
        \caption{Decision tree for the home battery}
        \label{fig:house3_bess}
    \end{subfigure}
    \hfill
    \begin{subfigure}[b]{\widthfigdecisiontreereal}
        \centering
        \includegraphics[width=\linewidth]{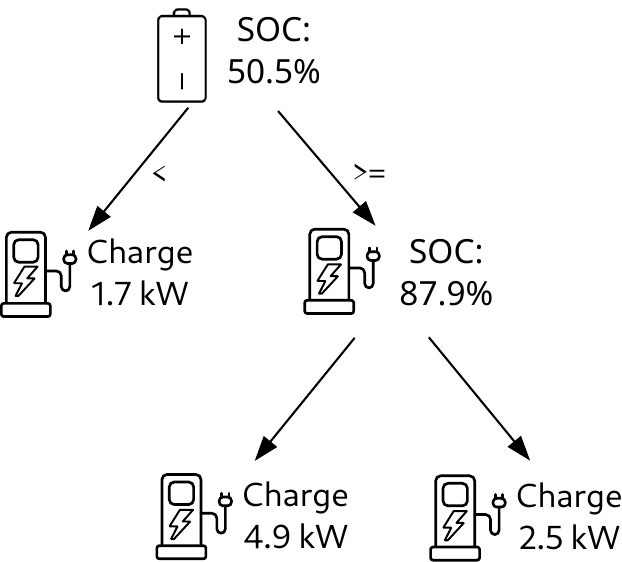}
        \caption{Decision tree for the EV charger}
        \label{fig:house3_ev}
    \end{subfigure}
    \caption{The TreeC EMS of house 3.}
    \label{fig:house3_decision_trees}
\end{figure}

\begin{figure}[H]
    \centering
    \begin{subfigure}[b]{\widthfigdecisiontreereal}
        \centering
        \includegraphics[width=\textwidth]{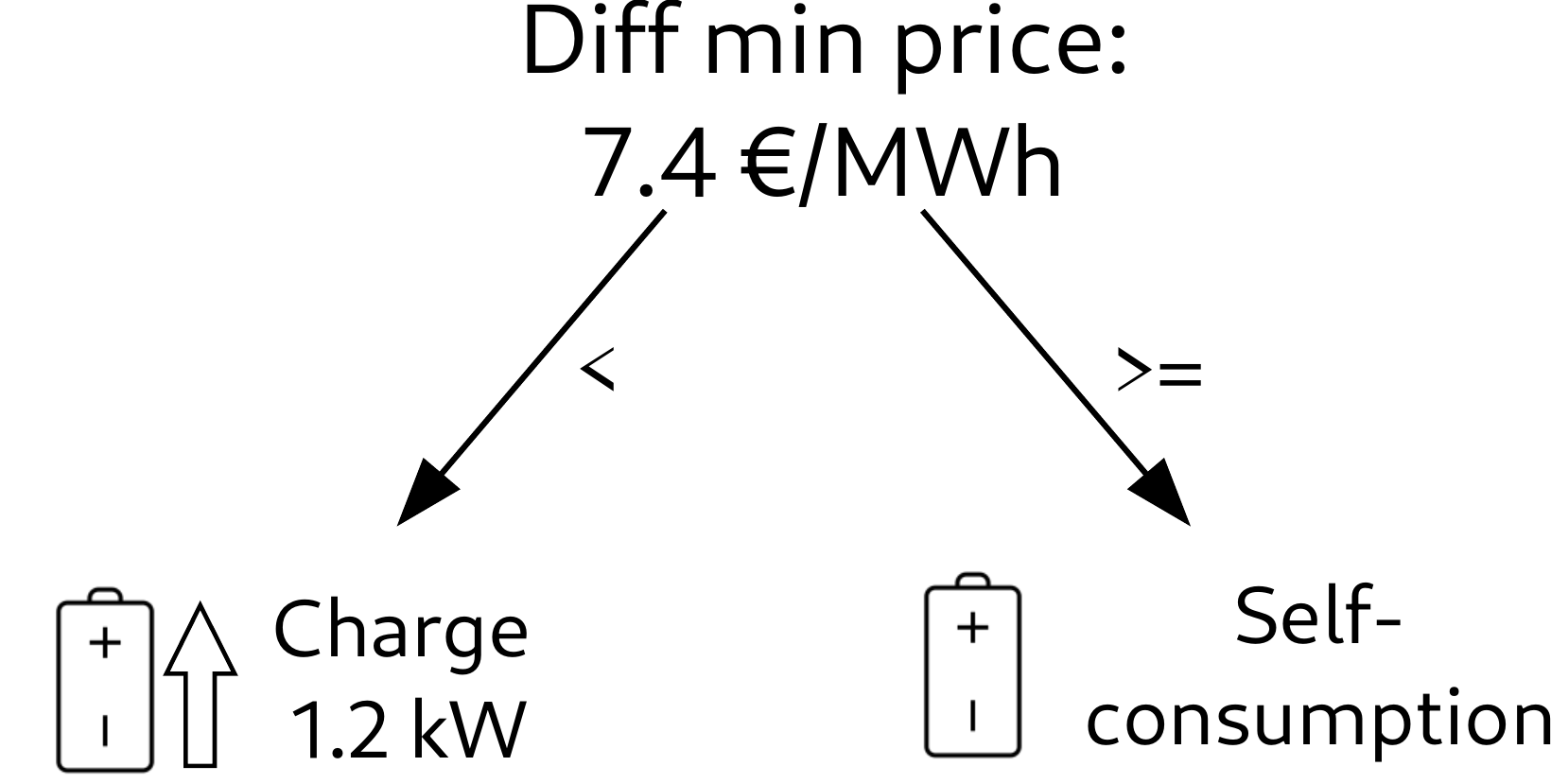}
        \caption{Decision tree for the home battery}
        \label{fig:house4_bess}
    \end{subfigure}
    \hfill
    \begin{subfigure}[b]{\widthfigdecisiontreereal}
        \centering
        \includegraphics[width=\linewidth]{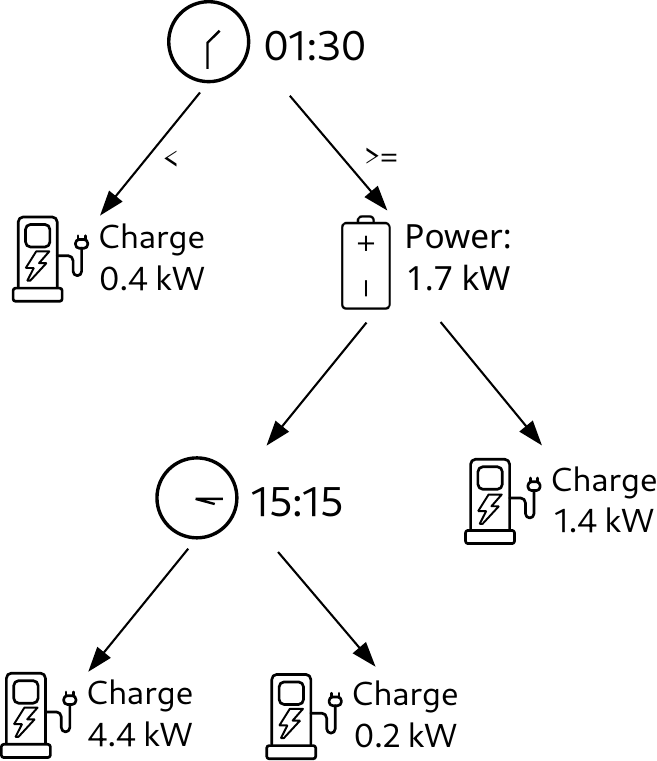}
        \caption{Decision tree for the EV charger}
        \label{fig:house4_ev}
    \end{subfigure}
    \caption{The TreeC EMS of house 4.}
    \label{fig:house4_decision_trees}
\end{figure}

\subsection{Experiment execution}

The experiment was conducted from 2024/04/11 at 15:00 to 2024/06/17 at 15:00 to obtain 48 experimental days fit for comparison.
In total, 19 experimental days were not used due to home batteries not reaching 100\% at the switching time, missing data or circuit breaker trips.
The reasons for these failures were the following:

\begin{itemize}
\item Server restart interrupting the execution program (2 days)
\item Battery of house 1 stopped responding until manual intervention (2 days)
\item Battery of house 4 stopped responding when fully charged, a bug fix was applied to limit the SOC to 95\% (1 day)
\item The bug fix for house 4 did not fix the issue. A new fix was applied to perform self-consumption above 95\% SOC (2 days)
\item A circuit breaker from the dynamic loads circuit of house 2 tripped because the breaking capacity was too low for high loads; a circuit breaker with a higher breaking capacity was used instead (6 days)
\item A previously uncaught error caused the experiment program to stop (2 days)
\item The main circuit breaker of house 3 blew because the apparent power of the house was too high for too long (2 days)
\item Missing grid measurements due to a malfunction of the grid meter of house 1 (2 days).
\end{itemize}

Some changes have been made to the experiment during its execution.
As listed above, we encountered an unexpectedly high amount of reactive power from the dynamic loads within our hardware setup (\cref{fig:setup_diagram}), causing a circuit breaker trip. 
This was because the safety layer was formulated using active power rates, essentially assuming a perfect power factor of 1. 
Therefore, halfway through the experiment, we incorporated the reactive power in the constraints - imposing limits on the grid's apparent power instead to limit the maximum main circuit breaker current correctly.
Extra measurements were taken on the main grid breaker, and the grid limit of 9.2 kVA apparent power corresponds to approximately 8.7 kW active power. 
The maximum grid power in simulation and the MPC constraints was therefore changed from 9.2 kW to 8.7 kW.
The correction was made on the 2024/5/21 at 15:00.

\newcommand{\figexample}{\begin{figure}[H]
    \centering
    \includegraphics[width=\textwidth]{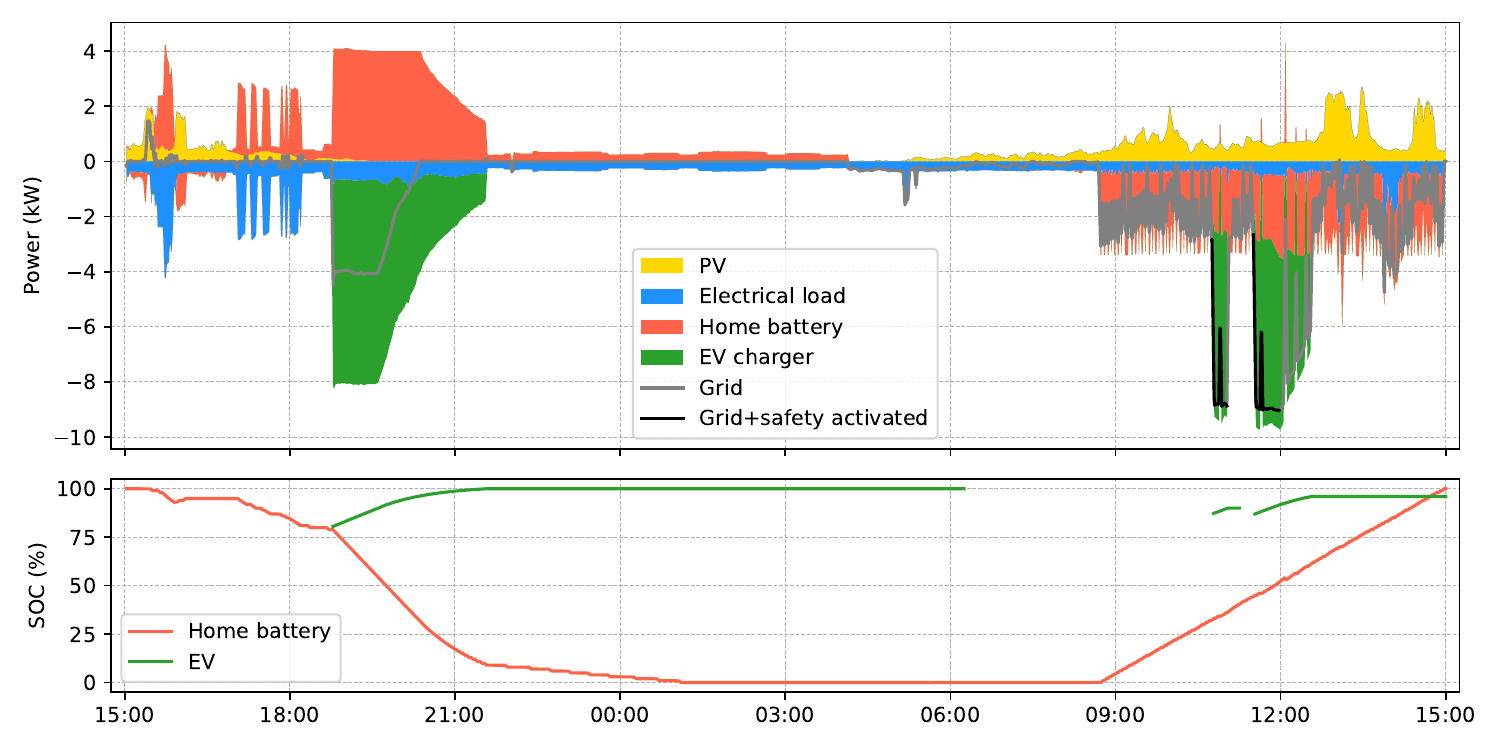}
    \caption{
    Representation of the experiment day starting on the 30th of May at 15:00. The upper plot shows the power profiles of the different assets of house 3 controlled by the RBC EMS. The grid power is represented in grey when the safety layer is not activated and in black when it is. The bottom plot shows the SOC of the home battery and EV. The enforced charging behaviour of the home battery is executed at the end of the experiment day to reach the 100\% SOC goal at 15:00. 
    }
    \label{fig:RBC_example}
\end{figure}
}

Two other minor changes were made during the experiment.
The EV SOC input used by the TreeC EMS had a value 100 times too low. This was corrected and had an impact on only one charging session without affecting the performance much. 
The error was also reproduced in the simulation (Corrected 2024/4/15 15:00:00).
The maximum SOC for the battery of house 4 was limited to 95\% as described in the bullet points above (First correction 2024/04/22 15:00:00, second correction 2024/04/30 15:00:00).

\Cref{fig:RBC_example} shows the RBC EMS controlling house 3 for an experiment day. 
The figure shows the power profiles of the different assets for this experiment day, the SOC of the EV and home battery, the enforced charging behaviour at the end of the day for the home battery and two periods when the safety layer corrected the EV and home battery powers.

\ifprintedversion
\figexample
\else
    \figexample
\fi

\section{Results and discussion}
\label{sec:results}
\subsection{EMS performance comparison}

\newcommand{\figcomparison}{
\begin{figure}[H]
    \centering
    \includegraphics[width=0.8\textwidth]{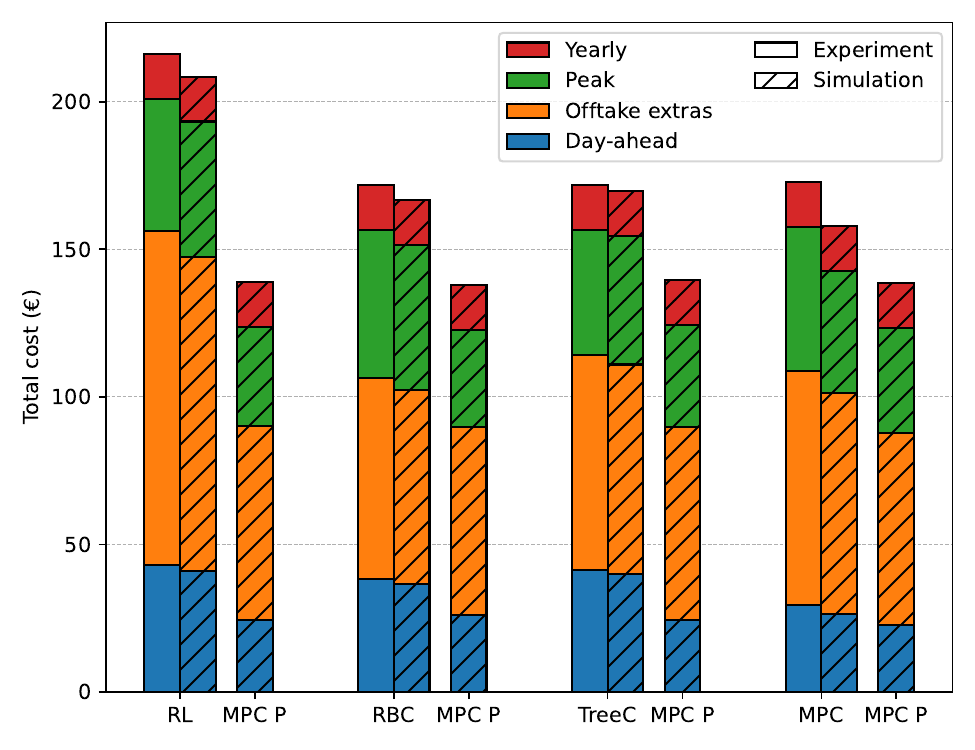}
    \caption{Costs of electricity for each EMS over the 48-day experiment from the real measurements and simulation.
    The total cost of electricity is divided into the four different costs from \cref{eq:tot_cost}.
    In addition to EMSs costs, the simulated costs of an MPC EMS with perfect forecast is also shown with the label MPC P.
    The MPC P costs are simulated using the same set of 48 days and house combinations as the EMS on its left and serve as benchmarks.
    }
    \label{fig:comparison_results}
\end{figure}

}

The RBC, TreeC and MPC EMSs obtained similar performances in the experiment as shown in \cref{fig:comparison_results}.
The RL EMS obtained a worse performance than the other EMSs, which is expected as the RL performed more exploratory actions than the other EMSs.
For the three other EMSs, them being close is unexpected but is explained in the following sections.

No EMS is close in performance to the MPC P benchmarks, indicating better implementations or methods could still be found.
The MPC P EMSs obtained scores between 137.9\euro{}\ and 139.5\euro{}\ even though they were evaluated on different day house combinations (i.e. the same day house combinations than the EMS for which they serve as a benchmark). 
The score difference is small, which indicates that the comparison procedure is fair.

\ifprintedversion
\else
    \figcomparison
\fi

\ifprintedversion
    \figcomparison
    \fi

\newcommand{\tabimportexport}{
\begin{table}[H]
    \caption{Electrical energy metrics for each EMS over the 48-day experiment from the real measurements. Net consumption is defined as the difference between grid-imported and grid-exported energy.}
    \label{tab:import_export_res}
    \centering
    \begin{tabular}{l||c|c|c|c}
    \textbf{Result} & \textbf{RL} & \textbf{RBC} & \textbf{TreeC} & \textbf{MPC}\\
    \hline
    Grid imported [kWh]       & 991.7  & 595.3 & 638.6 & 694.9 \\
    Grid exported [kWh]         & 458.3    & 91.3   & 142.2   & 190.5\\
    Net consumption [kWh]        & 533.4    & 504.0   & 496.3   & 504.4
    \end{tabular}
    
\end{table}
}

\ifprintedversion
\tabimportexport
\else
\fi

The results of \cref{tab:import_export_res} show that the RL EMS imported more electricity from the grid compared to the other EMSs, and it also has a higher net consumption. A higher net consumption is most probably caused by a higher use of the battery, which would result in more efficiency energy losses.

\ifprintedversion
\else
\tabimportexport
\fi

\subsubsection{Rule-based control}

The strategy of the RBC is to do as much self-consumption as possible, which means trying not to take electricity off the grid as shown in \cref{tab:import_export_res}.
This strategy is reflected in a comparatively low offtake extras cost, which is proportionally the highest cost category for all EMSs.

\subsubsection{Reinforcement learning}

The RL EMS obtained the worst score, mainly due to a higher offtake extras cost (distribution and taxes) as can be seen in \cref{fig:comparison_results} - which is not unexpected as we observed early-stage learning despite the pre-training, given the new (real) experience tuples.
Moreover, even though the RL agents were pre-trained offline with a 3-month-long historical dataset, each agent-house combination only had an experimental runtime of 1.7 weeks (12 days). When comparing to relevant simulated case studies, more required training time is also observed. For example, \citet{ceustersModelpredictiveControlReinforcement2021} reported 95\% MPC-like performance of the RL EMS after \(\sim18.6\) weeks and only matched the MPC EMS after \(\sim2\) years - also using 15-minute time steps.

\subsubsection{TreeC}

\ifprintedversion
\newcommand{\widthfigtreerbc}{0.75\linewidth}
\else
\newcommand{\widthfigtreerbc}{0.8\linewidth}
\fi

\newcommand{\fighousetreerbc}{
\begin{figure}[H]
    \centering
    \includegraphics[width=\widthfigtreerbc]{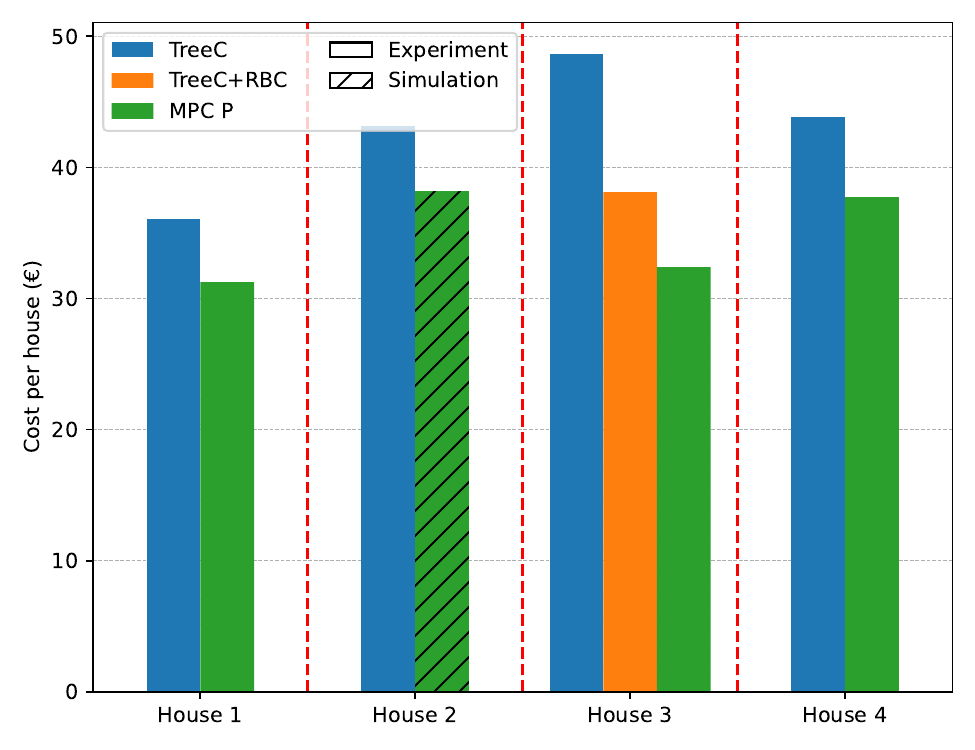}
    \caption{Comparison for each house of the total cost obtained with the TreeC EMS and the benchmark MPC P EMS. 
    For house 3, the TreeC+RBC EMS is added to the comparison and represents the TreeC EMS using the self-consumption control strategy of the RBC EMS to control the home battery.}
    \label{fig:house3_tree_rbc}
\end{figure}
}

\ifprintedversion
\fighousetreerbc

\fi

The TreeC EMS obtained the lowest peak cost but couldn't obtain as low offtake extras and day-ahead costs as the other MPC and RBC EMSs.
Looking at the decision trees obtained in \cref{fig:house1_decision_trees,fig:house2_decision_trees,fig:house3_decision_trees,fig:house4_decision_trees}, the low peak cost is explainable by lower charging setpoints of the EVs.

\ifprintedversion
\else
\fighousetreerbc
\fi

Intuitively, it is surprising that the TreeC EMS did not perform better than the RBC EMS.
\Cref{fig:house3_tree_rbc} shows that the benchmark costs of the MPC P EMS are 10-15\% lower than the TreeC EMS costs for houses 1,2 and 4.
For house 3, though, the MPC P EMS cost is 33\% lower than the TreeC EMS cost, showing a much larger difference than for the other houses.
Looking at the decision trees obtained from the TreeC training (see \cref{fig:house1_decision_trees,fig:house2_decision_trees,fig:house3_decision_trees,fig:house4_decision_trees}), the main difference between houses 1,2 and 4 and house 3 is the home battery strategy.
Houses 1,2 and 4 charge their battery when the electricity price is very low and otherwise do self-consumption.
House 3, on the other hand, does self-consumption until midnight and then charges the battery until the switching time at 15:00.
This strategy could well be valid in the training months of January to March when there is less PV production but during the experiment months of April to June, the strategy would lead to more offtake from the grid and, therefore, offtake extras costs compared to the self-consumption strategy of the RBC EMS.
When replacing the TreeC home battery strategy of house 3 with the self-consumption strategy of the RBC EMS in simulation, the MPC P cost is now only 15\% lower for that house as shown in \cref{fig:house3_tree_rbc}.
The TreeC+RBC policy for house 3 would have reduced the total cost of the TreeC EMS by ~10.5 \euro{}, making it better than the RBC EMS over the total cost of the 4 houses (see \cref{fig:comparison_results} for original total costs).

Overall, this shows the importance of using data from a training period that is representative of the experiment or test period.
In this case, the training data was not representative enough of the experiment, which led to the TreeC EMS obtaining a bad control strategy for the home battery of house 3.

\subsubsection{Model predictive control}
\label{sec:mpc_res}

The MPC EMS obtained the lowest day-ahead costs but couldn't obtain as low of a peak and offtake extras costs as the RBC and TreeC EMSs.
The low day-ahead cost is expected as the MPC knows the future Belgian day-ahead prices of electricity and can optimize accordingly.
Intuitively, it is surprising that the MPC did not optimize the other costs efficiently.

\begin{figure}[H]
    \centering
    \includegraphics[width=0.8\textwidth]{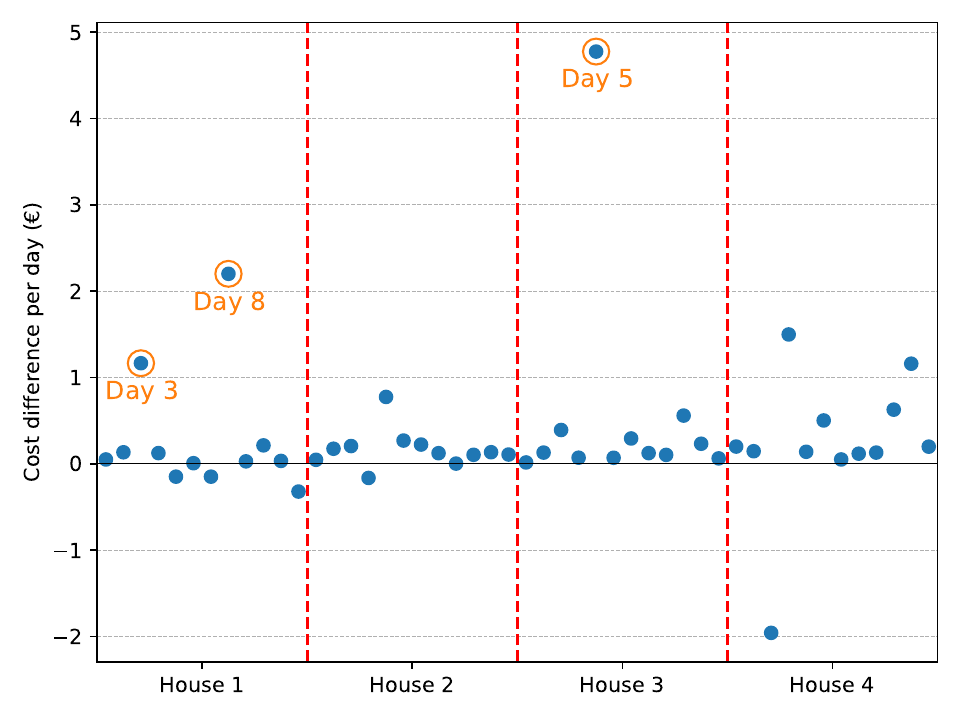}
    \caption{Difference in costs between the experiment and the simulation for the MPC EMS for each day.
    }
    \label{fig:diff_mpc}
\end{figure}

\Cref{fig:comparison_results} shows a bigger cost difference between simulation and reality than the other EMSs and obtained lower peak, offtake extras and day-ahead costs.
Certain days had cost differences exceeding 1 \euro{}, as seen in \cref{fig:diff_mpc}.
A deeper analysis was performed to understand these differences on three days that had this cost difference for houses 1 and 3. 
House 4 also had two days when the simulation results were worse than the experiment by more than 1 \euro{}.
However, on one day, the simulation outperformed the experiment by approximately 2 \euro{}, which mitigates the impact of the two worse days.

In all three instances, the main driver behind the cost differences is the peak cost.

\begin{itemize}
    \item Day 5 of house 3, a difference of 4.8 \euro{}: On the first time step of the day, a new short charging session forced the EV charger to charge at maximum power. 
    In the simulation, this peak was avoided by discharging the home battery at the same time, but this did not occur in the experiment.
    The MPC EMS in the experiment decided on the setpoint for the home battery and EV charger without receiving the information that an EV session would start at that time step and, therefore, did not set an appropriate home battery setpoint.
    This is a coding error that would happen when an EV session starts at the first time step of the day.
    \item Day 8 of house 1, a difference of 2.2 \euro{}: The peak of the day was higher in the experiment than in the simulation because the MPC set a higher charging power for the EV charger than in the simulation at midnight.
    The MPC EMS forecasted that there would be some PV production at that time because it received the PV production of the previous day in a wrong order.
    This wrong order was due to the device failing to take the grid measurements.
    The device failed to collect data from the \nth{29} of May at 10:45 to the \nth{30} of May at 06:45, and the investigated day started on the \nth{30} of May at 15:00.
    The forecaster of the MPC requested the data from the previous 24 hours, but due to the missing data and the formatting functions not sorting by default on time; the data was received starting from 06:45 until the time of request and then from 24 hours before the time of request until 06:45.
    This mixed order caused the forecaster to predict a PV production between 300W and 400W during nighttime and at a time when the MPC EMS decided to charge the EV at power close to the previous peak, thus obtaining a higher peak in the experiment than in the simulation for the day.
    \item Day 3 of house 1, difference of 1.7 \euro{}: Close to the end switch time of the investigated experiment day, the MPC EMS managed in simulation to charge the home battery at 100\% SOC and complete an EV session without increasing the peak cost. In contrast, in the experiment, the peak was increased due to bad management of the EV session and home battery 100\% SOC switching time constraint.
    The error came from a difference in the experiment and simulation MPC. Whenever the forecaster would predict a departure time later than 15:00, the simulation MPC would set the departure time to 15:00 instead. This correction was mistakenly not applied in the experiment MPC.
\end{itemize}

These errors were reproduced in simulation, and the total cost went from 157.8 \euro{} to 165 \euro{}. 
This brings the simulation cost closer to the experimental total cost of 172.8 \euro{} by half of the original difference.
A better implementation of the MPC EMS in the experiment would improve its performance.
On the other hand, this shows that the peak cost can penalize the performance of the MPC if there are some errors in the implementation.
EMSs based on less complex models and inputs, such as the RBC and TreeC EMS, did not experience similar errors.
Also, it would have been difficult to notice these errors without a simulation reproducing the MPC behaviour.

\subsection{Difference experiment vs simulations}

The following section analyses differences between simulation and experiment in addition to the ones already pointed out specifically for the MPC EMS in \cref{sec:mpc_res}.

The experiment runs in real-time while the simulation approximates this behaviour on a 15-minute time step.
In the experiment, the grid sensor measures both imported and exported energy for the same time step, while in simulation, there is only one net consumption value per time step.
The net consumption is the difference between the imported and exported energy from the grid.
Calculating the cost for one time step using both the imported and exported energy instead of the net consumption is always more expensive because offtake extras costs get removed when doing the energy difference for net consumption.
\Cref{tab:energy_diff} shows the score obtained in the experiment for each EMS when using the import and export energy values as well as the score obtained when using the net consumption to understand if the approximation of the simulation has a big impact on the results.

\begin{table}[H]
    \caption{Costs of electricity obtained for each EMS in the experiment when taking the export and import energy values and when calculating net consumption.}
    \centering
    \begin{tabular}{l||c|c|c}
    \textbf{EMS} & \begin{tabular}{@{}c@{}}\textbf{Experiment} \\ \textbf{(import/export)}\end{tabular} & \textbf{ } \begin{tabular}{@{}c@{}}\textbf{Experiment} \\ \textbf{(net consumption)}\end{tabular} & \begin{tabular}{@{}c@{}}\textbf{Simulation} \\ \textbf{(net consumption)}\end{tabular}\\
    \hline
    RL & 216.1\euro{} & 207.2\euro{} &  208.6\euro{} \\
    RBC & 171.7\euro{}  & 171.2\euro{} & 166.6\euro{} \\
    TreeC & 171.7\euro{} & 170.7\euro{} & 169.9\euro{} \\
    MPC & 172.8\euro{} & 170.1\euro{} & 157.8\euro{} \\
    \end{tabular}
    
    \label{tab:energy_diff}
\end{table}

The cost difference between import/export and net consumption is larger for the RL and MPC EMSs than for the TreeC and RBC EMSs.
A major behavioural difference between the MPC and RL EMSs with the TreeC and RBC EMSs is that the former control the home battery's setpoint while the latter often or exclusively do self-consumption.
Doing self-consumption allows the battery to adapt to real-time changes and has no or very little imported and exported energy when the battery is not full or empty.
Giving a fixed setpoint to the home battery for a whole time step, on the other hand, does not allow the battery to adapt to real-time changes and can lead to both imported and exported energy during the same time step.

A solution to this problem could be to require the home battery to maintain the grid power at a certain setpoint instead of keeping a fixed battery power for the whole time step.
Further simulation and experimentation would be needed to confirm if this is a good solution.

Both the MPC differences from \cref{sec:mpc_res} and the grid energy exchange calculation difference from this section explain some differences between the experiment and the simulation.
Other differences observed between experiment and simulation but more difficult to quantify in terms of cost include:
\begin{itemize}
    \item Battery capacities and efficiencies not being exactly the same as on the specification sheet.
    \item Batteries 1 and 2 having a clear minimum SOC of 5\%.
    \item Battery 4 having a variable minimum SOC, an upped soc of 95\% to avoid the battery not responding and a recurrent jump from ~85\% to 95\% SOC due to internal recalibration.
    \item The real-time safety layer of the experiment is approximated on a 15-minute time step in the simulation.
    \item The hybrid inverter's power limit is not modelled in the simulation.
\end{itemize}
With the MPC errors of \cref{sec:mpc_res} reproduced in simulation, the difference in total cost between experiment and simulation ranges from 1\% to 4.5\% for all EMSs. When calculating the experiment's total cost based on the net consumption, this range goes down to between  0.5\% to 3\% difference.
The remaining difference between the experiment and simulation can be explained by the points listed above and some other unnoticed differences but, in total, they do not account for a large difference in costs. 
As stated in \cref{sec:hardware}, the electrical loads and possible power losses of the system are not separately measured; therefore, the simulation uses data combining the two.
The possible power losses could depend on the home battery and EV usage, which would not be reflected in the simulation and needs to be further investigated in future work.
Nonetheless, the simple battery model (see \cref{eq:batt_model}) and the capacity and efficiency parameters provided by the battery manufacturers (see \cref{tab:batt_spec}) were sufficiently accurate to perform meaningful simulations.

\section{Conclusion}
\label{sec:conclusion}
This chapter presented the results of the real-world experimental validation of a novel machine learning method using a metaheuristic algorithm to generate an interpretable control policy modelled as a decision tree (i.e., TreeC \cite{ruddickTreeCMethodGenerate2025}). 
This, together with the comparison between an MPC, a RL, an RBC and an explainable Tree-based EMS on any case study, simulated or otherwise, using a novel evaluation procedure, aimed to be as fair as possible. We come to the following conclusions:

\begin{itemize}
    \item The RBC, TreeC and MPC EMSs obtained a similar economic performance with only a 0.6\% difference in cost between the three. The results highlight the importance of representative training data for the TreeC EMS and expert implementation for the MPC EMS to obtain better performances in real implementations.
    \item A pre-trained, yet model-free, RL agent still has a considerably long training time (even with hyperparameters chosen for a higher learning rate) and, therefore, monetary training cost that would need to be set off against the extra (if any) cost of other methods (e.g. modelling time).
    \item With the MPC experiment errors reproduced in simulation, the difference in costs between experiment and simulation ranges from 1\% to 4.5\% for all EMSs.
    This difference is in part explained by the simulation using a single net consumption value per time step for the energy exchanged with the grid, while the experiment measures both imported and exported energy per time step. 
    Overall, the simulations rather accurately represented reality in terms of cost and were essential to finding the MPC errors and TreeC training data issues.
\end{itemize}

While this chapter validates the results using real hardware, the testing occurred in a controlled environment. The next logical step involves testing the EMSs with real users to introduce genuine user needs and uncertainties into the system.

\mode<all> %

    \chapter[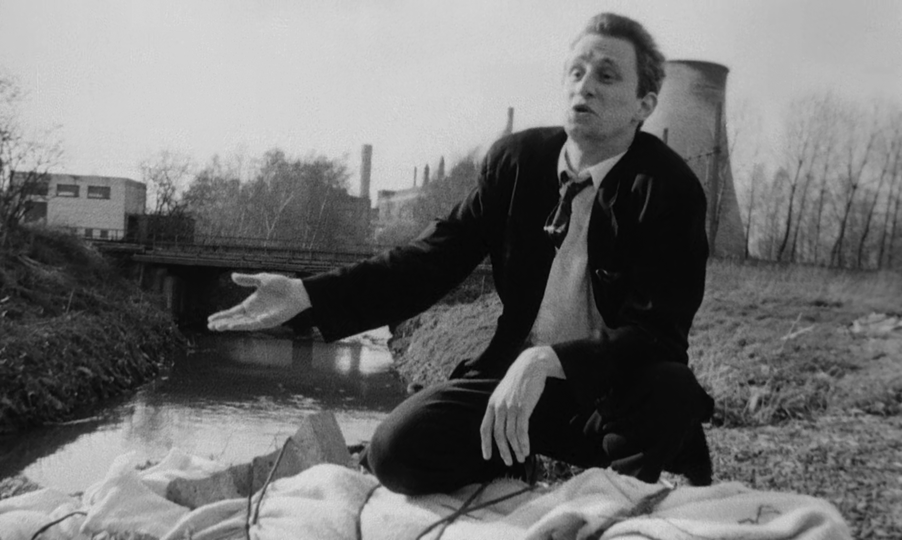]{Analysis of EMSs for electrified Belgian dwellings\label{ch:ems_relevance}}

\section{Introduction}

\begin{figure}
    \centerline{\includegraphics[width=0.6\linewidth]{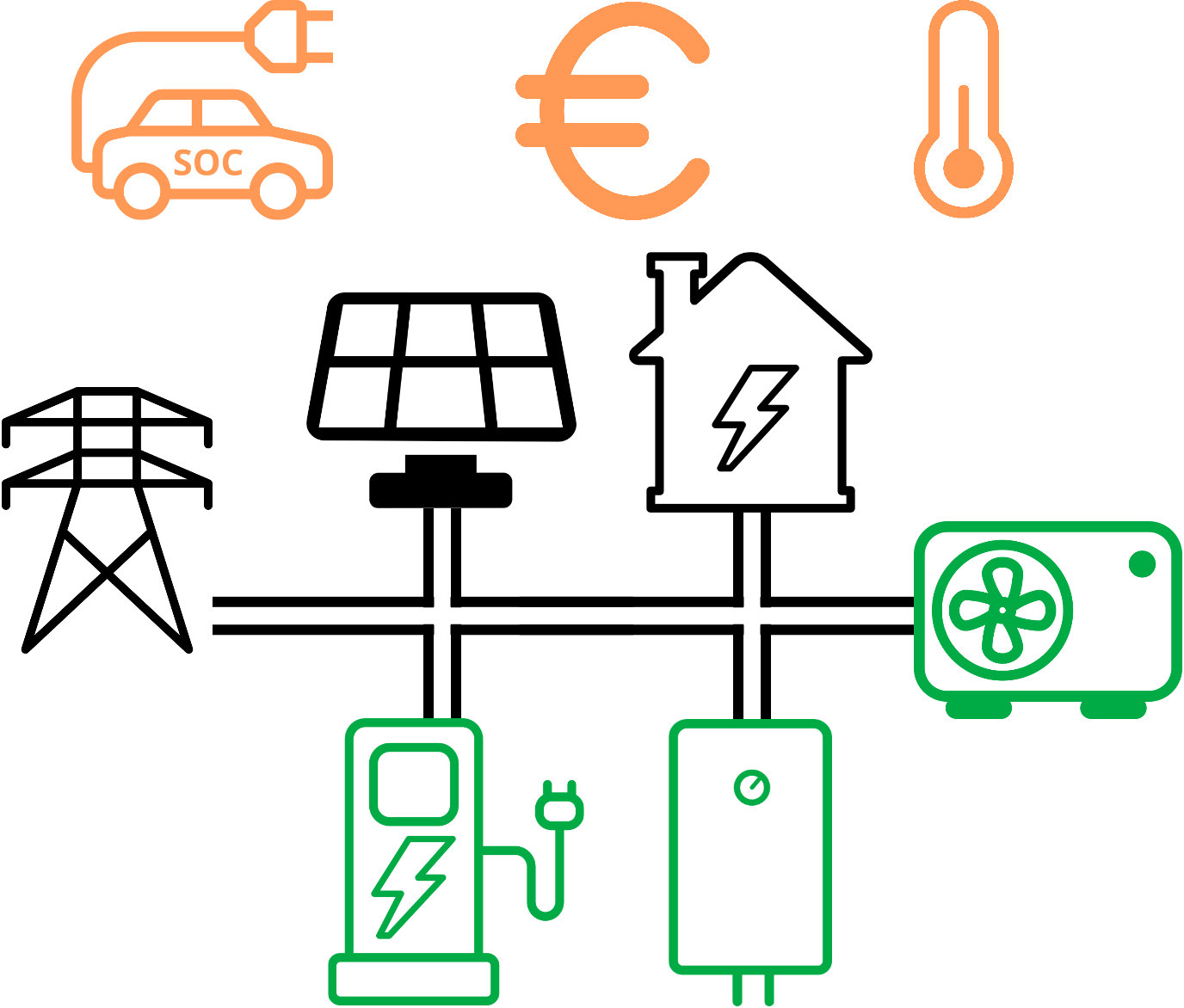}}
    \caption{Representation of the energy grid analysed in this chapter. The optimized metrics indicated in orange are EV charge completion, electricity cost, and thermal comfort. The controlled assets indicated in green are the EV charger, water heater, and heat pump.}
    \label{fig:relevance_scheme}
\end{figure}

An effective management of energy assets in residential buildings is becoming increasingly relevant in Belgium.
This is driven by the increasing electrification of our energy systems, the adoption of dynamic electricity pricing, and the implementation of grid peak tariffs in the residential sector \cite{ruddickRealworldValidationSafe2024}.
As these systems and regulations become more complex, the question arises: What are the benefits of implementing an advanced energy management system in electrified dwellings ? 

This chapter aims to answer this question by simulating a set of representative Belgian dwellings which have been electrified and are subject to dynamic pricing and grid peak tariffs.
The electrification of the dwellings is achieved by replacing the traditional heating systems with air-to-air heat pumps, adding domestic electric vehicle chargers due to the replacement of petrol cars with electric vehicles, and replacing gas water heaters with electric water heaters.
Other dwelling characteristics such as the presence of photovoltaic panels, insulation level, and construction type are kept close to current Belgium situation according to the most recent statistics we could find.
Real-world datasets are used for base electricity consumption, domestic water usage, electric vehicle charging sessions, photovoltaic production, weather data, and electricity tariffs. Two years of data (2022 for training, 2023 for evaluation) are used, reflecting a realistic scenario where a household adopts an energy management system (EMS)
 in January 2023 based on data from the previous year.
Home batteries are excluded from the analysis, as their adoption remains limited in Belgium. According to available statistics, about 3\% of dwellings had a home battery in 2024 \cite{vangastelHardeCijfersVorig2025}. Replacing gas boilers with heat pumps and petrol cars with electric vehicles is essential for reducing fossil fuel use, while the case for home batteries is less evident.

This study compares three distinct EMS strategies: rule-based control (RBC), which mimics the default operation of residential energy assets; model predictive control (MPC), a optimization-based approach; and TreeC, a machine learning-based method that models EMS strategy as decision trees.
In addition to these realistic EMSs, a model predictive control EMS with perfect forecast (MPC P) is also compared to show a maximum performance an EMS could technically achieve.

Buying a home EMS through a professional is advertised as costing between 280\,\euro{} and 680\,\euro{} with some additional subscription costs ranging between 0\,\euro{} and 360\,\euro{} per year \cite{maakjemeterslimAlleEMSProducten2025}. An EMS can be implemented cheaper using for example Home Assistant and the additional EMHASS module \cite{homeassistantHomeAssistant2025,davidusb-geekEnergyManagementHome2025} for which the basic hardware costs close to 100\,\euro{} but usually requires some technical knowledge to set up and configure.
To these basic costs can be added the maintenance, more expensive compatible equipment and additional necessary sensors.

This chapter addresses the following research questions for a representative set of electrified Belgian dwellings:

\begin{itemize}
    \item What are the benefits of EMSs like TreeC and MPC compared to a simple RBC in terms of cost savings, thermal comfort and EV charging satisfaction ?
    \item How much can algorithmic improvements realistically improve these benefits ?
    \item What is the influence of different dwelling characteristics on cost savings ?
\end{itemize}

\section{Building a representative set of Belgian dwellings}

This section describes the methodology for constructing a set of 500 representative Belgian dwellings. Each dwelling is first assigned a set of characteristics (such as region, number of rooms, and insulation level) based on the latest available Belgian statistics (see \cref{sec:dwelling_characteristics}). Using these characteristics, a simulation model is created for each dwelling, incorporating real-world data sources (see \cref{sec:thermal_data_and_models,sec:consumption_profiles,sec:ev_charger,sec:pv_panels,sec:hot_water}). This process yields 500 unique simulation models, each reflecting its own set of characteristics and data inputs. The overall methodology is illustrated by the flow chart in \cref{fig:dwelling_flow_chart}.

\newcommand{\hconnect}[2]{%
    \draw[thick,->] (#1.east) -- ++(0.564,0) |- (#2.west);
}
\newcommand{\vconnect}[2]{%
    \draw[thick,->] ([xshift=-1.3cm]#1.south) |- (#2.west);
}
\newcommand{\hshortconnect}[2]{%
    \draw[thick,->] (#1.east) -- ++(0.15,0) |- (#2.west);
}
\newcommand{\hlongconnect}[2]{%
    \draw[thick,->] (#1.east) -- ++(3.765,0) |- (#2.west);
}

\newlength{\defaultwidth}
\setlength{\defaultwidth}{2.9cm}

\newlength{\xshiftsubcategory}
\setlength{\xshiftsubcategory}{\dimexpr\defaultwidth/7\relax}

\newlength{\middlewidth}
\setlength{\middlewidth}{4.4cm}

\newlength{\rightwidth}
\setlength{\rightwidth}{3.2cm}

\newlength{\pvoutputwidth}
\setlength{\pvoutputwidth}{\dimexpr\middlewidth-1cm\relax}

\newlength{\hpxmlwidth}
\setlength{\hpxmlwidth}{\dimexpr\middlewidth-0.3cm\relax}

\newlength{\energypluswidth}
\setlength{\energypluswidth}{\dimexpr\middlewidth/2\relax}

\newlength{\belowleftrow}
\setlength{\belowleftrow}{0.5cm}

\begin{figure}[H]
    \centering
    \begin{tikzpicture}[
        node distance=0.8cm,
        every node/.style={draw, rounded corners, font=\scriptsize, align=center, fill=blue!20, minimum width=\defaultwidth, minimum height=0.9cm},
        every edge/.style={draw,->,thick},
    ]
    \node (all) at (0,0) {Applied to all \\dwellings};
    \node[below=\belowleftrow of all] (region) {Region \\ 
    \Cref{tab:stats_region}};
    \node[below=\belowleftrow of region, xshift=\xshiftsubcategory] (pv) {PV \\ \Cref{tab:stats_pv}};
    \node[below=\belowleftrow of pv] (garage) {Garage and car \\ \Cref{tab:stats_home_charger}};
    \node[below=\belowleftrow of garage] (central) {Central heating \\ \Cref{tab:stats_heating}};
    \node[below=\belowleftrow of central] (construction) {Construction type \\ \Cref{tab:stats_construction_type}};
    \node[below=\belowleftrow of construction] (insulation) {Insulation level \\ \Cref{tab:stats_insulation}};
    \node[below=\belowleftrow of insulation] (rooms) {Num. rooms \\ \Cref{tab:stats_rooms}};

    \node[below=\belowleftrow of rooms, xshift=\xshiftsubcategory] (surface) {Surface area \\ \Cref{tab:stats_surface_area}};
    \node[below=\belowleftrow of surface] (bedrooms) {Num. bedrooms \\ \Cref{tab:stats_bedrooms}};

    \node[right=1.315cm of all, minimum width=\middlewidth, fill=orange!20] (elecprof) {\textbf{Electric consumption}\\
    Profile data: \\ 
    \citet{schlemmingerDatasetElectricalSinglefamily2022}};
    \node[right=0.9cm of pv,yshift=0.85cm, minimum width=\middlewidth, fill=orange!20] (pvprof) {\textbf{PV} \\ Profile data: \\
      \includegraphics[width=\pvoutputwidth
      ]{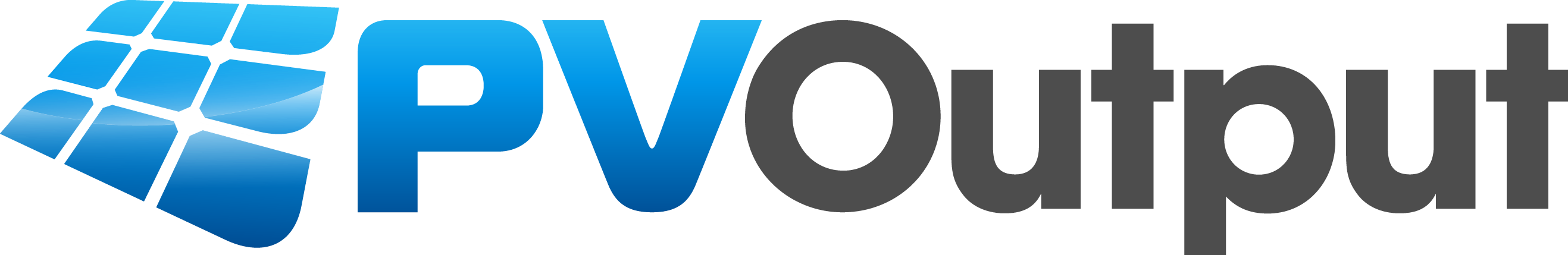}};
    \node[right=0.9cm of garage, minimum width=\middlewidth, fill=orange!20] (evprof) {\textbf{EV and charger} \\ 
    EV profile data: 
    \\ \citet{sorensenElectricVehicleCharging2024} \\
    Battery model: \\ \Cref{eq:ev_charging_reliability}};

    \node[below=0.5cm of evprof, minimum width=\middlewidth, fill=orange!20] (hpxml) {Inputs into: \\\includegraphics[width=\hpxmlwidth]{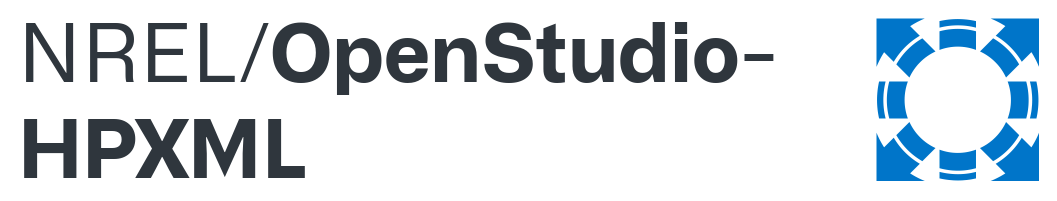}\\ Output: \\Thermal model (.idf)};

    \node[below=0.5cm of hpxml, minimum width=\middlewidth, fill=orange!20] (heatmodel) {\textbf{Thermal and heat pump}\\
    Thermal model (.idf)\\
    +\\
    Brussels weather data (.epw) \\
    +\\
    \includegraphics[width=\energypluswidth]{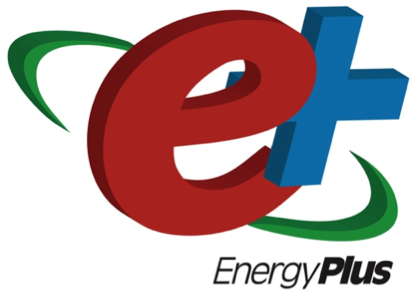}
    };

    \node[below=0.5cm of heatmodel, minimum width=\middlewidth, fill=orange!20] (hot_water) {Average hot water use from \\num. bedrooms \cite{meirelesDomesticHotWater2022}
     \\
    +\\
    Water consumption data:\\
    \citet{athanasiouDAIADOpenWater2014}\\ Output: \\
    Hot water profile};
    \node[below=0.5cm of hot_water, minimum width=\middlewidth, fill=orange!20] (dhwmodel) {\textbf{Domestic hot water}\\
    Hot water profile\\
    +\\
    Water heater model:\\ \Cref{eq:hot_water_params,eq:tank_temp_change}};
    
    \node[right=0.4cm of evprof, minimum width=\rightwidth, fill=green!20] (evcontrol) {\textbf{EV charger}\\
    Power setpoint};
    \node[below=3.5cm of evcontrol, minimum width=\rightwidth, fill=green!20] (heatcontrol) {\textbf{Heat pump}
    \\ Heating and \\cooling 
    temperature \\setpoints};
    \node[below=3.5cm of heatcontrol, minimum width=\rightwidth, fill=green!20] (dhwcontrol) {\textbf{Water heater}\\ Power setpoint};

    \path (evcontrol) -- (heatcontrol) coordinate[midway] (midctrl);

    \node[right=2cm of midctrl,yshift=-1cm, minimum width=0.5cm, fill=red!20] (dwsim) {Dwelling \\ simulation};

    \hconnect{all}{elecprof}
    \hconnect{region}{pvprof}
    \vconnect{region}{pv}
    \vconnect{region}{garage}
    \vconnect{region}{central}
    \vconnect{region}{construction}
    \vconnect{region}{insulation}
    \vconnect{region}{rooms}

    \hshortconnect{pv}{pvprof}
    \hconnect{garage}{evprof}
    \hconnect{central}{hpxml}
    \hconnect{construction}{hpxml}
    \hconnect{insulation}{hpxml}
    \vconnect{rooms}{surface}
    \hshortconnect{surface}{hpxml}
    \draw[thick, ->] (hpxml) -- (heatmodel);
    
    \vconnect{rooms}{bedrooms}
    \hshortconnect{bedrooms}{hot_water}
    \draw[thick, ->] (hot_water) -- (dhwmodel);

    \hlongconnect{elecprof}{dwsim}
    \hlongconnect{pvprof}{dwsim}
    \hshortconnect{evprof}{evcontrol}
    \hshortconnect{heatmodel}{heatcontrol}
    \hshortconnect{dhwmodel}{dhwcontrol}

    \hshortconnect{evcontrol}{dwsim}
    \hshortconnect{heatcontrol}{dwsim}
    \hshortconnect{dhwcontrol}{dwsim}

    \end{tikzpicture}
    \caption{Flow chart of this study's methodology to simulate a dwelling. Blue nodes represent the sampled characteristics of the dwelling (\cref{sec:dwelling_characteristics}), orange nodes represent the data and simulation models used for each part of the dwelling (\cref{sec:thermal_data_and_models,sec:consumption_profiles,sec:ev_charger,sec:pv_panels,sec:hot_water}), green nodes represent the possible control actions of the assets (\cref{sec:asset_management}), and the red node is the dwelling simulation that combines everything together.}
    \label{fig:dwelling_flow_chart}
\end{figure}
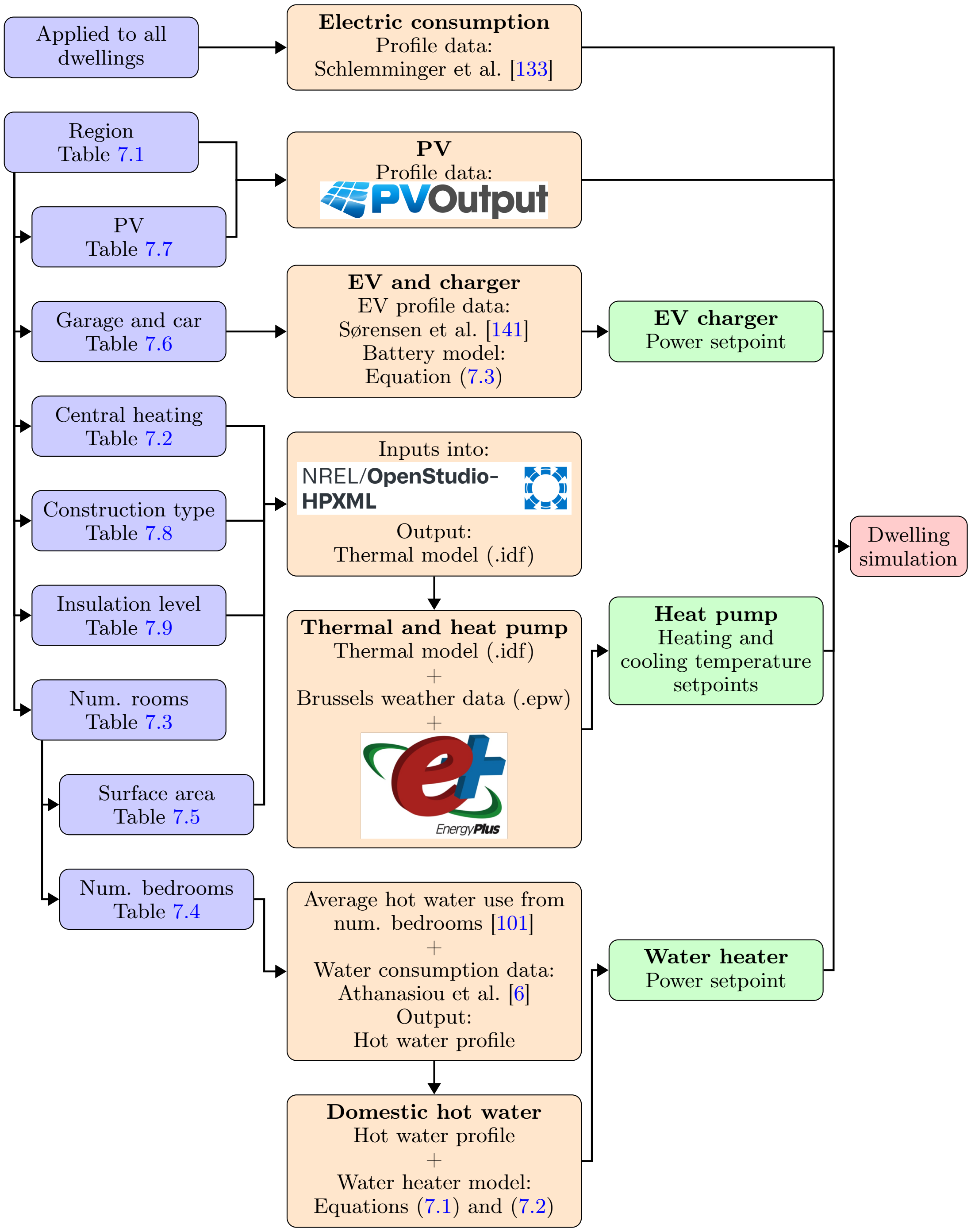

\subsection{Dwelling characteristics}
\label{sec:dwelling_characteristics}

Different statistics were collected to build the set of 500 representative Belgian dwellings.
This section goes through each of them giving the source and explanation when necessary. In the following tables, each cell shows the percentage of all dwellings that have both the characteristic specified by the row and the characteristic specified by the column.

To assign characteristics to the 500 representative dwellings, the number of dwellings with each characteristic is calculated by multiplying the relevant percentage by 500. For example, according to \cref{tab:stats_region}, 10.6\% of dwellings are in Brussels, 57.4\% in Flanders, and 32.0\% in Wallonia, resulting in 53 dwellings in Brussels, 287 in Flanders, and 160 in Wallonia. This establishes the initial pool of unassigned characteristics.

Assignment proceeds sequentially, starting with the most independent characteristic (region), and moving towards the most dependent (number of bedrooms and surface area) as illustrated by \cref{fig:dwelling_flow_chart}. For each dwelling, characteristics are sampled randomly from the remaining pool, using any already assigned characteristics as conditional constraints. This process continues until all dwellings have a complete set of characteristics, and the overall distribution matches the target statistics.

The percentage for the number of dwellings per region (\cref{tab:stats_region}), the presence of central heating (\cref{tab:stats_heating}) and the number of rooms (\cref{tab:stats_rooms}) are all obtained from the 2021 Belgian Census which gives the total number of dwellings fitting each characteristic without need for extrapolation \cite{statbelCensus2021Nombre2021,statbelCensus2021Nombre2021a}.

The number of bedrooms data from \cref{tab:stats_bedrooms} comes from an extrapolation of \citet{meirelesDomesticHotWater2022}. 
The source gives the number of bedrooms of 9\,206 Belgian apartments. 
To fit better to the global Belgian situation with 66\% houses (see \cref{tab:stats_construction_type}), the total proportion of 0 and 1 bedroom dwellings was reduced by 25\% and the total proportion of 3 to 6 bedroom dwellings was increased by 47.3\% to match the reduction of the 0 and 1 bedroom dwellings.
The correlation between the number of bedrooms and the number of rooms is established using common sense with the total bedroom and total room  proportions  as a constraint (see \cref{tab:stats_rooms}).
\citet{meirelesDomesticHotWater2022} indicates that the number of bedrooms in a dwelling is a good indicator of average water consumption. 
The number of bedrooms is used solely to determine the appropriate hot water consumption profile to assign to the dwelling, as detailed in \cref{sec:hot_water}.

The surface area data from \cref{tab:stats_surface_area} comes from tables IV.24 and IV.28 of \citet{vannesteLogementBelgique2007}.
Table IV.24 gives the number of dwellings with a certain surface area per region. 
Table IV.28 gives the number of dwellings with a certain number of rooms per region.
The correlation between the number of rooms and the surface area is determined to ensure that the final proportions per region closely match those presented in tables IV.24 and IV.28 of \citet{vannesteLogementBelgique2007}.

The garage and car ownership data from \cref{tab:stats_home_charger} comes from page 68 of \citet{hubertMobiliteQuotidienneBelges2003}.
The source gives the proportion of dwellings that have a garage in each region and the proportion of people with a garage that have at least one car.
\Cref{tab:stats_home_charger} shows the combination of these two proportions, it is used to determine the proportion of dwellings that would have a EV charger in each region.

The photovoltaic data from \cref{tab:stats_pv} comes from different sources for the different regions.
The number of photovoltaic installations in Brussels and Flanders comes from \citet{brugelEnergiesRenouvelablesRegion2024} and \citet{vlaamsenergie-enklimaatagentschapZonnepanelenVlaanderen2024} respectively. For Wallonia, the proportion of dwellings with a photovoltaic installation is given by \citet{lallemand16MenagesWallons2023}.

The construction type data in \cref{tab:stats_construction_type} combines two sources: \citet{statbelCensus2021Nombre2021b}, which provides the proportion of houses and apartments in each region, and Figure IV.13 from \citet{vannesteLogementBelgique2007}, which details the breakdown of terraced, semi-detached, and detached constructions by region.

The insulation data in \cref{tab:stats_insulation} is based on \citet{recticelinsulationBarometreLisolation20202020}, which reports the proportions of dwellings with roof, wall, and ground insulation in each region. However, the source does not specify the exact overlap between these insulation levels (e.g., how many dwellings have both wall and roof insulation, or all three types). To construct the final table, the following assumptions were made: dwellings with ground insulation also have wall and roof insulation, and dwellings with wall insulation also have roof insulation. This approach ensures that the total proportions match the source data while providing a reasonable estimate of the distribution of combined insulation levels.

\begin{table}[H]
    \centering
    \caption{Distribution of dwellings per region \cite{statbelCensus2021Nombre2021}}
    \begin{tabular}{|c | c | c|}
    \hline
    Brussels & Flanders & Wallonia \\
    \hline
    10.6\% & 57.4\% & 32.0\% \\
    \hline
    \end{tabular}
    \label{tab:stats_region}
    \end{table}

    \begin{table}[H]
    \centering
    \caption{Distribution of dwellings with central heating \cite{statbelCensus2021Nombre2021}}
    \begin{tabular}{|c | c | c | c||c|}
    \hline
     & Brussels & Flanders & Wallonia & Total \\
    \hline
    With central heating & 10.1\% & 51.9\% & 27.4\% & 89.4\% \\
    \hline
    Without central heating & 0.5\% & 5.5\% & 4.6\% & 10.6\% \\
    \hline
    \end{tabular}
    \label{tab:stats_heating}
    \end{table}

    \begin{table}[H]
    \centering
    \caption{Distribution of dwellings per number of rooms \cite{statbelCensus2021Nombre2021a}}
    \begin{tabular}{|c | c | c | c||c|}
    \hline
     & Brussels & Flanders & Wallonia & Total \\
    \hline
    1 room & 0.3\% & 0.5\% & 0.4\% & 1.2\% \\
    \hline
    2 rooms & 0.8\% & 1.6\% & 1.0\% & 3.4\% \\
    \hline
    3 rooms & 2.1\% & 4.6\% & 2.5\% & 9.2\% \\
    \hline
    4 rooms & 3.2\% & 10.0\% & 5.1\% & 18.3\% \\
    \hline
    5 rooms & 1.8\% & 14.0\% & 7.4\% & 23.2\% \\
    \hline
    6 rooms & 0.9\% & 12.6\% & 7.1\% & 20.6\% \\
    \hline
    7 rooms & 0.5\% & 7.3\% & 4.1\% & 11.9\% \\
    \hline
    8 rooms & 0.4\% & 3.6\% & 2.3\% & 6.3\% \\
    \hline
    9+ rooms & 0.6\% & 3.2\% & 2.1\% & 5.9\% \\
    \hline
    \end{tabular}
    \label{tab:stats_rooms}
    \end{table}

    \begin{table}[H]
    \centering
    \caption{Distribution of dwellings per number of bedrooms \cite{meirelesDomesticHotWater2022}}
    \setlength{\tabcolsep}{4pt}
    \small
    \begin{tabular}{|c | c | c | c | c | c | c | c | c | c||c|}
    \hline
    \begin{tabular}[c]{@{}c@{}}Number of\\rooms\end{tabular} & 1 & 2 & 3 & 4 & 5 & 6 & 7 & 8 & 9+ & Total \\
    \hline
    0 bedroom & 0.4\% & 1.2\% & 2.5\% & 2.5\% & 2.0\% & 0.3\% & - & - & - & 8.9\% \\
    \hline
    1 bedroom & 0.8\% & 2.2\% & 5.6\% & 5.6\% & 4.6\% & 1.2\% & - & - & - & 20.0\% \\
    \hline
    2 bedrooms & - & - & 1.1\% & 8.2\% & 12.3\% & 10.3\% & 5.1\% & 2.1\% & 2.0\% & 41.1\% \\
    \hline
    3 bedrooms & - & - & - & 2.0\% & 4.0\% & 6.5\% & 5.1\% & 2.2\% & 2.1\% & 21.9\% \\
    \hline
    4 bedrooms & - & - & - & - & 0.3\% & 2.2\% & 1.4\% & 1.3\% & 1.0\% & 6.2\% \\
    \hline
    5 bedrooms & - & - & - & - & - & 0.1\% & 0.3\% & 0.6\% & 0.5\% & 1.5\% \\
    \hline
    6 bedrooms & - & - & - & - & - & - & - & 0.1\% & 0.3\% & 0.4\% \\
    \hline
    \end{tabular}
    \label{tab:stats_bedrooms}
    \end{table}

    \begin{table}[H]
    \centering
    \caption{Distribution of dwellings per surface area \cite{vannesteLogementBelgique2007}}
    \setlength{\tabcolsep}{4pt}
    \small
    \begin{tabular}{|c | c | c | c | c | c | c | c | c | c||c|}
    \hline
    \begin{tabular}[c]{@{}c@{}}Number of\\rooms\end{tabular} & 1 & 2 & 3 & 4 & 5 & 6 & 7 & 8 & 9+ & Total \\
    \hline
    <35m² & 1.1\% & 1.7\% & 0.9\% & - & - & - & - & - & - & 3.7\% \\
    \hline
    35-54m² & 0.1\% & 1.0\% & 2.8\% & 5.5\% & 4.6\% & 3.1\% & - & - & - & 17.1\% \\
    \hline
    55-84m² & - & 0.7\% & 3.7\% & 7.3\% & 7.0\% & 5.1\% & 1.2\% & - & - & 25.0\% \\
    \hline
    85-104m² & - & - & 1.8\% & 3.7\% & 7.0\% & 6.2\% & 3.6\% & 0.5\% & 0.5\% & 23.3\% \\
    \hline
    105-124m² & - & - & - & 1.8\% & 2.3\% & 4.1\% & 3.6\% & 1.6\% & 1.5\% & 14.9\% \\
    \hline
    >124m² & - & - & - & - & 2.3\% & 2.1\% & 3.5\% & 4.2\% & 3.9\% & 16.0\% \\
    \hline
    \end{tabular}
    \label{tab:stats_surface_area}
    \end{table}

    \begin{table}[H]
    \centering
    \caption{Distribution of dwellings with a garage and a car \cite{hubertMobiliteQuotidienneBelges2003}}
    \begin{tabular}{|c | c | c | c||c|}
    \hline
     & Brussels & Flanders & Wallonia & Total \\
    \hline
    With garage and car & 3.8\% & 42.7\% & 22.4\% & 68.9\% \\
    \hline
    Without garage or car & 6.8\% & 14.7\% & 9.6\% & 31.1\% \\
    \hline
    \end{tabular}
    \label{tab:stats_home_charger}
    \end{table}

    \begin{table}[H]
    \centering
    \caption{Distribution of dwellings with photovoltaic installation \cite{brugelEnergiesRenouvelablesRegion2024,vlaamsenergie-enklimaatagentschapZonnepanelenVlaanderen2024,lallemand16MenagesWallons2023}}
    \begin{tabular}{|c | c | c | c||c|}
    \hline
     & Brussels & Flanders & Wallonia & Total \\
    \hline
    With photovoltaic & 0.2\% & 19.2\% & 6.4\% & 25.8\% \\
    \hline
    Without photovoltaic & 10.4\% & 38.2\% & 25.6\% & 74.2\% \\
    \hline
    \end{tabular}
    \label{tab:stats_pv}
    \end{table}

    \begin{table}[H]
    \centering
    \caption{Distribution of dwellings per construction type \cite{vannesteLogementBelgique2007}}
    \begin{tabular}{|c | c | c | c||c|}
    \hline
     & Brussels & Flanders & Wallonia & Total \\
    \hline
    Terraced house & 1.3\% & 11.7\% & 8.4\% & 21.4\% \\
    \hline
    Semi-detached house & 0.2\% & 10.6\% & 5.2\% & 16.0\% \\
    \hline
    Detached house & 0.1\% & 18.6\% & 10.1\% & 28.8\% \\
    \hline
    Apartment unit & 9.0\% & 16.5\% & 8.3\% & 33.8\% \\
    \hline
    \end{tabular}
    \label{tab:stats_construction_type}
    \end{table}

    \begin{table}[H]
    \centering
    \caption{Distribution of dwellings by insulation level \cite{recticelinsulationBarometreLisolation20202020}}
    \begin{tabular}{|c | c | c | c||c|}
    \hline
     & Brussels & Flanders & Wallonia & Total \\
    \hline
    No insulation & 1.3\% & 7.8\% & 11.5\% & 20.6\% \\
    \hline
    Roof insulation & 5.6\% & 16.5\% & 15.5\% & 37.6\% \\
    \hline
    Wall and roof insulation & - & 10.7\% & 0.2\% & 10.9\% \\
    \hline
    Ground and roof insulation & 0.1\% & - & - & 0.1\% \\
    \hline
    Ground, wall and roof insulation & 3.6\% & 22.4\% & 4.8\% & 30.8\% \\
    \hline
    \end{tabular}
    \label{tab:stats_insulation}
    \end{table}

\subsection{Dwelling thermal and heat pump models}
\label{sec:thermal_data_and_models}

The thermal models are created using the Openstudio-HPXML
workflow \cite{horowitzOpenStudioHPXMLWorkflow2024}. This workflow takes as inputs the various characteristics of the dwelling, in this case construction type, insulation level and surface area, and returns and .idf file containing a detailed thermal model of the dwelling with a properly sized heat pump that can then be used for Energyplus \cite{crawleyEnergyPlusCreatingNewgeneration2001} simulations (see \cref{fig:dwelling_flow_chart}).

The roof and ground are insulated with an R-value of 6.4 m²K/W and the walls with a value of 7.3 m²K/W following the values from \citet{recticelinsulationBarometreLisolation20202020}.
The uninsulated R-values are averages of typical Belgian building constructions: 0.5 m²K/W for the walls, 0.3 m²K/W for the roof and 0.4 m²K/W for the ground, as reported by \citet{energie+CoefficientTransmissionThermique2007}.

The windows U-values correspond to single glazing (5.8 W/m²K), double glazing (2.8 W/m²K) and triple glazing (0.6 W/m²K) values from \citet{energuide.beDoubleGlazingTriple2025}.

The air tightness values are taken from \citet{lavergeAirtightnessAssessmentSingle2014} and correspond to the median air tightness values for Belgian houses:
\begin{itemize}
    \item Constructed between 1990 and 1995 (9.3 ACH\textsubscript{50})
    \item Constructed between 2007 and 2010 and randomly sampled (5.5 ACH\textsubscript{50})
    \item Constructed between 2006 and 2013 and sampled from engaged owners that reported their air tightness levels (2.9 ACH\textsubscript{50})
    \item With high energy performance  (1.4 ACH\textsubscript{50})
\end{itemize}

The link between the insulation level, the window U-values and the air tightness is shown in \cref{tab:window_and_tightness}.

\begin{table}[H]
    
    \centering
    \caption{Window U-values and air tightness values}
    \begin{tabular}{|c | c | c |}
    \hline
    Insulation level & Window U-value & Air tightness \\
    \hline
    No insulation & 5.8 W/m²K & 9.3 ACH\textsubscript{50} \\
    \hline
    Roof insulation & 2.8 W/m²K & 5.5 ACH\textsubscript{50} \\
    \hline
    Double insulation & 2.8 W/m²K & 2.9 ACH\textsubscript{50} \\
    \hline
    Ground, wall and roof insulation & 0.6 W/m²K & 1.4 ACH\textsubscript{50} \\
    \hline
    \end{tabular}
    \label{tab:window_and_tightness}
\end{table}

Since only ranges of surface areas are given in the statistics from \cref{tab:stats_surface_area}, each range is assigned a specific value as input to the Openstudio-HPXML workflow.
The surface area is set to 25m\textsuperscript{2} for the dwellings of <35m\textsuperscript{2}, 150m\textsuperscript{2} for the dwellings of >124m\textsuperscript{2} and the average of the upper and lower bounds for the other surface area categories (e.g. 70m\textsuperscript{2} for the 55-84m\textsuperscript{2} range).

The dwelling's heating and cooling are managed by an air-to-air heat pump. This type of heat pump was selected due to its higher sales volume compared to air-to-water and geothermal heat pumps in Belgium \cite{eurobserver22ndReportState2023}.
The heat pump is modelled with a primary heating coil and a primary cooling coil, both operating via a compressor. Additionally, it includes a backup electric resistance heating coil that activates only when the primary heating coil cannot meet the heating demand. The heat pump is also modelled with a resistive defrost heater to prevent ice formation.
The heat pump is controllable by setting a heating and cooling set point temperature.

The historical weather data for Brussels is used in all simulations and was downloaded using the Weather Data Downloader from \citet{oikolabWeatherDataDownloader2025}. The obtained weather data file is in .epw format.

The simulation is conducted in EnergyPlus using the generated .idf thermal model file and the .epw Brussels weather data file as inputs. \Cref{sec:heat_pump_control} details how the heat pump is controlled during the simulation.

\subsection{Domestic hot water}
\label{sec:hot_water}

The domestic hot water system is modelled as a storage tank following the equations from \citet{dolanDevelopmentMonteCarlo1996} with a hot water consumption profile extrapolated from the \citet{athanasiouDAIADOpenWater2014} dataset.

The hot water tank is modelled using the equations presented in \cref{eq:hot_water_params,eq:tank_temp_change}:
\begin{equation}
\begin{aligned}
    &C = \rho^w c^w V  \\
    &G = \frac{SA}{R} \\
    &B_t =  \rho^w c^w F_t \\
    &R'_{t} = \frac{1}{G + B_t} \\
    &D_{t} = \exp\left(-\frac{\Delta t}{R'_{t} C}\right)
\end{aligned}
\label{eq:hot_water_params}
\end{equation}

where:

\begin{tabular}{ll}
    $C$ [J/K]               & tank heat capacity \\
    $\rho^w$ [kg/m$^3$]     & water density (1000 kg/m$^3$) \\
    $c^w$ [J/kg/K]          & specific heat capacity of water (4186 J/kg/K) \\
    $V$ [m$^3$]             & tank volume \\
    $G$, $B_t$, $R'_{t}$, $D_t$ & intermediate parameters and variables \\
    $SA$ [m$^2$]            & tank surface area \\
    $R$ [m$^2$K/W]          & tank insulation $R$-value (used: 10 m$^2$K/W) \\
    $F_t$ [m$^3$/s]         & water flow rate at time $t$ \\
    $\Delta t$ [s]          & time step duration \\
\end{tabular}

\ifprintedversion
\fi

\begin{equation}
    \ifprintedversion
    \small
    \fi
    T^{tank}_{t+1} = D_{t} \underbrace{T^{tank}_{t}}_{\text{init temp}} + \left( \underbrace{G R'_{t} T^{room}}_{\text{loss to room}}+ \underbrace{B_t R'_{t} T^{inlet}_{t}}_{\text{loss inlet water}} + \underbrace{\eta^{heater} P_t R'_{t}}_{\text{gain heater}} \right)(1 - D_{t})
\label{eq:tank_temp_change}
\end{equation}

where:

\begin{tabular}{ll}
    $T^{tank}_t$ [$^\circ$C] & mean tank temperature at time $t$ \\
    $T^{room}$ [$^\circ$C]   & ambient room temperature (used: 15 $^\circ$C) \\
    $T^{inlet}_t$ [$^\circ$C] & inlet (mains) water temperature at time $t$ \\
    $\eta^{heater}$ [--]             & heater efficiency (used: 1) \\
    $P_t$ [W]               & heater power at time $t$ \\
\end{tabular}

The \citet{athanasiouDAIADOpenWater2014} dataset comprises two and a half years of hourly water consumption data from 1\,007 households in Alicante.
\citet{meirelesDomesticHotWater2022} studied the water consumption of 9\,206 Belgian apartments and gave a month dependant ratio between total water consumption and hot water consumption. The mean of the monthly ratios is 32.3\% with a standard deviation of 2.5\% between the months.
The water consumption profiles from the \citet{athanasiouDAIADOpenWater2014} dataset are converted to hot water consumption profiles by multiplying each month's data by the corresponding monthly hot-to-total water ratio.

The study also gives statistical links between water consumption and the number of bedrooms in the apartments. 
A probability distribution is created linking the number of bedrooms to an estimated average hot water consumption in order to link a dwelling with a hot water consumption profile.

To sample the daily hot water consumption given a number of bedrooms in a dwelling, we use an inverse Gaussian distribution. 
An inverse Gaussian distribution is defined by its mean, shape and location parameters.
Each bedroom count gets its own distribution, with the mean value taken directly from \citet{meirelesDomesticHotWater2022}. The study also provides overall statistics (median, standard deviation, skewness, kurtosis) for hot water use across all apartments.

\newcommand{\hotdistrfig}{\begin{figure}[H]
    \centering
    \includegraphics[width=0.8\textwidth]{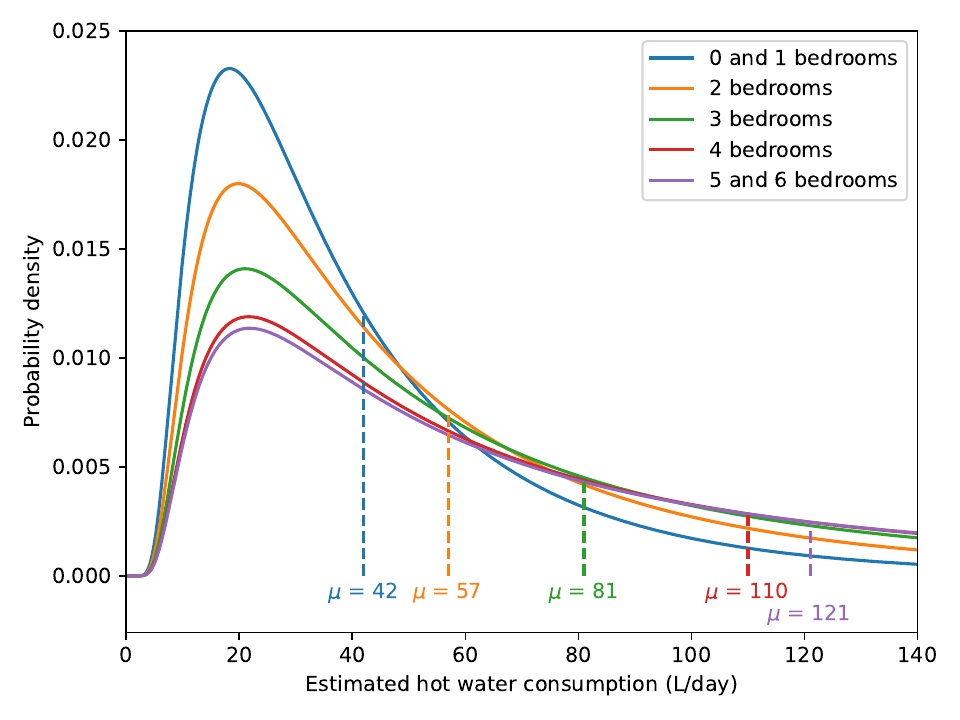}
    \caption{Probability density function of the estimated average daily hot water consumption per number of bedrooms in a dwelling. The mean value of the distribution is indicated by the vertical dashed line.}
    \label{fig:hot_water_distribution}
\end{figure}
}

\ifprintedversion
\hotdistrfig
\fi

We assume the same shape and location parameters for all distributions. These parameters are fitted so the combination of distributions matches the median value reported in the original dataset. The obtained shape and location parameters are 67.68 and 0 respectively. The resulting probability density functions for each bedroom count are shown in \cref{fig:hot_water_distribution}.

The mean and median of the combination of distributions is therefore the same as the original dataset but the standard deviation (77.2 for dataset vs 65.5 for distribution), skewness (61.3 for dataset vs 5.4 for distribution) and kurtosis (11.3 for dataset vs 66.6 for distribution) are still different. Knowing the exact distribution for each bedroom count would require the original data which is not available.

\ifprintedversion
\else
\hotdistrfig
\fi

The hot water consumption profile for a dwelling is determined by sampling from the inverse Gaussian distribution that matches the number of bedrooms in the dwelling. The profile selected is the one with the average daily hot water consumption closest to the sampled value.

In addition to the hot water consumption profile, the model from \citet{dolanDevelopmentMonteCarlo1996} need a temperature for the incoming cold water from the mains.
\citet{fuentesReviewDomesticHot2018} established that a sinusoidal function, which provides the mains water temperature as a function of the month is a good approximation. 
The upper and lower bounds of the sinusoidal function are set to 11\,°C and 16\,°C respectively \cite{v.sebastienMachinesLaverConsommentelles2020}.

The water heater is modelled as an electric resistance heater with a heater efficiency ($\eta^{heater}$) of 1. The volume of the storage tank is three times the expected average daily hot water consumption determined by the number of bedrooms to account for peak consumption periods (so 146 litres for a dwelling with 1 bedroom and 363 litres for a dwelling with 6 bedrooms).

\subsection{EV charger}
\label{sec:ev_charger}

The dwellings sampled to have a garage and a car (see \cref{tab:stats_home_charger}) are equipped with an EV charger for electric vehicle.
Each simulation uses a real user charging profile, which is a record of every home charging session information that occurred over the year. These profiles are sourced from \citet{sorensenElectricVehicleCharging2024}.
This dataset includes the arrival time, departure time and charged energy of each charging session of electric vehicles from 267 users charging at 12 residential locations in Norway.
The dataset also provides assumed arrival and departure state of charge, battery sizes and maximum charging powers for each charging session \cite{sorensenMethodGeneratingComplete2023}.
All charging profiles containing more than 600 days of data were included in this study, resulting in a total of 16 profiles.
Each dwelling with an EV charger is randomly assigned one of these 16 charging profiles.

\begin{table}[H]
    \centering
    \caption{Summary statistics of the 16 EV charging profiles used in this study.}
    \begin{tabular}{|l|c|c|}
        \hline
        Statistic & Mean & Std. dev. \\
        \hline
        Total energy charged per profile per year & 2\,414 kWh & 1\,464 kWh \\
        Number of sessions in a year per profile & 193.6 & 131.2 \\
        Session duration across all profiles & 12h16min & 11h12min \\
        Session SOC charged across all profiles  & 31.7\% & 20.0\% \\
        Std. dev. of session duration per profile & 9h46min & 5h32min \\
        Std. dev. of session SOC charged per profile & 17.3\% & 2.6\% \\
        \hline
    \end{tabular}
    \label{tab:ev_charging_profile_stats}
\end{table}

\Cref{tab:ev_charging_profile_stats} summarizes the variability in the EV charging profiles. 

On average, each EV charges by 2\,414 kWh per year at home. Assuming an EV efficiency of 0.191 kWh/km \cite{evdatabaseEnergyConsumptionFull2025}, this equates to an annual driving distance of 12\,638 km. This figure is close to the average yearly travelled distance by Belgian cars of 14\,420 km \cite{carpassRapportAnnuel20242025}, especially considering that not all users charge their EV exclusively at home.

On average, each profile has about 194 charging sessions per year, but this number varies widely between users (standard deviation: 131). The typical charging session lasts 12 hours and 16 minutes, with a standard deviation of 11 hours and 12 minutes, showing that some sessions are much shorter or longer than others. The average state of charge (SOC) charged per session is 31.7\%, but again, this varies a lot (standard deviation: 20\%). Even when looking at individual profiles, the average standard deviation within a user's sessions remains high at 9 hours and 46 minutes for session duration, and at 17.3\% for charged SOC. 

In summary, the average charged energy per year across the users is typical for Belgian drivers, and EV charging behaviour is highly variable both between users and from session to session for the same user. This highlights potential challenges in predicting this behaviour for energy management systems.

The model of the charged electric vehicle is defined by \cref{eq:ev_charging_reliability}:

\begin{equation}
    \label{eq:ev_charging_reliability}
    \text{SOC}_{t+1} = \text{SOC}_{t} + \frac{\eta^{charge} \times \text{P}^{\text{charge}}_{t} \times \Delta t}{\text{E}^{\text{batt}}}
\end{equation}

where:

\begin{tabular}{ll}
    $\text{SOC}_{t}$ [--]           & State of charge of the EV battery at time $t$ \\
    $\text{P}^{\text{charge}}_{t}$ [kW] & Power of the charger at time $t$ \\
    $\Delta t$ [h]                  & Time step duration \\
    $\text{E}^{\text{batt}}$ [kWh]    & Energy capacity of the battery \\
    $\eta^{\text{charge}}$ [--]     & Charging efficiency \\
\end{tabular}

The maximum charging power provided in the dataset for the 16 charging profiles is 3.7 kW. We are aware that home chargers in Belgium generally have a higher maximum power (7.4 kW or 11 kW) but we chose to keep the dataset's maximum charging power because the maximum power of the charger could influence the user's charging profile unpredictably (e.g. shorter connection times).

The charging efficiency $\eta^{\text{charge}}$ is set to 88\% and the departure SOC is always 95\% which are the assumptions made by \citet{sorensenMethodGeneratingComplete2023} for the dataset.

\subsection{Photovoltaic panels}
\label{sec:pv_panels}

The PV panels are simulated using real Belgian PV production data available on a 5-minute granularity from the PVoutput.org website \cite{pvoutput.orgPVOutputorg2025}.
A total of 380 PV profiles were downloaded from the website, representing all installations with a capacity below 10 kW that recorded measurements for at least 95\% of the days in both 2022 and 2023.
Each dwelling with PV panels is randomly assigned a PV profiles from its corresponding region out of the 380 profiles.

\subsection{Electricity consumption}
\label{sec:consumption_profiles}

The electricity consumptions are simulated using real house electricity consumption profiles available on a 2-minute granularity for houses located in Hamelin, Germany \cite{schlemmingerDatasetElectricalSinglefamily2022}.
The heating and cooling electricity consumption is measured separately from the rest of the electricity consumption, this study only uses the rest of the electricity consumption. 

Out of the 38 profiles available in the dataset, this study used the 22 profiles with data available for at least 98\% of the time steps for the years 2019 and 2020. 
Since all missing data must be imputed for the simulation, this threshold ensures that only a minimal amount of data requires imputation.

The 22 consumption profiles have an average annual electricity consumption of 2\,833 kWh, with a standard deviation of 862 kWh. For comparison, the average Flemish household consumed 2\,534 kWh according to Vlaamse Nutsregulator, while the European average reported by Eurostat is 3\,500 kWh \cite{vlaamsenutsregulatorEnergieverbruik2024}. These profiles are therefore reasonably representative of typical household yearly consumption, with the added benefit of being based on real, high-resolution measurements.

\subsection{Electricity prices}

The real-time pricing used in this study is based on a one-year dynamic contract of a major Belgian energy supplier, as if it were taken in January 2023\footnote{Full contract available on the paper's GitHub repository at 
\url{https://github.com/EVERGi/should_belgians_install_advanced_ems_paper}}.
The total cost is composed of four costs with different pricing mechanisms: the day-ahead cost, the offtake extras cost, the peak cost and the fixed yearly cost.

The day-ahead cost depends on the prices of \cref{eq:injection_price_usefull,eq:offtake_price_usefull}.

\begin{equation}
    Vo_t= 
\begin{cases}
    (Vd_t+0.002\frac{\SI{}{\EUR}}{kWh}) \times 1.06& \text{if } (Vd_t+0.002\frac{\SI{}{\EUR}}{kWh})\geq 0\\
    (Vd_t+0.002\frac{\SI{}{\EUR}}{kWh}),              & \text{otherwise}
\end{cases}
\label{eq:injection_price_usefull}
\end{equation}

\begin{equation}
    Vi_t=Vd_t\frac{\SI{}{\EUR}}{kWh}
\label{eq:offtake_price_usefull}
\end{equation}

where:

\begin{tabular}{ll}
    $Vo_t$ [\euro/kWh] & Offtake price at time step $t$ \\
    $Vi_t$ [\euro/kWh] & Injection price at time step $t$ \\
    $Vd_t$ [\euro/kWh] & Belgian day-ahead hourly price \\
\end{tabular}

\Cref{tab:day_ahead_price_stats} summarizes the Belgian day-ahead electricity prices ($Vd_t$) from 2020 to 2024. The table highlights significant year-to-year variation, with 2022 standing out as an exceptional year due to unusually high prices. In contrast, 2023 falls within the typical range observed over recent years.

\begin{table}[H]
    \centering
    \caption{Belgian day-ahead electricity prices for 2020–2024. All values are in \euro{}/kWh.}
    \begin{tabular}{|c|c|c|c|c|}
        \hline
        Year & Mean & Std. dev. & Min & Max \\
        \hline
        2020 & 0.032 & 0.017 & -0.115 & 0.200 \\
        2021 & 0.104 & 0.079 & -0.070 & 0.620 \\
        2022 & 0.245 & 0.135 & -0.100 & 0.871 \\
        2023 & 0.097 & 0.046 & -0.120 & 0.330 \\
        2024 & 0.070 & 0.0430 & -0.140 & 0.566 \\
        \hline
    \end{tabular}
    \label{tab:day_ahead_price_stats}
\end{table}

A value-added tax of 6\% is added to the offtake price when it is positive.
Extra costs that occur for energy taken from the grid for transport, distribution and taxes are calculated together and labelled as offtake extras cost.
The offtake price for all these additional offtake costs $Vf$ is 0.079 \euro/kWh. 

The monthly peak offtake cost $Cp_m$ is calculated by \cref{eq:peak_cost_usefull}.
\begin{equation}
Cp_m = Vp \times \max_{{t\in B_m}} (\frac{Eo_t}{\Delta t}, 2.5 kW)
\label{eq:peak_cost_usefull}
\end{equation}

where:

\begin{tabular}{ll}
    $Vp$ [\euro/kW] & Peak price (used: 3.66 \euro/kW) \\
    $Eo_t$ [kWh] & Energy taken from the grid at time step $t$ \\
    $B_m$ & Set of time steps in month $m$ \\
    $\Delta t$ [h] & Length of time step \\
\end{tabular}

This cost adds a penalty to the highest offtake power measured over any 15 minute time step within each month. 
The offtake energy is divided by the length of a time step to obtain the power in kW.
A minimum monthly peak power of 2.5 kW is enforced in the cost, even if the measured peak power is lower.

The monthly peak offtake cost, otherwise called capacity tariff, is a pricing mechanism applied to all residential consumers equipped with a digital meter in the Belgian region of Flanders.
The goal is to incentivize consumers to lower their peak offtake power from the grid.

Finally, part of the electric bill is a fixed yearly cost $Cy$ of 115.23\,\euro{} per year.

The total electricity cost $Ctot$ is calculated by \cref{eq:tot_cost_usefull}.

\begin{equation}
Ctot=\underbrace{\sum_{t\in T}{(Eo_tVo_t-Ei_tVi_t)}}_{\text{day-ahead cost}}+\underbrace{\sum_{t\in T}Eo_tVf}_{\text{offtake extras cost}}+\underbrace{\sum_{m \in M}{Cp_m}}_{\text{peak cost}}+\underbrace{Cy}_{\text{yearly cost}}
\label{eq:tot_cost_usefull}
\end{equation}

where:

\begin{tabular}{ll}
    $T$ & Set of all time steps \\
    $M$ & Set of all months \\
\end{tabular}

\section{Asset management}
\label{sec:asset_management}

The three controllable assets, when present in the dwelling, are the air-to-air heat pump, the water heater and the electric vehicle charger.

Every simulation is run for a period of one year with a time step of 15 minutes as this is the time period used to calculate the peak electricity cost in Belgium (see \cref{eq:peak_cost_usefull}).
The EMS can change at each time step the cooling and heating temperature set points of the heat pump, the electric power of the water heater and the charge power of the electric vehicle charger.

\subsection{Heat pump control}
\label{sec:heat_pump_control}

The heat pump operates within a default comfort range.
If the indoor temperature is outside the comfort range, the heat pump will bring the indoor temperature back to the comfort range. 
The EMS can set a heating or cooling temperature set point within this comfort range.

Based on different studies, \citet{peetersThermalComfortResidential2009} established usable comfort ranges for the different rooms of a residential building.
The comfort range depends on the reference external temperature $T^{e,ref}_d$ which is based on the minimum and maximum temperatures of the current and the last three days as shown in \cref{eq:te_ref}.

\begin{equation}
    T^{e,ref}_d = \frac{T^{avg}_{d} + 0.8 \times T^{avg}_{d-1} + 0.4 \times T^{avg}_{d-2} + 0.2 \times T^{avg}_{d-3}}{2.4}
\label{eq:te_ref}
\end{equation}

where:

\begin{tabular}{ll}
    $T^{avg}_{d}$ & Average of the maximum and minimum temperature on day $d$ \\
\end{tabular}

To avoid relying on weather forecasts for $T^{avg}_{d}$, this study uses $T^{avg}_{d-1}$ as a substitute in the calculation. The mean absolute difference in the resulting comfort range is only 0.18\,°C (standard deviation 0.24\,°C), showing that this simplification has minimal impact while reducing forecast dependency.

For each room a neutral temperature $T^n_d$ is calculated based on the reference external temperature $T^{e,ref}_d$.
The neutral temperature represents the temperature at which most people would feel comfortable in the room.
From $T^n_d$ a comfort range is calculated using the width of the comfort band $w$ and the band asymmetry parameter $\alpha$.
This study uses a width $w$ of 5 °C and an asymmetry parameter $\alpha$ of 0.7 which corresponds to a 10\% predicted percentage of dissatisfied (PPD) users \cite{peetersThermalComfortResidential2009}.

For the bedroom $T^n_d$ is calculated as shown in \cref{eq:bedroom_temp}.

\begin{equation}
    T^n_d = 
    \begin{cases} 
        16^\circ C & \text{if } T^{e,ref}_d < 0^\circ C \\
        0.23 \times T^{e,ref}_d + 16^\circ C & \text{if } 0^\circ C \leq T^{e,ref}_d < 12.6^\circ C \\
        0.77 \times T^{e,ref}_d + 9.18^\circ C & \text{if } 12.6^\circ C \leq T^{e,ref}_d < 21.8^\circ C \\
        26^\circ C & \text{if } T^{e,ref}_d \geq 21.8^\circ C
    \end{cases}
    \label{eq:bedroom_temp}
\end{equation}

The comfort range for the bedroom is calculated as shown in \cref{eq:range_sleep}.

\begin{equation}
    \begin{aligned}
        &T^{low}_t = \max(16^\circ C, T^n_d - w \times (1 - \alpha) )\\
        &T^{up}_t = \min(26^\circ C, T^n_d + w \times \alpha) 
    \end{aligned}
\label{eq:range_sleep}
\end{equation}

where: 

\begin{tabular}{ll}
    $T^{low}_t$ & Lower bound of the comfort range at time step $t$ \\
    $T^{up}_t$ & Upper bound of the comfort range at time step $t$ \\
\end{tabular}

For the rooms of the house other than the bedroom and the bathroom, $T^n_d$ is calculated as shown in \cref{eq:other_temp}.

\begin{equation}
    T^n_d = 
    \begin{cases} 
        20.4^\circ C + 0.06 \times T^{e,ref}_d & \text{if } T^{e,ref}_d < 12.5^\circ C \\
        16.63^\circ C + 0.36 \times T^{e,ref}_d & \text{if } T^{e,ref}_d \geq 12.5^\circ C
    \end{cases}
    \label{eq:other_temp}
\end{equation}

The comfort range for the rooms of the house other than the bedroom and the bathroom are calculated as shown in \cref{eq:range_other}.

\begin{equation}
    \begin{aligned}
&T^{low}_t = \max(18^\circ C, T^n_d - w \times (1 - \alpha))\\
&T^{up}_t = T^n_d + w \times \alpha 
    \end{aligned}
    \label{eq:range_other}
\end{equation}

\begin{figure}[H]
\centering
\includegraphics[width=0.8\textwidth]{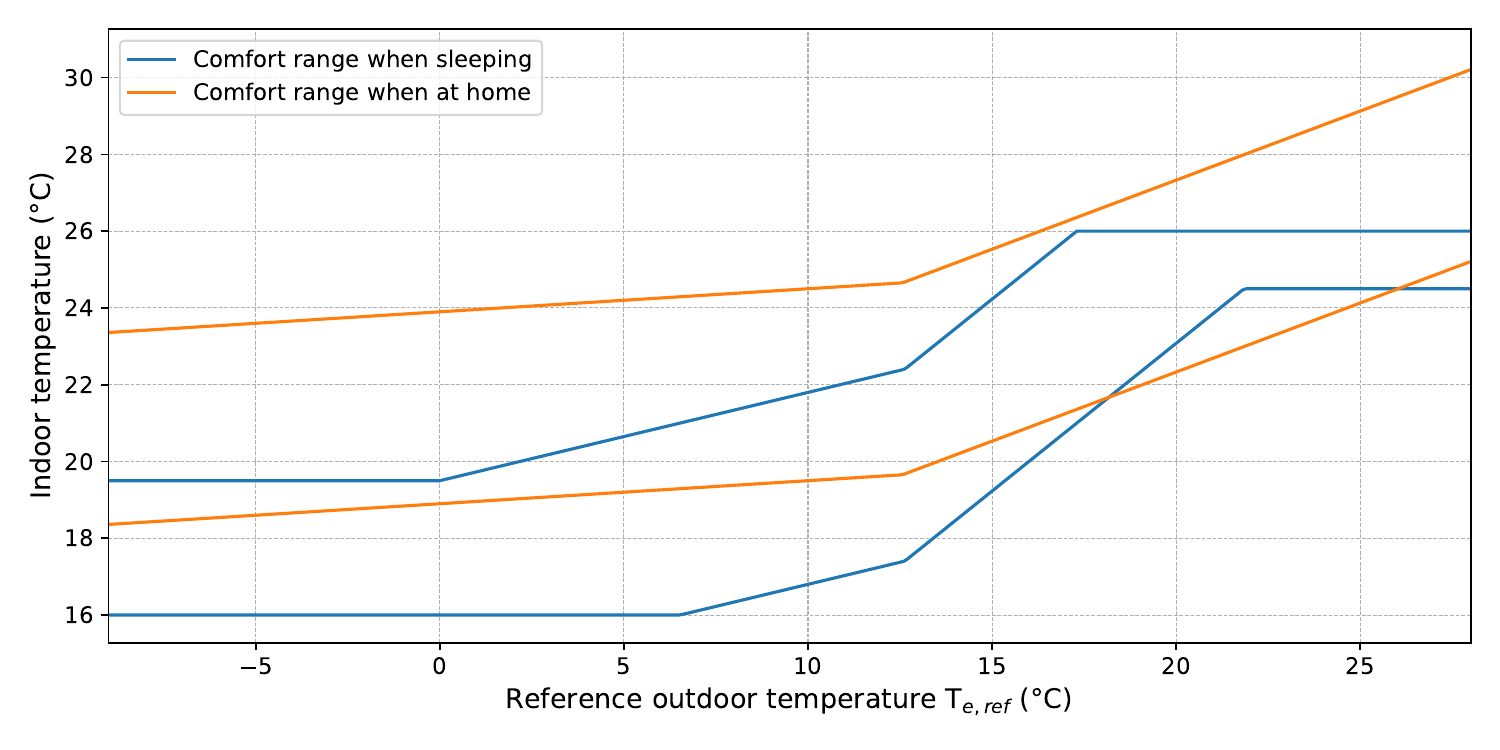}
\caption{
    Illustration of the comfort temperature ranges for sleeping and home occupancy profiles as a function of the reference external temperature $T^{e,ref}_d$. The plot covers $T^{e,ref}_d$ values from $-9\,^\circ$C to $28\,^\circ$C, representing the coldest and warmest reference temperatures observed in Belgium during the 21st century.
}
\label{fig:comfort_range}
\end{figure}

\Cref{fig:comfort_range} illustrates how the comfort temperature bounds $T^{low}_t$ and $T^{up}_t$ vary with the reference external temperature $T^{e,ref}_d$ for both sleeping and home occupancy profiles. For an example of how the comfort range evolve throughout a weekend, check \cref{fig:example_weekend}.

\citet{aertsDiscreteOccupancyProfiles2013} defines a set of occupancy profiles out of 3\,474 Belgian households.
The paper finds 7 daily occupancy profiles representative of people's behaviour which cover 79\% of the evaluated households.
The occurrence of each profile depends on whether the day is a weekday, Saturday, or Sunday.
In these profiles, the household is either at home, absent or sleeping.

Each dwelling is assigned one of the 7 representative occupancy profiles for the weekdays, Saturdays, and Sundays based on the distribution provided by \citet{aertsDiscreteOccupancyProfiles2013}.
The occupancy profiles are used to determine the comfort range of the dwelling at each time of the day.
When at home, the comfort range is defined by \cref{eq:range_other}, when sleeping by \cref{eq:range_sleep} and when absent the range is set to $T^{low}_t=5^\circ C$ and $T^{up}_t=40^\circ C$.

\subsection{Water heater control}

The hot water tank has a temperature range from 55 °C to 65 °C that it should maintain.
These values are the minimum and maximum domestic hot water temperatures from the European, Australian, Canada, Japan and Unites States standards \cite{fuentesReviewDomesticHot2018}. 
If the water temperature goes below 55 °C, the water heater will heat the water at maximum power for one 15 minute time step. Whenever the water temperature is above 65 °C, the water heater will not heat the water.
Otherwise, the water can be heated between 0 kW and the maximum power of the water heater.

\subsection{Electric vehicle charger control}
\label{sec:ev_charger_control}

The electric vehicle charger is controlled by setting a power between 0 kW and 3.7 kW.
Once the required departure state of charge is reached, charging stops.
Currently, most EVs do not provide information about their battery's state of charge \cite{sorensenMethodGeneratingComplete2023}. Therefore, in this work, the EMSs are assumed to operate without knowledge of the battery's state of charge.

For dwellings with an EV charger, two types of simulations are run. In the first, the EMS fully controls EV charging, allowing to assess how well each EMS charges the EV (see \cref{fig:avg_soc_final,fig:cummul_freq_soc}). In the second, charging is enforced so the EV always reaches full charge before departure, ensuring a fair cost comparison across EMSs (see \cref{fig:total_opex,fig:total_opex_diff,fig:dwelling_features_results,fig:opex_diff_separate}).

In the enforced scenario, if the EV is not fully charged near departure, the charger switches to 3.7 kW to guarantee a full charge, simulating a driver’s request for a complete charge before leaving.

\section{Energy management systems}

\subsection{Rule-based control}
\label{sec:relevance_rbc}
The rule-based control EMS uses simple, predefined strategies for each controllable asset, reflecting typical default operation modes of these assets:

\begin{itemize}
    \item \textbf{Heat pump:} The heat pump setpoint is determined by obtaining the comfort range for the next 1 hour and 30 minutes. The controller gets the highest lower bound ($T^{low}_t$) and the lowest upper bound ($T^{up}_t$) during this period and sets the heating and cooling setpoints to these values.
    \item \textbf{Domestic water heater:} The water heater operates at maximum power for 15 minutes whenever the tank temperature drops below 55\,°C.
    \item \textbf{Electric vehicle charger:} The EV is always charged at the maximum power.
\end{itemize}

The 1 hour and 30 minute time horizon was selected based on the final results to ensure that the rule-based controller achieves a thermal discomfort level similar to the other EMSs (see \cref{fig:discomfort_level}).

\subsection{Model predictive control}

Model predictive control (MPC) is a control strategy that optimizes the operation of controllable assets over a prediction horizon by solving an optimization problem at each time step.
The MPC uses a mathematical model of the assets and predicts all necessary parameters such as weather, electricity prices, and energy consumption of non-controllable assets to define the set points for the controllable assets.

This study implements two MPC approaches: a realistic MPC (MPC) and a perfect forecast MPC (MPC P).
The realistic MPC is a proven technology control algorithm for EMSs. The perfect forecast MPC serves as an unrealistically performant EMS to indicate an upper EMS performance limit.

The realistic MPC uses realistic forecasts for most parameters and uses a prediction horizon that extends up to when day-ahead prices ($Vd_t$) are available, that is until midnight of the current day if before 14:00 Central European Time or until midnight of the following day if after 14:00, resulting in a horizon ranging from 10 hours and 15 minutes to 34 hours.
MPC P uses a perfect forecast for all parameters and has a fixed prediction horizon of 48 hours.

Both MPC approaches are implemented using the Gurobi solver \cite{gurobioptimizationllcGurobiOptimizerReference2024}.

\subsubsection{Mathematical model}

The MPC model is formulated as a mixed-integer linear programming problem.

The dwelling's thermal model is approximated using the resistor-capacitor analogy for thermal modelling \cite{madsenEstimationContinuoustimeModels1995,drgonaAllYouNeed2020}. \Cref{eq:mpc_thermal_model} shows the implementation of this model in the MPC. EnergyPlus does not allow to control the power of the heat pump directly so the control setpoint used is the temperature of the model at the next time step ($T^{in}_{t+1}$).

\begin{equation}
C^{dwell}\frac{T^{in}_{t+1}-T^{in}_{t}}{\Delta t} = \frac{T^{out}_{t}-T^{in}_{t}}{TR}+Q^{pump}_{t} + gAQ^{sun}_{t}
\label{eq:mpc_thermal_model}
\end{equation}

where:

\begin{tabular}{ll}
    $C^{dwell}$ [J/K] & Thermal capacitance of the dwelling \\
    $T^{in}_{t}$ [K] & Indoor temperature at time step $t$ \\
    $T^{out}_{t}$ [K] & Outdoor temperature at time step $t$ \\   
    $TR$ [K/W] & Thermal resistance of the dwelling \\
    $Q^{pump}_{t}$ [W] & Thermal power of the heat pump at time step $t$ \\
    $gA$ [m²] & Solar aperture \\
    $Q^{sun}_{t}$ [W/m] & Solar irradiance at time step $t$ \\
\end{tabular}

The primary heating and cooling coils and the backup heating coil are modelled linearly using a fixed efficiency conversion between the electric power the thermal power as shown in \cref{eq:mpc_heat_pump}. 

\begin{equation}
Q^{pump}_{t} = \eta^{heat}P^{heat}_{t}-\eta^{cool}P^{cool}_{t}+\eta^{backup}P^{backup}_{t}
\label{eq:mpc_heat_pump}
\end{equation}

where:

\begin{tabular}{ll}
    $\eta^{coil}$ [--] & Efficiency of the heat, cool or backup coil 
     \\
    $P^{coil}_{t}$ [W] & Electric power of the heat, cool or backup coil at time step $t$ \\
\end{tabular}

The total electric power of the heat pump $P_{t}^{pump}$ is modelled as the sum of the electric power of the primary heating and cooling coils and the backup heating coil as shown in \cref{eq:mpc_heat_pump_total}.

\begin{equation}
P^{pump}_{t} = P^{heat}_{t}+P^{cool}_{t}+P^{backup}_{t}
\label{eq:mpc_heat_pump_total}
\end{equation}

The parameters for the thermal capacitance ($C^{dwell}$), thermal resistance ($TR$), solar aperture ($gA$), and coil efficiencies ($\eta^{heat}$, $\eta^{cool}$, $\eta^{backup}$) are calibrated by fitting the MPC model to the detailed simulation results for each dwelling over the year 2022. 

When using the accurate simulation values for $T_t^{out}$, $T_t^{in}$, $Q_t^{pump}$ and $Q_t^{sun}$, the calibrated model from \cref{eq:mpc_thermal_model} predicts the indoor temperature for the next time step $T_{t+1}^{in}$ with a mean absolute error of 0.12\,°C (standard deviation 0.14\,°C) across all dwellings. This outperforms a naive baseline that assumes the indoor temperature remains constant, which yields a mean absolute error of 0.16\,°C (standard deviation 0.31\,°C).
These results indicate that the simulation's thermal dynamics are more complex than those captured by the simplified RC model used in the MPC.

The maximum electric power of each heat pump coil is constrained by the maximum capacities defined by the Openstudio-HPXML workflow for the respective dwelling.
The MPC model does not include the ventilation consumption and the defrosting coil consumption present in the simulation model.

The indoor temperature $T^{in}_{t}$ is constrained to be within the comfort range defined in \cref{sec:heat_pump_control} for the predicted horizon.

The water heater's MPC model closely mirrors the simulation model by implementing \cref{eq:tank_temp_change} and enforcing heating at maximum power for 15 minutes whenever the tank temperature falls below 55\,°C.
The main difference is that, in the MPC model, the tank temperature is strictly limited to a maximum of 65\,°C, rather than allowing it to exceed this value and simply disabling heating in the following time step.

The electric vehicle charger is modelled using \cref{eq:ev_charging_reliability} for the electric vehicle charging and the charging session is constrained to charge a given amount of energy in a given time period.
This exactly matches the simulation model of the charger.

The total electric power of the dwelling is the sum of the different assets as shown in \cref{eq:mpc_total_power}.

\begin{equation}
    P^{total}_{t} = P^{pump}_{t}+P^{water}_{t}+P^{EV}_{t}-P^{PV}_{t}+P^{load}_{t}
\label{eq:mpc_total_power}
\end{equation}

Overall, the MPC model provides a close approximation of the simulation model.

\subsubsection{Objective function}

To calculate the objective function, the energy offtake $Eo_t$ and injection $Ei_t$ are calculated using the total electric power $P^{total}_{t}$ following \cref{eq:mpc_energy_offtake,eq:mpc_energy_injection}.

\begin{equation}
    Eo_t = 
    \begin{cases}
        P^{total}_{t} \Delta t, & \text{if } P^{total}_{t} \geq 0 \\
        0, & \text{otherwise}
    \end{cases}
    \label{eq:mpc_energy_offtake}
\end{equation}
\begin{equation}
    Ei_t = 
    \begin{cases}
        -P^{total}_{t} \Delta t, & \text{if } P^{total}_{t} < 0 \\
        0, & \text{otherwise}
    \end{cases}
    \label{eq:mpc_energy_injection}
\end{equation}

Model predictive control uses an adapted version from \cref{eq:tot_cost_usefull} as its objective function without the yearly cost, which cannot be optimized, and where the peak cost increase is directly reflected on the prediction horizon (see \cref{eq:mpc_obj_relevance}).

\begin{equation}
O^{MPC}=\underbrace{\sum_{t\in H}{(Eo_tVo_t-Ei_tVi_t)}}_{\text{day-ahead cost}}+\underbrace{\sum_{t\in H}Eo_tVf}_{\text{offtake extras cost}}+\underbrace{Vp \times \max_{{t\in H}} (\frac{Eo_t}{\Delta t}, Pp)}_{\text{peak cost}}
\label{eq:mpc_obj_relevance}
\end{equation}

where:

\begin{tabular}{ll}
    $H$ & Set of time steps in the prediction horizon \\
    $Pp$ [kW] & Previous grid peak power for that month (minimum 2.5 kW) \\
\end{tabular}

\subsubsection{Forecast}

The forecasted variables used by the MPC's model are:

\begin{itemize}
    \item The outdoor temperature ($T^{out}_{t}$)
    \item The solar irradiance ($Q^{sun}_{t}$)
    \item The hot water flow ($F_t$)
    \item The arrival time of the EV
    \item The departure time of the EV
    \item The charged energy of the EV
    \item The PV production ($P^{PV}_{t}$)
    \item The base load consumption ($P^{load}_{t}$)
\end{itemize}

MPC P uses the actual future values of all relevant variables over its 48-hour prediction horizon.

For the realistic MPC, all non EV related variables in the next hour are predicted using the most recent value, while subsequent time steps are predicted using values from the preceding 24 hours at the same time of the day.

Predicting the arrival time of an EV is particularly challenging due to the high variability in user behaviour.
Therefore, the MPC only assumes that an EV arrives when a new charging session starts. For each new charging session, the departure time and charged energy are predicted using k-Nearest Neighbor (k-NN) \cite{coverNearestNeighborPattern1967} regression as it showed good performance in \citet{genovForecastingFlexibilityCharging2024} to predict departure time and charged energy of EVs.
The forecasting features used are the arrival time and week day, the charged energy from the previous charging session, the total connection time of the previous charging session, and the time since the start of the previous charging session. The model is trained on the 2022 charging profile data of the dwelling the MPC is controlling and the k-NN model is retrained with one additional charging session data every time a new charging session is finished during simulation.

To account for prediction errors, if the MPC incorrectly predicts that the EV is fully charged but it remains connected and not fully charged, the MPC will schedule charging at 3.7 kW for the next 2 hours to ensure additional charging. Conversely, if the MPC overestimates the departure time and the enforced charging behaviour was triggered at the previous time step (see \cref{sec:ev_charger_control}), then the MPC assumes the EV will be charged at maximum power for the next 2 hours.

\subsection{TreeC}

\subsubsection{Training process}

The training of TreeC is performed by evaluating the performance of different decision tree based EMSs on the simulation of the year 2022. The enforced charging mechanism described in \cref{sec:ev_charger_control} is not used during the TreeC training simulations.

Two changes have been made to the original TreeC method described in \cite{ruddickTreeCMethodGenerate2025}:

\begin{itemize}
\item The covariance matrix adaptation evolution strategy was replaced by the metaheuristic algorithm particle swarm optimization \cite{kennedyParticleSwarmOptimization1995} as it performed better in \cite{ruddickRealworldValidationSafe2024}.
This study used the particle swarm optimization algorithm implemented in the Python library pygmo \cite{biscaniParallelGlobalMultiobjective2020} with default parameters which corresponds to the canonical particle swarm optimization algorithm described in \cite{poliParticleSwarmOptimization2007}.
\item After training, the best decision tree is pruned by iteratively removing leaves whose removal does not degrade the objective value by more than 5\,\euro{} compared to the unpruned tree. This threshold is selected to balance performance and interpretability: it is low enough to avoid sacrificing meaningful cost savings, yet high enough to eliminate leaves that contribute minimally to overall performance.

The least used leaves are tested for removal first. 
\end{itemize}

The TreeC EMS is trained in simulation with the particle swarm optimization algorithm over 300 generations and a population of 200 individuals.
This means 60\,000 decision tree based EMSs are evaluated in simulation on the 2022 training data and the best one is then selected.
The population size of 200 was chosen as it obtained better results for the real-world and non-unimodal benchmarks of the IEEE Competitions in Evolutionary Computation compared to the more historically implemented population sizes of 20-50 individuals \cite{piotrowskiPopulationSizeParticle2020}.
To reduce the risk of converging to a local optimum, five independent EMSs are trained and pruned for each dwelling. The EMS with the best performance among the five is then selected for use in the experiment.

The TreeC training was performed on a high performance computer
composed of Intel\textregistered{} Xeon\textregistered{} Gold, AMD EPYC\texttrademark{} 9384X and AMD EPYC\texttrademark{}  9535 processors. Each training used 100 cores and evaluated 300\ 000 EMSs (60\ 000 per metaheuristic optimization $\times$ 5 to avoid local optimums). On average a training took 5 days and 20 hours to complete with each EMS evaluation taking on average 34 seconds on a single core corresponding to the simulation of the dwelling for one year.

\subsubsection{Decision tree model}

The training can select from up to 23 different inputs to use as splitting features in the decision trees.
The inputs related to specific assets are not used in the dwellings where the asset is not present.

The inputs are:
\begin{itemize}
\item The time of the day
\item The day of the week
\item The cosine of the day of the year, normalized so that $-1$ corresponds to December 21\textsuperscript{st} (winter solstice) and $+1$ to June 21\textsuperscript{st} (summer solstice)
\item $P^{asset}_{t-1}$, the average power of each asset (heat pump, water heater, the EV charger, base load, PV) and total power for the previous time step (6 inputs)
\item $T^{in}_{t}$, the indoor temperature
\item $T^{low}_{t}$, the lower bound of the comfort range for the next time step, in 30 minutes and in 2 hours (3 inputs)
\item $T^{tank}_{t}$, the water tank temperature
\item $T^{out}_{t}$, the outdoor temperature
\item $Q^{sun}_{t}$, the solar irradiance
\item $Vd_t$, the day-ahead price
\item $\max_{{h\in H}}(Vd_h)-Vd_t$ the difference between the maximum and the current price for the next 6, 12 and 24 hours with $H$ being the set of future time steps (3 inputs)
\item $Vd_t-\min_{{h\in H}}(Vd_h)$, the difference between the current price and the minimum price for the next 6, 12 and 24 hours with $H$ being the set of future time steps (3 inputs)
\end{itemize}

The TreeC EMS can control the heat pump through three types of setpoints: comfort rule-based, temperature setpoint and adjustment to indoor temperature.
The comfort rule-based setpoint in TreeC follows the same logic as in the rule-based controller: it sets the heat pump setpoints for cooling and heating to match the comfort range over a specified time window (see \cref{sec:relevance_rbc}). Unlike the rule-based controller, however, TreeC select the length of this time window anywhere from 0 to 12 hours.
The temperature setpoint sets the heat pump setpoints for cooling and heating to a fixed temperature between 5,\,°C and 40\,°C. 
The adjustment to indoor temperature setpoint sets the heat pump's heating and cooling setpoints relative to the current indoor temperature $T^{in}_{t}$. For heating, the setpoint is defined as $T^{in}_{t}$ plus a fixed offset; for cooling, as $T^{in}_{t}$ minus a fixed offset. This fixed offset can be selected within the range of $-2\,^\circ$C to $5\,^\circ$C.

The TreeC EMS control the water heater and EV charger by setting directly the power of the asset.

\subsubsection{Objective function}

The TreeC EMS's goal is to minimize the total electricity cost, keep the indoor temperature within the comfort range, and charge the EV sufficiently before departure.
The total electricity cost $Ctot$ is calculated using \cref{eq:tot_cost_usefull}.

Discomfort is measured in Kh (Kelvin-hours), which represents how long and by how much the indoor temperature deviates from the comfort range. For example, being 2 °C above the comfort range for 1 hour results in 2 Kh of discomfort.

The discomfort above the upper bound of the comfort range $\delta^{up}_t$ and below the lower bound $\delta^{low}_t$ are calculated using \cref{eq:discomfort_cost_up,eq:discomfort_cost_low}.

\begin{equation}
     \delta^{up}_t = \int_{0}^{\Delta t} \max\left( \frac{T_{t+1} - T_{t}}{\Delta t} x + (T_t - T^{up}_t),\, 0K \right) dx
     \label{eq:discomfort_cost_up}
\end{equation}

\begin{equation}    
    \delta^{low}_t = \int_{0}^{\Delta t} \max\left( \frac{T_{t} - T_{t+1}}{\Delta t} x + (T^{low}_t - T_t),\, 0K \right) dx
    \label{eq:discomfort_cost_low}
\end{equation}

where:

\begin{tabular}{ll}
    $T_{t}$ [K] & Indoor temperature at time step $t$ \\
    $\Delta t$ [h] & Length of the time step \\
    $T^{up}_t$ [K] & Upper bound of the comfort range at time step $t$ \\
    $T^{low}_t$ [K] & Lower bound of the comfort range at time step $t$ \\
\end{tabular}

The total discomfort cost $C\delta$ is calculated by \cref{eq:discomfort_cost_usefull}.

\begin{equation}
    C\delta = 
   max(0\SI{}{\EUR},
        w1\times(\sum_{t \in T} (\delta^{up}_t + \delta^{low}_t)-30 Kh))
    \label{eq:discomfort_cost_usefull}
\end{equation}

where:

\begin{tabular}{ll}
    $C\delta$ [\euro] & Discomfort cost \\
    $w_1$ [\euro/Kh] & Penalty weight for discomfort cost (5 \euro/Kh) \\
\end{tabular}

A discomfort penalty is only applied if the total annual discomfort exceeds 30 Kh, which corresponds to a low level of discomfort over the year. For any discomfort above this threshold, a penalty of 5\,\euro{}/Kh is imposed, strongly incentivizing the EMS to maintain thermal comfort.

The cost for missed EV charging energy $Ccharg$ is calculated using \cref{eq:charg_cost_usefull}.

\begin{equation}
    Ccharg = \sum_{s \in S} 
    \begin{cases}
        w_2 \times E^{miss}_s, & \text{if } E^{miss}_s \leq E^{5\%} \\
        w_2 \times E^{5\%} + w_3 \times (E^{miss}_s - E^{5\%}), & \text{otherwise}
    \end{cases}
    \label{eq:charg_cost_usefull}
\end{equation}

where:

\begin{tabular}{ll}
    $Ccharg$ [\euro] & Penalty cost for missed EV charging energy \\
    $S$ & Set of all charging sessions \\
    $E^{miss}_s$ [kWh] & Missed energy in session $s$ \\
    $E^{5\%}$ [kWh] & Energy corresponding to charging the EV by 5\% SOC \\
    $w_2$ [\euro/kWh] & Penalty weight for missed energy up to $E^{5\%}$ (0.5 \euro/kWh) \\
    $w_3$ [\euro/kWh] & Penalty weight for missed energy exceeding $E^{5\%}$ (2 \euro/kWh) \\
\end{tabular}

A missed energy event occurs when the simulated EMS charges less energy than what was originally delivered in the dataset for a given charging session. In other words, it is the shortfall between the original and the actually delivered energy by the EMS. 

Missing up to 5\% of the EV's SOC in a charging session is undesirable but not critical, so a moderate penalty of 0.5\,\euro/kWh is applied reflecting an exceptionally high price of electricity (see \cref{tab:day_ahead_price_stats}). However, missing more than 5\% SOC is considered a significant problem, and a much higher penalty of 2\,\euro/kWh is imposed to strongly discourage such outcomes.

The total objective function $O$ of the TreeC EMS is calculated by summing the different costs from \cref{eq:tot_cost_usefull,eq:discomfort_cost_usefull,eq:charg_cost_usefull} as shown in \cref{eq:treec_obj_usefull}.

\begin{equation}
    O^{TreeC} = Ctot +  C\delta +  Ccharg
    \label{eq:treec_obj_usefull}
\end{equation}

\subsubsection{Interpretable decision tree}

\begin{figure}[H]
    \centering
    \begin{subfigure}{0.48\textwidth}
        \includegraphics[width=\linewidth]{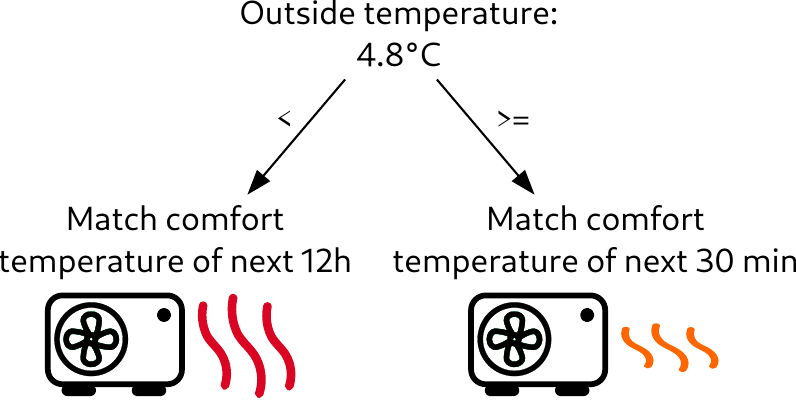}
        \caption{Heating setpoint
        }
        \label{fig:heating_decision_tree}
    \end{subfigure}
    \hfill
    \begin{subfigure}{0.48\textwidth}
        \includegraphics[width=\linewidth]{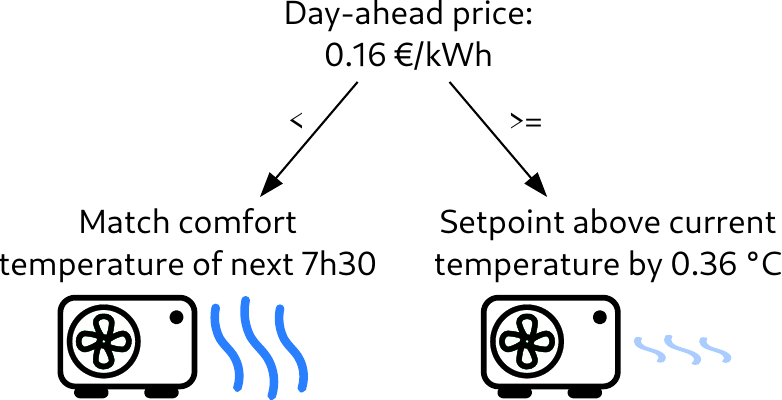}
        \caption{Cooling setpoint}
        \label{fig:cooling_decision_tree}
    \end{subfigure}
    \vskip\baselineskip
    \begin{subfigure}{0.42\textwidth}
        \includegraphics[width=\linewidth]{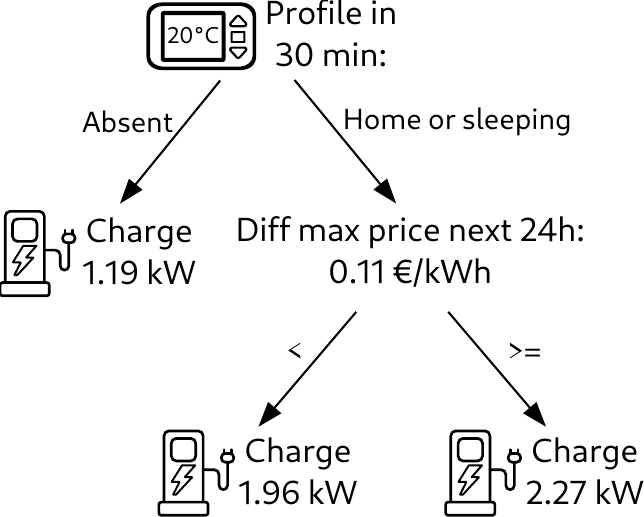}
        \caption{EV charger power setpoint}
        \label{fig:charger_decision_tree}
    \end{subfigure}
    \hfill
    \begin{subfigure}{0.54\textwidth}
        \includegraphics[width=\linewidth]{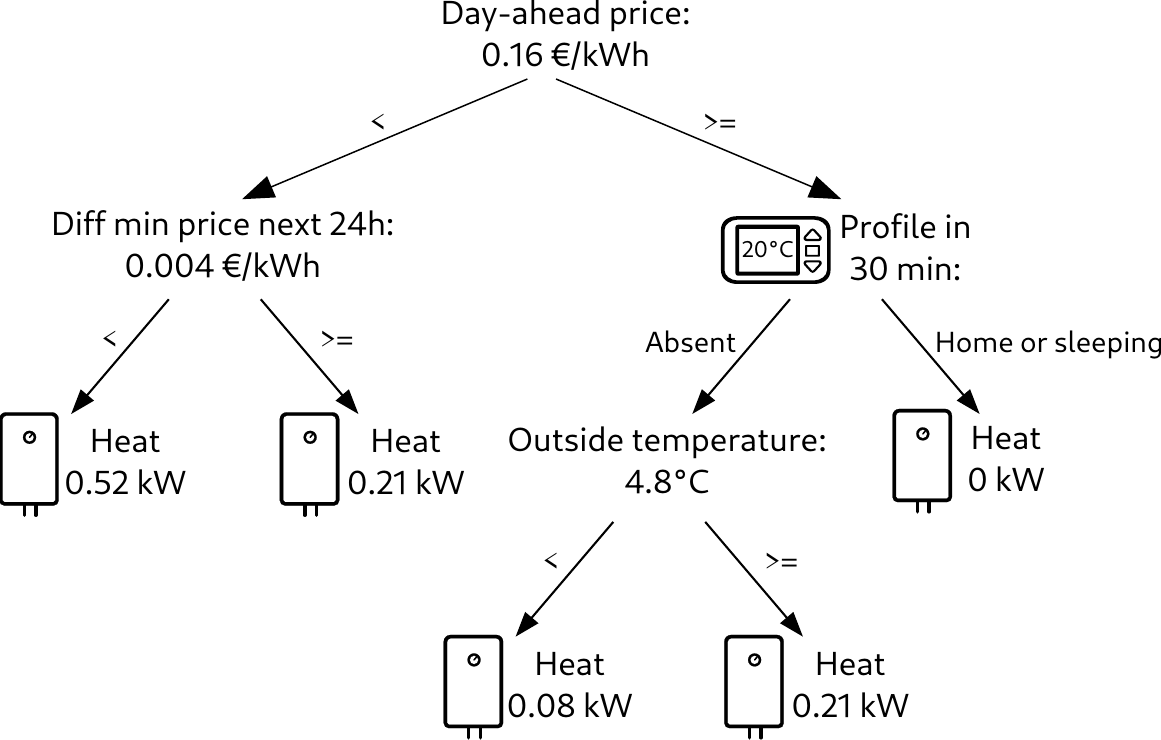}
        \caption{Water heater power setpoint}
        \label{fig:water_heater_decision_tree}
    \end{subfigure}
    \caption{Decision trees for each controllable asset in the TreeC EMS for dwelling 1.}
    \label{fig:complete_dwelling}
\end{figure}

As an example, the TreeC EMS obtained after training for dwelling 1 is presented  in \cref{fig:complete_dwelling} as it is the first dwelling with all possible assets present (PV, heat pump, water heater and EV charger).

The decision tree is interpretable for each of the set points it controls.
For the heating setpoint in \cref{fig:heating_decision_tree}, when the outdoor temperature is cold (below 4.8 °C), the EMS is precautious and matches the comfort range for the next 12 hours to not incur discomfort, otherwise it matches it for the next 30 minutes as this is more energy efficient. For the cooling setpoint in \cref{fig:cooling_decision_tree}, when the day-ahead price is considered low enough (below 0.16\,\euro/kWh), the EMS matches the comfort range 7h30 minutes ahead, otherwise it just avoids that the indoor temperature increases too fast. For the EV charger in \cref{fig:charger_decision_tree}, the EMS charges between 1.96 kW and 2.27 kW when the occupancy profile is marked as home or sleeping in 30 minutes and at a lower power of 1.19 kW then occupancy profile is absent. The EV charger never charges at full power to avoid peaks. For the water heater in \cref{fig:water_heater_decision_tree}, the EMS heats the water more when the day-ahead price is considered low enough and doesn't heat at all when the occupancy profile is marked as home or sleeping in 30 minutes. 

The figures and explanation highlight the interpretability of the TreeC EMS and also illustrate how it can coordinate the operation of different assets. For example, the EV charger tends to charge more when the water heater is not heating, helping to avoid simultaneous power peaks. This is achieved by both controllers using the same occupancy profile feature.

\section{Results}

\begin{figure}[H]
    \centering
    \includegraphics[width=1.0\textwidth]{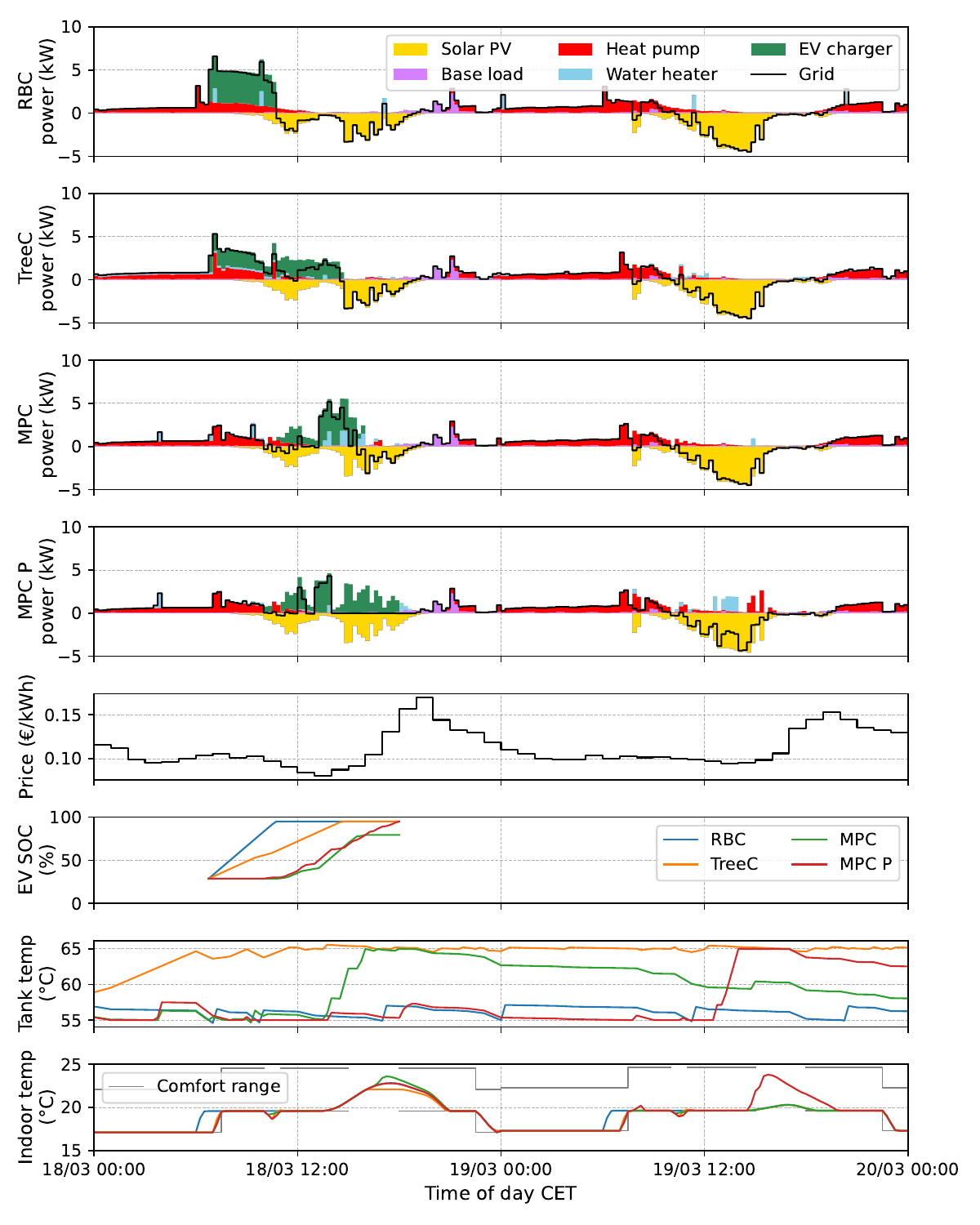}
    \caption{Illustration of the operation of the different EMSs over the weekend of 18-19 March 2023 for dwelling 1.
    The top four plots show the stacked power consumption and production of each asset, with the total grid connection power indicated in black.
    The four lower plots show the day-ahead electricity prices, the state of charge of the connected EV, the water heater's tank temperature, and the indoor temperature for each EMS.
    }
    
    \label{fig:example_weekend}
\end{figure}

\Cref{fig:example_weekend} shows how the different EMSs operate during the last weekend of winter for dwelling 1. This dwelling is a detached house in Flanders with all assets present: heat pump, water heater, EV charger, and PV system. The water heater has a 171 litre tank, the house is partially insulated (roof only), and the PV system has a capacity of 5 kWp.
This dwelling was chosen because it is the first with all assets present, and the selected weekend highlights an unsuccessful EV charging session under the MPC EMS. 
In this example, MPC incorrectly forecasted that the EV would remain connected until 12:30 on the 19\textsuperscript{th} of March, whereas in reality, the EV departed at 18:00 on the 18\textsuperscript{th} of March. Because the day-ahead prices, and therefore injection prices, before 18:00 on the 18\textsuperscript{th} of March were higher than those before 12:30 on the 19\textsuperscript{th} of March, MPC planned to inject excess PV generation on the 18\textsuperscript{th} March and use the excess PV on 19 March to charge the EV. 
Mathematically this all makes sense, but since the prediction was incorrect, the EV was not fully charged by the time of departure contrary to the other EMSs.
RBC charges to the maximum power, TreeC charges at a lower power to avoid peaks but still high enough to ensure the EV is fully charged and MPC P charges the EV using all excess PV production and cheap day ahead prices.

For the domestic hot water, TreeC heats the water tank over almost the entire weekend because the price of electricity is mostly below 0.16\,\euro/kWh (see \cref{fig:water_heater_decision_tree}), MPC and MPC P make more use of the excess PV production to heat the water tank, and RBC keeps the water tank around the lower 55 °C bound.

For heating, all EMSs generally maintain the indoor temperature near the lower bound of the comfort range except for moments where the sun and outdoor temperatures heat the house naturally. MPC and MPC P take advantage of excess PV production to provide additional heating at certain times. RBC starts heating earlier in the morning than the other EMSs, which results in slightly lower energy efficiency.

\begin{figure}[H]
    \centering
    \begin{subfigure}{0.48\textwidth}
        \includegraphics[width=\linewidth]{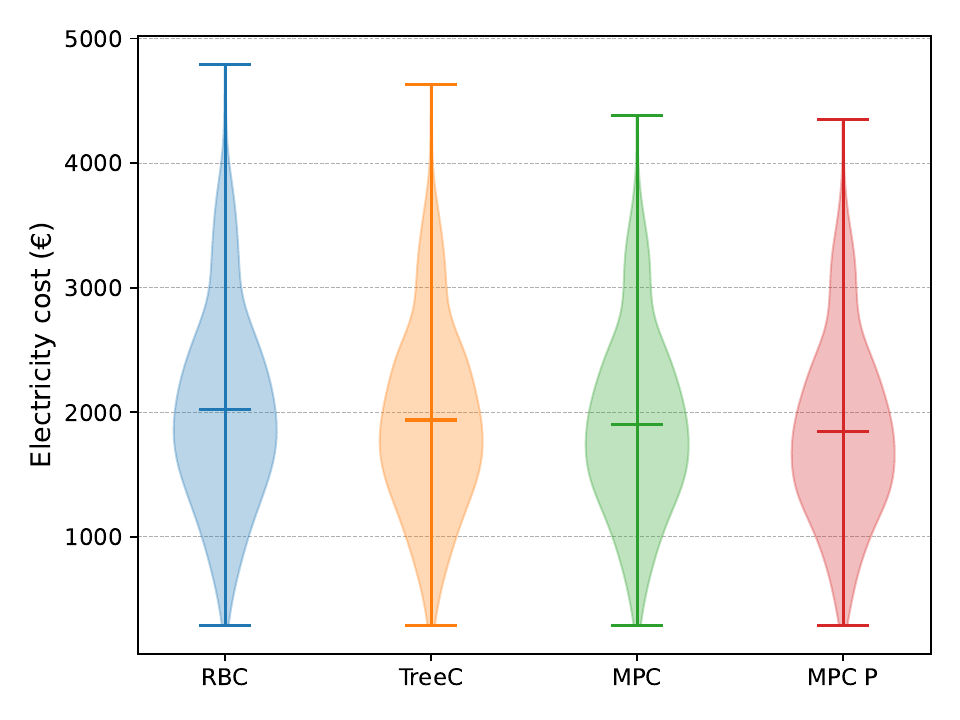}
        \caption{Yearly cost}
        \label{fig:total_opex}
    \end{subfigure}
    \hfill
    \begin{subfigure}{0.48\textwidth}
        \includegraphics[width=\linewidth]{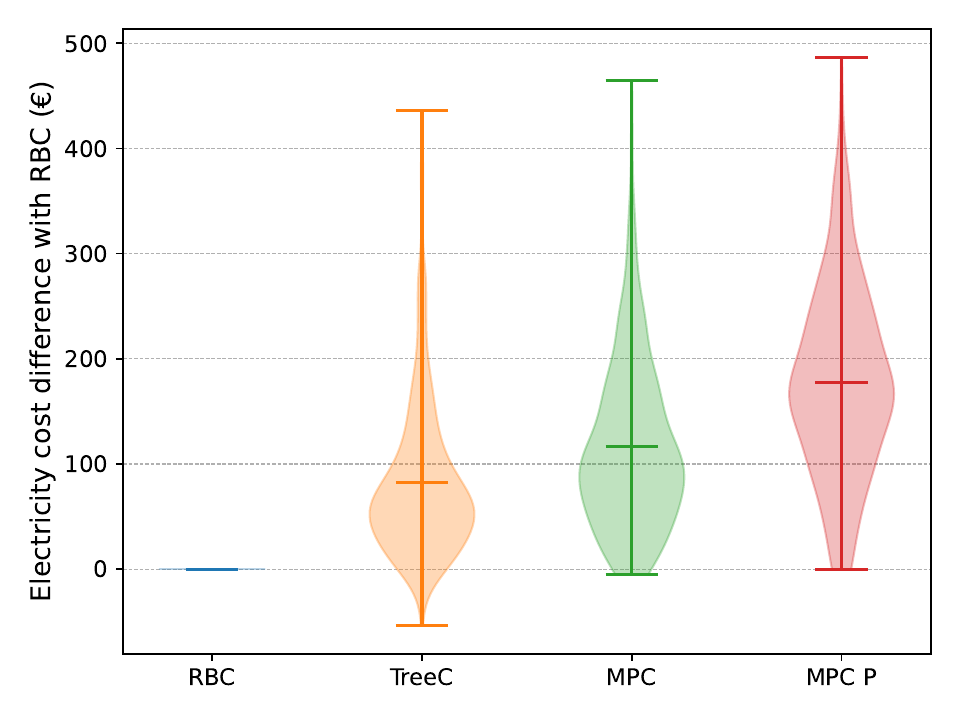}
        \caption{Difference with RBC}
        \label{fig:total_opex_diff}
    \end{subfigure}
    \begin{subfigure}{0.48\textwidth}
        \includegraphics[width=\linewidth]{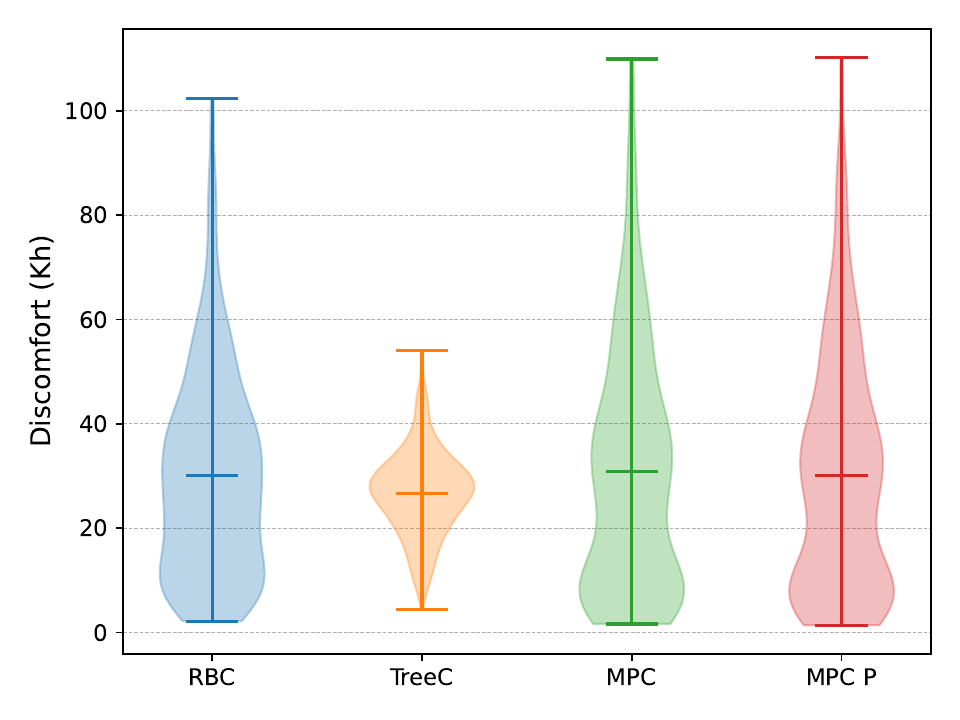}
        \caption{Thermal discomfort}
        \label{fig:discomfort_level}
    \end{subfigure}
    \hfill
    \begin{subfigure}{0.48\textwidth}
        \includegraphics[width=\linewidth]{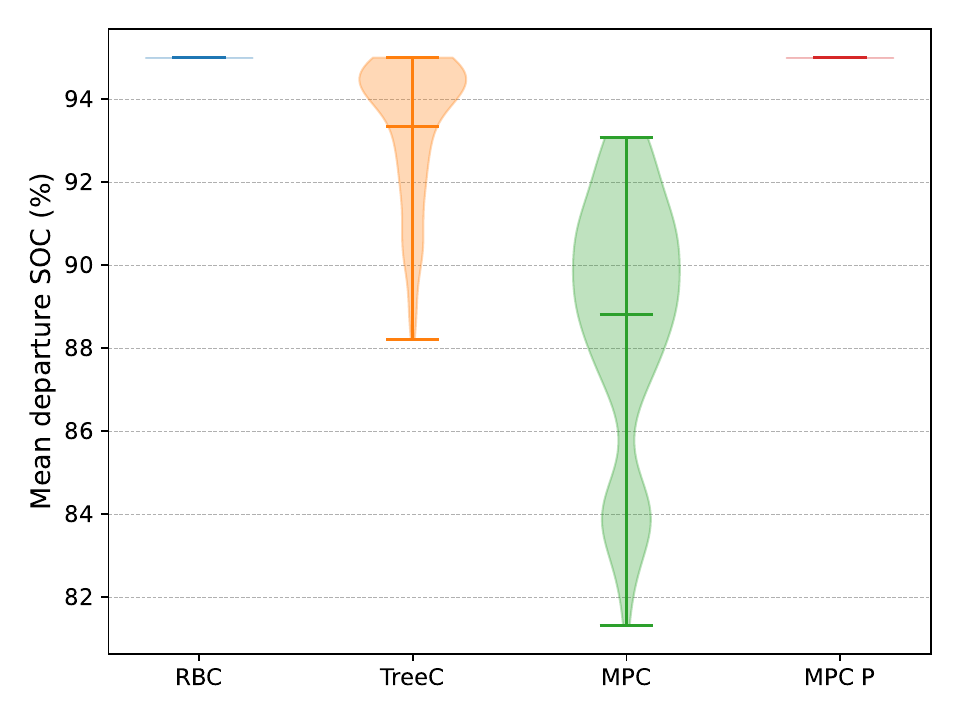}
        \caption{EV charge completion}
        \label{fig:avg_soc_final}
    \end{subfigure}
    
    \caption{Comparison of key performance metrics for the different EMSs: (a) total yearly cost, (b) difference in electricity cost with the RBC EMS, (c) thermal discomfort, and (d) mean departure state of charge for EV charging sessions.
    For each EMS, violin plots (a) and (b) include 500 dwellings, (c) includes the 447 dwellings with a heat pump, and (d) includes the 344 dwellings with an EV charger. Each violin plot shows the distribution of results across all dwellings. The width of the violin at each value reflects the relative frequency of dwellings with that result. The horizontal dashes within each violin indicate the minimum, mean, and maximum values for each EMS.
    }
    \label{fig:performance_comparison}
\end{figure}

Results in \cref{fig:performance_comparison} show different metrics to compare EMSs across the 500 dwellings.
\Cref{fig:total_opex} shows the total yearly cost for each dwelling. The results indicate a hierarchy in performance: the RBC EMS has the highest costs, followed by TreeC, then MPC, with MPC P achieving the lowest costs overall. There is considerable variation in total costs across dwellings, reflecting the impact of factors such as insulation quality, and the presence of EV chargers or PV systems on overall energy consumption.

To provide a clearer comparison of the EMS performance relative to RBC, \cref{fig:total_opex_diff} presents the reduction in total yearly cost for each EMS compared to the RBC EMS. The results show that TreeC achieves a mean cost reduction of 82.5\,\euro{}, MPC achieves 116.6\,\euro{}, and the upper bound MPC P achieves 177.4\,\euro{}.
TreeC and MPC rarely achieve a worse cost than RBC, which is likely counteracted by an increase in thermal comfort.

\Cref{fig:discomfort_level} shows the thermal discomfort for each EMS with all EMSs achieving similar mean discomfort levels between 26 and 31 Kh. TreeC manages to keep all dwellings below 55 Kh discomfort while the other EMSs have some dwellings with discomforts up to 111 Kh.
Overall even 111 Kh can be considered acceptable especially since some dwellings are not insulated at all and this only represents 111 hours in the year when the indoor temperature is 1 °C above or below the comfort range with the rest of the year being within the comfort range.

\ifprintedversion
    \newcommand{\widthfigcummul}{0.9\textwidth}
\else
    \newcommand{\widthfigcummul}{1.0\textwidth}
\fi

\begin{figure}[H]
    \centering
    \includegraphics[width=\widthfigcummul]{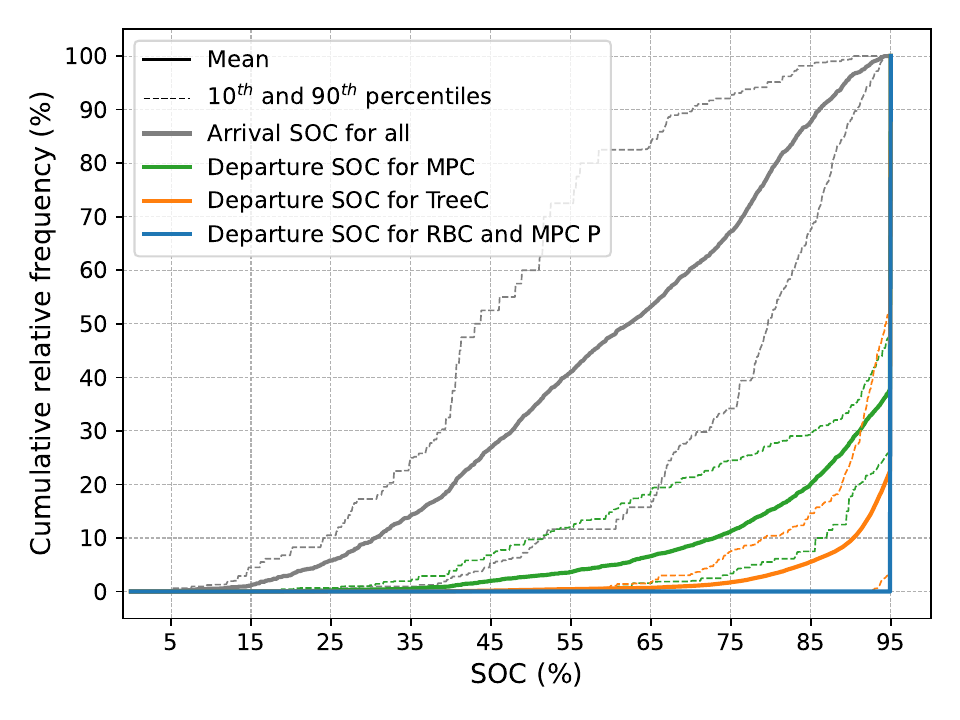}
    \caption{Cumulative distribution of EV charging session SOC at arrival (grey) and departure (coloured) for each EMS. 
        Thick lines show the mean cumulative relative frequency of SOC across all dwellings with an EV charger, representing the percentage of sessions at or below the given SOC. 
        Thin dashed lines indicate the 80\% interpercentile range (10\textsuperscript{th} to 90\textsuperscript{th} percentiles), highlighting variability between dwellings.
        This figure demonstrates how well each EMS ensures EVs are sufficiently charged before departure.
    }
    \label{fig:cummul_freq_soc}
\end{figure}

\Cref{fig:avg_soc_final} shows the mean departure SOC for the EV charging sessions across all dwellings with an EV charger. The original dataset assumed the departure charging SOC to be 95\% for all sessions so it is normal that RBC and MPC P always achieve this value.
For this particular metric TreeC performs better than MPC achieving a mean of 93.3\% departure SOC compared to 88.8\% for MPC.
This shows that TreeC is able to charge the EV more effectively than MPC but the average departure SOC does not always give a complete representation of the charging performance.

To give a better representation, \cref{fig:cummul_freq_soc} presents the cumulative distribution of EV SOC at departure for each EMS. The results show that, with MPC, the SOC falls on average below 85\% in 20\% of charging sessions and below 65\% in 6.6\% of sessions. In contrast, TreeC only drops below these thresholds in 5.5\% and 0.7\% of sessions, respectively. We think that TreeC generally maintains an acceptable charging level for most users while the MPC EMS may leave a significant share of drivers dissatisfied. Developing more refined metrics for driver charging satisfaction would be valuable for future studies to better quantify this aspect.

\ifprintedversion
    \newcommand{\widthfigdwellfeat}{0.48\textwidth}
\else
    \newcommand{\widthfigdwellfeat}{0.48\textwidth}
    
\fi

\newcommand{\dwellfeatfig}{
\ifprintedversion
\begin{figure}[H]
\else
\begin{figure}[ht]
\fi
    \centering
    
    \begin{subfigure}{\widthfigdwellfeat}
        \includegraphics[width=\linewidth]{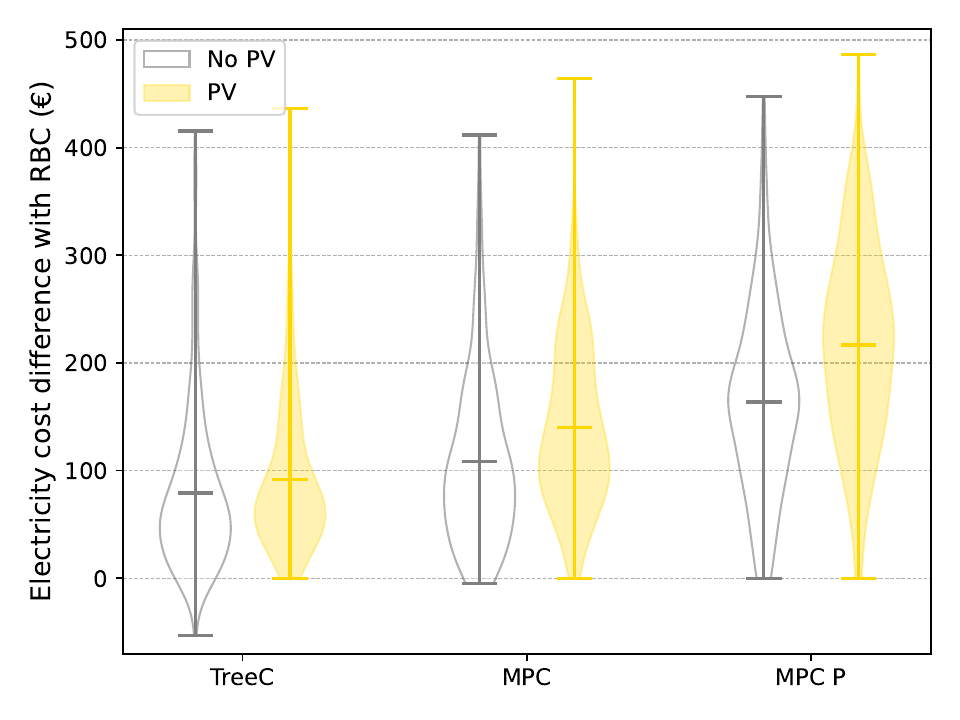}
        \caption{PV}
        \label{fig:pv_results}
    \end{subfigure}
    \hfill
    \begin{subfigure}{\widthfigdwellfeat}
        \includegraphics[width=\linewidth]{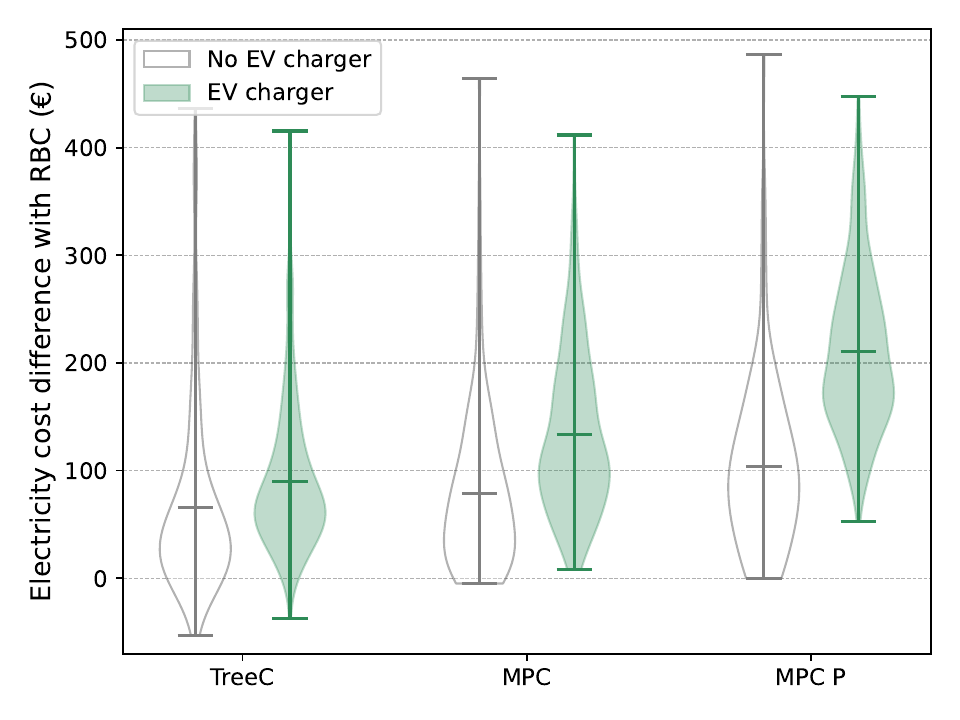}
        \caption{EV charger}
        \label{fig:home_charger_results}
    \end{subfigure}
    \begin{subfigure}{\widthfigdwellfeat}
        \includegraphics[width=\linewidth]{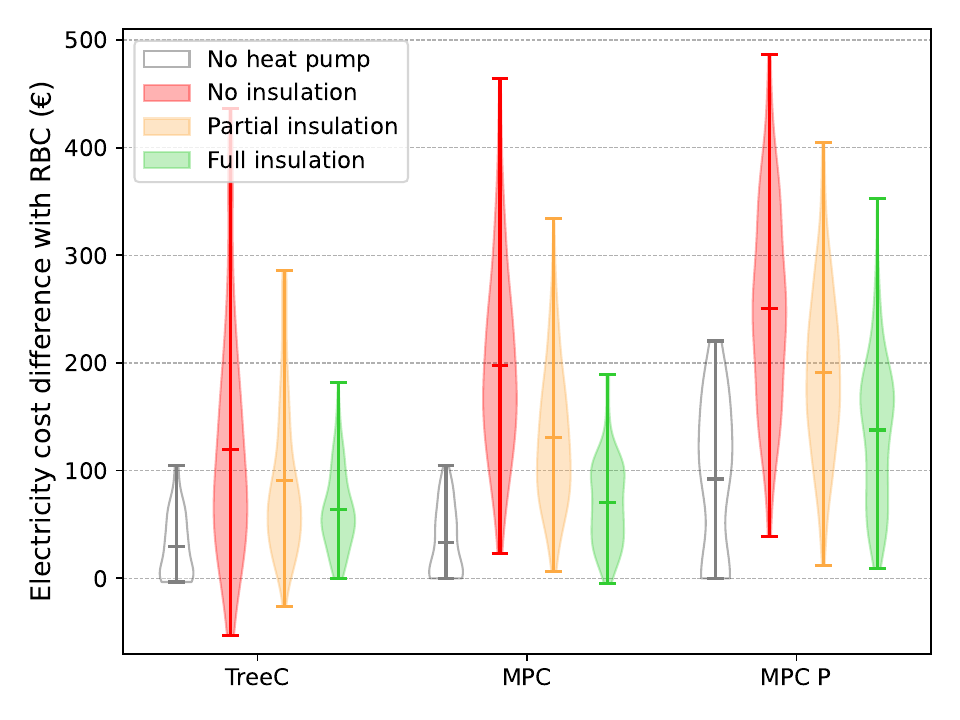}
        \caption{Insulation}
        \label{fig:insulation_results}
    \end{subfigure}
    \hfill
    \begin{subfigure}{\widthfigdwellfeat}
        \includegraphics[width=\linewidth]{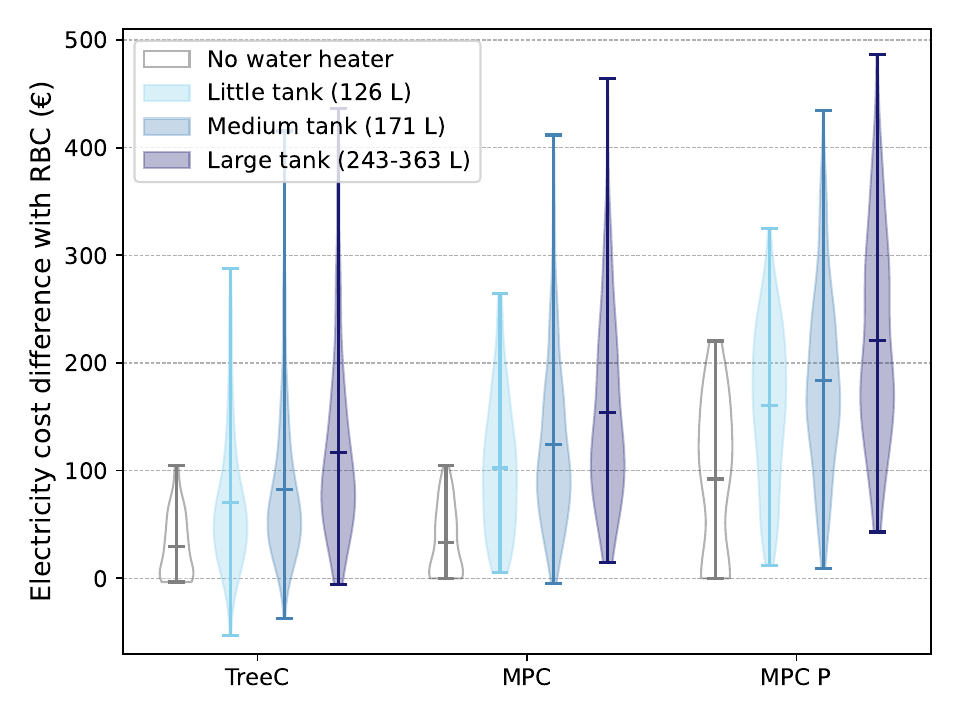}
        \caption{Tank volume}
        \label{fig:water_heater_volume_results}
    \end{subfigure}
    \begin{subfigure}{\widthfigdwellfeat}
    \includegraphics[width=\linewidth]{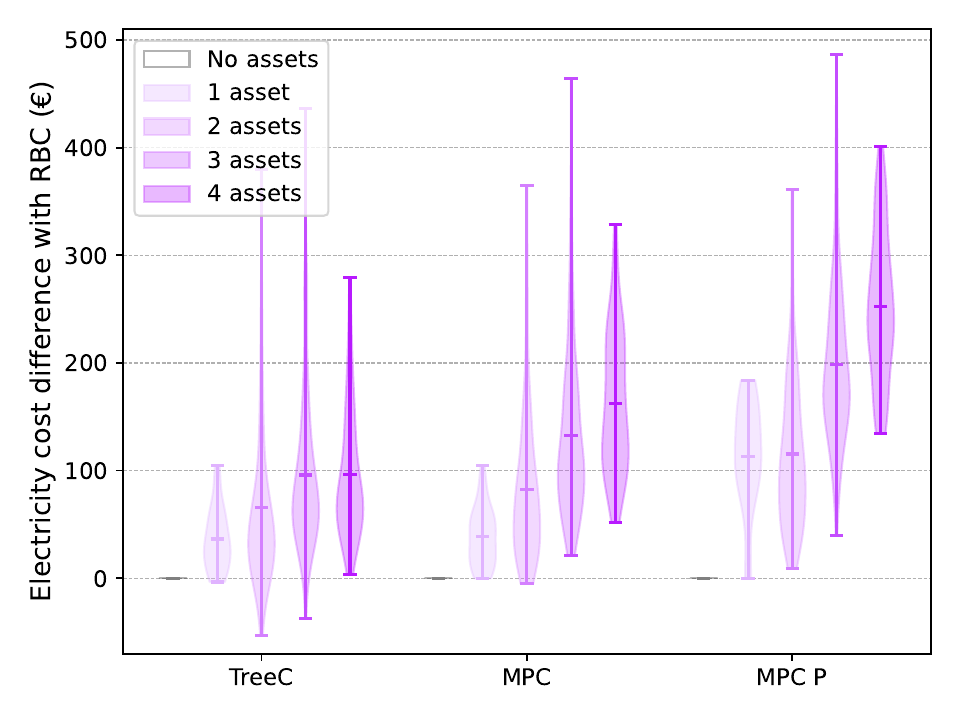}
    \caption{Asset number}
    \label{fig:good_assets_results}
    \end{subfigure}
    \caption{Comparison of EMS performance for different dwelling features: (a) PV presence, (b) EV charger presence, (c) insulation level, and (d) water heater's tank volume (e) number of assets.}
    \label{fig:dwelling_features_results}
\end{figure}
}

\ifprintedversion
\else    
\dwellfeatfig
\fi

\Cref{fig:dwelling_features_results} shows the impact of different dwelling features on the EMS performance. Most results are as expected: \cref{fig:pv_results} shows that an EMS makes better improvements when the dwelling has a PV system, \cref{fig:home_charger_results} shows the same for EV chargers, \cref{fig:insulation_results,fig:water_heater_volume_results} shows the same for central heating and \cref{fig:good_assets_results} shows that generally the more assets a dwelling has, the better the EMS performs.

Interestingly \cref{fig:insulation_results} shows that more insulation has negative impact on the EMSs improvement compared to RBC. This is certainly because dwellings with more insulation have a lower energy consumption and therefore the EMSs have less opportunity to reduce costs.
For the water tank size, \cref{fig:water_heater_volume_results} shows a positive correlation between the tank size and the EMS performance.

\ifprintedversion
\dwellfeatfig
\fi

\ifprintedversion
\begin{figure}[H]
\else
\begin{figure}[ht]
\fi
    \centering
    \includegraphics[width=0.8\textwidth]{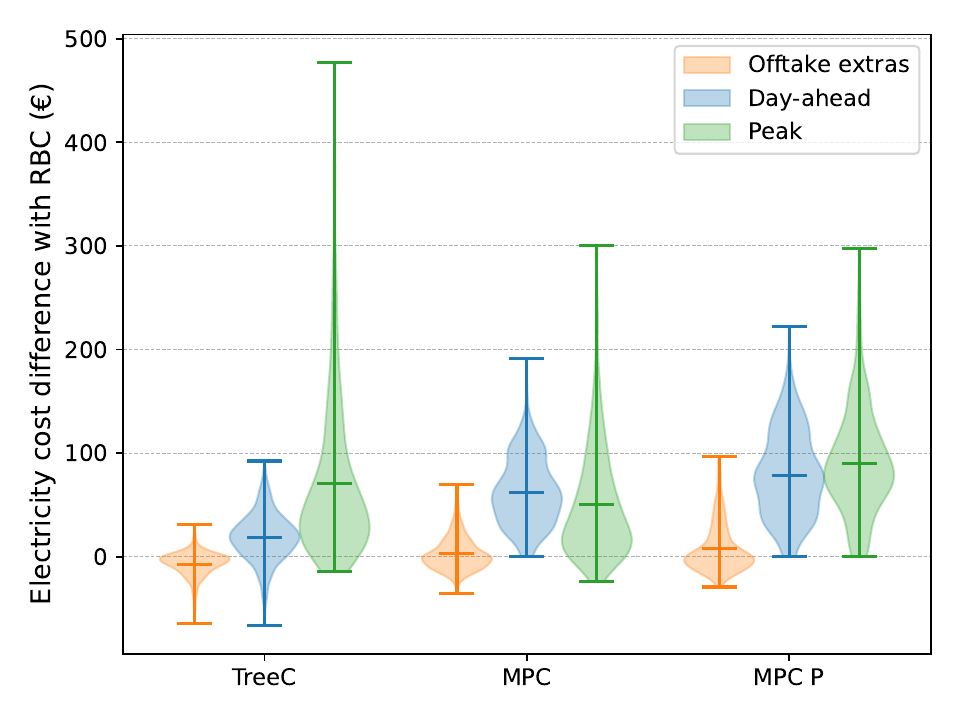}
    \caption{Breakdown of cost savings relative to the RBC EMS for each cost component defined in \cref{eq:tot_cost_usefull}. Offtake extras costs improvement reflects better energy efficiency, day-ahead costs improvement reflects better hourly price optimization, and peak costs improvement reflects reductions in peak grid power. Each box plot summarizes the results across the 500 dwellings.}
    \label{fig:opex_diff_separate}
\end{figure}

To understand better how the EMSs achieve cost reductions, \cref{fig:opex_diff_separate} breaks down the savings by cost component. 

Only minimal savings are achieved in the offtake extras costs related to energy efficiency. This is expected, as the RBC is already quite energy efficient for water and space heating. Efficiency gains can be realized by directly consuming electricity from PV, but 74.2\% of dwellings do not have a PV system.

MPC and MPC P achieve significant cost saving in the day-ahead costs, this is the main cost component that gives these methods their edge over TreeC. This can be explained by MPC based techniques having perfect hourly knowledge of the future day-ahead prices and being able to schedule assets based on these prices more effectively than TreeC which relies on incomplete knowledge of the future prices and a less sophisticated decision tree control model.

For the peak cost component, TreeC slightly outperforms MPC. During training, TreeC learns to minimize peak costs by coordinating asset operation and selecting lower power setpoints, thereby avoiding large simultaneous peaks. While MPC also attempts to coordinate assets to reduce peaks, prediction errors can occur due to forecast or model inaccuracies and result in unexpected peaks, which may explain its lower performance compared to TreeC in this area.

\ifprintedversion
\begin{figure}[H]
\else
\begin{figure}[ht]
\fi
    \centering
    \includegraphics[width=1.0\textwidth]{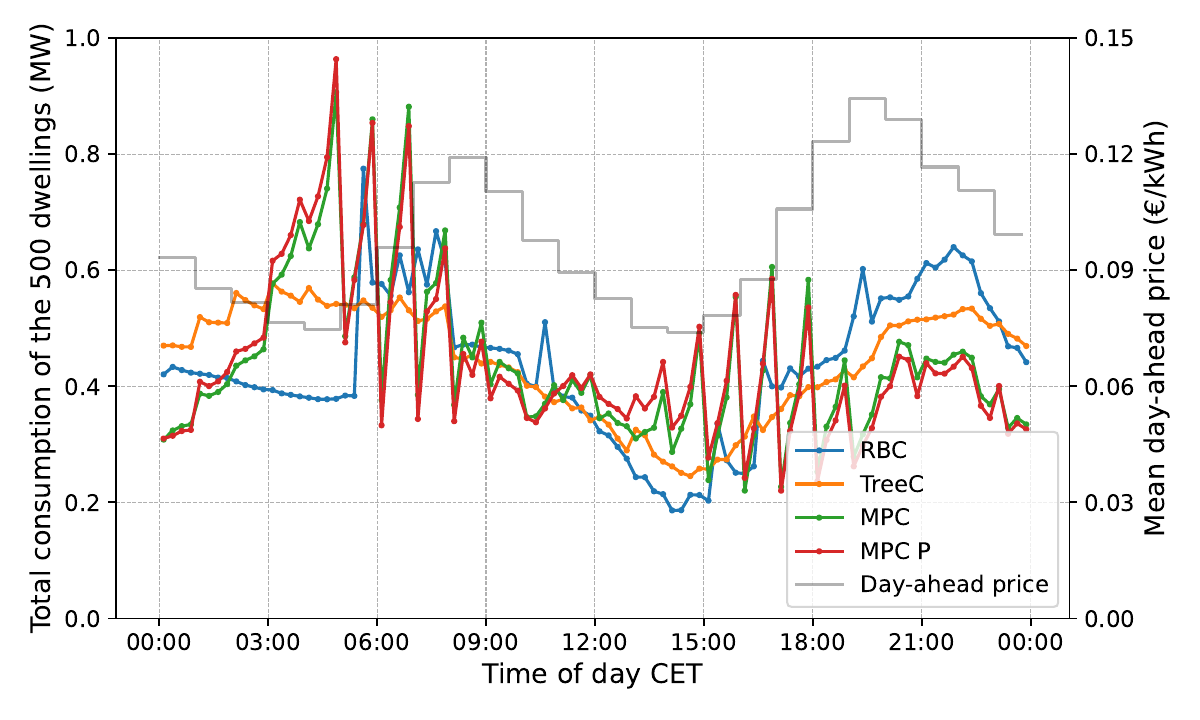}
    \caption{Yearly mean of the total grid power consumption of the 500 dwellings for each time step of the day. Each EMS is represented as well as the mean day-ahead prices for the year 2023.
    }
    \label{fig:average_daily_grid_power}
\end{figure}

Another relevant aspect to consider is how the peak and hourly day-ahead costs influence the total grid power consumption of the 500 dwellings. \Cref{fig:average_daily_grid_power} shows on 24 hours the mean total grid power for each EMS. The MPC and MPC P EMSs show almost identical average behaviour. Overall TreeC and the MPC EMSs consume more than RBC during hours with lower day ahead prices, and all manage reduce the evening RBC total peak. 
However, the MPC EMSs show a clear peak for the last time step of the hours between 04:00 and 08:00 and between 14:00 and 18:00 which are the times the dwelling heat up to reach a higher comfort temperature. This is because the MPC EMSs calculates that it is better to heat the dwelling at the last time step of the hour to gain advantage of the low day-ahead price of that hour and at the same time loose as little as possible of the heat bought at that price. Buying at an earlier time step of that hour would result in more of the bought heat being lost to the exterior. 
The TreeC EMSs didn't learn this behaviour and therefore has a much smoother total grid power consumption profile. 

The peaks caused by the MPC EMSs are problematic for the grid because: 1) they can cause local congestion, requiring costly grid reinforcement, and 2) the sharp, short-lived spikes in aggregate demand increase intra-hour variability, making grid balancing more challenging for operators. This undermines the intent of hourly pricing signals, which are designed to encourage demand shifting across entire hours rather than concentrating consumption at specific moments within those hours.

\section{Discussion}

When analysing all the results, TreeC and MPC have their strengths and weaknesses.
TreeC is more interpretable (see \cref{fig:complete_dwelling}), achieves higher EV charge completion rates, and results in a smoother total grid power consumption profile when all dwellings are equipped with the same EMS. However, MPC obtains more cost reductions, which is arguably the most important metric for evaluating EMS performance.

To improve the cost reduction of TreeC, future work should focus on improving its ability to reduce the day-ahead costs. This can be done by retraining every month the EMS to adapt to the changing day-ahead prices, adding more price related inputs that can be selected by the decision trees, or using more sophisticated action on the leaf nodes of the decision trees. Improving the training time of TreeC through reduced simulation speed, improved training methodology, or transfer learning is also necessary to make it more practical for real-world applications.

\begin{table}[H]
    \centering
    \caption{Mean absolute error (MAE) of two forecasts for the EV charging sessions. Values show the mean and standard deviation of MAE across all dwellings with an EV charger for 2023.}
    \begin{tabular}{|l|l|c|}
        \hline
        Forecast method & Predicted value & MAE \\
         & & (Mean  $\pm$ Standard deviation) \\
        \hline
        K-NN regression & Connection time & 7h14min $\pm$ 4h33min \\
        Repeat last & Connection time & 9h14min $\pm$ 5h7min \\
        K-NN regression & SOC demand  & 12.8\% $\pm$ 3.4\% \\ 
        Repeat last & SOC demand  & 16.4\% $\pm$ 5.4\% \\
        \hline
    \end{tabular}
    \label{tab:ev_charging_stats}
\end{table}

The results for MPC P indicate that improved forecasting can enhance both cost savings and EV charge completion. However, accurately predicting EV user behaviour and charging sessions remains challenging due to the high variability observed in \cref{tab:ev_charging_profile_stats} and the substantial mean absolute errors (MAE) for both k-NN regression and "repeat last" forecasts in \cref{tab:ev_charging_stats} exceeding 7 hours for connection time and 12\% for SOC demand. To address this, future work should consider multi-scenario or stochastic MPC approaches that explicitly account for the uncertainty in EV charging behaviour, as conventional forecasting methods struggle with this aspect.

The mean annual savings compared to the RBC EMS are 82.5\,\euro{} for TreeC, 116.6\,\euro{} for MPC, and 177.4\,\euro{} for MPC P. These savings are relatively modest compared to the cost of buying, installing, maintaining, and subscription fees of current home EMS solutions on the market (between 280\,\euro{} and 680\,\euro{} to buy and 0\,\euro{} and 360\,\euro{} per year for subscription fees \cite{maakjemeterslimAlleEMSProducten2025}). Specific cases can justify an EMS implementation as the results do show some cost reduction going up to 464\,\euro{} are possible, but these are not the norm.

Different factors could increase the commercial attractiveness of advanced EMSs:

\begin{itemize}
\item More volatility in electricity prices. Having more extreme low and high prices would increase the potential savings of shifting consumption towards low price periods.
\item Allowing participation in balancing markets. This would create additional revenue streams for households providing flexibility to the grid.
\item A wider adoption of home batteries. The additional flexibility provided by batteries could increase the potential savings from an advanced EMS.
\item Wider adoption of standardized protocols in large electrical appliances. Standardization would lower EMS integration costs by making it easier to connect and control devices, potentially allowing simple plug-and-play EMS installation.
\end{itemize}

The code and data for this study are open-source, enabling other researchers to benchmark their EMS solutions in the same context. All materials are available at:
\url{https://github.com/EVERGi/should_belgians_install_advanced_ems_paper}
However, further algorithmic improvements alone are unlikely to deliver greater benefits than those demonstrated by MPC P.

\section{Conclusion}

This chapter presented a comparison of three EMSs for electrified residential buildings in Belgium: a rule-based controller, a model predictive control EMS, and a decision tree-based EMS called TreeC. The study evaluated the performance across one year simulations of 500 dwellings with characteristics representative of current Belgian dwellings but where the fossil fuel based assets for heating, hot water and car transportation are replaced by electric equivalents. 

On average, TreeC and MPC achieved cost reductions of 82.5\,\euro{} and 116.6\,\euro{}, respectively, compared to the RBC EMS, which had a mean total cost of 2021\,\euro{}.
MPC obtained better cost reduction but worse EV charge completion with an average of 88.8\% SOC at departure compared to 93.3\% for TreeC with a full charge set at 95\% SOC.
The results show that TreeC and MPC performed better than RBC in terms of cost for dwellings, with PV systems, with EV chargers, larger water tanks, and more assets in general which is expected, it also worked better for dwellings with less insulation which is a less intuitive result.
The total power consumption of the 500 dwellings together obtained large volatility from one time step to the next when all dwellings use MPC, while TreeC obtained a smoother total power consumption profile. 

The main recommended improvement for TreeC is to enhance its ability to reduce hourly day-ahead costs. This could be achieved by retraining the EMS more frequently to adapt to changing electricity over time or by making the decision tree model more responsive to hourly price signals. For MPC, future work should focus on addressing the uncertainty in EV charging sessions, for example by adopting stochastic or multi-scenario MPC approaches.

In most cases, the cost savings achieved by advanced EMSs are modest relative to the investment and effort required for their implementation, when compared to the default operation of assets (RBC EMS). Further algorithmic improvements are unlikely to deliver significantly greater benefits than a mean cost reduction of 177.4\,\euro{} achieved by the MPC with perfect forecast (MPC P) EMS.
Increasing the potential cost savings, and thus the attractiveness of advanced EMSs, could be achieved by modifying electricity pricing structures, enabling participation in flexibility markets, reducing EMS investment and maintenance costs, or integrating home batteries.

    \chapter[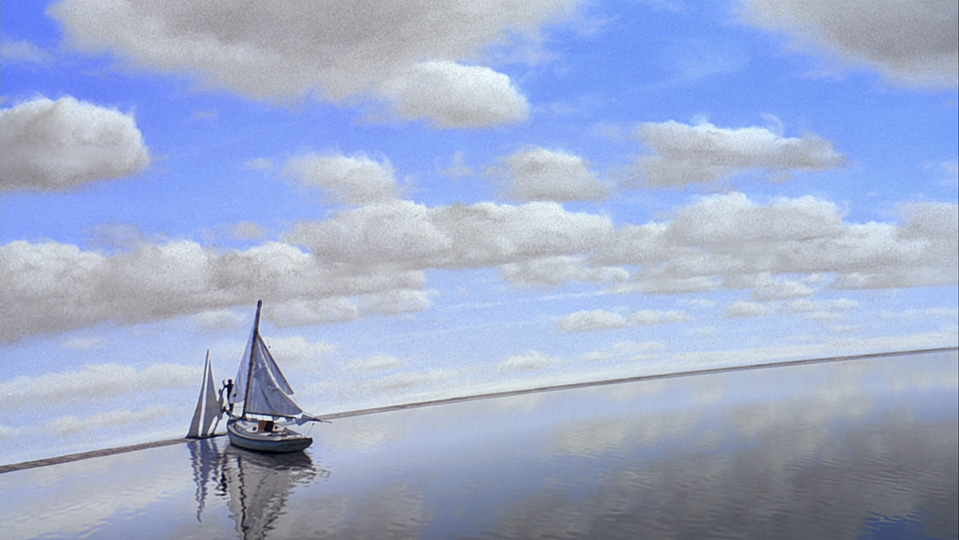]{Conclusions, recommendations, and future work\label{ch:conclusions}}

\section{Answers to the research questions}

As residential systems become more electrified, the need for effective energy management systems (EMS) becomes increasingly important. 
This thesis has contributed to improving the research knowledge in this field guided by five main \textbf{research questions}: 

\begin{enumerate}
    \item Can an energy management system method use interpretable control models, improve its control model by training on data, and be versatile ?
    \item Is such an energy management system performant meaning that it obtains performances closer to proven technology such as model predictive control than to baseline methods such as rule-based control in a variety of cases ?
    \item Can such an energy management system be implemented in a real case and how does it perform compared to baseline and proven technology methods ?
    \item What are the tangible benefits for residents of deploying energy management systems in electrified dwellings in Belgium?
    \item Can we define a new machine learning paradigm that categorizes algorithms used to solve control problems but do not fit in the three traditional machine learning categories ?
\end{enumerate}

The first research question was addressed by developing the TreeC energy management system. As a machine learning-based approach, TreeC is capable of improving its decision-making model through data-driven training. Its decision tree structure ensures interpretability, allowing users to understand and trust its decisions. 
Furthermore, TreeC demonstrated versatility by being successfully implemented across four simulated case studies and one real-world experiment involving different types of systems, objectives, and constraints without requiring any case specific adaptation of the method itself.

TreeC's interpretability provides several key advantages over other energy management methods:

\begin{itemize}
    \item Enables easier identification of potential errors in the learned model before deployment
    \item Facilitates the analysis of unexpected behaviours during operation
    \item Highlights the most important inputs and actions for efficient system control
    \item Simplifies the transfer of learned knowledge to similar systems
\end{itemize}

For all the cases in this thesis, TreeC has successfully generated interpretable energy management systems as shown in \cref{fig:anm_trees,fig:boptest_tree,fig:house1_decision_trees,fig:house2_decision_trees,fig:house3_decision_trees,fig:house4_decision_trees,fig:complete_dwelling}.

The second research question was answered by evaluating TreeC's performance against proven technology methods in two public benchmarks (\textit{Gym-ANM ANM6-Easy} and \textit{BOPTEST hydronic heat pump}) and one competition (Citylearn 2022). 

In the ANM6-Easy benchmark, TreeC achieved a grid security score of 19.2, closely matching the performance of model predictive control with perfect forecasts (14.4) and reinforcement learning (16.2). All three methods effectively maintained grid security and prevented collapse, significantly outperforming model predictive control with constant forecasts, which scored 127.7.

In the BOPTEST benchmark, TreeC performed closely to the best energy management system for each of the six scenarios, with an average electricity cost only 0.03\,\euro{}/m\textsuperscript{2} higher and total discomfort just 0.42\,Kh greater.
These results are comparable to the 0.04\,\euro{}/m\textsuperscript{2} and 2.91\,Kh achieved by model predictive control with perfect forecast variants, and substantially better than the 0.27\,\euro{}/m\textsuperscript{2} and 24.41\,Kh of reinforcement learning, as well as the 0.21\,\euro{}/m\textsuperscript{2} and 6.05\,Kh of rule-based control.

In the CityLearn 2022 competition, TreeC ranked 7\textsuperscript{th} out of 42 submissions, surpassed by three model predictive control solutions, one holistic learning approach, a hybrid rule-based and reinforcement learning solution, and a supervised learning-based solution. The winning entry combined multi-scenario stochastic model predictive control with reinforcement learning.

In summary, TreeC consistently obtained similar performances to proven technology methods across a variety of benchmarks. While it did not consistently surpass other approaches in terms of raw performance, its high level of interpretability offers a distinct advantage for practical deployment and analysis, making it a compelling choice when raw performance is not the sole criterion.

The third research question was addressed by implementing TreeC in a real-world experiment involving four homes, and comparing its performance to rule-based control, model predictive control, and reinforcement learning.

TreeC achieved an electricity cost of 171.7\,\euro{}, matching the performance of rule-based control (171.7\,\euro{}) and model predictive control (172.8\,\euro{}), while outperforming reinforcement learning (216.1\,\euro{}). In this scenario, neither TreeC nor model predictive control achieved cost savings compared to rule-based control. Further analysis showed the importance of using a representative training set for TreeC to improve real-world performance. For model predictive control, the absence of cost savings was mainly due to inaccurate input data, which are more prevalent in real-world settings and to which this method was more sensitive than TreeC and rule-based control.

This experiment demonstrated that TreeC can be successfully implemented on real hardware, but also underscored the practical challenges involved in real-world deployment.

The fourth research question was investigated using a large-scale scenario simulating 500 representative electrified Belgian dwellings. This scenario evaluated the tangible benefits of deploying TreeC and model predictive control, compared to rule-based control, in dwellings subject to hourly electricity pricing and a capacity tariff.

TreeC and model predictive control achieved average annual cost savings of 82.5\,\euro{} and 116.6\,\euro{} per dwelling, respectively, compared to rule-based control. Both methods maintained thermal comfort and ensured electric vehicles were charged on average to 93.3\% and 88.8\%, respectively, out of a maximum of 95\% charge. However, given the modest financial savings relative to the costs of implementing and maintaining commercial home energy management systems, these solutions may not be economically viable for most electrified households.

Model predictive control with perfect forecast was also evaluated to give an upper bound on the potential savings achieving average savings of 177.4\,\euro{} per dwelling. This shows some potential for algorithmic improvement of EMS solutions but also highlights that even with perfect knowledge of the future, the savings remain limited in this scenario.

Furthermore, the results indicated that if all dwellings simultaneously implemented the same algorithm, model predictive control would lead to significantly higher peaks in aggregated electricity consumption than TreeC, potentially causing congestion and balancing challenges for the grid.

In addition to the four main research questions, this thesis also addressed two additional research questions:

The fifth research question resulted in the proposition of a new machine learning paradigm termed holistic learning. Unlike reinforcement learning, which relies on intermediate rewards to update its model, holistic learning focuses solely on the final outcome of a sequence of actions during training. This approach inherently captures the delayed effects of actions (such as battery charging or the thermal inertia of buildings) which are prevalent in energy management problems. Methods that fit this paradigm include metaheuristic optimization used to  generate fuzzy logic models, neuroevolution, TreeC among others. While fuzzy logic methods have already been applied to energy management, neuroevolution, commonly used in robotics, has not yet been explored in this context. Grouping these approaches under the holistic learning paradigm may facilitate clearer referencing and comparison to other approaches in future research.

\section{Recommendations}

Drawing on the findings and experiences presented throughout this thesis, this section provides targeted recommendations for key stakeholders in the energy management ecosystem. These suggestions are informed by the strengths, limitations, and practical insights gained from the research, with the aim of supporting more effective development, deployment, and adoption of energy management systems in both scientific and real-world contexts.

For \textbf{developers using machine learning}:
\begin{itemize}
    \item \textbf{Avoid using non-interpretable models when possible}: Neural network based methods are used in many applications, even in some where an interpretable model would work just as well as it has been shown in this thesis. The reason is that the libraries to train the neural networks are more mature and easier to use but this leads to the loss of the benefits that come with interpretability. Hopefully more people using interpretable models will lead to more mature libraries in the future.
    \item \textbf{Explore holistic learning methods for your control tasks}: Although reinforcement learning remains the predominant machine learning 
    paradigm for control problems, it can be limited in situations where the reward function is difficult to define at each time step, such as when actions have very delayed effects (e.g., battery charging). Holistic learning addresses this challenge by optimizing based on the final outcome, inherently accounting for delayed effects in the model update process. This approach can work better in certain control tasks like shown in this thesis.
\end{itemize}

For \textbf{scientists working on energy management systems}:
\begin{itemize}
    \item \textbf{Benchmarks and competitions are important}: The benchmarks and competitions used in this thesis have been crucial to compare fairly the different methods against the work done by others. Other fields of research such as computer vision and natural language processing have a more mature ecosystem with great results. We should try to approach this level of maturity to improve our field.
    \item \textbf{Stochastic model predictive control seems very promising}: The winning solution of the Citylearn 2022 competitions was based on multi-scenario stochastic model predictive control. It seems a very promising method to deal with uncertainty in energy management problems. 
    In terms of pure performance, I expect it will be very hard to beat this method.
\end{itemize}

For \textbf{potential users of energy management systems}:
\begin{itemize}
    \item \textbf{Be precautious before investing in energy management systems}: Some commercial home energy management systems are advertised as being able to save 50\% of your electricity bill. This is unlikely as illustrated by the 500 electrified dwelling case so be sure to check what are the realistic savings you can expect before investing. If you have some technical skills and want to dedicate time into this, have a look into Home Assistant which an extremely popular open-source home automation platform with energy management capabilities.
\end{itemize}

For \textbf{policy makers}:
\begin{itemize}
    \item \textbf{Support open source and standards}: This work would not have been possible without all the open source code effort made by the community around the world. Not having to code or buy a software that someone else has already made publicly available really drives innovation and allows achieving more in less time with fewer expenses. The same goes for standards, for the real experiment every battery and photovoltaic inverter had different protocols for communication and control. Standardizing this would allow implementing energy management systems much easier which is necessary to make them affordable enough for people to use.
    \item \textbf{Promote realistic expectations and transparent communication about new energy technologies}: When introducing innovations such as energy management systems and energy communities, ensure that the advertized benefits accurately reflect real-world outcomes. Overstating financial savings can lead to disappointment and distrust. 
    For example, many people may expect to save significantly more than 100\,\euro{} per year with an electrified dwelling and dynamic tariffs, but the results presented in this thesis indicate that such savings are often not the norm. Similarly, encouraging collective purchase of photovoltaics through energy communities may not be beneficial for most individuals, as solar generation typically coincides with periods of low electricity prices and low consumption, and organizing a community requires considerable effort. Additionally, energy contracts examined in this thesis impose a 120\,\euro{} fee per participant in the community. It is important to avoid discouraging well-intentioned individuals from solutions that promote shifting electricity consumption and collective investment in renewable energy sources by providing clear and accurate information.
    \end{itemize}

\section{Future work}

This thesis has laid some important groundwork for the development of future research on energy management systems.
The key directions for future research based on this thesis are:

\begin{itemize}
    \item \textbf{Improving the training time of TreeC}: The training time of TreeC is a significant limitation. This could be done be coding the simulation environments with a faster programming language such as C++ or Rust, and by using previously performant EMS decision trees to initialize the training of new decision trees.
    \item \textbf{Improving the performance of TreeC}: TreeC did not achieve the best performance in all cases. Exploring how to improve its decision tree model to perform better in certain cases while keeping it interpretable is important for its future development. 
    \item \textbf{Expand real-world deployments of TreeC with users}: Implementing TreeC in real-world settings provides valuable insights into the practical challenges of deployment and operation. Increasing the number of real-world applications will help to improve the method especially regarding the interraction with users. This would provide an opportunity to further evaluate the benefits of TreeC's interpretability with users, as well as its robustness against unexpected user behaviors and real-world variability. One promising direction is to develop a Home Assistant integration for TreeC (similar to EMHASS \cite{davidusb-geekEnergyManagementHome2025}), enabling broader adoption and facilitating further testing.
    \item \textbf{Establish the 500 electrified dwellings scenario as a benchmark}: The 500-dwelling case presented in this thesis has strong potential as a standardized benchmark for energy management systems. To realize this, further work is needed to make it more user-friendly, clearly define acceptable levels of thermal comfort and electric vehicle charging, and implement a public leaderboard to facilitate transparent comparison of different methods.
    \item \textbf{Leverage the 500 electrified dwellings scenario to explore broader energy management topics}: Beyond algorithmic comparisons, the 500-dwelling scenario offers a valuable platform for investigating additional aspects such as the impact of various energy tariffs, different degrees of electrification, and widespread adoption of home batteries. These analyses can provide deeper insights into the practical implications and potential benefits of advanced energy management strategies in realistic settings.
\end{itemize}

\section{Publications and code}

\subsection{Peer-reviewed journal}
\begin{itemize}
    \item J. Ruddick, L. Ramirez Camargo, M. A. Putratama, M. Messagie, and T. Coosemans. “TreeC: A Method to Generate Interpretable Energy Management Systems Using a Metaheuristic Algorithm”. In: Knowledge-Based Systems (2025). \\
    Code: \url{https://github.com/EVERGi/treec-paper-results}\\
    Thesis sections: \Cref{sec:treec} and \cref{ch:simul_cases} \cref{sec:ANM6-Easy,sec:boptest}
    \item J. Ruddick, G. Ceusters, G. Van Kriekinge, E. Genov, C. De Cauwer, T. Coosemans, and M. Messagie. “Real-World Validation of Safe Reinforcement Learning, Model Predictive Control and Decision Tree-Based Home Energy Management Systems”. In: Energy and AI (2024). \\
    Code: \url{https://github.com/EVERGi/real_validation_saferl_treec_paper} \\
    Thesis chapter: \Cref{ch:summary_cases}
\end{itemize}

\subsection{Peer-reviewed conference}
\begin{itemize}
    \item J. Ruddick, E. Genov, L. R. Camargo, T. Coosemans, and M. Messagie. “Evolutionary Scheduling of University Activities Based on Consumption Forecasts to Minimise Electricity Costs”. In: 2022 IEEE Congress on Evolutionary Computation (CEC) (2022) \\ 
    Code: \url{https://github.com/EVERGi/predict-then-schedule-university-activities}\\
\end{itemize}

\subsection{Submitted}
\begin{itemize}
    \item J. Ruddick, T. Coosemans, M. Messagie, and C. De Cauwer. “Should Belgians install an advanced energy management system in their electrified dwellings ?”. To: BuildSys 2026 \\
    Code: \url{https://github.com/EVERGi/should_belgians_install_advanced_ems_paper} \\
\end{itemize}

\subsection{Co-author publications}

\begin{itemize}
    \item C. Bergmeir et al. “Predict+Optimize Problem in Renewable Energy Scheduling”. In: IEEE Access (2025).
    \item E. Genov, J. Ruddick, C. Bergmeir, M. Vafaeipour, T. Coosemans, S. Garcia, and M. Messagie. Balancing Forecast Accuracy and Switching Costs in Online Optimization of Energy Management Systems. In: Expert Systems with Applications (2025).
\end{itemize}

\mode<all> %

\backmatter
    \printbibliography     %
        \ifprintedversion
    \else
    \makebackcoverVUBdigital
    \fi

    \end{document}